\documentclass[12pt,preprint]{aastex}

\newcommand{\kms}{\ensuremath{\textrm{km~s}^{-1}}}
\newcommand{\flunits}{\textrm{erg~s\ensuremath{^{-1}}~cm\ensuremath{^{-2}}~\AA\ensuremath{^{-1}}}}
\newcommand{\hubbleunits}{\textrm{km~s}\ensuremath{^{-1}}~\textrm{Mpc}\ensuremath{^{-1}}}
\newcommand{\ha}{\textrm{H}\ensuremath{\alpha}}
\newcommand{\hb}{\textrm{H}\ensuremath{\beta}}
\newcommand{\hg}{\textrm{H}\ensuremath{\gamma}}
\newcommand{\hd}{\textrm{H}\ensuremath{\delta}}
\newcommand{\nii}{[\textrm{N}~\textsc{ii}]}
\newcommand{\oii}{[\textrm{O}~\textsc{ii}]}
\newcommand{\oiii}{[\textrm{O}~\textsc{iii}]}
\newcommand{\hii}{\textrm{H}~\textsc{ii}}
\newcommand{\halam}{\textrm{H}\ensuremath{\alpha~\lambda6563}}
\newcommand{\hblam}{\textrm{H}\ensuremath{\beta~\lambda4861}}
\newcommand{\oilam}{[\textrm{O}~\textsc{i}]~\ensuremath{\lambda6300}}
\newcommand{\oiilam}{\oii~\ensuremath{\lambda3727}}
\newcommand{\oiiilam}{[\textrm{O}~\textsc{iii}]~\ensuremath{\lambda5007}}
\newcommand{\oiidoublet}{[\textrm{O}~\textsc{ii}]~\ensuremath{\lambda\lambda3726,3729}}
\newcommand{\oiiidoublet}{[\textrm{O}~\textsc{iii}]~\ensuremath{\lambda\lambda4959,5007}}
\newcommand{\siidoublet}{[\textrm{S}~\textsc{ii}]~\ensuremath{\lambda\lambda6716,6731}}
\newcommand{\niidoublet}{[\textrm{N}~\textsc{ii}]~\ensuremath{\lambda\lambda6548,6584}}
\newcommand{\ewha}{\textrm{EW}(\textrm{H}\ensuremath{\alpha})}

\newcommand{\ebv}{E(\bv)}
\newcommand{\av}{\ensuremath{A_{V}}}
\newcommand{\rv}{\ensuremath{R_{V}}}
\newcommand{\mb}{\ensuremath{M_{B}}} 
\newcommand{\lfir}{\ensuremath{L({\rm FIR})}}
\newcommand{\lb}{\ensuremath{L(B)}}
\newcommand{\lbsun}{\lb\ensuremath{_{\sun}}}
\newcommand{\lblbsun}{\lb/\lbsun}

\newcommand{\noprint}[1]{}
\newcommand{\figsetstart}{{\bf Fig. Set} }
\newcommand{\figsetend}{}
\newcommand{\figsetgrpstart}{}
\newcommand{\figsetgrpend}{}
\newcommand{\figsetnum}[1]{{\bf #1.}}
\newcommand{\figsettitle}[1]{ {\bf #1} }
\newcommand{\figsetgrpnum}[1]{\noprint{#1}}
\newcommand{\figsetgrptitle}[1]{\noprint{#1}}
\newcommand{\figsetplot}[1]{\noprint{#1}}
\newcommand{\figsetgrpnote}[1]{\noprint{#1}}

\slugcomment{Accepted to ApJS}
\shortauthors{Moustakas \& Kennicutt}
\shorttitle{Integrated Spectrophotometry of Galaxies} 

\begin{document}

\title{An Integrated Spectrophotometric Survey of Nearby Star-Forming
  Galaxies}

\author{John Moustakas\altaffilmark{1} \& Robert~C. Kennicutt,
  Jr.\altaffilmark{1,2}} 
\altaffiltext{1}{Steward Observatory, University of Arizona, 933 N
  Cherry Ave., Tucson, AZ 85721, USA;
  \mbox{jmoustakas@as.arizona.edu}} 
\altaffiltext{2}{Current address: Institute of Astronomy, University 
  of Cambridge, Madingley Road, Cambridge CB3 0HA, UK;
  \mbox{robk@ast.cam.ac.uk}}  

\setcounter{footnote}{2}

%

\begin{abstract}
We present integrated optical spectrophotometry for a sample of $417$
nearby galaxies.  Our observations consist of spatially integrated,
${\rm S/N}=10-100$ spectroscopy between $3600$ and $6900$~\AA{} at
$\sim8$~\AA{} FWHM resolution.  In addition, we present nuclear
($2\farcs5\times2\farcs5$) spectroscopy for $153$ of these objects.
Our sample targets a diverse range of galaxy types, including
starbursts, peculiar galaxies, interacting/merging systems, dusty,
infrared-luminous galaxies, and a significant number of normal
galaxies.  We use population synthesis to model and subtract the
stellar continuum underlying the nebular emission lines.  This
technique results in emission-line measurements reliably corrected for
stellar absorption.  Here, we present the integrated and nuclear
spectra, the nebular emission-line fluxes and equivalent widths, and a
comprehensive compilation of ancillary data available in the
literature for our sample.  In a series of subsequent papers we use
these data to study optical star-formation rate indicators, nebular
abundance diagnostics, the luminosity-metallicity relation, the dust
properties of normal and starburst galaxies, and the star-formation
histories of infrared-luminous galaxies.
\end{abstract}

\keywords{atlases --- galaxies: fundamental parameters --- galaxies:
ISM --- galaxies: starburst --- galaxies: stellar content ---
techniques: spectroscopic} 


\section{INTRODUCTION}\label{sec:intro}

Integrated optical spectrophotometry provides a powerful means of
investigating the physical drivers of galaxy evolution.  Optical
spectral diagnostics may be used to constrain the star-formation rate,
star-formation history, stellar mass, chemical abundance, and dust
content of galaxies \citep[e.g.,][]{kauffmann03a, panter03, tremonti04,
brinchmann04}.  Consequently, tremendous effort has gone into
obtaining optical spectroscopy of both nearby and distant galaxies.
\citet[hereafter K92]{kenn92a} conducted the first systematic analysis
of the integrated spectroscopic properties of $55$ nearby galaxies
along the Hubble sequence, including a small number of peculiar
objects.  Optical spectroscopy for the centers of $45$ starburst
galaxies was presented by \citet{storchi95} and \citet{mcquade95},
providing comprehensive spectral coverage from the ultraviolet
\citep{kinney93} to the near-infrared \citep{calzetti97a, wu02}.
Previously, spectroscopic studies of nearby galaxies targeted just
their nuclear or high surface-brightness regions, whose physical
properties in general are not representative of the whole galaxy.  The
K92 spectral atlas also provided the first spatially unbiased
observations of local galaxies which could be compared directly
against observations of distant galaxies.  Subsequently,
\citet{jansen00a,jansen00b} presented the Nearby Field Galaxy Survey
(NFGS), an integrated spectrophotometric and imaging survey of a
representative sample of $196$ nearby field galaxies.
\citet{gavazzi04} have extended the sample of galaxies with integrated
spectroscopy to include normal galaxies in higher-density
environments, including the Virgo and Coma clusters \citep[see
also][]{gavazzi02}.  However, all these published surveys share a
common limitation for high-redshift applications, namely the
relatively small representation of starbursts, interacting/merging
systems, infrared-luminous galaxies, and other peculiar types that
represent only a few percent of the $z=0$ population, but which become
substantial contributors to the observed high-redshift populations.
Consequently, there exists a need for a large integrated
spectrophotometric survey that samples a more diverse range of galaxy
types.

Large fiber-optic redshift surveys of nearby galaxies are
revolutionizing our quantitative understanding of the physical drivers
of galaxy evolution.  The Sloan Digital Sky Survey
\citep[SDSS;][]{york00} and the 2dF Galaxy Redshift Survey
\citep[2dFGRS;][]{colless01} have obtained optical spectra for nearly
$10^{6}$ galaxies, providing the opportunity to study the stellar and
emission-line properties of nearby galaxies with unprecendented
statistical precision.  However, these surveys suffer two important
limitations for high-redshift applications: incomplete spatial
coverage by the spectroscopic aperture, and a magnitude-limited
selection criterion that targets primarily the most luminous
present-day galaxies.  Incomplete spatial coverage, or aperture bias,
may be particularly severe since many physical properties of galaxies
vary with galactocentric radius \citep[e.g., stellar populations,
metallicity, extinction, etc.; see, e.g.,][]{kewley05}.  For example,
\citet{tremonti04} estimate that aperture bias in the SDSS results in
chemical abundance measurements that over-estimate the integrated
abundance by at least $25\%$.  Furthermore, these surveys primarily
target the most luminous present-day galaxies, which in general will
not be representative of galaxies at an earlier stage in their
evolutionary history.

To address these limitations of existing surveys, we have obtained
high signal-to-noise (${\rm S/N}=10-100$), integrated optical
($3600-6900$~\AA) spectrophotometry for $417$ nearby galaxies.  Our
survey targets a broad range of galaxy types with ongoing star
formation, including ultraviolet- and infrared-luminous starbursts,
galaxies with enhanced nuclear or circumnuclear star formation,
peculiar galaxies, and interacting/merging systems, and a large number
of normal galaxies.  Although our sample is incomplete in a strict
magnitude-limited sense, the primary objective of our survey was to
achieve wide coverage of the physical parameter space spanned by the
$z=0$ population of star-forming galaxies, focusing on the population
of nearby starburst galaxies that likely dominate at high redshift.
We utilize the drift-scanning technique developed by K92 to obtain
spatially integrated spectroscopy at intermediate ($\sim8$~\AA{} FWHM)
spectral resolution.  Drift-scanning consists of moving a narrow
long-slit back-and-forth across the optical extent of a nearby galaxy
that may subtend several arcminutes on the sky.  This observational
technique results in a luminosity-weighted integrated spectrum
analogous to traditional (spatially fixed) spectroscopy of distant
galaxies.

In \S\ref{sec:data} we present our spectroscopic sample.  We describe
the sample selection, characterize the properties of the sample,
describe our observations and data reductions, quantify the
spectrophotometric accuracy of the data, and present the final
integrated spectral atlas.  In \S\ref{sec:fitting} we describe {\tt
iSPEC1d}, a spectral synthesis and emission-line fitting code we
developed to achieve accurate removal of the stellar continuum
underlying the nebular emission lines, and in \S\ref{sec:measure} we
provide flux and equivalent width measurements of the strong nebular
emission lines.  Finally, in \S\ref{sec:summary} we summarize our
results.  To compute distances and absolute magnitudes we adopt a
Hubble constant $H_{0}=70$~\hubbleunits{} \citep{spergel03,
freedman01}.

\section{THE DATA}\label{sec:data}

\subsection{Sample Selection}\label{sec:sample}

We draw our sample of galaxies from four primary sources.  We select
one subset from the First Byurakan Survey of ultraviolet-excess
galaxies \citep{markarian89}.  We primarily target the subsamples
studied by \citet{huchra77} and \citet{balzano83} to weight our
selection in favor of starburst galaxies and against active galactic
nuclei (AGN).  The resulting sample of $\sim125$ galaxies contains a
diverse collection of galaxy types, including blue compact galaxies
with strong emission lines, starburst nuclei, and normal massive
spiral and irregular galaxies with abnormally high SFRs.

An ultraviolet-selected starburst sample such as the one above may be
biased toward galaxies with low dust content.  Therefore, in order to
extend our coverage to dusty starbursts we observe a second subsample
of infrared galaxies from the IRAS Warm Galaxy Survey [WGS;
$S_{\nu}(60~\micron) / S_{\nu}(100~\micron)>0.25$] and Bright Galaxy
Survey (BGS; $L_{\mathrm IR}>10^{11}~L_{\sun}$) \citep{kim95,
veilleux95}.  To extend the baseline in infrared luminosity we include
several ultra-luminous infrared galaxies \citep[ULIRGs; $L_{\mathrm
IR}>10^{12}~L_{\sun}$;][]{sanders96} from \citet{veilleux99}.  This
selection in terms of infrared luminosity and color results in a
diverse morphological and spectroscopic sample of $\sim100$ galaxies,
including relatively normal (but unusually massive) star-forming
spirals, edge-on luminous galaxies, AGN-dominated galaxies, and
interacting/merging systems.

To increase the sampling of ``normal'' galaxies and to ensure a more
representative range of luminosities, inclinations, and types, we
select another subset of galaxies from two volume-limited samples.
These observations include $\sim75$ galaxies from the $11$~HUGS survey
(R.~C. Kennicutt et~al., 2005a, in preparation), an H$\alpha$ and
ultraviolet imaging survey of galaxies within the $11$~Mpc local
volume, and $53$ star-forming galaxies in the Ursa Major cluster.
This nearby cluster \citep[$d=19.8$~Mpc;][]{freedman01} is a
dynamically young system comprised of many gas-rich spiral galaxies
\citep[see][and subsequent papers]{tully96}.

Finally, we incorporate into our sample unpublished integrated
spectrophotometry for $35$ morphologically disturbed galaxies/systems
from the Ph.~D. thesis by Anne Turner \citep{turner98}.  These objects
are selected from the \citet{arp66} atlas of peculiar galaxies based
on visual evidence of recent interactions or merging in the form of
tidal tails, bridges, or shells.  These data have been obtained using
the same instrumental setup and observing technique as the main
survey, and have been re-reduced to ensure consistency with our new
observations (see \S\ref{sec:redux}).

In addition to the integrated spectrophotometric observations, during
the course of our survey we also obtained nuclear
($2\farcs5\times2\farcs5$) spectroscopy for $\sim30\%$ of the sample.
The list of objects with both nuclear and integrated spectra are a
heterogenous subset since the nuclear spectra were not the focus of
the primary survey.  However, we present these data in this paper
because they may be useful for a variety of astrophysical
applications, such as investigating the connection between the global
versus the nuclear/circumnuclear properties of nearby galaxies.

\subsection{Sample Properties}\label{sec:properties}

To increase the archival value of our new spectroscopic observations,
we compile multi-wavelength broadband photometry and other relevant
galaxy properties using the Nasa Extragalactic Database
(NED\footnote{\url{http://nedwww.ipac.caltech.edu}}), the Lyon-Meudon
Extragalactic Database
(LEDA\footnote{\url{http://leda.univ-lyon1.fr}}), and the SIMBAD
Astronomical
Database\footnote{\url{http://simbad.u-strasbg.fr/Simbad}}.
Table~\ref{table:general_properties} presents these global properties
in the following thirteen columns, which we describe in more detail
below: (1) unique object identification number; (2) galaxy name; (3)
other common galaxy name or names; (4) right ascension; (5)
declination; (6) heliocentric velocity; (7) foreground Galactic
reddening \citep{schlegel98}; (8) morphological type; (9) major axis
diameter at the $25$~mag~arcsec$^{-2}$ isophote; (10) minor axis
diameter at the same isophotal diameter; (11) galaxy position angle;
(12) distance; and (13) distance reference.

We adopt coordinates and heliocentric redshifts for our sample from
NED, SIMBAD, or LEDA, as noted in
Table~\ref{table:general_properties}.  We obtain major- and minor-axis
diameters from the Third Reference Catalog of Bright Galaxies
\citep[hereafter RC3]{devac91}, supplemented with additional
measurements from LEDA, the ESO/Uppsala Survey of the ESO(B) Atlas
\citep{lauberts82}, the Uppsala General Catalog of Galaxies
\citep[UGC;][]{nilson73}, the Morphological Catalogue of Galaxies
\citep[MCG;][]{vorontsov62}, and the NED ``Basic Data,'' listed in
order of preference.  We obtain position angles from the same set of
references, supplemented by measurements from either the 2MASS Large
Galaxy Atlas \citep[LGA;][]{jarrett03}, or the 2MASS Extended Source
Catalog \citep[XSC;][]{jarrett00}, in that order.  We adopt
morphological types from the RC3, LEDA, or from our own classification
if no other type is available.


Due to the proximity of our sample ($\sim1/4$ of our galaxies are
nearer than the Virgo cluster), peculiar velocities can be a
significant pertubation on the observed radial velocity.  Therefore,
we search the literature for direct distance estimates to as many
galaxies as possible.  These distances are based on a variety of
techniques, although we give particular preference to the Cepheid
period-luminosity relation and the magnitude of the tip of the
red-giant branch.  Our direct distances rely predominantly on the
large compilations by \citet{freedman01}, \citet{karachentsev04},
\citet{tonry01}, and \citet{shapley01b}, listed in order of
preference.  We use the Cepheid-based distances in \citet{freedman01}
corrected for metallicity \citep[see also][]{sakai04}.  We assign
members of the Ursa Major cluster \citep[as defined by][]{tully96} a
common distance of $19.8\pm1.6$~Mpc, based on the revised Cepheid
distance scale \citep{freedman01}.  Finally, we compute distances to
all other galaxies using the multi-attractor infall model developed by
\citet{mould00}.  This model accounts for linear infall onto the Virgo
cluster, the Great Attractor, and the Shapley concentration, which we
assume to be independent pertubations on the measured radial velocity
of the galaxy.  For our calculations we adopt the model parameters
listed in Table~A1 and the set of assumptions enumerated in \S2 of
\citet{mould00}.  In particular, galaxies that lie within the
cone-of-influence of each attractor are forced to the velocity of the
attractor (in the Local Group frame-of-reference).  We find good
statistical agreement between the distances using this model and the
direct distances.  We also inter-compare distances based the
\citet{mould00} infall model to the distances in \citet{tully88} and
again find good statistical agreement and no gross systematic
differences.

We compile optical and infrared photometry for our sample using NED's
batch photometry retrieval system.  We tabulate total broadband $UBV$
magnitudes, uncorrected for Galactic extinction and inclination
effects, predominantly from the RC3, adopting the RC3 $m_{B}$
measurements if $B_{T}$ magnitudes are unavailable.  For objects
without RC3 photometry we use LEDA $UBV$ magnitudes
\citep{prugniel98}.  We take total J- ($1.2~\micron$), H-
($1.6~\micron$), and K$_{\rm s}$-band ($2.2$~\micron) magnitudes from
the 2MASS/LGA \citep{jarrett03} or the 2MASS/XSC \citep{jarrett00}, in
that order.  The 2MASS total magnitudes are derived by extrapolating
the observed J-band surface brightness profile to $\sim4$ disk scale
lengths and integrating in each band \citep{jarrett00}.  Finally, we
tabulate infrared fluxes at $12$, $25$, $60$, and $100$~\micron{} for
our sample.  We obtain these measurements from the IRAS large optical
galaxy catalog \citep{rice88}, the Bright Galaxy Survey
\citep{soifer89}, or the Faint Source Catalog \citep{moshir90}, ranked
in order of preference \citep[following][]{bell03}.  We correct our
$UBVJHK_{\rm s}$ magnitudes for foreground Galactic extinction using
the \ebv{} value listed in Table~\ref{table:general_properties} and
the \citet{odonnell94} Milky Way extinction curve assuming
$\rv\equiv\av/\ebv=3.1$.  Table~\ref{table:photometric_properties}
lists these photometric data for our sample.

\begin{figure}
\epsscale{0.9}
\plotone{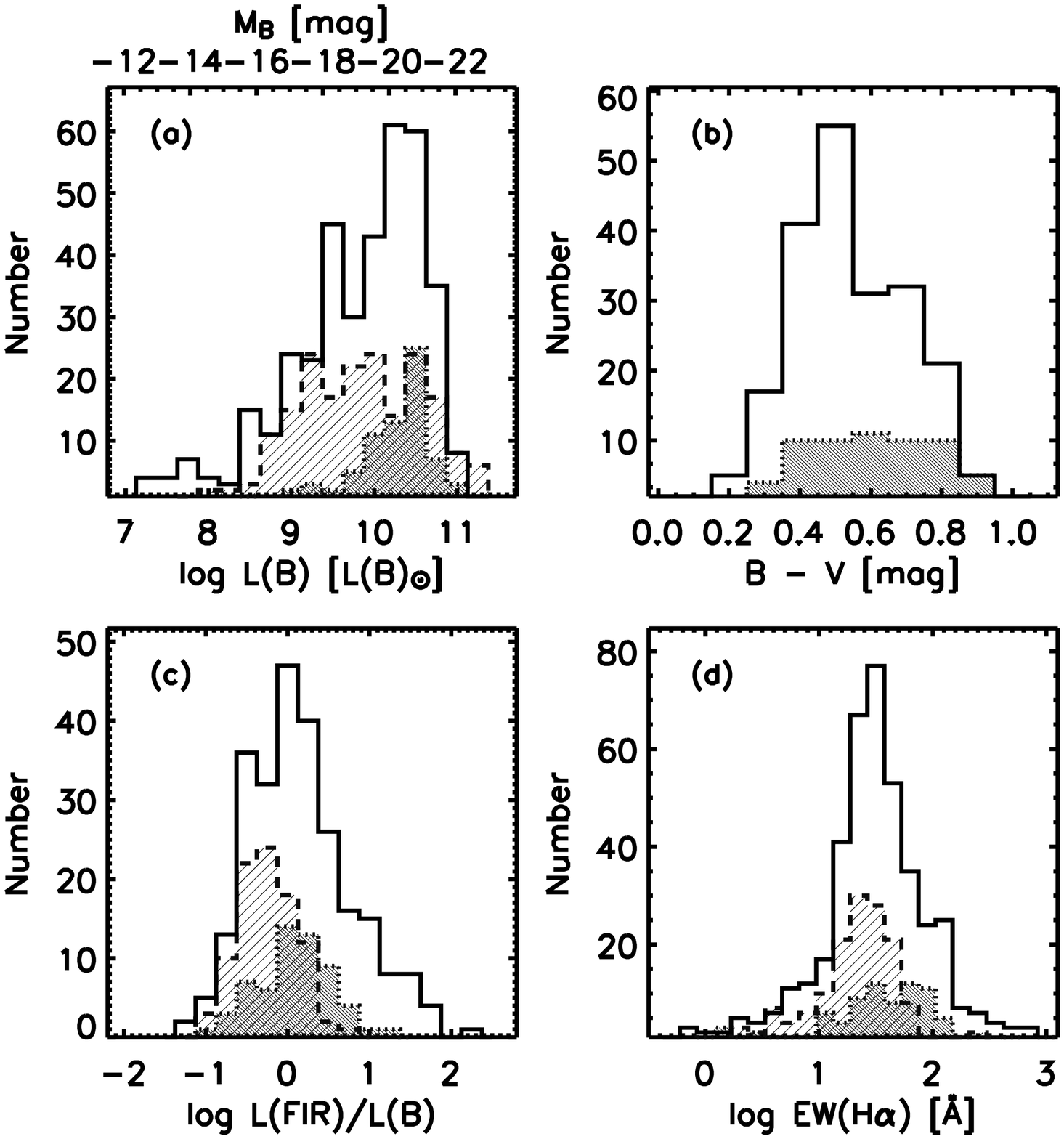}
\caption{Comparison of the distribution of physical properties for our
survey (\emph{solid-line} open histogram), the K92 spectral atlas
\citep[\emph{dotted-line}, dark grey histogram;][]{kenn92a, kenn92b},
and the NFGS \citep[\emph{dashed-line}, light grey
histogram;][]{jansen00a, jansen00b}.  (\emph{a}) B-band luminosity,
\lb, where we use $M_{\sun, B}=+5.42$~mag to convert between
\lblbsun{} and \mb; (\emph{b}) \bv{} color; (\emph{c}) far-infrared
(FIR) to B-band luminosity ratio, \lfir/\lb; and (\emph{d}) equivalent
width of \halam, \ewha.  In panel (\emph{d}), \ewha{} for our sample
(see \S\ref{sec:measure}) and the NFGS (see \citet{jansen00a}) have
been corrected for underlying stellar absorption; no absorption
corrections have been applied to the K92 measurements, and, following
\citet{kenn92b}, we adopt a constant $\nii/\ha=0.5$ ratio to convert
from ${\rm EW}(\ha+\nii)$ to \ewha. \label{fig:distributions}}
\end{figure}

Figure~\ref{fig:distributions} illustrates the broad range of physical
properties spanned by our galaxy sample.  In panels (\emph{a}),
(\emph{b}), (\emph{c}), and (\emph{d}), respectively, we plot the
distribution of $B$-band luminosity, \bv{} color, far-infrared (FIR)
to $B$-band luminosity ratio, and equivalent width (EW) of \ha{}
(\emph{solid-line}, open histogram).  For comparison, we show the
distributions of the same properties for two other widely used
integrated spectrophotometric surveys, the K92 spectral atlas
\citep[\emph{dotted-line}, dark grey histogram;][]{kenn92a, kenn92b}
and the NFGS \citep[\emph{dashed-line}, light grey
histogram;][]{jansen00a, jansen00b}.  The optical and infrared
photometry for all three samples have been collected from the
literature in the same way, as described in detail above.
Unfortunately, $V$-band photometry is only available for $20\%$ of the
NFGS; therefore, we do not plot its \bv{} color distribution in
Figure~\ref{fig:distributions}\emph{b}.  We compute the FIR
luminosity, $\lfir\equiv L(40-120~\micron)$, using the $60$ and
$100$~\micron{} IRAS fluxes according the formula in \citet{helou88}.
The \ha{} EWs for our sample (see \S\ref{sec:measure};
Table~\ref{table:int_lineEW}) and the NFGS \citep{jansen00a} have been
corrected for underlying stellar absorption, but no corrections have
been applied to the K92 data.  In addition, for the K92 measurements,
we follow \citet{kenn92b} and adopt a constant $\nii/\ha=0.5$ ratio to
convert from ${\rm EW}(\ha+\nii)$ to \ewha.

Figure~\ref{fig:distributions}\emph{a} shows that our galaxy sample
spans a wide range of blue luminosity, which ensures broad coverage of
physical quantities that correlate with luminosity, such as
metallicity, excitation, and dust content.  Compared to the NFGS, a
representative survey of nearby galaxies, our sample includes many
more luminous ($\mb\lesssim-18$~mag) galaxies, in addition to a
relatively large percentage of faint ($\mb>-16$~mag) dwarf galaxies
($8\%$ of our sample, compared to $0.8\%$ in the NFGS).  The K92 atlas
includes predominantly luminous galaxies, as discussed by
\citet{jansen01}.  The observed \bv{} color distribution of our sample
(Fig.~\ref{fig:distributions}\emph{b}) varies by one magnitude,
indicating a broad range of star-formation histories, from blue,
actively star-forming galaxies, to more quiescent (or very dusty)
galaxies with red \bv{} colors.  The range in color is similar to the
K92 atlas, although our selection of ultraviolet-excess galaxies (see
\S\ref{sec:sample}) incorporates many more galaxies bluer than
$\bv\simeq0.3$~mag.  Figure~\ref{fig:distributions}\emph{c} plots the
\lfir/\lb{} distribution for the three samples.  This ratio
characterizes the relative amount of energy absorbed by dust and
re-emitted into the infrared.  Because we include a large number of
infrared-luminous galaxies, \lfir/\lb{} in our sample varies by more
than a factor of $\sim3500$ with a median ratio of $0.07$~dex.  By
comparison, the \lfir/\lb{} distributions of the optically selected
K92 and NFGS samples are narrower and do not include as many extremely
dusty galaxies [$\lfir/\lb>10$].  Finally,
Figure~\ref{fig:distributions}\emph{d} compares the \ewha{}
distributions of the three samples.  The \ewha{} is proportional to
the birthrate parameter, or the ratio of the current to the
past-average star-formation rate \citep{kenn94}.  Nearly $30\%$ of our
galaxies have $\ewha>50$~\AA, compared to $10\%$ in the NFGS, and
$40\%$ in the K92 survey.

%
%


\subsection{Observations}\label{sec:obs}

We obtain our spectrophotometric observations at the $2.3$-meter Bok
telescope on Kitt Peak using the Boller \& Chivens (B\&C)
spectrograph.  The B\&C spectrograph is equipped with a Loral
$1200\times800$ square-pixel back-illumated CCD with $15~\micron$
pixels.  The $400$~line~mm$^{-1}$ reflection grating, blazed at
$\sim5200$~\AA, affords nearly one octave of spectral coverage between
$3600$ and $\sim6900$~\AA{} with $2.75$~\AA{} pixels at a full width
at half-maximum (FWHM) resolution of $\sim8$~\AA{} through a
$2\farcs5$ slit.  The physical scale at the detector, binned by $2$
pixels, is $1\farcs66$~pixel$^{-1}$.  The usable slit length is
$\sim3\farcm3$, enabling adequate sky subtraction in all but the
largest galaxies in our sample.  An UV-$36$ blue-blocking filter
eliminates contamination from second-order light; the transmittance
through the UV-$36$ filter is $50\%$ at $3600$~\AA{} and $90\%$ at
$4000$~\AA.  All the spectra presented in this atlas have been
obtained using the identical instrumental setup and observing
technique, which ensures data uniformity.

The calibration data for each night include bias frames, dome flats,
and evening twilight sky flats.  Helium-argon (HeAr) comparison lamps
taken during the course of the night at a variety of positions on the
sky facilitate wavelength calibration.  We observe spectrophotometric
standard stars from the
CALSPEC\footnote{\url{http://www.stsci.edu/instruments/observatory/cdbs/calspec.html}}
\citep{bohlin01} and \citet{massey88} star lists at low airmass
($<1.2$) and near the parallactic angle $4-8$ times during the night
using both $2\farcs5$ and $4\farcs5$ slits.  We observe more than
$\sim75\%$ of our integrated spectra and $\sim85\%$ of our nuclear
spectra during clear sky conditions.  In \S\ref{sec:redux} we discuss
our flux-calibration procedure and quantify the relative and absolute
spectrophotometric accuracy of our observations.

We implement the drift-scanning technique developed by K92 to obtain
integrated spectrophotometry of our sample at the spectral resolution
afforded by a $2\farcs5$ long-slit \citep[see
also][]{jansen00b,gavazzi04}.  Using this method we scan perpendicular
to the slit $10-20$ times over the optical extent of the galaxy during
a single exposure using dedicated telescope control software.  By
scanning the galaxy multiple times we average over any short-term
variations in atmospheric transparency.  The choice of scan length and
exposure time vary with the size and surface brightness of each
object.  Our scan lengths range from $15\arcsec-800\arcsec$ and total
exposure times vary from $15-120$ minutes split into two or more
consecutive exposures to facilitate cosmic-ray rejection (see
\S\ref{sec:redux}).  The effective time $t_{\rm eff}$ spent on a
fixed spatial location of the galaxy is given by

\begin{equation}
t_{\rm eff} = t \, \left(\frac{2\farcs5}{\Delta_{\rm scan}}\right),  
\label{eq:teff}
\end{equation}

\noindent where $t$ is the total exposure time and $\Delta_{\rm scan}$
is the diameter of the drift-scan perpendicular to the slit in
arcseconds.  Drift-scanning affords two specific advantages over
pointed long-slit spectroscopy.  First, it is insensitive to the
effects of atmospheric refraction because in general $\Delta_{\rm
scan}\gg\Delta(\theta)$, where $\Delta(\theta)$ is the amount of
wavelength-dependent atmospheric refraction in arcseconds as a
function of the parallactic angle, $\theta$ \citep{filippenko82}.  And
second, drift-scanned spectroscopy recovers any light lost due to
wavelength-dependent variations in seeing because, once again,
$\Delta_{\rm scan}\gg\Delta_{\rm seeing}$, where $\Delta_{\rm seeing}$
is the FWHM of the seeing disk in arcseconds as a function of
wavelength.

In addition to our integrated spectra we have obtained nuclear
spectroscopy for $\sim30\%$ of our sample.  These observations are
typically based on $1-2$ five-minute exposures using a fixed
$2\farcs5\times2\farcs5$ entrance aperture.  For our nuclear spectra
we always rotate the slit along the parallactic angle to minimize
atmospheric refraction effects \citep{filippenko82}.  Although many of
these nuclear spectra have been obtained during clear or mostly clear
observing conditions, we cannot guarantee their absolute
spectrophotometric accuracy due to variations in seeing, pointing
accuracy, and transparency.

Table~\ref{table:journal} presents a summary of our integrated
spectrophotometric observations in the following ten columns: (1)
galaxy identification number from
Table~\ref{table:general_properties}; (2) galaxy name; (3) diameter of
the drift scan perpendicular to the slit; (4) extraction aperture
diameter along the slit; (5) slit position angle; (6) total exposure
time; (7) flag indicating whether the spectrum was obtained during
clear or non-photometric conditions; (8) uncertainty in the absolute
spectrophotometric accuracy of the spectrum (see
\S\ref{sec:accuracy}); and (9) remarks regarding the individual galaxy
or spectral extraction.

\subsection{Reductions}\label{sec:redux}

We reduce our spectroscopic observations using {\tt iSPEC2d}, a generalized
long-slit data reduction and analysis software package written in IDL.
{\tt iSPEC2d} offers several advantages over some existing spectroscopic
data reduction packages such as error propagation, bad pixel tracking,
automated wavelength calibration, two-dimensional sky subtraction,
minimal pixel resampling (interpolation), and web-page visualizations
to facilitate inspection of each step in the reduction procedure.

First we repair dead pixels and bad columns, subtract the overscan
noise, trim, subtract the bias frame, and divide by the master flat
field.  We construct the flat field in the spectral dimension by
fitting an high-order cubic spline to a median-averaged dome flat.  We
average several twilight sky flats and model the residual illumination
pattern with a low-order polynomial.  We find that flat-fielded
exposures of the evening sky are spatially constant at the $1-2\%$
level.  For each observation we process a corresponding
two-dimensional error map, which we construct using the known
read-noise and gain of the detector and assuming Poisson statistics.

We measure the two-dimensional mapping between pixels and wavelength
using our HeAr lamps.  The wavelength residuals are typically
$0.2-0.4$~\AA{} based on $20-50$ arc lines.  Next, before coadding, we
compare the spatial profiles of sequential science exposures to solve
for any small shifts (typically $<2$ pixels) that may arise from
back-lash at the turn-around points of the drift-scan.  We remove
residual cosmic rays using L.~A.~COSMIC \citep{dokkum01}, which
identifies hot pixels by convolving the data with a Laplacian
edge-detection kernel.  We experimented with a variety of cosmic-ray
rejection schemes and found L.~A.~COSMIC to be the most reliable.
Finally, we examine our cosmic-ray rejected images to verify that
neither narrow emission lines nor bright sky lines are inadvertently
affected.

Next, we interactively define sky apertures for each object and
sky-subtract using the corresponding two-dimensional wavelength map.
We fit the over-sampled sky spectrum with an high-order cubic b-spline
in the spectral dimension and a first- or second-order polynomial in
the spatial dimension, selecting $2.75$~\AA{} as the spacing between
adjacent b-spline knots.  \citet{kelson03} discuss in detail the
advantages of two-dimensional sky subtraction.  We estimate the
sky-subtraction uncertainty by computing the error in the mean sky
level at each column (wavelength), and propagate this uncertainty into
the corresponding error map.  We test our simplified assumption of
using the error in the mean sky value as the sky-subtraction
uncertainty by sky-subtracting $500$ Monte Carlo realizations of
several hundred galaxy and standard-star spectra spanning a range of
sky apertures and exposure times.  For each object we measure the
standard deviation of the two-dimensional sky model at each column and
compare it to our ``simple'' uncertainty estimate.  On average, we
find the ratio of the two noise estimates to be unity.  

We flux-calibrate our observations using standard stars from an entire
run, ranging from $1-6$ consecutive nights.  We obtain the mean
sensitivity function for each run by fitting an high-order b-spline
fit to $5-40$ standard stars observed during clear sky conditions and
good seeing.  We adopt the CALSPEC absolute spectral energy
distributions \citep{bohlin01}, binned in $10$~\AA{} bandpasses, or
the \citet{massey88} spectrophotometry binned every $50$~\AA{} if no
CALSPEC calibration is available.  We use the nominal Kitt Peak
extinction curve to correct for atmospheric extinction and reddening
as a function of airmass.  In \S\ref{sec:accuracy} we describe our
flux calibration procedure in more detail and quantify the relative
and absolute spectrophotometric accuracy of our data.

We extract one-dimensional spectra using interactively defined
apertures enclosing the full spatial profile of the galaxy, or a fixed
$2\farcs5$ aperture for our nuclear spectra.  We account for the small
($<1\%$) spatial tilt in our data by fitting a linear function to the
flux-weighted trace of the galaxy profile.  We carefully treat
interacting/merging systems with overlapping spatial profiles by
extracting spectra for each individual object and for the composite
system.  We either exclude contaminating foreground stars from the
extraction aperture if they are near the edge of the galaxy, or
subtract them from the two-dimensional spectrum using the following
two-step procedure.  First, we fit a Gaussian function to the spatial
profile of the star averaged every $\sim100$ columns to measure the
mean variation in FWHM and center.  We use low-order polynomial
interpolation to estimate these quantities at every column.  Next,
measure the star's spectral energy distribution by fitting a Gaussian
profile to the stellar profile at every column by constraining the
width and center and solving for the amplitude.  This procedure yields
a robust two-dimensional model of the contaminating star, which we
then subtract from the data before extracting the galaxy spectrum
normally.  We apply a small correction (typically $\sim0.5$~\AA) to
each object's wavelength solution using $10-20$ night sky lines, and
repair sky-subtraction residuals near [\ion{O}{1}]~$\lambda5577$,
Na~D~$\lambda5889$, [\ion{O}{1}]~$\lambda6300$, and
[\ion{O}{1}]~$\lambda6336$ for display purposes.  However, we do not
repair sky-subtraction residuals that fall near the Na~D interstellar
absorption line or the \oilam{} nebular emission line in the
lowest-redshift galaxies in our sample.  Finally, we correct each
spectrum for Galactic reddening and extinction using the
\citet{schlegel98} dust maps and the \citet{odonnell94} Milky Way
extinction curve assuming $R_{\rm V} = 3.1$.  The mean (median)
reddening for the full sample is $0.05\pm0.07$~mag ($0.03$~mag), with
$<2\%$ of our sample suffering more than $0.2$~mag of foreground
Galactic extinction.

\subsection{Spectrophotometric Accuracy}\label{sec:accuracy} 

In this section we discuss the relative and absolute
spectrophotometric accuracy of our observations.  Variations in
seeing, atmospheric transparency, and pointing prevent the nuclear
spectra from being tied to an absolute spectrophotometric scale.
However, because our integrated spectra enclose all, or nearly all the
optical light of the galaxy, observations made during photometric
conditions may be fluxed on an absolute spectrophotometric scale.
Drift-scanned long-slit spectroscopy, therefore, is analogous to
large-aperture rectangular photometry with the added bonus of yielding
spectral information.  Note that we do not correct our integrated
spectra for any light missed by the drift scan.  

\begin{figure}
\epsscale{1.0} 
\plotone{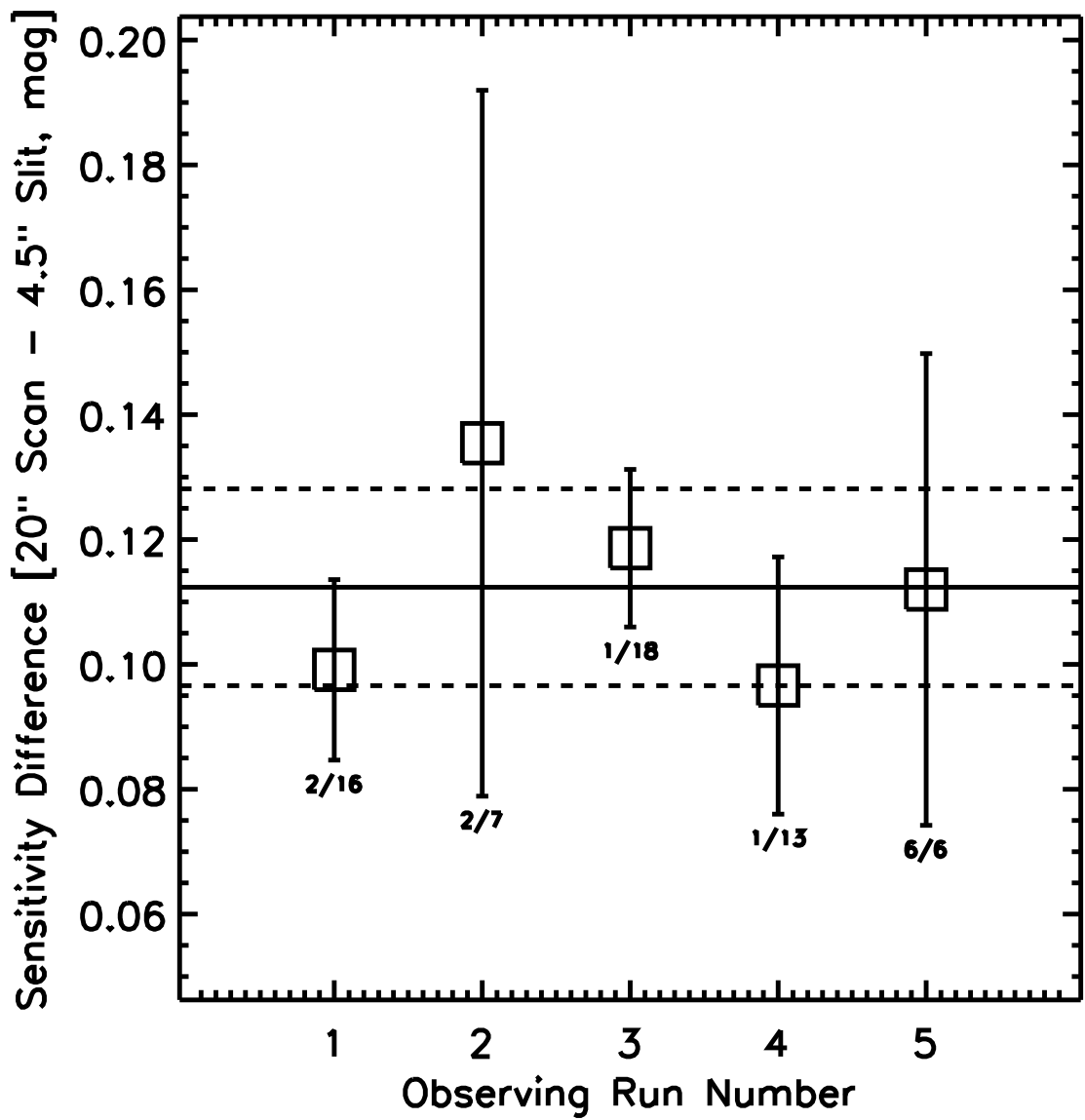}
\caption{Measured magnitude difference near $\sim5500$~\AA{} between
the $20\arcsec$ drift-scanned and $4\farcs5$-slit sensitivity
functions for $5$ separate observing runs.  This figure quantifies the
correction needed to account for light missed by the $4\farcs5$ slit
due to seeing.  The unweighted mean magnitude correction is
$0.11\pm0.02$~mag.  The pair of numbers below each measurement, for
example $2/7$, give the numbers of standard stars used to construct
the drift-scanned sensitivity function ($2$) and the $4\farcs5$
sensitivity function ($7$), respectively.  The error bar for each
point is the quadrature sum of the observed uncertainty in each
contributing sensitivity function.
\label{fig:zeropoint}}
\end{figure}

We flux-calibrate our spectra using standard stars observed with the
$4\farcs5$ slit during good seeing and clear skies.  Assuming
comparable seeing for all the standard-star observations, the scatter
in the observed sensitivity function relative to the mean quantifies
the variation in atmospheric transparency during the course of the
observing run.  The mean (median) scatter in our sensitivity functions
near $\sim5500$~\AA{} is $0.07^{+0.07}_{-0.05}$~mag ($0.06$~mag).
Unfortunately, even with the $4\farcs5$ slit we still miss some of the
light of the standards.  To quantify this effect, we conduct
$20\arcsec$ drift scans of several standard stars during photometric
conditions in addition to the nominal $4\farcs5$ slit observations.
Figure~\ref{fig:zeropoint} compares the sensitivity difference between
the drift-scanned and $4\farcs5$ slit observations near
$\sim5500$~\AA, illustrating that $0.11\pm0.02$~mag of light is lost
through the $4\farcs5$ slit.  We attribute this $\sim10\%$ light-loss
to seeing: assuming a Gaussian point-spread function, a
$4\farcs5$-wide slit encloses just $\sim90\%$ of the energy in the
median seeing of our observations, $1\farcs4$ (FWHM).  We add this
scalar zero-point shift to every sensitivity function before
flux-calibrating, and propagate the $0.02$~mag error in the correction
into our absolute spectrophotometric error budget.  Finally, we also
include a $2\%$ uncertainty in the absolute standard-star calibrations
\citep{bohlin01}.  We give the final absolute spectrophotometric
uncertainty of each integrated spectrum, excluding systematic
uncertainties arising from missing light or systematic sky-subtraction
errors, in column~(9) of Table~\ref{table:journal}.

We conduct both internal and external comparisons to further assess
the spectrophotometric quality of our observations.  First, we compare
multiple observations of the same galaxy.  Twenty-three objects in our
integrated galaxy sample have been observed two or more times and
extracted using identical, or nearly identical rectangular apertures.
We find that the $1\sigma$ relative error is $\pm3.3\%$ with no
wavelength dependence.  

\begin{figure}
\epsscale{0.9}
\plotone{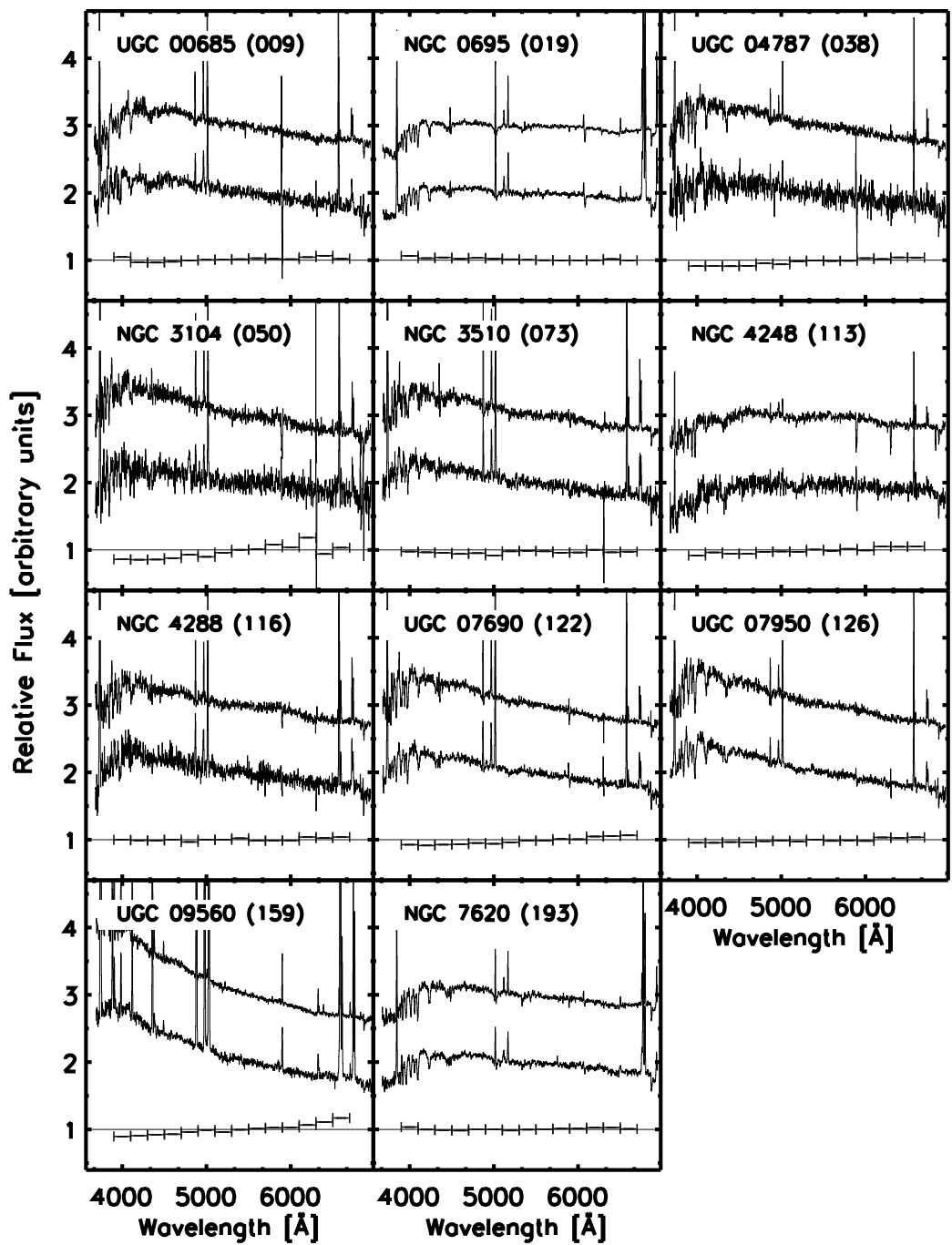}
\caption{Comparison of the $11$ galaxies in common with the Nearby
Field Galaxy Survey \citep[NFGS;][]{jansen00b}.  Each object is
identified by the galaxy name listed in
Table~\ref{table:general_properties} and the corresponding NFGS
identification number in parenthesis.  We normalize the spectra to the
mean flux around $5500\pm50$~\AA{} and offset each NFGS spectrum one
unit upwards for clarity.  The ratio of the two spectra are plotted in
$200$-\AA{} wide bins to illustrate the differences in continuum
shape.
\label{fig:nfgs_compare_spectra}}
\end{figure}

Next, we intercompare $11$ galaxies in common with the NFGS, both
corrected using the same Galactic reddening value.  We normalize each
pair of spectra to the mean flux around $5500\pm50$~\AA{} and compute
their ratio in $200$~\AA-wide bins to characterize the variation in
continuum shape.  We show this comparison in
Figure~\ref{fig:nfgs_compare_spectra} with the NFGS spectrum offset
upwards by one unit for clarity.  The correspondence between our
observations and the NFGS is very good, bearing in mind that in
general they are based on different drift-scan lengths, slit position
angles, and extraction apertures.  Excluding two objects, NGC~3104 and
UGC~09560, we find an average relative error of $\pm3.7\%$, with a
weak wavelength dependence in the sense that our spectra are bluer
than the NFGS by $\pm3\%$ between $4000$~\AA{} and $6500$~\AA.
Table~\ref{table:nfgs_compare} compares our emission-line equivalent
width measurements of \oiilam, \oiiilam, \hblam{}, and \halam{} for
these $11$ objects (described in detail in \S\ref{sec:fitting})
against the \citet{jansen00b} measurements.  The spectral differences
for NGC~3104 may be due to a bright foreground star, which has been
subtracted from our data but may not have been subtracted from the
NFGS spectrum.  For UGC~09560 aperture mismatch may be responsible for
the observed discrepancy.  Our spectroscopic aperture misses
$\sim10\%$ of the light while the NFGS aperture encloses the whole
galaxy.  Alternatively, the NFGS spectrum of UGC~09560 may be
suffering from second-order contamination since a blue-blocking filter
was not used for the NFGS observations \citep{jansen00b}.  Despite
these discrepancies, however, we conclude that, on average, our
observations agree very well with the NFGS.

\begin{figure}
\epsscale{0.9}
\plotone{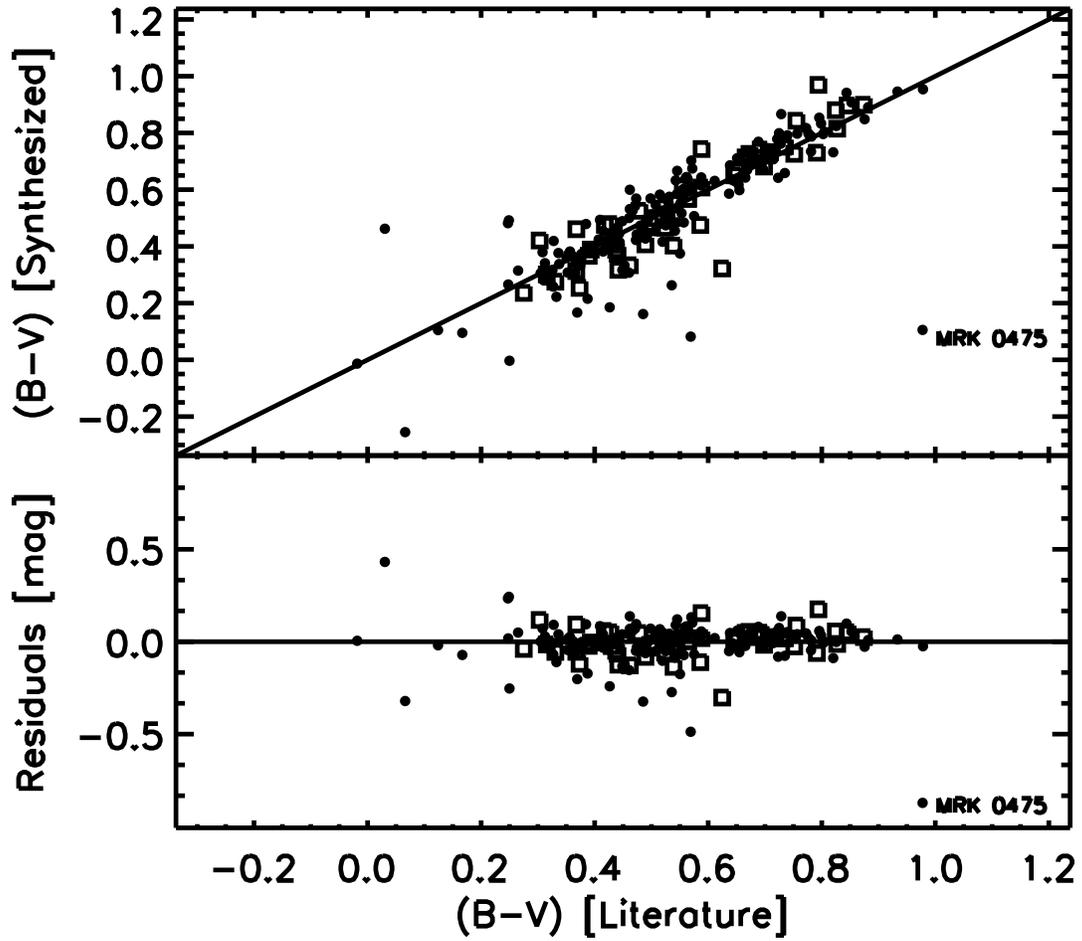}
\caption{Comparison of our synthesized \bv{} colors against the
literature.  We plot galaxies observed during clear and
non-photometric conditions using filled circles and open squares,
respectively.  We label Mrk~0475, the largest significant outlier, in
both panels and discuss it in \S\ref{sec:accuracy}.
\label{fig:compare_colors}} 
\end{figure}

Another test of our relative and absolute spectrophotometry compares
magnitudes synthesized directly from our spectra against published
broadband photometry (see \S\ref{sec:sample} and
Table~\ref{table:photometric_properties}).  We synthesize
spectrophotometric $B$- and $V$-band magnitudes using the
\citet{bessell90} filter response function and the \citet{lejeune97}
theoretical spectrum of Vega tied to the \citet{hayes85} V-band Vega
zero-point.  In Figure~\ref{fig:compare_colors} we plot our
synthesized \bv{} colors against measurements from the literature.  We
identify one significant outlier in this comparison, Mrk~0475, a
nearby irregular starburst galaxy.  The measurements from the
literature indicate that the \bv{} color for Mrk~0475 is $\sim1$~mag,
which is unrealistically red for this type of object.  The $1\sigma$
residual scatter including all the points is $0.11$~mag, or $0.06$~mag
if we exclude fourteen $3\sigma$ outliers, with no systematic offset.
Given the typical measurement uncertainty in the photometry from the
literature, $\sim0.2$~mag, we conclude that our \bv{} colors are
statistically consistent with the published \bv{} colors.

\begin{figure}
\epsscale{0.9}
\plotone{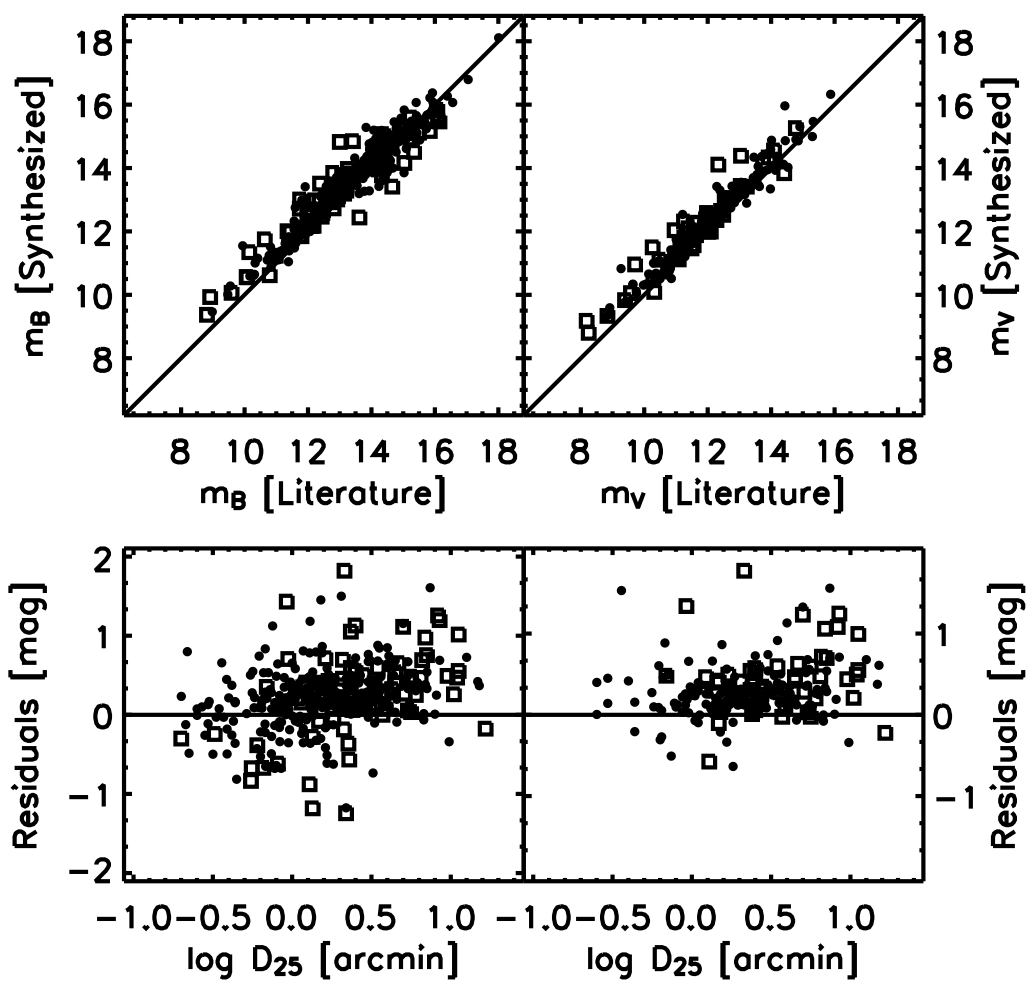}
\caption{Comparison of apparent magnitudes synthesized directly from
our spectra against literature measurements in the B- (\emph{left})
and V-band (\emph{right}).  The bottom panel shows the magnitude
residuals versus $D_{25}$, the $25$~mag~arcsec$^{-2}$ isophotal
diameter of the major axis in arcminutes.  Points are coded as in
Figure~\ref{fig:compare_colors}.
\label{fig:compare_mags}}
\end{figure}

Figure~\ref{fig:compare_mags} compares published $B$- and $V$-band
magnitudes against our synthesized photometry.  Several caveats should
be kept in mind for this comparison.  First, the photometry from the
literature has been taken from the RC3 and LEDA
(\S\ref{sec:properties}), which give total magnitudes extrapolated to
infinity, whereas our synthesized magnitudes typically only include
the galaxy light within the $25$~mag~arcsec$^{-2}$ isophote.
Unfortunately, these catalogs do not publish the aperture correction
used to transform from isophotal to total magnitude, which prevents us
from making a more direct comparison.  Second, the fraction of light
enclosed by our spectroscopic aperture is generally $<100\%$, in
particular for the largest galaxies in our sample.  Therefore, we
anticipate that our synthesized photometry will be systematically
fainter than the measurements from the literature, with a possible
residual systematic that correlates with galaxy size.

In Figure~\ref{fig:compare_mags} (\emph{left}) we compare the apparent
$B$ magnitudes, $m_{B}$.  The photometric observations exhibit a
$0.39$~mag residual scatter, or $0.32$~mag excluding eight $3\sigma$
outliers.  Not surprisingly, the scatter in our non-photometric
observations is much larger, $0.54$~mag, or a factor of $\sim3.5$.
Below this comparison we plot the magnitude residuals versus $D_{25}$,
the $25$~mag~arcsec$^{-2}$ isophotal diameter of the major axis.  We
find the residuals correlated with $D_{25}$ in the sense that our
synthesized magnitudes miss a larger fraction of the optical light
with increasing galaxy size.  The median residual increases from
$0.0-0.1$~dex for the smallest objects in our sample
($D_{25}\lesssim0\farcm5$), to $0.3-0.4$~dex for objects larger than
$\sim4\farcm5$.  In Figure~\ref{fig:compare_mags} (\emph{right}) we
compare our V-band apparent magnitudes against measurements from the
literature.  The $1\sigma$ V-band residuals for objects observed
during clear conditions is $0.35$~mag, or $0.23$~mag after clipping
seven significant outliers.  Galaxies observed during cloudy
conditions exhibit a scatter of $0.39$~mag excluding one extreme
outlier.  We also find the median V-band residuals correlated with
galaxy size, increasing by a factor of two from the smallest to the
largest galaxies in our sample.  In conclusion, given all the
uncertainties, we find good overall agreement between our synthesized
magnitudes and the published broadband photometry.  Without isophotal
magnitudes for the galaxies in our sample, however, it is very
difficult to estimate the light fraction enclosed by our spectroscopic
aperture, or to test externally our absolute spectrophotometric
calibration.

\begin{figure}
\epsscale{0.9}
\plotone{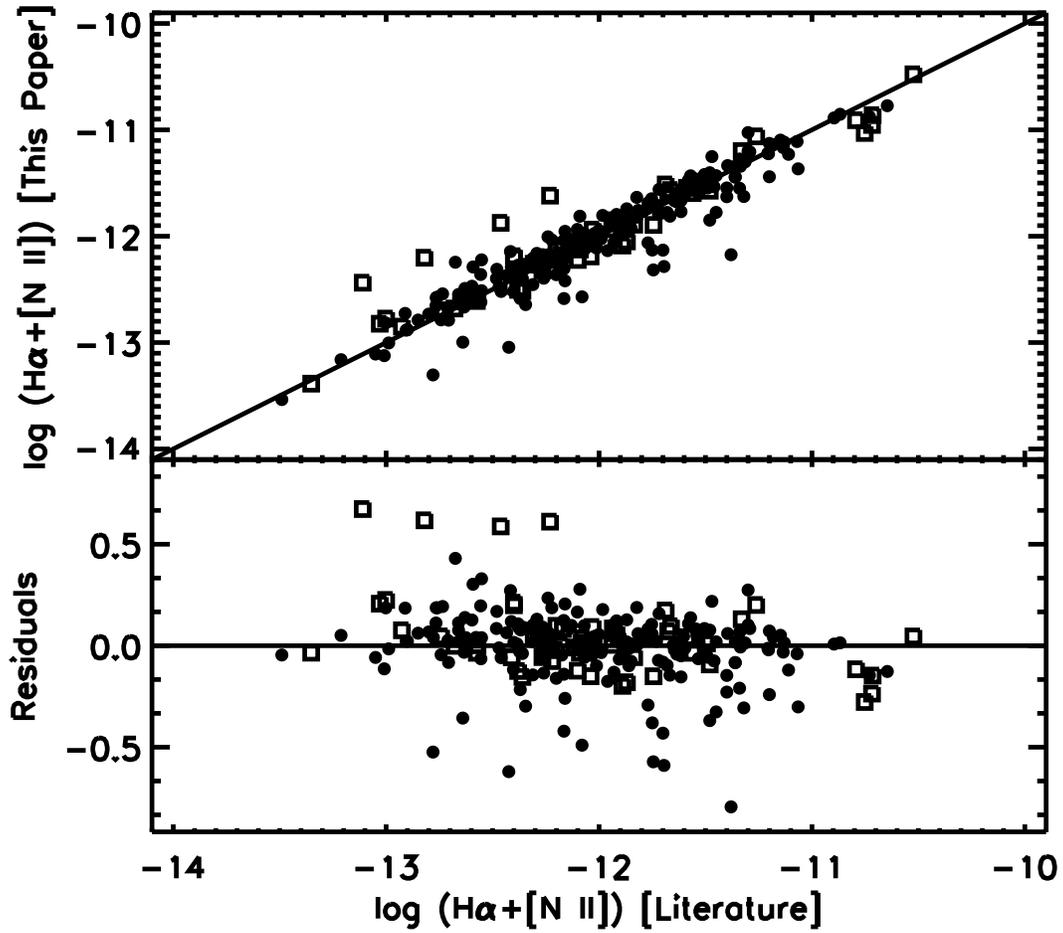}
\caption{Comparison of our integrated spectrophotometric
\ha+\niidoublet{} emission-line fluxes in \flunits{} against
measurements in the literature using the symbols defined in
Figure~\ref{fig:compare_colors}.}
\label{fig:compare_halpha_flux}
\end{figure}

\begin{figure}
\epsscale{0.9}
\plotone{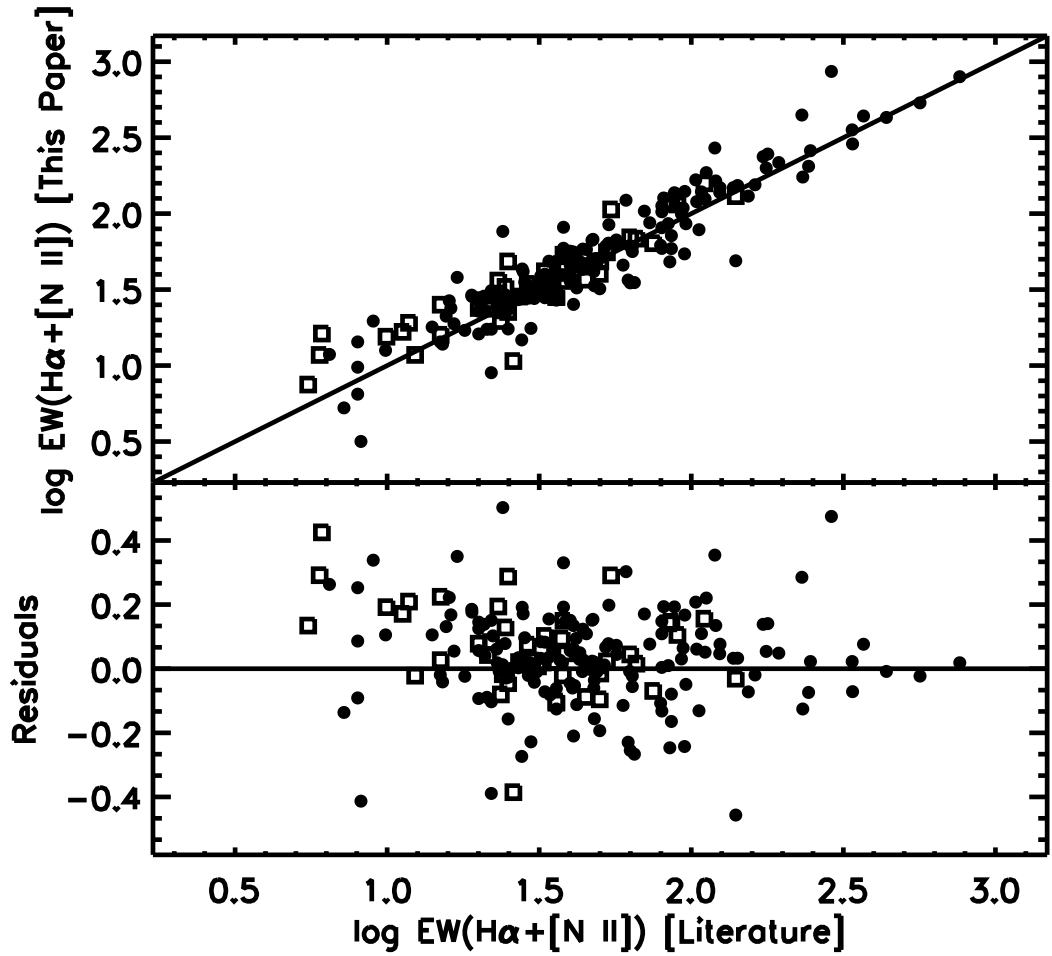}
\caption{Comparison of \ha+\niidoublet{} equivalent width measurements
in Angstroms from our integrated spectrophotometry against the
literature values with the individual points coded as in
Figure~\ref{fig:compare_colors}.}
\label{fig:compare_halpha_ew}
\end{figure}

Finally, we compare our integrated spectrophotometric
\ha+\niidoublet{} emission-line flux and equivalent-width measurements
against published values compiled and homogenized from the literature
(R.~C. Kennicutt et~al., 2005b, in preparation).  As described in
detail in \S\ref{sec:fitting}, we carefully subtract the stellar
continuum from our spectra using population synthesis before measuring
the nebular emission lines; consequently, we account for underlying
stellar absorption in our \ha{} measurements.  The data from the
literature are based on narrow-band imaging supplemented with
photoelectric measurements, and have been corrected for underlying
stellar absorption but not for foreground Galactic reddening.  Using a
$15\arcsec$ search radius we find $252$ overlapping galaxies with
\ha{} and \nii{} fluxes detected at more than $3\sigma$ in both
samples, and $206$ objects with well-measured equivalent widths.  In
Figure~\ref{fig:compare_halpha_flux} we compare our emission-line
fluxes against data from the literature, coding points according to
whether they have been observed during photometric or non-photometric
conditions.  Overall, we find excellent agreement among the two
measurements.  Only considering galaxies observed during photometric
conditions and removing twelve $3\sigma$ outliers, we find a $1\sigma$
residual scatter of $0.12$~dex, or $32\%$, and a median (mean)
systematic offset of just $0.02$~dex ($0.01$~dex).  In
Figure~\ref{fig:compare_halpha_ew} we compare our \ha+\niidoublet{}
equivalent width measurements against the corresponding measurements
from the literature, both corrected for stellar absorption.  After
removing nine significant outliers, we find a median (mean) systematic
difference of $0.03$~dex ($0.04$~dex) and a residual scatter of
$0.12$~dex, or $\sim32\%$.

To summarize, based on a variety of comparisons we find that the
spectrophotometric quality of our observations is on average very
good.  From both internal and external comparisons, we find that our
relative spectrophotometric uncertainty is $3-4\%$ between
$\sim3600$~\AA{} and $\sim6900$~\AA.  The absolute spectrophotometric
accuracy of our data is more difficult to estimate due to the absence
of matched-aperture broadband photometry.  Nevertheless, neglecting
light missed by our extraction apertures, the systematic uncertainty
in our absolute spectrophometry is of order $10-30\%$, and potentially
much larger on a case-by-case basis.  Because of our inability to
estimate this systematic uncertainty accurately, it has not been
included in column~(9) of Table~\ref{table:journal}.  However, the
absence of a systematic offset between our spectrophotometric
\ha+\niidoublet{} fluxes and measurements based on narrow-band imaging
indicates that, on average, our observations include all the
star-forming regions in our sample of galaxies.

\subsection{Presentation of the Atlas}\label{sec:presentation}

In Figure~Set~8 we present our integrated spectrophotometric atlas.
We plot each spectrum as normalized $f_{\lambda}(\lambda)$ versus rest
wavelength in Angstroms.  We choose the normalization factor for each
object to achieve a balance between showing the full range in flux and
illustrating the finer details of the continuum.  Adjacent to each
spectrum we provide a visualization of the galaxy using a
logarithmically scaled Digitized Sky Survey image.  Each image shows
our rectangular spectroscopic aperture as a \emph{solid} line and the
$25$~mag~arcsec$^{-2}$ isophotal size of the galaxy from
Table~\ref{table:general_properties}, where available, as a
\emph{dashed} ellipse.  The horizontal solid line in the lower-left
corner of each image represents $30\arcsec$.

\section{CONTINUUM \& EMISSION-LINE FITTING}\label{sec:fitting}

\subsection{Motivation}\label{sec:motivation}

Typically the integrated spectrum of a galaxy in the optical is a
luminosity-weighted sum of an emission-line spectrum superposed on a
stellar continuum/absorption-line spectrum.  To first order,
therefore, we can use population synthesis to model the stellar
continuum and subtract it from the data to isolate the nebular
emission-line spectrum.  This technique, including a simple
prescription for dust reddening, offers a powerful way to study the
integrated spectral properties of galaxies.

Mathematically, spectral synthesis fitting is a bounded (non-negative)
least squares problem.  The objective is to find the linear
combination of stellar continuum models that optimally reproduces the
integrated absorption-line spectrum of the galaxy.  High-resolution
population synthesis models with broad spectral coverage are ideally
suited for this application \citep[e.g.,][]{leborgne04, bruzual03}.  In
general, however, degeneracies between age, metallicity, and dust
reddening, as well as computational difficulties such as finding a
local versus the global minimum, complicate the interpretation of the
best-fitting absorption-line spectrum.  A better constrained objective
is to use population synthesis modeling simply as a means of measuring
the nebular emission-line fluxes free from the systematic effects of
underlying stellar absorption, without attempting to interpret
physically the model continuum.

Most studies correct only the Balmer emission lines for stellar
absorption, typically assuming a constant $2$~\AA{}
\citep[e.g.,][]{mccall85}.  However, this value and the assumption of
a constant correction only apply to spectroscopy of individual \hii{}
regions, where young ($\lesssim20$~Myr) stellar populations dominate
the underlying continuum.  \citet{rosa-gonzalez02} demonstrate that
the \oiidoublet{} doublet also suffers from significant stellar
absorption in galaxies with strong Balmer breaks, while the other
forbidden nebular lines suffer weak metal-line absorption.  In
general, the Balmer absorption-line equivalent widths in the
integrated spectra of star-forming galaxies vary with wavelength
according to the particular star-formation history of each object.
Furthermore, the total absorption-line equivalent width, which is
independent of spectral resolution, is larger than the absorption
correction, or the fraction of the total absorption underlying the
nebular emission line.  Balmer lines measured from lower spectral
resolution spectra suffer larger amounts of underlying stellar
absorption.  To minimize many of these uncertainties, studies
frequently restrict their sample to a minimum \ha{} or \hb{}
emission-line equivalent width, which unfortunately introduces a bias
toward galaxies with larger current to past-average star-formation
rates \citep{kenn94}.  By comparison, spectral synthesis modeling
accurately determines the underlying stellar continuum for each
individual galaxy, enabling even low equivalent-width nebular lines to
be measured with high precision.  In the following sections we present
{\tt iSPEC1d}, our implementation of the technique described above.
For similar recent implementations of stellar continuum subtraction we
refer the reader to the following papers: \citet{reichardt01},
\citet{panter03}, \citet{bruzual03}, \citet{tremonti04},
\citet{brinchmann04}, \citet{cid05}, and \citet{savaglio05}.

\subsection{Choice of Templates}\label{sec:templates}

We choose our continuum templates from the \citet[hereafter
BC03]{bruzual03} population synthesis models.  We select the BC03
models computed using the `Padova 1994' stellar evolutionary tracks
\citep[and references therein]{girardi96}, the STELIB stellar library
\citep{leborgne03}, and the \citet{salpeter55} initial mass function
(IMF) integrated between $0.1~\mathcal{M}_{\sun}$ and
$100~\mathcal{M}_{\sun}$.  At each metallicity BC03 provide $221$
spectral energy distributions at unevenly spaced ages ranging from
$0$~Myr to $20$~Gyr.  The FWHM resolution of the models is $3$~\AA{}
between $3200$~\AA{} and $9500$~\AA{} with $1$~\AA{} binning.

After extensive experimentation we select eight instantaneous-burst,
solar-metallicity models corresponding to the following ages: $0$,
$5$, $25$, $102$, $255$, $641$, $1434$, and $12000$~Myr
(Fig.~\ref{fig:templates}).  Clearly, there exists a degeneracy
between selecting young, solar-metallicity models, versus more evolved
metal-poor simple stellar populations.  However, we emphasize that the
primary motivation for fitting the stellar continuum is to correct
quantitatively and self-consistently the Balmer emission lines for
stellar absorption, and to enable measurement of weak nebular emission
lines.  In this context, eight solar-metallicity templates adequately
reproduce the stellar continua of our galaxies.  Furthermore, our
emission-line fluxes do not change significantly if we select fewer or
more templates, or if we allow the stellar metallicity to vary from
solar.  In a future paper we will investigate the stellar
metallicities and detailed star formation histories of the galaxies in
our sample using more generalized population synthesis fitting
techniques \citep[e.g.,][]{panter03, bruzual03, cid05}.

\subsection{The Fitting Algorithm}\label{sec:algorithm}

We begin with a flux-calibrated galaxy spectrum $S(\lambda)$ as a
function of wavelength $\lambda$, a variance spectrum
$\sigma^2(\lambda)$, and a set of $N$ single metallicity population
synthesis templates $\mathcal{T}^{i}(\lambda)$ ($i=1,...,N$).  Let
$\xi_{S}$ be the FWHM resolution of the data due to instrumental and
stellar velocity broadening.  Because our instrumental resolution,
$\sim8$~\AA{} FWHM, or $\sim435~\kms$ at $5500$~\AA, prohibits us from
measuring the stellar velocity dispersion $\sigma_{*}$ directly, we
assume a fixed $\sigma_{*}=100$~\kms{} for the whole sample.  We
obtain only minimal improvement in the quality of the continuum
subtraction if we adopt a velocity dispersion that varies with the
galaxy luminosity based on, for example, the Tully-Fisher relation.
Computationally, our algorithm searches for the optimal non-negative
linear combination of $\mathcal{T}^{i}(\lambda)$ that best reproduces
the observed absorption-line spectrum, subtracts it from the data, and
then simultaneously fits all the nebular emission lines using
physically motivated constraints on the Gaussian line profiles.  We
summarize this procedure in more detail next.

\begin{enumerate}

\item{First, we shift all the observed quantities to the rest-frame
using the redshifts in Table~\ref{table:general_properties}: $\lambda
= \lambda_{\rm obs}/(1+z)$, $S = S_{\rm obs}(1+z)$, $\sigma^{2}_{S} =
\sigma^{2}_{S_{\rm obs}}(1+z)^{2}$, and $\xi_{S} = \xi_{S_{\rm
obs}}/(1+z)$.  Consequently, all measured quantities such as
equivalenth widths are returned in the rest frame-of-reference.}

\item{Next, we resample the fitting templates onto the rest wavelength
spacing and pixel size of the data using high-order spline
interpolation.  We broaden the templates to the spectral resolution of
the data using a Gaussian convolution kernel:

\begin{equation}
\mathcal{T}_{\mathrm B}^{i}(\lambda) =
\frac{1}{\sqrt{2\pi}\sigma(\lambda)}
\int_{\lambda_{\rm min}}^{\lambda_{\rm max}}
{\rm d}{\lambda}^{\prime}\, \mathcal{T}^{i}(\lambda^{\prime})
\exp\left[-\frac{(\lambda-\lambda^{\prime})^2}
{2\sigma^2(\lambda^{\prime})}\right],
\end{equation}

\noindent where 

\begin{equation}
\sigma(\lambda) \equiv
\frac{\sqrt{\xi_{S}^2(\lambda) -
\xi_{\mathcal{T}}^2(\lambda)}}{2\sqrt{2\ln2}}   
\end{equation}

\noindent is the Gaussian broadening function width and
$\xi_{\mathcal{T}}=3$~\AA{} is the FWHM resolution of the templates.} 

\item{We mask all pixels within $\pm20$~\AA{} of the nebular emission
lines, the Na~D~$\lambda\lambda5890,5896$ interstellar absorption
doublet, and the [\ion{O}{1}]~$\lambda5577$ night sky line, as well as
the atmospheric telluric absorption band at $6850-6960$~\AA.}

\item{The best-fitting model spectrum $\mathcal{F}(\lambda)$ is
given by

\begin{equation}
\mathcal{F}(\lambda) = \sum_{i=1}^{N} A_{i}\,\mathcal{T}_{\mathrm 
B}^{i}(\lambda)\,10^{-0.4\,\alpha_{i}\,{\mathrm E}(B-V)\,
k(\lambda)} \label{eq:model} 
\end{equation}

\noindent where the $A_{i}$ are the non-negative coefficients we
require, $k(\lambda)$ is an assumed reddening curve, and the product
$\alpha_{i}\,\ebv$ controls the reddening of each template (described
in \S\ref{sec:ebvmodel}).  To find the $A_{i}$ coefficients we
minimize the $\chi^2$ statistic using MPFIT\footnote{A
Levenberg-Marquardt least-squares minimization routine available at
\url{http://cow.physics.wisc.edu/\~{}craigm/idl}.}, where

\begin{equation}
\chi^2 \equiv \sum_{j=1}^{M} \frac{[S(\lambda_{j})-
\mathcal{F}(\lambda_{j})]^2} {\sigma^2[S(\lambda_{j})]}
\label{eq:chi2}
\end{equation}

\noindent and $M$ is the number of unmasked pixels in the continuum.}

\item{We iterate steps $1$ through $4$ twice, but after the first
iteration we improve on the initial absorption-line redshift of the
galaxy by cross-correlating the best-fitting model spectrum
$\mathcal{F}(\lambda)$ with the observed spectrum $S(\lambda)$.  This
step accounts for small errors in the input redshift or wavelength
array, although we do not allow the redshift to vary by more than
$\pm500$~\kms{}.  In addition, based on some simple simulations, to
prevent spurious shifts we demand a median ${\rm S/N}>10$ in the
continuum before updating the redshift.  On the second iteration we
also input the best-fitting coefficients from the previous iteration
as an initial guess to ensure rapid convergence.}

\item{Finally, we subtract the best-fitting absorption-line spectrum
from the data to obtain a pure emission-line spectrum,
$\mathcal{E}(\lambda) \equiv S(\lambda) - \mathcal{F}(\lambda)$.  We
model $\mathcal{E}(\lambda)$ as a sum of $n_{\mathrm line}$ Gaussian
functions at the rest wavelengths of the nebular emission lines of
interest.  The velocity width of each emission line is given by
$\sigma_{\rm line} = \sqrt{\sigma^{2}_{\rm gas} + \sigma^{2}_{\rm
S}}$, where $\sigma_{\rm gas}$ is the intrinsic velocity dispersion of
the gas in \kms{} and $\sigma_{\rm S} =
1.273\times10^{5}\,\xi_{T}/\lambda_{\rm line}$ is the instrumental
resolution in \kms{} and $\lambda_{\rm line}$ is the wavelength of the
emission line.  We tie all the Balmer emission lines together by
constraining their $\sigma_{\rm gas}$ and redshift.  We apply the same
constraints separately to the forbidden emission lines.  This
procedure reduces the number of free emission-line parameters from
$3\times n_{\mathrm line}$ to $n_{\mathrm line}+2$ and succeeds
remarkably well at fitting weak emission lines since their profiles
are tied to stronger lines.}

\end{enumerate}

The final outputs from {\tt iSPEC1d} are the best-fitting
absorption-line spectrum and the integrated fluxes of the nebular
emission lines, self-consistently corrected for underlying stellar
absorption.  In Figure~\ref{fig:example_specfit} we show four examples
of our spectral synthesis modeling results.  In the left panels we
plot the integrated spectrum and the best-fitting stellar continuum
for NGC~0034, NGC~4651, NGC~3893, and NGC~0337.  For clarity we do not
plot the fits to the nebular emission lines.  In the right panels we
show an expanded view of the data and our model fits centered on the
\hb{} nebular emission line.  In all these objects \hb{} is severely
affected by underlying stellar absorption.  After continuum
subtraction, we measure \hb{} emission-line EWs of $1.2$, $3.7$,
$5.7$, and $10.4$~\AA, from top to bottom, respectively.  The EWs of
the absorption corrections are $5.9$, $3.9$, $4.6$, and $5.2$~\AA,
respectively.  Frequently, population synthesis cannot be used to
model the integrated stellar continua of galaxies, for example, if the
spectra have not been flux-calibrated, or if the signal-to-noise ratio
is too low.  In these cases, a statistical Balmer absorption
correction is normally adopted.  However, as discussed in
\S\ref{sec:motivation}, in general, the appropriate correction varies
across the Balmer-line sequence according to the particular
star-formation history of each galaxy, and depends on the spectral
resolution of the data, being larger for lower-resolution spectra.  In
our integrated sample the average \hb{} and \ha{} stellar absorption
corrections are $4.4\pm0.63$~\AA{} and $2.8\pm0.38$~\AA.  For
comparison, \citet{kobulnicky99a} find that the \hb{} absorption
correction in their sample and in the K92 atlas ranges from $1-6$~\AA,
with a mean of $3\pm2$~\AA.  



\subsection{Reddening Model}\label{sec:ebvmodel}

Virtually all of the galaxies in our sample require continuum
reddening to achieve reasonable fits to their integrated or nuclear
spectrum.  Given our poor understanding of the effects of dust
attenuation on the integrated spectral properties of galaxies
\citep[e.g.,][]{witt00}, we opt for a simple, physically motivated
model.  For our sample we adopt the \citet{charlot00} galactic
attenuation curve, parameterized as
$k(\lambda)=\rv\,(\lambda/5500~{\rm \AA})^{-0.7}$, where $\rv=5.9$.
Below, we discuss the effect of selecting a different $k(\lambda)$. 


Based on some simple arguments we expect the young stellar populations
to suffer a larger amount of dust extinction relative to the old
stellar populations.  Because of their short lifetimes, early-type
stars are located near their dusty, natal star-forming regions,
whereas late-type stars are more uniformly dispersed throughout the
galaxy where interstellar dust extinction is less severe
\citep{calzetti94, mayya96, charlot00, zaritsky02}.  To parameterize
this observational constraint we define

\begin{equation}
\ebv_{\rm old}\equiv\alpha\ebv_{\rm young},
\label{eq:alpha}
\end{equation}

\noindent where $\ebv_{\rm young}$ and $\ebv_{\rm old}$ correspond to
the attenuation of the stellar populations younger and older than
$10$~Myr, respectively, and $0\leq\alpha\leq1$ is a constant.  The
time boundary between young and old stellar populations roughly
corresponds to the dispersal time for giant molecular clouds
\citep[e.g.,][]{blitz80}.  Computationally, we solve for the
best-fitting parameters $\ebv_{\rm young}$ and $\alpha$, and use
equation~(\ref{eq:alpha}) to obtain $\ebv_{\rm old}$.  We also tested
a one-parameter reddening model, fixing $\alpha\equiv1$ for all ages
and solving for a scalar reddening value, \ebv.  

In general we achieve better continuum fits with our two-parameter
reddening model, which is not surprising given the added flexibility.
We also find, however, that the optimal reddening parameters are
highly degenerate with the best-fitting star-formation history.  For
example, we obtain comparable $\chi^{2}$ values with a highly reddened
young stellar population as with an unreddened, evolved stellar
population.  This effect is well-known as the age-reddening
degeneracy.  Ultimately, since we are not attempting to constrain the
star-formation histories of our sample, only to achieve robust stellar
absorption corrections, we adopt the one-parameter reddening model as
our default, and post-pone a more in-depth analysis of these
degeneracies to a future study.

How physical are the derived continuum reddening values?  In
Figure~\ref{fig:ebv_correlation} we explore this question by plotting
the derived continuum reddening, \ebv, versus the nebular reddening as
derived from the \ha/\hb{} Balmer decrement for our nuclear and
integrated spectra.  Using a Spearman rank correlation test we find
these independent variables correlated at $>10\sigma$ significance.
For reference we overplot lines of constant $\alpha=(0.5,0.25,0.1)$,
assuming that the nebular reddening traces the obscuration of the
young stellar populations.  We find that, on average, the stellar
continuum suffers a fractional amount of dust extinction relative to
the nebular lines.  For the integrated spectra the median (mean) ratio
is $0.34$ ($0.32\pm0.21$), and $0.45$ ($0.40\pm0.21$) for the nuclear
spectra.  For comparison, based on an analysis of the centers
($10\arcsec\times20\arcsec$) of a sample of $\sim40$ starburst
galaxies, \citet{calzetti97b} find an average ratio of $0.44\pm0.03$
\citep[see also][]{calzetti94, storchi94, calzetti00}.

Finally, we explore the effect of choosing a different $k(\lambda)$ on
the quality of the continuum fits and on the derived emission-line
fluxes and equivalent widths (see \S\ref{sec:measure}).  To conduct
this test we refit our integrated spectral atlas with a standard Milky
Way extinction curve \citep[$\rv=3.1$;][]{odonnell94}, and with the
Small Magellanic Cloud (SMC) extinction curve
\citep[$\rv=2.74$;][]{gordon03}.  We compare these results against the
values we obtain using our default \citet{charlot00} attenuation
curve.  We find no significant changes in the quality of the continuum
fits using either the Milky Way or SMC extinction curves: the standard
deviation of the ratio of the derived $\chi^{2}$ values is $1.4\%$ and
$1.9\%$, respectively, with no systematic differences.  The residuals
of the best-fitting \ebv{} values exhibit no systematic differences,
and a scatter of just $\pm3\%$ based on either the Milky Way or SMC
extinction curves.  Finally, we find no systematic residuals among the
emission-line fluxes and equivalent widths measured using any of the
three parameterizations of $k(\lambda)$.  Moreover the standard
deviation of the residuals are always $<1.5\%$, which is well within
the typical measurement uncertainty of the data.

\section{EMISSION-LINE MEASUREMENTS}\label{sec:measure} 

After continuum subtraction the residual spectrum is a pure
emission-line spectrum readily modeled using multi-Gaussian profile
fitting.  In our integrated and nuclear spectra we simultaneously fit
the \ha, \hb, \hg, and \hd{} Balmer emission lines and the following
forbidden nebular emission lines: \oiidoublet, \oiiidoublet, \oilam,
\niidoublet, and \siidoublet.  Because the \oii{} doublet is
unresolved at our spectral resolution we sum the individual Gaussian
components to form the \oiilam{} emission line.  In addition to the
velocity width and redshift constraints described in
\S\ref{sec:algorithm}, we also constrain the \nii\ doublet ratio,
$\nii \lambda6584/\lambda6548 \simeq 3$ \citep{storey00}, since these
lines are moderately blended with \ha{} at our spectral resolution.
Furthermore, we do not allow the emission-line redshift to differ from
the continuum redshift by more than $\pm1000$~\kms{} and we forbid the
intrinsic velocity width from exceeding $500$~\kms.

In Tables~\ref{table:int_lineflux} and \ref{table:nuc_lineflux} we
present the emission-line fluxes, $1\sigma$ flux errors, and $1\sigma$
upper limits for the strong nebular emission lines in our integrated
and nuclear spectra, respectively.  These measurements have been
corrected for foreground Galactic extinction using the reddening
values in Table~\ref{table:general_properties} and the
\citet{odonnell94} Milky Way extinction curve, and for underlying
stellar absorption as described in \S\ref{sec:algorithm}; they have
not been corrected for reddening intrinsic to the galaxy.  We compute
upper limits for undetected (${\rm S/N}<1$) emission lines by
measuring the standard deviation of the continuum near the line,
$\delta_{\rm lc}$, and determining the total (instrumental plus
intrinsic) velocity width at that wavelength, $\sigma_{\rm line}$.
The intrinsic velocity width can be estimated because the emission
lines are tied together.  The $1\sigma$ upper limit is then given by
$\sqrt{2\pi}\,\sigma_{\rm line}\,\delta_{\rm lc}$, assuming that the
undetected line is a Gaussian function whose sigma-width is
$\sigma_{\rm line}$ Angstroms.

The errors given in Tables~\ref{table:int_lineflux} and
\ref{table:nuc_lineflux} underestimate the true error because they
only incorporate statistical sources of uncertainty due to Poisson
noise, read noise, sky subtraction, and sensitivity function division.
We conduct two experiments to estimate the systematic uncertainty in
our flux measurements.  First, we compare the measured \oiiidoublet{}
doublet ratio to the theoretical ratio: $\oiii~\lambda5007/\lambda4959
\simeq 3$ \citep{storey00}.  We find that for ${\rm
EW}(\oiii~\lambda4959)\gtrsim3$~\AA{} the measured ratio agrees with
the theoretical value within $\pm4\%$.  In our second test we select
the \oiiilam{} emission line, which is well-isolated and exhibits a
wide range of intensities, and compare the Gaussian-integrated flux
against the total flux within $\pm3~\sigma_{\rm line}$ of the line
center, where $\sigma_{\rm line}$ is the total (instrumental plus
intrinsic) Gaussian line-width.  Once again we derive a mean scatter
of $\pm4\%$ with no systematic dependence on the emission-line
equivalent width.  In conclusion we recommend adding a $4\%$
systematic error in quadrature to the uncertainties given in
Tables~\ref{table:int_lineflux} and \ref{table:nuc_lineflux}.

The equivalent width measurements and $1\sigma$ upper limits for the
strong nebular lines in our integrated and nuclear spectra,
respectively, are given in Tables~\ref{table:int_lineEW} and
\ref{table:nuc_lineEW}.  In these tables we adopt the convention that
emission-line equivalent widths are positive.  As discussed above the
equivalent widths errors only reflect the statistical uncertainty in
the measurement.  To determine the equivalent width of a line we
divide its flux by the continuum at the line-center, neglecting any
small-scale variations in the continuum over the line-profile.  We
compute the local continuum as the average flux in two bandpasses
centered on the rest wavelength of the line, $\lambda_{0}$, given by
[$\lambda_{0}+3\,\sigma_{\rm line}$, $\lambda_{0}+10\,\sigma_{\rm
line}$] and [$\lambda_{0}-10\,\sigma_{\rm line}$,
$\lambda_{0}-3\,\sigma_{\rm line}$], where $\sigma_{\rm line}$ is the
total (intrinsic plus instrumental) Gaussian width of the line in
Angstroms.  This measurement is made on the emission-line subtracted
spectrum $S(\lambda) - \mathcal{E}(\lambda)$ to ensure that nearby
emission lines do not bias the local continuum measurement.  The
equivalent widths of the Balmer emission lines must be treated
separately from the forbidden lines due to the strong underlying
stellar absorption.  For each Balmer line we define red and blue
pseudo-continuum bandpasses, given in Table~\ref{table:balmers},
analogous to the Lick indices \citep{worthey94a, worthey97, trager98}.
Finally, we interpolate the mean continuum flux in the red and blue
bandpasses at the line-center and divide by the emission-line flux to
derive the equivalent width.

\section{SUMMARY}\label{sec:summary}

We conclude by briefly highlighting a few key aspects of our survey
with respect to sample selection, data quality, and analysis methods,
and by describing a handful of the scientific investigations we are
currently pursuing with these data.  The strengths of our survey are
the size ($417$ galaxies) and diversity of the sample, the
signal-to-noise ($10-100$) and relative spectrophotometric accuracy
($\sim4\%$) of the observations, and the quantitative methods we have
developed to measure the nebular emission lines.  Our sample includes
a wide range of galaxy types, from low-luminosity dwarf starburst
galaxies such as I~Zw~018 (UGCA~166) and NGC~1569; large disk galaxies
such Messier~51a (NGC~5194) and NGC~2903; interacting/merging systems
at various stages of their dynamical interaction such as the Mice
(NGC~4676) and the Antennae (Arp~244); and luminous- and
ultra-luminous infrared galaxies such as Arp~220, to name just a few
specific examples.  The spectrophotometric observations we present
constitute a homogenous data set obtained using the same telescope,
instrumental setup, and observing technique, which minimizes or
eliminates systematic errors in the data acquisition and data
processing.  Furthermore, we have reduced the data using a new
long-slit data reduction software package ({\tt iSPEC2d}), which
enables us to propagate carefully the uncertainties in the data
through every stage of the reductions.  Finally, we measure the
nebular emission lines precisely and free from the systematic effects
of stellar absorption using population synthesis modeling and
subtraction of the stellar continuum ({\tt iSPEC1d}).

The integrated spectra we have obtained are spatially unbiased,
thereby making them ideal for studying the global spectrophotometric
properties of nearby galaxies.  In addition, our observations may
serve as a unique reference sample for interpreting the properties of
spatially unresolved, distant galaxies.  In a series of subsequent
papers we will be using these data to study and calibrate rest-frame
optical star-formation rate indicators \citep{moustakas05b},
emission-line abundance diagnostics, the physical sources of scatter
in the luminosity-metallicity relation, the geometry and dust content
of normal and starburst galaxies, and the global star-formation
histories of infrared-luminous galaxies.

\acknowledgements

The authors would like to thank Christy Tremonti for contributing
invaluable technical expertise during the course of this project, and
for constructive comments on an earlier draft of the paper.
J.~M. would also like to acknowledge discussions, encouragement, and
other helpful contributions from Sanae Akiyama, Eric Bell, Daniel
Eisenstein, Doug Finkbeiner, Karl Gordon, Joannah Hinz, Janice Lee,
Andy Marble, Leonidas Moustakas, Chien Peng, Amy Stutz, and Dennis
Zaritsky.  The anonymous referee's suggestions also helped improve the
clarity of the paper.  The data reduction software developed during
the course of this project benefited enormously from IDL routines
written by David Schlegel, Scott Burles, Doug Finkbeiner, and Craig
Markwardt, and from the IDL Astronomy User's Library, which is
maintained by Wayne Landsman at the Goddard Space Flight Center.
Funding for this project has been provided by NSF grant AST-0307386,
NASA grant NAG5-8326, and a SINGS grant, provided by NASA through JPL
contract 1224769.  This research has made use of NASA's Astrophysics
Data System Bibliographic Services, the VizieR catalogue access tool,
the SIMBAD database, operated at CDS, Strasbourg, France, the LEDA
database, and the NASA/IPAC Extragalactic Database, which is operated
by the Jet Propulsion Laboratory, California Institute of Technology,
under contract with the National Aeronautics and Space Administration.

{\it Facility:} \facility{Bok (Boller \& Chivens spectrograph)}

\clearpage

\clearpage

\clearpage

\clearpage

\clearpage

\clearpage

\clearpage

\clearpage


\figsetstart
\figsetnum{8}
\figsettitle{Spectral Visualizations}
\figsetgrpstart
\figsetgrpnum{8.1}
\figsetgrptitle{Visualization of NGC 0023.}
\figsetplot{\includegraphics[scale=0.7,angle=90]{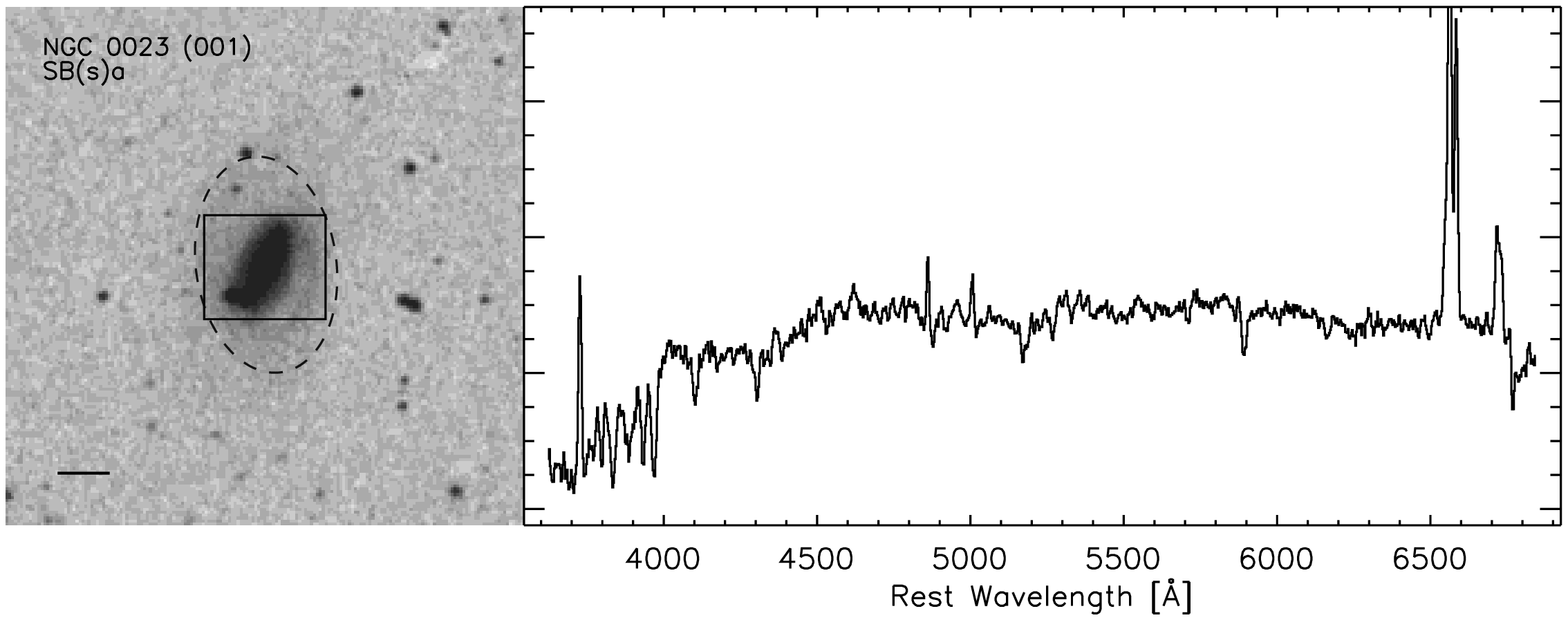}}
\figsetgrpnote{Presentation of our integrated spectral atlas.  We plot the spectrum as $f_{\lambda}(\lambda)$ versus rest wavelength, normalized in a way that attempts to achieve a balance between showing the full range in flux and illustrating the finer details of the continuum.  Our integrated spectrum is accompanied by a Digitized Sky Survey image which illustrates our rectangular spectroscopic aperture as a \emph{solid} outline and the $25$~mag~arcsec$^{-2}$ isophotal size of the galaxy as a \emph{dashed} ellipse.  The image legend gives the galaxy name, the unique identification number in parenthesis, and the morphological type as listed in Table~\ref{table:general_properties}.  The horizontal solid line in the lower-left corner of each image represents $30\arcsec$.}
\figsetgrpend
 
\figsetgrpstart
\figsetgrpnum{8.2}
\figsetgrptitle{Visualization of NGC 0034.}
\figsetplot{\includegraphics[scale=0.7,angle=90]{f8_2.eps}}
\figsetgrpnote{Presentation of our integrated spectral atlas.  We plot the spectrum as $f_{\lambda}(\lambda)$ versus rest wavelength, normalized in a way that attempts to achieve a balance between showing the full range in flux and illustrating the finer details of the continuum.  Our integrated spectrum is accompanied by a Digitized Sky Survey image which illustrates our rectangular spectroscopic aperture as a \emph{solid} outline and the $25$~mag~arcsec$^{-2}$ isophotal size of the galaxy as a \emph{dashed} ellipse.  The image legend gives the galaxy name, the unique identification number in parenthesis, and the morphological type as listed in Table~\ref{table:general_properties}.  The horizontal solid line in the lower-left corner of each image represents $30\arcsec$.}
\figsetgrpend
 
\figsetgrpstart
\figsetgrpnum{8.3}
\figsetgrptitle{Visualization of ARP 256 N.}
\figsetplot{\includegraphics[scale=0.7,angle=90]{f8_3.eps}}
\figsetgrpnote{Presentation of our integrated spectral atlas.  We plot the spectrum as $f_{\lambda}(\lambda)$ versus rest wavelength, normalized in a way that attempts to achieve a balance between showing the full range in flux and illustrating the finer details of the continuum.  Our integrated spectrum is accompanied by a Digitized Sky Survey image which illustrates our rectangular spectroscopic aperture as a \emph{solid} outline and the $25$~mag~arcsec$^{-2}$ isophotal size of the galaxy as a \emph{dashed} ellipse.  The image legend gives the galaxy name, the unique identification number in parenthesis, and the morphological type as listed in Table~\ref{table:general_properties}.  The horizontal solid line in the lower-left corner of each image represents $30\arcsec$.}
\figsetgrpend
 
\figsetgrpstart
\figsetgrpnum{8.4}
\figsetgrptitle{Visualization of ARP 256.}
\figsetplot{\includegraphics[scale=0.7,angle=90]{f8_4.eps}}
\figsetgrpnote{Presentation of our integrated spectral atlas.  We plot the spectrum as $f_{\lambda}(\lambda)$ versus rest wavelength, normalized in a way that attempts to achieve a balance between showing the full range in flux and illustrating the finer details of the continuum.  Our integrated spectrum is accompanied by a Digitized Sky Survey image which illustrates our rectangular spectroscopic aperture as a \emph{solid} outline and the $25$~mag~arcsec$^{-2}$ isophotal size of the galaxy as a \emph{dashed} ellipse.  The image legend gives the galaxy name, the unique identification number in parenthesis, and the morphological type as listed in Table~\ref{table:general_properties}.  The horizontal solid line in the lower-left corner of each image represents $30\arcsec$.}
\figsetgrpend
 
\figsetgrpstart
\figsetgrpnum{8.5}
\figsetgrptitle{Visualization of ARP 256 S.}
\figsetplot{\includegraphics[scale=0.7,angle=90]{f8_5.eps}}
\figsetgrpnote{Presentation of our integrated spectral atlas.  We plot the spectrum as $f_{\lambda}(\lambda)$ versus rest wavelength, normalized in a way that attempts to achieve a balance between showing the full range in flux and illustrating the finer details of the continuum.  Our integrated spectrum is accompanied by a Digitized Sky Survey image which illustrates our rectangular spectroscopic aperture as a \emph{solid} outline and the $25$~mag~arcsec$^{-2}$ isophotal size of the galaxy as a \emph{dashed} ellipse.  The image legend gives the galaxy name, the unique identification number in parenthesis, and the morphological type as listed in Table~\ref{table:general_properties}.  The horizontal solid line in the lower-left corner of each image represents $30\arcsec$.}
\figsetgrpend
 
\figsetgrpstart
\figsetgrpnum{8.6}
\figsetgrptitle{Visualization of NGC 0095.}
\figsetplot{\includegraphics[scale=0.7,angle=90]{f8_6.eps}}
\figsetgrpnote{Presentation of our integrated spectral atlas.  We plot the spectrum as $f_{\lambda}(\lambda)$ versus rest wavelength, normalized in a way that attempts to achieve a balance between showing the full range in flux and illustrating the finer details of the continuum.  Our integrated spectrum is accompanied by a Digitized Sky Survey image which illustrates our rectangular spectroscopic aperture as a \emph{solid} outline and the $25$~mag~arcsec$^{-2}$ isophotal size of the galaxy as a \emph{dashed} ellipse.  The image legend gives the galaxy name, the unique identification number in parenthesis, and the morphological type as listed in Table~\ref{table:general_properties}.  The horizontal solid line in the lower-left corner of each image represents $30\arcsec$.}
\figsetgrpend
 
\figsetgrpstart
\figsetgrpnum{8.7}
\figsetgrptitle{Visualization of NGC 0151.}
\figsetplot{\includegraphics[scale=0.7,angle=90]{f8_7.eps}}
\figsetgrpnote{Presentation of our integrated spectral atlas.  We plot the spectrum as $f_{\lambda}(\lambda)$ versus rest wavelength, normalized in a way that attempts to achieve a balance between showing the full range in flux and illustrating the finer details of the continuum.  Our integrated spectrum is accompanied by a Digitized Sky Survey image which illustrates our rectangular spectroscopic aperture as a \emph{solid} outline and the $25$~mag~arcsec$^{-2}$ isophotal size of the galaxy as a \emph{dashed} ellipse.  The image legend gives the galaxy name, the unique identification number in parenthesis, and the morphological type as listed in Table~\ref{table:general_properties}.  The horizontal solid line in the lower-left corner of each image represents $30\arcsec$.}
\figsetgrpend
 
\figsetgrpstart
\figsetgrpnum{8.8}
\figsetgrptitle{Visualization of NGC 0157.}
\figsetplot{\includegraphics[scale=0.7,angle=90]{f8_8.eps}}
\figsetgrpnote{Presentation of our integrated spectral atlas.  We plot the spectrum as $f_{\lambda}(\lambda)$ versus rest wavelength, normalized in a way that attempts to achieve a balance between showing the full range in flux and illustrating the finer details of the continuum.  Our integrated spectrum is accompanied by a Digitized Sky Survey image which illustrates our rectangular spectroscopic aperture as a \emph{solid} outline and the $25$~mag~arcsec$^{-2}$ isophotal size of the galaxy as a \emph{dashed} ellipse.  The image legend gives the galaxy name, the unique identification number in parenthesis, and the morphological type as listed in Table~\ref{table:general_properties}.  The horizontal solid line in the lower-left corner of each image represents $30\arcsec$.}
\figsetgrpend
 
\figsetgrpstart
\figsetgrpnum{8.9}
\figsetgrptitle{Visualization of NGC 0178.}
\figsetplot{\includegraphics[scale=0.7,angle=90]{f8_9.eps}}
\figsetgrpnote{Presentation of our integrated spectral atlas.  We plot the spectrum as $f_{\lambda}(\lambda)$ versus rest wavelength, normalized in a way that attempts to achieve a balance between showing the full range in flux and illustrating the finer details of the continuum.  Our integrated spectrum is accompanied by a Digitized Sky Survey image which illustrates our rectangular spectroscopic aperture as a \emph{solid} outline and the $25$~mag~arcsec$^{-2}$ isophotal size of the galaxy as a \emph{dashed} ellipse.  The image legend gives the galaxy name, the unique identification number in parenthesis, and the morphological type as listed in Table~\ref{table:general_properties}.  The horizontal solid line in the lower-left corner of each image represents $30\arcsec$.}
\figsetgrpend
 
\figsetgrpstart
\figsetgrpnum{8.10}
\figsetgrptitle{Visualization of NGC 0232.}
\figsetplot{\includegraphics[scale=0.7,angle=90]{f8_10.eps}}
\figsetgrpnote{Presentation of our integrated spectral atlas.  We plot the spectrum as $f_{\lambda}(\lambda)$ versus rest wavelength, normalized in a way that attempts to achieve a balance between showing the full range in flux and illustrating the finer details of the continuum.  Our integrated spectrum is accompanied by a Digitized Sky Survey image which illustrates our rectangular spectroscopic aperture as a \emph{solid} outline and the $25$~mag~arcsec$^{-2}$ isophotal size of the galaxy as a \emph{dashed} ellipse.  The image legend gives the galaxy name, the unique identification number in parenthesis, and the morphological type as listed in Table~\ref{table:general_properties}.  The horizontal solid line in the lower-left corner of each image represents $30\arcsec$.}
\figsetgrpend
 
\figsetgrpstart
\figsetgrpnum{8.11}
\figsetgrptitle{Visualization of NGC 0244.}
\figsetplot{\includegraphics[scale=0.7,angle=90]{f8_11.eps}}
\figsetgrpnote{Presentation of our integrated spectral atlas.  We plot the spectrum as $f_{\lambda}(\lambda)$ versus rest wavelength, normalized in a way that attempts to achieve a balance between showing the full range in flux and illustrating the finer details of the continuum.  Our integrated spectrum is accompanied by a Digitized Sky Survey image which illustrates our rectangular spectroscopic aperture as a \emph{solid} outline and the $25$~mag~arcsec$^{-2}$ isophotal size of the galaxy as a \emph{dashed} ellipse.  The image legend gives the galaxy name, the unique identification number in parenthesis, and the morphological type as listed in Table~\ref{table:general_properties}.  The horizontal solid line in the lower-left corner of each image represents $30\arcsec$.}
\figsetgrpend
 
\figsetgrpstart
\figsetgrpnum{8.12}
\figsetgrptitle{Visualization of NGC 0245.}
\figsetplot{\includegraphics[scale=0.7,angle=90]{f8_12.eps}}
\figsetgrpnote{Presentation of our integrated spectral atlas.  We plot the spectrum as $f_{\lambda}(\lambda)$ versus rest wavelength, normalized in a way that attempts to achieve a balance between showing the full range in flux and illustrating the finer details of the continuum.  Our integrated spectrum is accompanied by a Digitized Sky Survey image which illustrates our rectangular spectroscopic aperture as a \emph{solid} outline and the $25$~mag~arcsec$^{-2}$ isophotal size of the galaxy as a \emph{dashed} ellipse.  The image legend gives the galaxy name, the unique identification number in parenthesis, and the morphological type as listed in Table~\ref{table:general_properties}.  The horizontal solid line in the lower-left corner of each image represents $30\arcsec$.}
\figsetgrpend
 
\figsetgrpstart
\figsetgrpnum{8.13}
\figsetgrptitle{Visualization of IC 0051.}
\figsetplot{\includegraphics[scale=0.7,angle=90]{f8_13.eps}}
\figsetgrpnote{Presentation of our integrated spectral atlas.  We plot the spectrum as $f_{\lambda}(\lambda)$ versus rest wavelength, normalized in a way that attempts to achieve a balance between showing the full range in flux and illustrating the finer details of the continuum.  Our integrated spectrum is accompanied by a Digitized Sky Survey image which illustrates our rectangular spectroscopic aperture as a \emph{solid} outline and the $25$~mag~arcsec$^{-2}$ isophotal size of the galaxy as a \emph{dashed} ellipse.  The image legend gives the galaxy name, the unique identification number in parenthesis, and the morphological type as listed in Table~\ref{table:general_properties}.  The horizontal solid line in the lower-left corner of each image represents $30\arcsec$.}
\figsetgrpend
 
\figsetgrpstart
\figsetgrpnum{8.14}
\figsetgrptitle{Visualization of IC 1586.}
\figsetplot{\includegraphics[scale=0.7,angle=90]{f8_14.eps}}
\figsetgrpnote{Presentation of our integrated spectral atlas.  We plot the spectrum as $f_{\lambda}(\lambda)$ versus rest wavelength, normalized in a way that attempts to achieve a balance between showing the full range in flux and illustrating the finer details of the continuum.  Our integrated spectrum is accompanied by a Digitized Sky Survey image which illustrates our rectangular spectroscopic aperture as a \emph{solid} outline and the $25$~mag~arcsec$^{-2}$ isophotal size of the galaxy as a \emph{dashed} ellipse.  The image legend gives the galaxy name, the unique identification number in parenthesis, and the morphological type as listed in Table~\ref{table:general_properties}.  The horizontal solid line in the lower-left corner of each image represents $30\arcsec$.}
\figsetgrpend
 
\figsetgrpstart
\figsetgrpnum{8.15}
\figsetgrptitle{Visualization of MRK 0960.}
\figsetplot{\includegraphics[scale=0.7,angle=90]{f8_15.eps}}
\figsetgrpnote{Presentation of our integrated spectral atlas.  We plot the spectrum as $f_{\lambda}(\lambda)$ versus rest wavelength, normalized in a way that attempts to achieve a balance between showing the full range in flux and illustrating the finer details of the continuum.  Our integrated spectrum is accompanied by a Digitized Sky Survey image which illustrates our rectangular spectroscopic aperture as a \emph{solid} outline and the $25$~mag~arcsec$^{-2}$ isophotal size of the galaxy as a \emph{dashed} ellipse.  The image legend gives the galaxy name, the unique identification number in parenthesis, and the morphological type as listed in Table~\ref{table:general_properties}.  The horizontal solid line in the lower-left corner of each image represents $30\arcsec$.}
\figsetgrpend
 
\figsetgrpstart
\figsetgrpnum{8.16}
\figsetgrptitle{Visualization of NGC 0278.}
\figsetplot{\includegraphics[scale=0.7,angle=90]{f8_16.eps}}
\figsetgrpnote{Presentation of our integrated spectral atlas.  We plot the spectrum as $f_{\lambda}(\lambda)$ versus rest wavelength, normalized in a way that attempts to achieve a balance between showing the full range in flux and illustrating the finer details of the continuum.  Our integrated spectrum is accompanied by a Digitized Sky Survey image which illustrates our rectangular spectroscopic aperture as a \emph{solid} outline and the $25$~mag~arcsec$^{-2}$ isophotal size of the galaxy as a \emph{dashed} ellipse.  The image legend gives the galaxy name, the unique identification number in parenthesis, and the morphological type as listed in Table~\ref{table:general_properties}.  The horizontal solid line in the lower-left corner of each image represents $30\arcsec$.}
\figsetgrpend
 
\figsetgrpstart
\figsetgrpnum{8.17}
\figsetgrptitle{Visualization of NGC 0309.}
\figsetplot{\includegraphics[scale=0.7,angle=90]{f8_17.eps}}
\figsetgrpnote{Presentation of our integrated spectral atlas.  We plot the spectrum as $f_{\lambda}(\lambda)$ versus rest wavelength, normalized in a way that attempts to achieve a balance between showing the full range in flux and illustrating the finer details of the continuum.  Our integrated spectrum is accompanied by a Digitized Sky Survey image which illustrates our rectangular spectroscopic aperture as a \emph{solid} outline and the $25$~mag~arcsec$^{-2}$ isophotal size of the galaxy as a \emph{dashed} ellipse.  The image legend gives the galaxy name, the unique identification number in parenthesis, and the morphological type as listed in Table~\ref{table:general_properties}.  The horizontal solid line in the lower-left corner of each image represents $30\arcsec$.}
\figsetgrpend
 
\figsetgrpstart
\figsetgrpnum{8.18}
\figsetgrptitle{Visualization of UGC 00591.}
\figsetplot{\includegraphics[scale=0.7,angle=90]{f8_18.eps}}
\figsetgrpnote{Presentation of our integrated spectral atlas.  We plot the spectrum as $f_{\lambda}(\lambda)$ versus rest wavelength, normalized in a way that attempts to achieve a balance between showing the full range in flux and illustrating the finer details of the continuum.  Our integrated spectrum is accompanied by a Digitized Sky Survey image which illustrates our rectangular spectroscopic aperture as a \emph{solid} outline and the $25$~mag~arcsec$^{-2}$ isophotal size of the galaxy as a \emph{dashed} ellipse.  The image legend gives the galaxy name, the unique identification number in parenthesis, and the morphological type as listed in Table~\ref{table:general_properties}.  The horizontal solid line in the lower-left corner of each image represents $30\arcsec$.}
\figsetgrpend
 
\figsetgrpstart
\figsetgrpnum{8.19}
\figsetgrptitle{Visualization of NGC 0337.}
\figsetplot{\includegraphics[scale=0.7,angle=90]{f8_19.eps}}
\figsetgrpnote{Presentation of our integrated spectral atlas.  We plot the spectrum as $f_{\lambda}(\lambda)$ versus rest wavelength, normalized in a way that attempts to achieve a balance between showing the full range in flux and illustrating the finer details of the continuum.  Our integrated spectrum is accompanied by a Digitized Sky Survey image which illustrates our rectangular spectroscopic aperture as a \emph{solid} outline and the $25$~mag~arcsec$^{-2}$ isophotal size of the galaxy as a \emph{dashed} ellipse.  The image legend gives the galaxy name, the unique identification number in parenthesis, and the morphological type as listed in Table~\ref{table:general_properties}.  The horizontal solid line in the lower-left corner of each image represents $30\arcsec$.}
\figsetgrpend
 
\figsetgrpstart
\figsetgrpnum{8.20}
\figsetgrptitle{Visualization of UGC 00685.}
\figsetplot{\includegraphics[scale=0.7,angle=90]{f8_20.eps}}
\figsetgrpnote{Presentation of our integrated spectral atlas.  We plot the spectrum as $f_{\lambda}(\lambda)$ versus rest wavelength, normalized in a way that attempts to achieve a balance between showing the full range in flux and illustrating the finer details of the continuum.  Our integrated spectrum is accompanied by a Digitized Sky Survey image which illustrates our rectangular spectroscopic aperture as a \emph{solid} outline and the $25$~mag~arcsec$^{-2}$ isophotal size of the galaxy as a \emph{dashed} ellipse.  The image legend gives the galaxy name, the unique identification number in parenthesis, and the morphological type as listed in Table~\ref{table:general_properties}.  The horizontal solid line in the lower-left corner of each image represents $30\arcsec$.}
\figsetgrpend
 
\figsetgrpstart
\figsetgrpnum{8.21}
\figsetgrptitle{Visualization of IC 1623 A.}
\figsetplot{\includegraphics[scale=0.7,angle=90]{f8_21.eps}}
\figsetgrpnote{Presentation of our integrated spectral atlas.  We plot the spectrum as $f_{\lambda}(\lambda)$ versus rest wavelength, normalized in a way that attempts to achieve a balance between showing the full range in flux and illustrating the finer details of the continuum.  Our integrated spectrum is accompanied by a Digitized Sky Survey image which illustrates our rectangular spectroscopic aperture as a \emph{solid} outline and the $25$~mag~arcsec$^{-2}$ isophotal size of the galaxy as a \emph{dashed} ellipse.  The image legend gives the galaxy name, the unique identification number in parenthesis, and the morphological type as listed in Table~\ref{table:general_properties}.  The horizontal solid line in the lower-left corner of each image represents $30\arcsec$.}
\figsetgrpend
 
\figsetgrpstart
\figsetgrpnum{8.22}
\figsetgrptitle{Visualization of IC 1623.}
\figsetplot{\includegraphics[scale=0.7,angle=90]{f8_22.eps}}
\figsetgrpnote{Presentation of our integrated spectral atlas.  We plot the spectrum as $f_{\lambda}(\lambda)$ versus rest wavelength, normalized in a way that attempts to achieve a balance between showing the full range in flux and illustrating the finer details of the continuum.  Our integrated spectrum is accompanied by a Digitized Sky Survey image which illustrates our rectangular spectroscopic aperture as a \emph{solid} outline and the $25$~mag~arcsec$^{-2}$ isophotal size of the galaxy as a \emph{dashed} ellipse.  The image legend gives the galaxy name, the unique identification number in parenthesis, and the morphological type as listed in Table~\ref{table:general_properties}.  The horizontal solid line in the lower-left corner of each image represents $30\arcsec$.}
\figsetgrpend
 
\figsetgrpstart
\figsetgrpnum{8.23}
\figsetgrptitle{Visualization of IC 1623 B.}
\figsetplot{\includegraphics[scale=0.7,angle=90]{f8_23.eps}}
\figsetgrpnote{Presentation of our integrated spectral atlas.  We plot the spectrum as $f_{\lambda}(\lambda)$ versus rest wavelength, normalized in a way that attempts to achieve a balance between showing the full range in flux and illustrating the finer details of the continuum.  Our integrated spectrum is accompanied by a Digitized Sky Survey image which illustrates our rectangular spectroscopic aperture as a \emph{solid} outline and the $25$~mag~arcsec$^{-2}$ isophotal size of the galaxy as a \emph{dashed} ellipse.  The image legend gives the galaxy name, the unique identification number in parenthesis, and the morphological type as listed in Table~\ref{table:general_properties}.  The horizontal solid line in the lower-left corner of each image represents $30\arcsec$.}
\figsetgrpend
 
\figsetgrpstart
\figsetgrpnum{8.24}
\figsetgrptitle{Visualization of MCG -03-04-014.}
\figsetplot{\includegraphics[scale=0.7,angle=90]{f8_24.eps}}
\figsetgrpnote{Presentation of our integrated spectral atlas.  We plot the spectrum as $f_{\lambda}(\lambda)$ versus rest wavelength, normalized in a way that attempts to achieve a balance between showing the full range in flux and illustrating the finer details of the continuum.  Our integrated spectrum is accompanied by a Digitized Sky Survey image which illustrates our rectangular spectroscopic aperture as a \emph{solid} outline and the $25$~mag~arcsec$^{-2}$ isophotal size of the galaxy as a \emph{dashed} ellipse.  The image legend gives the galaxy name, the unique identification number in parenthesis, and the morphological type as listed in Table~\ref{table:general_properties}.  The horizontal solid line in the lower-left corner of each image represents $30\arcsec$.}
\figsetgrpend
 
\figsetgrpstart
\figsetgrpnum{8.25}
\figsetgrptitle{Visualization of CGCG 436-030.}
\figsetplot{\includegraphics[scale=0.7,angle=90]{f8_25.eps}}
\figsetgrpnote{Presentation of our integrated spectral atlas.  We plot the spectrum as $f_{\lambda}(\lambda)$ versus rest wavelength, normalized in a way that attempts to achieve a balance between showing the full range in flux and illustrating the finer details of the continuum.  Our integrated spectrum is accompanied by a Digitized Sky Survey image which illustrates our rectangular spectroscopic aperture as a \emph{solid} outline and the $25$~mag~arcsec$^{-2}$ isophotal size of the galaxy as a \emph{dashed} ellipse.  The image legend gives the galaxy name, the unique identification number in parenthesis, and the morphological type as listed in Table~\ref{table:general_properties}.  The horizontal solid line in the lower-left corner of each image represents $30\arcsec$.}
\figsetgrpend
 
\figsetgrpstart
\figsetgrpnum{8.26}
\figsetgrptitle{Visualization of NGC 0474.}
\figsetplot{\includegraphics[scale=0.7,angle=90]{f8_26.eps}}
\figsetgrpnote{Presentation of our integrated spectral atlas.  We plot the spectrum as $f_{\lambda}(\lambda)$ versus rest wavelength, normalized in a way that attempts to achieve a balance between showing the full range in flux and illustrating the finer details of the continuum.  Our integrated spectrum is accompanied by a Digitized Sky Survey image which illustrates our rectangular spectroscopic aperture as a \emph{solid} outline and the $25$~mag~arcsec$^{-2}$ isophotal size of the galaxy as a \emph{dashed} ellipse.  The image legend gives the galaxy name, the unique identification number in parenthesis, and the morphological type as listed in Table~\ref{table:general_properties}.  The horizontal solid line in the lower-left corner of each image represents $30\arcsec$.}
\figsetgrpend
 
\figsetgrpstart
\figsetgrpnum{8.27}
\figsetgrptitle{Visualization of UGC 00903.}
\figsetplot{\includegraphics[scale=0.7,angle=90]{f8_27.eps}}
\figsetgrpnote{Presentation of our integrated spectral atlas.  We plot the spectrum as $f_{\lambda}(\lambda)$ versus rest wavelength, normalized in a way that attempts to achieve a balance between showing the full range in flux and illustrating the finer details of the continuum.  Our integrated spectrum is accompanied by a Digitized Sky Survey image which illustrates our rectangular spectroscopic aperture as a \emph{solid} outline and the $25$~mag~arcsec$^{-2}$ isophotal size of the galaxy as a \emph{dashed} ellipse.  The image legend gives the galaxy name, the unique identification number in parenthesis, and the morphological type as listed in Table~\ref{table:general_properties}.  The horizontal solid line in the lower-left corner of each image represents $30\arcsec$.}
\figsetgrpend
 
\figsetgrpstart
\figsetgrpnum{8.28}
\figsetgrptitle{Visualization of NGC 0520.}
\figsetplot{\includegraphics[scale=0.7,angle=90]{f8_28.eps}}
\figsetgrpnote{Presentation of our integrated spectral atlas.  We plot the spectrum as $f_{\lambda}(\lambda)$ versus rest wavelength, normalized in a way that attempts to achieve a balance between showing the full range in flux and illustrating the finer details of the continuum.  Our integrated spectrum is accompanied by a Digitized Sky Survey image which illustrates our rectangular spectroscopic aperture as a \emph{solid} outline and the $25$~mag~arcsec$^{-2}$ isophotal size of the galaxy as a \emph{dashed} ellipse.  The image legend gives the galaxy name, the unique identification number in parenthesis, and the morphological type as listed in Table~\ref{table:general_properties}.  The horizontal solid line in the lower-left corner of each image represents $30\arcsec$.}
\figsetgrpend
 
\figsetgrpstart
\figsetgrpnum{8.29}
\figsetgrptitle{Visualization of NGC 0523.}
\figsetplot{\includegraphics[scale=0.7,angle=90]{f8_29.eps}}
\figsetgrpnote{Presentation of our integrated spectral atlas.  We plot the spectrum as $f_{\lambda}(\lambda)$ versus rest wavelength, normalized in a way that attempts to achieve a balance between showing the full range in flux and illustrating the finer details of the continuum.  Our integrated spectrum is accompanied by a Digitized Sky Survey image which illustrates our rectangular spectroscopic aperture as a \emph{solid} outline and the $25$~mag~arcsec$^{-2}$ isophotal size of the galaxy as a \emph{dashed} ellipse.  The image legend gives the galaxy name, the unique identification number in parenthesis, and the morphological type as listed in Table~\ref{table:general_properties}.  The horizontal solid line in the lower-left corner of each image represents $30\arcsec$.}
\figsetgrpend
 
\figsetgrpstart
\figsetgrpnum{8.30}
\figsetgrptitle{Visualization of NGC 0615.}
\figsetplot{\includegraphics[scale=0.7,angle=90]{f8_30.eps}}
\figsetgrpnote{Presentation of our integrated spectral atlas.  We plot the spectrum as $f_{\lambda}(\lambda)$ versus rest wavelength, normalized in a way that attempts to achieve a balance between showing the full range in flux and illustrating the finer details of the continuum.  Our integrated spectrum is accompanied by a Digitized Sky Survey image which illustrates our rectangular spectroscopic aperture as a \emph{solid} outline and the $25$~mag~arcsec$^{-2}$ isophotal size of the galaxy as a \emph{dashed} ellipse.  The image legend gives the galaxy name, the unique identification number in parenthesis, and the morphological type as listed in Table~\ref{table:general_properties}.  The horizontal solid line in the lower-left corner of each image represents $30\arcsec$.}
\figsetgrpend
 
\figsetgrpstart
\figsetgrpnum{8.31}
\figsetgrptitle{Visualization of NGC 0660.}
\figsetplot{\includegraphics[scale=0.7,angle=90]{f8_31.eps}}
\figsetgrpnote{Presentation of our integrated spectral atlas.  We plot the spectrum as $f_{\lambda}(\lambda)$ versus rest wavelength, normalized in a way that attempts to achieve a balance between showing the full range in flux and illustrating the finer details of the continuum.  Our integrated spectrum is accompanied by a Digitized Sky Survey image which illustrates our rectangular spectroscopic aperture as a \emph{solid} outline and the $25$~mag~arcsec$^{-2}$ isophotal size of the galaxy as a \emph{dashed} ellipse.  The image legend gives the galaxy name, the unique identification number in parenthesis, and the morphological type as listed in Table~\ref{table:general_properties}.  The horizontal solid line in the lower-left corner of each image represents $30\arcsec$.}
\figsetgrpend
 
\figsetgrpstart
\figsetgrpnum{8.32}
\figsetgrptitle{Visualization of MRK 0360.}
\figsetplot{\includegraphics[scale=0.7,angle=90]{f8_32.eps}}
\figsetgrpnote{Presentation of our integrated spectral atlas.  We plot the spectrum as $f_{\lambda}(\lambda)$ versus rest wavelength, normalized in a way that attempts to achieve a balance between showing the full range in flux and illustrating the finer details of the continuum.  Our integrated spectrum is accompanied by a Digitized Sky Survey image which illustrates our rectangular spectroscopic aperture as a \emph{solid} outline and the $25$~mag~arcsec$^{-2}$ isophotal size of the galaxy as a \emph{dashed} ellipse.  The image legend gives the galaxy name, the unique identification number in parenthesis, and the morphological type as listed in Table~\ref{table:general_properties}.  The horizontal solid line in the lower-left corner of each image represents $30\arcsec$.}
\figsetgrpend
 
\figsetgrpstart
\figsetgrpnum{8.33}
\figsetgrptitle{Visualization of III Zw 035.}
\figsetplot{\includegraphics[scale=0.7,angle=90]{f8_33.eps}}
\figsetgrpnote{Presentation of our integrated spectral atlas.  We plot the spectrum as $f_{\lambda}(\lambda)$ versus rest wavelength, normalized in a way that attempts to achieve a balance between showing the full range in flux and illustrating the finer details of the continuum.  Our integrated spectrum is accompanied by a Digitized Sky Survey image which illustrates our rectangular spectroscopic aperture as a \emph{solid} outline and the $25$~mag~arcsec$^{-2}$ isophotal size of the galaxy as a \emph{dashed} ellipse.  The image legend gives the galaxy name, the unique identification number in parenthesis, and the morphological type as listed in Table~\ref{table:general_properties}.  The horizontal solid line in the lower-left corner of each image represents $30\arcsec$.}
\figsetgrpend
 
\figsetgrpstart
\figsetgrpnum{8.34}
\figsetgrptitle{Visualization of IC 1727.}
\figsetplot{\includegraphics[scale=0.7,angle=90]{f8_34.eps}}
\figsetgrpnote{Presentation of our integrated spectral atlas.  We plot the spectrum as $f_{\lambda}(\lambda)$ versus rest wavelength, normalized in a way that attempts to achieve a balance between showing the full range in flux and illustrating the finer details of the continuum.  Our integrated spectrum is accompanied by a Digitized Sky Survey image which illustrates our rectangular spectroscopic aperture as a \emph{solid} outline and the $25$~mag~arcsec$^{-2}$ isophotal size of the galaxy as a \emph{dashed} ellipse.  The image legend gives the galaxy name, the unique identification number in parenthesis, and the morphological type as listed in Table~\ref{table:general_properties}.  The horizontal solid line in the lower-left corner of each image represents $30\arcsec$.}
\figsetgrpend
 
\figsetgrpstart
\figsetgrpnum{8.35}
\figsetgrptitle{Visualization of IC 0162.}
\figsetplot{\includegraphics[scale=0.7,angle=90]{f8_35.eps}}
\figsetgrpnote{Presentation of our integrated spectral atlas.  We plot the spectrum as $f_{\lambda}(\lambda)$ versus rest wavelength, normalized in a way that attempts to achieve a balance between showing the full range in flux and illustrating the finer details of the continuum.  Our integrated spectrum is accompanied by a Digitized Sky Survey image which illustrates our rectangular spectroscopic aperture as a \emph{solid} outline and the $25$~mag~arcsec$^{-2}$ isophotal size of the galaxy as a \emph{dashed} ellipse.  The image legend gives the galaxy name, the unique identification number in parenthesis, and the morphological type as listed in Table~\ref{table:general_properties}.  The horizontal solid line in the lower-left corner of each image represents $30\arcsec$.}
\figsetgrpend
 
\figsetgrpstart
\figsetgrpnum{8.36}
\figsetgrptitle{Visualization of UGC 01281.}
\figsetplot{\includegraphics[scale=0.7,angle=90]{f8_36.eps}}
\figsetgrpnote{Presentation of our integrated spectral atlas.  We plot the spectrum as $f_{\lambda}(\lambda)$ versus rest wavelength, normalized in a way that attempts to achieve a balance between showing the full range in flux and illustrating the finer details of the continuum.  Our integrated spectrum is accompanied by a Digitized Sky Survey image which illustrates our rectangular spectroscopic aperture as a \emph{solid} outline and the $25$~mag~arcsec$^{-2}$ isophotal size of the galaxy as a \emph{dashed} ellipse.  The image legend gives the galaxy name, the unique identification number in parenthesis, and the morphological type as listed in Table~\ref{table:general_properties}.  The horizontal solid line in the lower-left corner of each image represents $30\arcsec$.}
\figsetgrpend
 
\figsetgrpstart
\figsetgrpnum{8.37}
\figsetgrptitle{Visualization of NGC 0694.}
\figsetplot{\includegraphics[scale=0.7,angle=90]{f8_37.eps}}
\figsetgrpnote{Presentation of our integrated spectral atlas.  We plot the spectrum as $f_{\lambda}(\lambda)$ versus rest wavelength, normalized in a way that attempts to achieve a balance between showing the full range in flux and illustrating the finer details of the continuum.  Our integrated spectrum is accompanied by a Digitized Sky Survey image which illustrates our rectangular spectroscopic aperture as a \emph{solid} outline and the $25$~mag~arcsec$^{-2}$ isophotal size of the galaxy as a \emph{dashed} ellipse.  The image legend gives the galaxy name, the unique identification number in parenthesis, and the morphological type as listed in Table~\ref{table:general_properties}.  The horizontal solid line in the lower-left corner of each image represents $30\arcsec$.}
\figsetgrpend
 
\figsetgrpstart
\figsetgrpnum{8.38}
\figsetgrptitle{Visualization of NGC 0695.}
\figsetplot{\includegraphics[scale=0.7,angle=90]{f8_38.eps}}
\figsetgrpnote{Presentation of our integrated spectral atlas.  We plot the spectrum as $f_{\lambda}(\lambda)$ versus rest wavelength, normalized in a way that attempts to achieve a balance between showing the full range in flux and illustrating the finer details of the continuum.  Our integrated spectrum is accompanied by a Digitized Sky Survey image which illustrates our rectangular spectroscopic aperture as a \emph{solid} outline and the $25$~mag~arcsec$^{-2}$ isophotal size of the galaxy as a \emph{dashed} ellipse.  The image legend gives the galaxy name, the unique identification number in parenthesis, and the morphological type as listed in Table~\ref{table:general_properties}.  The horizontal solid line in the lower-left corner of each image represents $30\arcsec$.}
\figsetgrpend
 
\figsetgrpstart
\figsetgrpnum{8.39}
\figsetgrptitle{Visualization of UGC 01385.}
\figsetplot{\includegraphics[scale=0.7,angle=90]{f8_39.eps}}
\figsetgrpnote{Presentation of our integrated spectral atlas.  We plot the spectrum as $f_{\lambda}(\lambda)$ versus rest wavelength, normalized in a way that attempts to achieve a balance between showing the full range in flux and illustrating the finer details of the continuum.  Our integrated spectrum is accompanied by a Digitized Sky Survey image which illustrates our rectangular spectroscopic aperture as a \emph{solid} outline and the $25$~mag~arcsec$^{-2}$ isophotal size of the galaxy as a \emph{dashed} ellipse.  The image legend gives the galaxy name, the unique identification number in parenthesis, and the morphological type as listed in Table~\ref{table:general_properties}.  The horizontal solid line in the lower-left corner of each image represents $30\arcsec$.}
\figsetgrpend
 
\figsetgrpstart
\figsetgrpnum{8.40}
\figsetgrptitle{Visualization of NGC 0750.}
\figsetplot{\includegraphics[scale=0.7,angle=90]{f8_40.eps}}
\figsetgrpnote{Presentation of our integrated spectral atlas.  We plot the spectrum as $f_{\lambda}(\lambda)$ versus rest wavelength, normalized in a way that attempts to achieve a balance between showing the full range in flux and illustrating the finer details of the continuum.  Our integrated spectrum is accompanied by a Digitized Sky Survey image which illustrates our rectangular spectroscopic aperture as a \emph{solid} outline and the $25$~mag~arcsec$^{-2}$ isophotal size of the galaxy as a \emph{dashed} ellipse.  The image legend gives the galaxy name, the unique identification number in parenthesis, and the morphological type as listed in Table~\ref{table:general_properties}.  The horizontal solid line in the lower-left corner of each image represents $30\arcsec$.}
\figsetgrpend
 
\figsetgrpstart
\figsetgrpnum{8.41}
\figsetgrptitle{Visualization of NGC 0784.}
\figsetplot{\includegraphics[scale=0.7,angle=90]{f8_41.eps}}
\figsetgrpnote{Presentation of our integrated spectral atlas.  We plot the spectrum as $f_{\lambda}(\lambda)$ versus rest wavelength, normalized in a way that attempts to achieve a balance between showing the full range in flux and illustrating the finer details of the continuum.  Our integrated spectrum is accompanied by a Digitized Sky Survey image which illustrates our rectangular spectroscopic aperture as a \emph{solid} outline and the $25$~mag~arcsec$^{-2}$ isophotal size of the galaxy as a \emph{dashed} ellipse.  The image legend gives the galaxy name, the unique identification number in parenthesis, and the morphological type as listed in Table~\ref{table:general_properties}.  The horizontal solid line in the lower-left corner of each image represents $30\arcsec$.}
\figsetgrpend
 
\figsetgrpstart
\figsetgrpnum{8.42}
\figsetgrptitle{Visualization of UGC 01561.}
\figsetplot{\includegraphics[scale=0.7,angle=90]{f8_42.eps}}
\figsetgrpnote{Presentation of our integrated spectral atlas.  We plot the spectrum as $f_{\lambda}(\lambda)$ versus rest wavelength, normalized in a way that attempts to achieve a balance between showing the full range in flux and illustrating the finer details of the continuum.  Our integrated spectrum is accompanied by a Digitized Sky Survey image which illustrates our rectangular spectroscopic aperture as a \emph{solid} outline and the $25$~mag~arcsec$^{-2}$ isophotal size of the galaxy as a \emph{dashed} ellipse.  The image legend gives the galaxy name, the unique identification number in parenthesis, and the morphological type as listed in Table~\ref{table:general_properties}.  The horizontal solid line in the lower-left corner of each image represents $30\arcsec$.}
\figsetgrpend
 
\figsetgrpstart
\figsetgrpnum{8.43}
\figsetgrptitle{Visualization of NGC 0877.}
\figsetplot{\includegraphics[scale=0.7,angle=90]{f8_43.eps}}
\figsetgrpnote{Presentation of our integrated spectral atlas.  We plot the spectrum as $f_{\lambda}(\lambda)$ versus rest wavelength, normalized in a way that attempts to achieve a balance between showing the full range in flux and illustrating the finer details of the continuum.  Our integrated spectrum is accompanied by a Digitized Sky Survey image which illustrates our rectangular spectroscopic aperture as a \emph{solid} outline and the $25$~mag~arcsec$^{-2}$ isophotal size of the galaxy as a \emph{dashed} ellipse.  The image legend gives the galaxy name, the unique identification number in parenthesis, and the morphological type as listed in Table~\ref{table:general_properties}.  The horizontal solid line in the lower-left corner of each image represents $30\arcsec$.}
\figsetgrpend
 
\figsetgrpstart
\figsetgrpnum{8.44}
\figsetgrptitle{Visualization of NGC 0922.}
\figsetplot{\includegraphics[scale=0.7,angle=90]{f8_44.eps}}
\figsetgrpnote{Presentation of our integrated spectral atlas.  We plot the spectrum as $f_{\lambda}(\lambda)$ versus rest wavelength, normalized in a way that attempts to achieve a balance between showing the full range in flux and illustrating the finer details of the continuum.  Our integrated spectrum is accompanied by a Digitized Sky Survey image which illustrates our rectangular spectroscopic aperture as a \emph{solid} outline and the $25$~mag~arcsec$^{-2}$ isophotal size of the galaxy as a \emph{dashed} ellipse.  The image legend gives the galaxy name, the unique identification number in parenthesis, and the morphological type as listed in Table~\ref{table:general_properties}.  The horizontal solid line in the lower-left corner of each image represents $30\arcsec$.}
\figsetgrpend
 
\figsetgrpstart
\figsetgrpnum{8.45}
\figsetgrptitle{Visualization of NGC 0943.}
\figsetplot{\includegraphics[scale=0.7,angle=90]{f8_45.eps}}
\figsetgrpnote{Presentation of our integrated spectral atlas.  We plot the spectrum as $f_{\lambda}(\lambda)$ versus rest wavelength, normalized in a way that attempts to achieve a balance between showing the full range in flux and illustrating the finer details of the continuum.  Our integrated spectrum is accompanied by a Digitized Sky Survey image which illustrates our rectangular spectroscopic aperture as a \emph{solid} outline and the $25$~mag~arcsec$^{-2}$ isophotal size of the galaxy as a \emph{dashed} ellipse.  The image legend gives the galaxy name, the unique identification number in parenthesis, and the morphological type as listed in Table~\ref{table:general_properties}.  The horizontal solid line in the lower-left corner of each image represents $30\arcsec$.}
\figsetgrpend
 
\figsetgrpstart
\figsetgrpnum{8.46}
\figsetgrptitle{Visualization of ARP 309.}
\figsetplot{\includegraphics[scale=0.7,angle=90]{f8_46.eps}}
\figsetgrpnote{Presentation of our integrated spectral atlas.  We plot the spectrum as $f_{\lambda}(\lambda)$ versus rest wavelength, normalized in a way that attempts to achieve a balance between showing the full range in flux and illustrating the finer details of the continuum.  Our integrated spectrum is accompanied by a Digitized Sky Survey image which illustrates our rectangular spectroscopic aperture as a \emph{solid} outline and the $25$~mag~arcsec$^{-2}$ isophotal size of the galaxy as a \emph{dashed} ellipse.  The image legend gives the galaxy name, the unique identification number in parenthesis, and the morphological type as listed in Table~\ref{table:general_properties}.  The horizontal solid line in the lower-left corner of each image represents $30\arcsec$.}
\figsetgrpend
 
\figsetgrpstart
\figsetgrpnum{8.47}
\figsetgrptitle{Visualization of NGC 0942.}
\figsetplot{\includegraphics[scale=0.7,angle=90]{f8_47.eps}}
\figsetgrpnote{Presentation of our integrated spectral atlas.  We plot the spectrum as $f_{\lambda}(\lambda)$ versus rest wavelength, normalized in a way that attempts to achieve a balance between showing the full range in flux and illustrating the finer details of the continuum.  Our integrated spectrum is accompanied by a Digitized Sky Survey image which illustrates our rectangular spectroscopic aperture as a \emph{solid} outline and the $25$~mag~arcsec$^{-2}$ isophotal size of the galaxy as a \emph{dashed} ellipse.  The image legend gives the galaxy name, the unique identification number in parenthesis, and the morphological type as listed in Table~\ref{table:general_properties}.  The horizontal solid line in the lower-left corner of each image represents $30\arcsec$.}
\figsetgrpend
 
\figsetgrpstart
\figsetgrpnum{8.48}
\figsetgrptitle{Visualization of NGC 0958.}
\figsetplot{\includegraphics[scale=0.7,angle=90]{f8_48.eps}}
\figsetgrpnote{Presentation of our integrated spectral atlas.  We plot the spectrum as $f_{\lambda}(\lambda)$ versus rest wavelength, normalized in a way that attempts to achieve a balance between showing the full range in flux and illustrating the finer details of the continuum.  Our integrated spectrum is accompanied by a Digitized Sky Survey image which illustrates our rectangular spectroscopic aperture as a \emph{solid} outline and the $25$~mag~arcsec$^{-2}$ isophotal size of the galaxy as a \emph{dashed} ellipse.  The image legend gives the galaxy name, the unique identification number in parenthesis, and the morphological type as listed in Table~\ref{table:general_properties}.  The horizontal solid line in the lower-left corner of each image represents $30\arcsec$.}
\figsetgrpend
 
\figsetgrpstart
\figsetgrpnum{8.49}
\figsetgrptitle{Visualization of NGC 0959.}
\figsetplot{\includegraphics[scale=0.7,angle=90]{f8_49.eps}}
\figsetgrpnote{Presentation of our integrated spectral atlas.  We plot the spectrum as $f_{\lambda}(\lambda)$ versus rest wavelength, normalized in a way that attempts to achieve a balance between showing the full range in flux and illustrating the finer details of the continuum.  Our integrated spectrum is accompanied by a Digitized Sky Survey image which illustrates our rectangular spectroscopic aperture as a \emph{solid} outline and the $25$~mag~arcsec$^{-2}$ isophotal size of the galaxy as a \emph{dashed} ellipse.  The image legend gives the galaxy name, the unique identification number in parenthesis, and the morphological type as listed in Table~\ref{table:general_properties}.  The horizontal solid line in the lower-left corner of each image represents $30\arcsec$.}
\figsetgrpend
 
\figsetgrpstart
\figsetgrpnum{8.50}
\figsetgrptitle{Visualization of NGC 0976.}
\figsetplot{\includegraphics[scale=0.7,angle=90]{f8_50.eps}}
\figsetgrpnote{Presentation of our integrated spectral atlas.  We plot the spectrum as $f_{\lambda}(\lambda)$ versus rest wavelength, normalized in a way that attempts to achieve a balance between showing the full range in flux and illustrating the finer details of the continuum.  Our integrated spectrum is accompanied by a Digitized Sky Survey image which illustrates our rectangular spectroscopic aperture as a \emph{solid} outline and the $25$~mag~arcsec$^{-2}$ isophotal size of the galaxy as a \emph{dashed} ellipse.  The image legend gives the galaxy name, the unique identification number in parenthesis, and the morphological type as listed in Table~\ref{table:general_properties}.  The horizontal solid line in the lower-left corner of each image represents $30\arcsec$.}
\figsetgrpend
 
\figsetgrpstart
\figsetgrpnum{8.51}
\figsetgrptitle{Visualization of NGC 0972.}
\figsetplot{\includegraphics[scale=0.7,angle=90]{f8_51.eps}}
\figsetgrpnote{Presentation of our integrated spectral atlas.  We plot the spectrum as $f_{\lambda}(\lambda)$ versus rest wavelength, normalized in a way that attempts to achieve a balance between showing the full range in flux and illustrating the finer details of the continuum.  Our integrated spectrum is accompanied by a Digitized Sky Survey image which illustrates our rectangular spectroscopic aperture as a \emph{solid} outline and the $25$~mag~arcsec$^{-2}$ isophotal size of the galaxy as a \emph{dashed} ellipse.  The image legend gives the galaxy name, the unique identification number in parenthesis, and the morphological type as listed in Table~\ref{table:general_properties}.  The horizontal solid line in the lower-left corner of each image represents $30\arcsec$.}
\figsetgrpend
 
\figsetgrpstart
\figsetgrpnum{8.52}
\figsetgrptitle{Visualization of NGC 1003.}
\figsetplot{\includegraphics[scale=0.7,angle=90]{f8_52.eps}}
\figsetgrpnote{Presentation of our integrated spectral atlas.  We plot the spectrum as $f_{\lambda}(\lambda)$ versus rest wavelength, normalized in a way that attempts to achieve a balance between showing the full range in flux and illustrating the finer details of the continuum.  Our integrated spectrum is accompanied by a Digitized Sky Survey image which illustrates our rectangular spectroscopic aperture as a \emph{solid} outline and the $25$~mag~arcsec$^{-2}$ isophotal size of the galaxy as a \emph{dashed} ellipse.  The image legend gives the galaxy name, the unique identification number in parenthesis, and the morphological type as listed in Table~\ref{table:general_properties}.  The horizontal solid line in the lower-left corner of each image represents $30\arcsec$.}
\figsetgrpend
 
\figsetgrpstart
\figsetgrpnum{8.53}
\figsetgrptitle{Visualization of NGC 1036.}
\figsetplot{\includegraphics[scale=0.7,angle=90]{f8_53.eps}}
\figsetgrpnote{Presentation of our integrated spectral atlas.  We plot the spectrum as $f_{\lambda}(\lambda)$ versus rest wavelength, normalized in a way that attempts to achieve a balance between showing the full range in flux and illustrating the finer details of the continuum.  Our integrated spectrum is accompanied by a Digitized Sky Survey image which illustrates our rectangular spectroscopic aperture as a \emph{solid} outline and the $25$~mag~arcsec$^{-2}$ isophotal size of the galaxy as a \emph{dashed} ellipse.  The image legend gives the galaxy name, the unique identification number in parenthesis, and the morphological type as listed in Table~\ref{table:general_properties}.  The horizontal solid line in the lower-left corner of each image represents $30\arcsec$.}
\figsetgrpend
 
\figsetgrpstart
\figsetgrpnum{8.54}
\figsetgrptitle{Visualization of NGC 1068.}
\figsetplot{\includegraphics[scale=0.7,angle=90]{f8_54.eps}}
\figsetgrpnote{Presentation of our integrated spectral atlas.  We plot the spectrum as $f_{\lambda}(\lambda)$ versus rest wavelength, normalized in a way that attempts to achieve a balance between showing the full range in flux and illustrating the finer details of the continuum.  Our integrated spectrum is accompanied by a Digitized Sky Survey image which illustrates our rectangular spectroscopic aperture as a \emph{solid} outline and the $25$~mag~arcsec$^{-2}$ isophotal size of the galaxy as a \emph{dashed} ellipse.  The image legend gives the galaxy name, the unique identification number in parenthesis, and the morphological type as listed in Table~\ref{table:general_properties}.  The horizontal solid line in the lower-left corner of each image represents $30\arcsec$.}
\figsetgrpend
 
\figsetgrpstart
\figsetgrpnum{8.55}
\figsetgrptitle{Visualization of NGC 1058.}
\figsetplot{\includegraphics[scale=0.7,angle=90]{f8_55.eps}}
\figsetgrpnote{Presentation of our integrated spectral atlas.  We plot the spectrum as $f_{\lambda}(\lambda)$ versus rest wavelength, normalized in a way that attempts to achieve a balance between showing the full range in flux and illustrating the finer details of the continuum.  Our integrated spectrum is accompanied by a Digitized Sky Survey image which illustrates our rectangular spectroscopic aperture as a \emph{solid} outline and the $25$~mag~arcsec$^{-2}$ isophotal size of the galaxy as a \emph{dashed} ellipse.  The image legend gives the galaxy name, the unique identification number in parenthesis, and the morphological type as listed in Table~\ref{table:general_properties}.  The horizontal solid line in the lower-left corner of each image represents $30\arcsec$.}
\figsetgrpend
 
\figsetgrpstart
\figsetgrpnum{8.56}
\figsetgrptitle{Visualization of NGC 1084.}
\figsetplot{\includegraphics[scale=0.7,angle=90]{f8_56.eps}}
\figsetgrpnote{Presentation of our integrated spectral atlas.  We plot the spectrum as $f_{\lambda}(\lambda)$ versus rest wavelength, normalized in a way that attempts to achieve a balance between showing the full range in flux and illustrating the finer details of the continuum.  Our integrated spectrum is accompanied by a Digitized Sky Survey image which illustrates our rectangular spectroscopic aperture as a \emph{solid} outline and the $25$~mag~arcsec$^{-2}$ isophotal size of the galaxy as a \emph{dashed} ellipse.  The image legend gives the galaxy name, the unique identification number in parenthesis, and the morphological type as listed in Table~\ref{table:general_properties}.  The horizontal solid line in the lower-left corner of each image represents $30\arcsec$.}
\figsetgrpend
 
\figsetgrpstart
\figsetgrpnum{8.57}
\figsetgrptitle{Visualization of UGC 02238.}
\figsetplot{\includegraphics[scale=0.7,angle=90]{f8_57.eps}}
\figsetgrpnote{Presentation of our integrated spectral atlas.  We plot the spectrum as $f_{\lambda}(\lambda)$ versus rest wavelength, normalized in a way that attempts to achieve a balance between showing the full range in flux and illustrating the finer details of the continuum.  Our integrated spectrum is accompanied by a Digitized Sky Survey image which illustrates our rectangular spectroscopic aperture as a \emph{solid} outline and the $25$~mag~arcsec$^{-2}$ isophotal size of the galaxy as a \emph{dashed} ellipse.  The image legend gives the galaxy name, the unique identification number in parenthesis, and the morphological type as listed in Table~\ref{table:general_properties}.  The horizontal solid line in the lower-left corner of each image represents $30\arcsec$.}
\figsetgrpend
 
\figsetgrpstart
\figsetgrpnum{8.58}
\figsetgrptitle{Visualization of NGC 1087.}
\figsetplot{\includegraphics[scale=0.7,angle=90]{f8_58.eps}}
\figsetgrpnote{Presentation of our integrated spectral atlas.  We plot the spectrum as $f_{\lambda}(\lambda)$ versus rest wavelength, normalized in a way that attempts to achieve a balance between showing the full range in flux and illustrating the finer details of the continuum.  Our integrated spectrum is accompanied by a Digitized Sky Survey image which illustrates our rectangular spectroscopic aperture as a \emph{solid} outline and the $25$~mag~arcsec$^{-2}$ isophotal size of the galaxy as a \emph{dashed} ellipse.  The image legend gives the galaxy name, the unique identification number in parenthesis, and the morphological type as listed in Table~\ref{table:general_properties}.  The horizontal solid line in the lower-left corner of each image represents $30\arcsec$.}
\figsetgrpend
 
\figsetgrpstart
\figsetgrpnum{8.59}
\figsetgrptitle{Visualization of MRK 0600.}
\figsetplot{\includegraphics[scale=0.7,angle=90]{f8_59.eps}}
\figsetgrpnote{Presentation of our integrated spectral atlas.  We plot the spectrum as $f_{\lambda}(\lambda)$ versus rest wavelength, normalized in a way that attempts to achieve a balance between showing the full range in flux and illustrating the finer details of the continuum.  Our integrated spectrum is accompanied by a Digitized Sky Survey image which illustrates our rectangular spectroscopic aperture as a \emph{solid} outline and the $25$~mag~arcsec$^{-2}$ isophotal size of the galaxy as a \emph{dashed} ellipse.  The image legend gives the galaxy name, the unique identification number in parenthesis, and the morphological type as listed in Table~\ref{table:general_properties}.  The horizontal solid line in the lower-left corner of each image represents $30\arcsec$.}
\figsetgrpend
 
\figsetgrpstart
\figsetgrpnum{8.60}
\figsetgrptitle{Visualization of NGC 1134.}
\figsetplot{\includegraphics[scale=0.7,angle=90]{f8_60.eps}}
\figsetgrpnote{Presentation of our integrated spectral atlas.  We plot the spectrum as $f_{\lambda}(\lambda)$ versus rest wavelength, normalized in a way that attempts to achieve a balance between showing the full range in flux and illustrating the finer details of the continuum.  Our integrated spectrum is accompanied by a Digitized Sky Survey image which illustrates our rectangular spectroscopic aperture as a \emph{solid} outline and the $25$~mag~arcsec$^{-2}$ isophotal size of the galaxy as a \emph{dashed} ellipse.  The image legend gives the galaxy name, the unique identification number in parenthesis, and the morphological type as listed in Table~\ref{table:general_properties}.  The horizontal solid line in the lower-left corner of each image represents $30\arcsec$.}
\figsetgrpend
 
\figsetgrpstart
\figsetgrpnum{8.61}
\figsetgrptitle{Visualization of UGC 02369 N.}
\figsetplot{\includegraphics[scale=0.7,angle=90]{f8_61.eps}}
\figsetgrpnote{Presentation of our integrated spectral atlas.  We plot the spectrum as $f_{\lambda}(\lambda)$ versus rest wavelength, normalized in a way that attempts to achieve a balance between showing the full range in flux and illustrating the finer details of the continuum.  Our integrated spectrum is accompanied by a Digitized Sky Survey image which illustrates our rectangular spectroscopic aperture as a \emph{solid} outline and the $25$~mag~arcsec$^{-2}$ isophotal size of the galaxy as a \emph{dashed} ellipse.  The image legend gives the galaxy name, the unique identification number in parenthesis, and the morphological type as listed in Table~\ref{table:general_properties}.  The horizontal solid line in the lower-left corner of each image represents $30\arcsec$.}
\figsetgrpend
 
\figsetgrpstart
\figsetgrpnum{8.62}
\figsetgrptitle{Visualization of UGC 02369.}
\figsetplot{\includegraphics[scale=0.7,angle=90]{f8_62.eps}}
\figsetgrpnote{Presentation of our integrated spectral atlas.  We plot the spectrum as $f_{\lambda}(\lambda)$ versus rest wavelength, normalized in a way that attempts to achieve a balance between showing the full range in flux and illustrating the finer details of the continuum.  Our integrated spectrum is accompanied by a Digitized Sky Survey image which illustrates our rectangular spectroscopic aperture as a \emph{solid} outline and the $25$~mag~arcsec$^{-2}$ isophotal size of the galaxy as a \emph{dashed} ellipse.  The image legend gives the galaxy name, the unique identification number in parenthesis, and the morphological type as listed in Table~\ref{table:general_properties}.  The horizontal solid line in the lower-left corner of each image represents $30\arcsec$.}
\figsetgrpend
 
\figsetgrpstart
\figsetgrpnum{8.63}
\figsetgrptitle{Visualization of UGC 02369 S.}
\figsetplot{\includegraphics[scale=0.7,angle=90]{f8_63.eps}}
\figsetgrpnote{Presentation of our integrated spectral atlas.  We plot the spectrum as $f_{\lambda}(\lambda)$ versus rest wavelength, normalized in a way that attempts to achieve a balance between showing the full range in flux and illustrating the finer details of the continuum.  Our integrated spectrum is accompanied by a Digitized Sky Survey image which illustrates our rectangular spectroscopic aperture as a \emph{solid} outline and the $25$~mag~arcsec$^{-2}$ isophotal size of the galaxy as a \emph{dashed} ellipse.  The image legend gives the galaxy name, the unique identification number in parenthesis, and the morphological type as listed in Table~\ref{table:general_properties}.  The horizontal solid line in the lower-left corner of each image represents $30\arcsec$.}
\figsetgrpend
 
\figsetgrpstart
\figsetgrpnum{8.64}
\figsetgrptitle{Visualization of NGC 1140.}
\figsetplot{\includegraphics[scale=0.7,angle=90]{f8_64.eps}}
\figsetgrpnote{Presentation of our integrated spectral atlas.  We plot the spectrum as $f_{\lambda}(\lambda)$ versus rest wavelength, normalized in a way that attempts to achieve a balance between showing the full range in flux and illustrating the finer details of the continuum.  Our integrated spectrum is accompanied by a Digitized Sky Survey image which illustrates our rectangular spectroscopic aperture as a \emph{solid} outline and the $25$~mag~arcsec$^{-2}$ isophotal size of the galaxy as a \emph{dashed} ellipse.  The image legend gives the galaxy name, the unique identification number in parenthesis, and the morphological type as listed in Table~\ref{table:general_properties}.  The horizontal solid line in the lower-left corner of each image represents $30\arcsec$.}
\figsetgrpend
 
\figsetgrpstart
\figsetgrpnum{8.65}
\figsetgrptitle{Visualization of NGC 1143.}
\figsetplot{\includegraphics[scale=0.7,angle=90]{f8_65.eps}}
\figsetgrpnote{Presentation of our integrated spectral atlas.  We plot the spectrum as $f_{\lambda}(\lambda)$ versus rest wavelength, normalized in a way that attempts to achieve a balance between showing the full range in flux and illustrating the finer details of the continuum.  Our integrated spectrum is accompanied by a Digitized Sky Survey image which illustrates our rectangular spectroscopic aperture as a \emph{solid} outline and the $25$~mag~arcsec$^{-2}$ isophotal size of the galaxy as a \emph{dashed} ellipse.  The image legend gives the galaxy name, the unique identification number in parenthesis, and the morphological type as listed in Table~\ref{table:general_properties}.  The horizontal solid line in the lower-left corner of each image represents $30\arcsec$.}
\figsetgrpend
 
\figsetgrpstart
\figsetgrpnum{8.66}
\figsetgrptitle{Visualization of ARP 118.}
\figsetplot{\includegraphics[scale=0.7,angle=90]{f8_66.eps}}
\figsetgrpnote{Presentation of our integrated spectral atlas.  We plot the spectrum as $f_{\lambda}(\lambda)$ versus rest wavelength, normalized in a way that attempts to achieve a balance between showing the full range in flux and illustrating the finer details of the continuum.  Our integrated spectrum is accompanied by a Digitized Sky Survey image which illustrates our rectangular spectroscopic aperture as a \emph{solid} outline and the $25$~mag~arcsec$^{-2}$ isophotal size of the galaxy as a \emph{dashed} ellipse.  The image legend gives the galaxy name, the unique identification number in parenthesis, and the morphological type as listed in Table~\ref{table:general_properties}.  The horizontal solid line in the lower-left corner of each image represents $30\arcsec$.}
\figsetgrpend
 
\figsetgrpstart
\figsetgrpnum{8.67}
\figsetgrptitle{Visualization of NGC 1144.}
\figsetplot{\includegraphics[scale=0.7,angle=90]{f8_67.eps}}
\figsetgrpnote{Presentation of our integrated spectral atlas.  We plot the spectrum as $f_{\lambda}(\lambda)$ versus rest wavelength, normalized in a way that attempts to achieve a balance between showing the full range in flux and illustrating the finer details of the continuum.  Our integrated spectrum is accompanied by a Digitized Sky Survey image which illustrates our rectangular spectroscopic aperture as a \emph{solid} outline and the $25$~mag~arcsec$^{-2}$ isophotal size of the galaxy as a \emph{dashed} ellipse.  The image legend gives the galaxy name, the unique identification number in parenthesis, and the morphological type as listed in Table~\ref{table:general_properties}.  The horizontal solid line in the lower-left corner of each image represents $30\arcsec$.}
\figsetgrpend
 
\figsetgrpstart
\figsetgrpnum{8.68}
\figsetgrptitle{Visualization of NGC 1156.}
\figsetplot{\includegraphics[scale=0.7,angle=90]{f8_68.eps}}
\figsetgrpnote{Presentation of our integrated spectral atlas.  We plot the spectrum as $f_{\lambda}(\lambda)$ versus rest wavelength, normalized in a way that attempts to achieve a balance between showing the full range in flux and illustrating the finer details of the continuum.  Our integrated spectrum is accompanied by a Digitized Sky Survey image which illustrates our rectangular spectroscopic aperture as a \emph{solid} outline and the $25$~mag~arcsec$^{-2}$ isophotal size of the galaxy as a \emph{dashed} ellipse.  The image legend gives the galaxy name, the unique identification number in parenthesis, and the morphological type as listed in Table~\ref{table:general_properties}.  The horizontal solid line in the lower-left corner of each image represents $30\arcsec$.}
\figsetgrpend
 
\figsetgrpstart
\figsetgrpnum{8.69}
\figsetgrptitle{Visualization of NGC 1266.}
\figsetplot{\includegraphics[scale=0.7,angle=90]{f8_69.eps}}
\figsetgrpnote{Presentation of our integrated spectral atlas.  We plot the spectrum as $f_{\lambda}(\lambda)$ versus rest wavelength, normalized in a way that attempts to achieve a balance between showing the full range in flux and illustrating the finer details of the continuum.  Our integrated spectrum is accompanied by a Digitized Sky Survey image which illustrates our rectangular spectroscopic aperture as a \emph{solid} outline and the $25$~mag~arcsec$^{-2}$ isophotal size of the galaxy as a \emph{dashed} ellipse.  The image legend gives the galaxy name, the unique identification number in parenthesis, and the morphological type as listed in Table~\ref{table:general_properties}.  The horizontal solid line in the lower-left corner of each image represents $30\arcsec$.}
\figsetgrpend
 
\figsetgrpstart
\figsetgrpnum{8.70}
\figsetgrptitle{Visualization of NGC 1275.}
\figsetplot{\includegraphics[scale=0.7,angle=90]{f8_70.eps}}
\figsetgrpnote{Presentation of our integrated spectral atlas.  We plot the spectrum as $f_{\lambda}(\lambda)$ versus rest wavelength, normalized in a way that attempts to achieve a balance between showing the full range in flux and illustrating the finer details of the continuum.  Our integrated spectrum is accompanied by a Digitized Sky Survey image which illustrates our rectangular spectroscopic aperture as a \emph{solid} outline and the $25$~mag~arcsec$^{-2}$ isophotal size of the galaxy as a \emph{dashed} ellipse.  The image legend gives the galaxy name, the unique identification number in parenthesis, and the morphological type as listed in Table~\ref{table:general_properties}.  The horizontal solid line in the lower-left corner of each image represents $30\arcsec$.}
\figsetgrpend
 
\figsetgrpstart
\figsetgrpnum{8.71}
\figsetgrptitle{Visualization of UGCA 073.}
\figsetplot{\includegraphics[scale=0.7,angle=90]{f8_71.eps}}
\figsetgrpnote{Presentation of our integrated spectral atlas.  We plot the spectrum as $f_{\lambda}(\lambda)$ versus rest wavelength, normalized in a way that attempts to achieve a balance between showing the full range in flux and illustrating the finer details of the continuum.  Our integrated spectrum is accompanied by a Digitized Sky Survey image which illustrates our rectangular spectroscopic aperture as a \emph{solid} outline and the $25$~mag~arcsec$^{-2}$ isophotal size of the galaxy as a \emph{dashed} ellipse.  The image legend gives the galaxy name, the unique identification number in parenthesis, and the morphological type as listed in Table~\ref{table:general_properties}.  The horizontal solid line in the lower-left corner of each image represents $30\arcsec$.}
\figsetgrpend
 
\figsetgrpstart
\figsetgrpnum{8.72}
\figsetgrptitle{Visualization of NGC 1345.}
\figsetplot{\includegraphics[scale=0.7,angle=90]{f8_72.eps}}
\figsetgrpnote{Presentation of our integrated spectral atlas.  We plot the spectrum as $f_{\lambda}(\lambda)$ versus rest wavelength, normalized in a way that attempts to achieve a balance between showing the full range in flux and illustrating the finer details of the continuum.  Our integrated spectrum is accompanied by a Digitized Sky Survey image which illustrates our rectangular spectroscopic aperture as a \emph{solid} outline and the $25$~mag~arcsec$^{-2}$ isophotal size of the galaxy as a \emph{dashed} ellipse.  The image legend gives the galaxy name, the unique identification number in parenthesis, and the morphological type as listed in Table~\ref{table:general_properties}.  The horizontal solid line in the lower-left corner of each image represents $30\arcsec$.}
\figsetgrpend
 
\figsetgrpstart
\figsetgrpnum{8.73}
\figsetgrptitle{Visualization of NGC 1359.}
\figsetplot{\includegraphics[scale=0.7,angle=90]{f8_73.eps}}
\figsetgrpnote{Presentation of our integrated spectral atlas.  We plot the spectrum as $f_{\lambda}(\lambda)$ versus rest wavelength, normalized in a way that attempts to achieve a balance between showing the full range in flux and illustrating the finer details of the continuum.  Our integrated spectrum is accompanied by a Digitized Sky Survey image which illustrates our rectangular spectroscopic aperture as a \emph{solid} outline and the $25$~mag~arcsec$^{-2}$ isophotal size of the galaxy as a \emph{dashed} ellipse.  The image legend gives the galaxy name, the unique identification number in parenthesis, and the morphological type as listed in Table~\ref{table:general_properties}.  The horizontal solid line in the lower-left corner of each image represents $30\arcsec$.}
\figsetgrpend
 
\figsetgrpstart
\figsetgrpnum{8.74}
\figsetgrptitle{Visualization of NGC 1377.}
\figsetplot{\includegraphics[scale=0.7,angle=90]{f8_74.eps}}
\figsetgrpnote{Presentation of our integrated spectral atlas.  We plot the spectrum as $f_{\lambda}(\lambda)$ versus rest wavelength, normalized in a way that attempts to achieve a balance between showing the full range in flux and illustrating the finer details of the continuum.  Our integrated spectrum is accompanied by a Digitized Sky Survey image which illustrates our rectangular spectroscopic aperture as a \emph{solid} outline and the $25$~mag~arcsec$^{-2}$ isophotal size of the galaxy as a \emph{dashed} ellipse.  The image legend gives the galaxy name, the unique identification number in parenthesis, and the morphological type as listed in Table~\ref{table:general_properties}.  The horizontal solid line in the lower-left corner of each image represents $30\arcsec$.}
\figsetgrpend
 
\figsetgrpstart
\figsetgrpnum{8.75}
\figsetgrptitle{Visualization of NGC 1385.}
\figsetplot{\includegraphics[scale=0.7,angle=90]{f8_75.eps}}
\figsetgrpnote{Presentation of our integrated spectral atlas.  We plot the spectrum as $f_{\lambda}(\lambda)$ versus rest wavelength, normalized in a way that attempts to achieve a balance between showing the full range in flux and illustrating the finer details of the continuum.  Our integrated spectrum is accompanied by a Digitized Sky Survey image which illustrates our rectangular spectroscopic aperture as a \emph{solid} outline and the $25$~mag~arcsec$^{-2}$ isophotal size of the galaxy as a \emph{dashed} ellipse.  The image legend gives the galaxy name, the unique identification number in parenthesis, and the morphological type as listed in Table~\ref{table:general_properties}.  The horizontal solid line in the lower-left corner of each image represents $30\arcsec$.}
\figsetgrpend
 
\figsetgrpstart
\figsetgrpnum{8.76}
\figsetgrptitle{Visualization of IRAS 03359+1523.}
\figsetplot{\includegraphics[scale=0.7,angle=90]{f8_76.eps}}
\figsetgrpnote{Presentation of our integrated spectral atlas.  We plot the spectrum as $f_{\lambda}(\lambda)$ versus rest wavelength, normalized in a way that attempts to achieve a balance between showing the full range in flux and illustrating the finer details of the continuum.  Our integrated spectrum is accompanied by a Digitized Sky Survey image which illustrates our rectangular spectroscopic aperture as a \emph{solid} outline and the $25$~mag~arcsec$^{-2}$ isophotal size of the galaxy as a \emph{dashed} ellipse.  The image legend gives the galaxy name, the unique identification number in parenthesis, and the morphological type as listed in Table~\ref{table:general_properties}.  The horizontal solid line in the lower-left corner of each image represents $30\arcsec$.}
\figsetgrpend
 
\figsetgrpstart
\figsetgrpnum{8.77}
\figsetgrptitle{Visualization of NGC 1421.}
\figsetplot{\includegraphics[scale=0.7,angle=90]{f8_77.eps}}
\figsetgrpnote{Presentation of our integrated spectral atlas.  We plot the spectrum as $f_{\lambda}(\lambda)$ versus rest wavelength, normalized in a way that attempts to achieve a balance between showing the full range in flux and illustrating the finer details of the continuum.  Our integrated spectrum is accompanied by a Digitized Sky Survey image which illustrates our rectangular spectroscopic aperture as a \emph{solid} outline and the $25$~mag~arcsec$^{-2}$ isophotal size of the galaxy as a \emph{dashed} ellipse.  The image legend gives the galaxy name, the unique identification number in parenthesis, and the morphological type as listed in Table~\ref{table:general_properties}.  The horizontal solid line in the lower-left corner of each image represents $30\arcsec$.}
\figsetgrpend
 
\figsetgrpstart
\figsetgrpnum{8.78}
\figsetgrptitle{Visualization of UGC 02982.}
\figsetplot{\includegraphics[scale=0.7,angle=90]{f8_78.eps}}
\figsetgrpnote{Presentation of our integrated spectral atlas.  We plot the spectrum as $f_{\lambda}(\lambda)$ versus rest wavelength, normalized in a way that attempts to achieve a balance between showing the full range in flux and illustrating the finer details of the continuum.  Our integrated spectrum is accompanied by a Digitized Sky Survey image which illustrates our rectangular spectroscopic aperture as a \emph{solid} outline and the $25$~mag~arcsec$^{-2}$ isophotal size of the galaxy as a \emph{dashed} ellipse.  The image legend gives the galaxy name, the unique identification number in parenthesis, and the morphological type as listed in Table~\ref{table:general_properties}.  The horizontal solid line in the lower-left corner of each image represents $30\arcsec$.}
\figsetgrpend
 
\figsetgrpstart
\figsetgrpnum{8.79}
\figsetgrptitle{Visualization of UGCA 090.}
\figsetplot{\includegraphics[scale=0.7,angle=90]{f8_79.eps}}
\figsetgrpnote{Presentation of our integrated spectral atlas.  We plot the spectrum as $f_{\lambda}(\lambda)$ versus rest wavelength, normalized in a way that attempts to achieve a balance between showing the full range in flux and illustrating the finer details of the continuum.  Our integrated spectrum is accompanied by a Digitized Sky Survey image which illustrates our rectangular spectroscopic aperture as a \emph{solid} outline and the $25$~mag~arcsec$^{-2}$ isophotal size of the galaxy as a \emph{dashed} ellipse.  The image legend gives the galaxy name, the unique identification number in parenthesis, and the morphological type as listed in Table~\ref{table:general_properties}.  The horizontal solid line in the lower-left corner of each image represents $30\arcsec$.}
\figsetgrpend
 
\figsetgrpstart
\figsetgrpnum{8.80}
\figsetgrptitle{Visualization of ESO 550-IG 025.}
\figsetplot{\includegraphics[scale=0.7,angle=90]{f8_80.eps}}
\figsetgrpnote{Presentation of our integrated spectral atlas.  We plot the spectrum as $f_{\lambda}(\lambda)$ versus rest wavelength, normalized in a way that attempts to achieve a balance between showing the full range in flux and illustrating the finer details of the continuum.  Our integrated spectrum is accompanied by a Digitized Sky Survey image which illustrates our rectangular spectroscopic aperture as a \emph{solid} outline and the $25$~mag~arcsec$^{-2}$ isophotal size of the galaxy as a \emph{dashed} ellipse.  The image legend gives the galaxy name, the unique identification number in parenthesis, and the morphological type as listed in Table~\ref{table:general_properties}.  The horizontal solid line in the lower-left corner of each image represents $30\arcsec$.}
\figsetgrpend
 
\figsetgrpstart
\figsetgrpnum{8.81}
\figsetgrptitle{Visualization of ESO 550-IG 025 N.}
\figsetplot{\includegraphics[scale=0.7,angle=90]{f8_81.eps}}
\figsetgrpnote{Presentation of our integrated spectral atlas.  We plot the spectrum as $f_{\lambda}(\lambda)$ versus rest wavelength, normalized in a way that attempts to achieve a balance between showing the full range in flux and illustrating the finer details of the continuum.  Our integrated spectrum is accompanied by a Digitized Sky Survey image which illustrates our rectangular spectroscopic aperture as a \emph{solid} outline and the $25$~mag~arcsec$^{-2}$ isophotal size of the galaxy as a \emph{dashed} ellipse.  The image legend gives the galaxy name, the unique identification number in parenthesis, and the morphological type as listed in Table~\ref{table:general_properties}.  The horizontal solid line in the lower-left corner of each image represents $30\arcsec$.}
\figsetgrpend
 
\figsetgrpstart
\figsetgrpnum{8.82}
\figsetgrptitle{Visualization of ESO 550-IG 025 S.}
\figsetplot{\includegraphics[scale=0.7,angle=90]{f8_82.eps}}
\figsetgrpnote{Presentation of our integrated spectral atlas.  We plot the spectrum as $f_{\lambda}(\lambda)$ versus rest wavelength, normalized in a way that attempts to achieve a balance between showing the full range in flux and illustrating the finer details of the continuum.  Our integrated spectrum is accompanied by a Digitized Sky Survey image which illustrates our rectangular spectroscopic aperture as a \emph{solid} outline and the $25$~mag~arcsec$^{-2}$ isophotal size of the galaxy as a \emph{dashed} ellipse.  The image legend gives the galaxy name, the unique identification number in parenthesis, and the morphological type as listed in Table~\ref{table:general_properties}.  The horizontal solid line in the lower-left corner of each image represents $30\arcsec$.}
\figsetgrpend
 
\figsetgrpstart
\figsetgrpnum{8.83}
\figsetgrptitle{Visualization of NGC 1569.}
\figsetplot{\includegraphics[scale=0.7,angle=90]{f8_83.eps}}
\figsetgrpnote{Presentation of our integrated spectral atlas.  We plot the spectrum as $f_{\lambda}(\lambda)$ versus rest wavelength, normalized in a way that attempts to achieve a balance between showing the full range in flux and illustrating the finer details of the continuum.  Our integrated spectrum is accompanied by a Digitized Sky Survey image which illustrates our rectangular spectroscopic aperture as a \emph{solid} outline and the $25$~mag~arcsec$^{-2}$ isophotal size of the galaxy as a \emph{dashed} ellipse.  The image legend gives the galaxy name, the unique identification number in parenthesis, and the morphological type as listed in Table~\ref{table:general_properties}.  The horizontal solid line in the lower-left corner of each image represents $30\arcsec$.}
\figsetgrpend
 
\figsetgrpstart
\figsetgrpnum{8.84}
\figsetgrptitle{Visualization of NGC 1560.}
\figsetplot{\includegraphics[scale=0.7,angle=90]{f8_84.eps}}
\figsetgrpnote{Presentation of our integrated spectral atlas.  We plot the spectrum as $f_{\lambda}(\lambda)$ versus rest wavelength, normalized in a way that attempts to achieve a balance between showing the full range in flux and illustrating the finer details of the continuum.  Our integrated spectrum is accompanied by a Digitized Sky Survey image which illustrates our rectangular spectroscopic aperture as a \emph{solid} outline and the $25$~mag~arcsec$^{-2}$ isophotal size of the galaxy as a \emph{dashed} ellipse.  The image legend gives the galaxy name, the unique identification number in parenthesis, and the morphological type as listed in Table~\ref{table:general_properties}.  The horizontal solid line in the lower-left corner of each image represents $30\arcsec$.}
\figsetgrpend
 
\figsetgrpstart
\figsetgrpnum{8.85}
\figsetgrptitle{Visualization of NGC 1614.}
\figsetplot{\includegraphics[scale=0.7,angle=90]{f8_85.eps}}
\figsetgrpnote{Presentation of our integrated spectral atlas.  We plot the spectrum as $f_{\lambda}(\lambda)$ versus rest wavelength, normalized in a way that attempts to achieve a balance between showing the full range in flux and illustrating the finer details of the continuum.  Our integrated spectrum is accompanied by a Digitized Sky Survey image which illustrates our rectangular spectroscopic aperture as a \emph{solid} outline and the $25$~mag~arcsec$^{-2}$ isophotal size of the galaxy as a \emph{dashed} ellipse.  The image legend gives the galaxy name, the unique identification number in parenthesis, and the morphological type as listed in Table~\ref{table:general_properties}.  The horizontal solid line in the lower-left corner of each image represents $30\arcsec$.}
\figsetgrpend
 
\figsetgrpstart
\figsetgrpnum{8.86}
\figsetgrptitle{Visualization of NGC 1800.}
\figsetplot{\includegraphics[scale=0.7,angle=90]{f8_86.eps}}
\figsetgrpnote{Presentation of our integrated spectral atlas.  We plot the spectrum as $f_{\lambda}(\lambda)$ versus rest wavelength, normalized in a way that attempts to achieve a balance between showing the full range in flux and illustrating the finer details of the continuum.  Our integrated spectrum is accompanied by a Digitized Sky Survey image which illustrates our rectangular spectroscopic aperture as a \emph{solid} outline and the $25$~mag~arcsec$^{-2}$ isophotal size of the galaxy as a \emph{dashed} ellipse.  The image legend gives the galaxy name, the unique identification number in parenthesis, and the morphological type as listed in Table~\ref{table:general_properties}.  The horizontal solid line in the lower-left corner of each image represents $30\arcsec$.}
\figsetgrpend
 
\figsetgrpstart
\figsetgrpnum{8.87}
\figsetgrptitle{Visualization of IRAS 05189-2524.}
\figsetplot{\includegraphics[scale=0.7,angle=90]{f8_87.eps}}
\figsetgrpnote{Presentation of our integrated spectral atlas.  We plot the spectrum as $f_{\lambda}(\lambda)$ versus rest wavelength, normalized in a way that attempts to achieve a balance between showing the full range in flux and illustrating the finer details of the continuum.  Our integrated spectrum is accompanied by a Digitized Sky Survey image which illustrates our rectangular spectroscopic aperture as a \emph{solid} outline and the $25$~mag~arcsec$^{-2}$ isophotal size of the galaxy as a \emph{dashed} ellipse.  The image legend gives the galaxy name, the unique identification number in parenthesis, and the morphological type as listed in Table~\ref{table:general_properties}.  The horizontal solid line in the lower-left corner of each image represents $30\arcsec$.}
\figsetgrpend
 
\figsetgrpstart
\figsetgrpnum{8.88}
\figsetgrptitle{Visualization of UGCA 114.}
\figsetplot{\includegraphics[scale=0.7,angle=90]{f8_88.eps}}
\figsetgrpnote{Presentation of our integrated spectral atlas.  We plot the spectrum as $f_{\lambda}(\lambda)$ versus rest wavelength, normalized in a way that attempts to achieve a balance between showing the full range in flux and illustrating the finer details of the continuum.  Our integrated spectrum is accompanied by a Digitized Sky Survey image which illustrates our rectangular spectroscopic aperture as a \emph{solid} outline and the $25$~mag~arcsec$^{-2}$ isophotal size of the galaxy as a \emph{dashed} ellipse.  The image legend gives the galaxy name, the unique identification number in parenthesis, and the morphological type as listed in Table~\ref{table:general_properties}.  The horizontal solid line in the lower-left corner of each image represents $30\arcsec$.}
\figsetgrpend
 
\figsetgrpstart
\figsetgrpnum{8.89}
\figsetgrptitle{Visualization of UGCA 116.}
\figsetplot{\includegraphics[scale=0.7,angle=90]{f8_89.eps}}
\figsetgrpnote{Presentation of our integrated spectral atlas.  We plot the spectrum as $f_{\lambda}(\lambda)$ versus rest wavelength, normalized in a way that attempts to achieve a balance between showing the full range in flux and illustrating the finer details of the continuum.  Our integrated spectrum is accompanied by a Digitized Sky Survey image which illustrates our rectangular spectroscopic aperture as a \emph{solid} outline and the $25$~mag~arcsec$^{-2}$ isophotal size of the galaxy as a \emph{dashed} ellipse.  The image legend gives the galaxy name, the unique identification number in parenthesis, and the morphological type as listed in Table~\ref{table:general_properties}.  The horizontal solid line in the lower-left corner of each image represents $30\arcsec$.}
\figsetgrpend
 
\figsetgrpstart
\figsetgrpnum{8.90}
\figsetgrptitle{Visualization of NGC 2139.}
\figsetplot{\includegraphics[scale=0.7,angle=90]{f8_90.eps}}
\figsetgrpnote{Presentation of our integrated spectral atlas.  We plot the spectrum as $f_{\lambda}(\lambda)$ versus rest wavelength, normalized in a way that attempts to achieve a balance between showing the full range in flux and illustrating the finer details of the continuum.  Our integrated spectrum is accompanied by a Digitized Sky Survey image which illustrates our rectangular spectroscopic aperture as a \emph{solid} outline and the $25$~mag~arcsec$^{-2}$ isophotal size of the galaxy as a \emph{dashed} ellipse.  The image legend gives the galaxy name, the unique identification number in parenthesis, and the morphological type as listed in Table~\ref{table:general_properties}.  The horizontal solid line in the lower-left corner of each image represents $30\arcsec$.}
\figsetgrpend
 
\figsetgrpstart
\figsetgrpnum{8.91}
\figsetgrptitle{Visualization of NGC 2146.}
\figsetplot{\includegraphics[scale=0.7,angle=90]{f8_91.eps}}
\figsetgrpnote{Presentation of our integrated spectral atlas.  We plot the spectrum as $f_{\lambda}(\lambda)$ versus rest wavelength, normalized in a way that attempts to achieve a balance between showing the full range in flux and illustrating the finer details of the continuum.  Our integrated spectrum is accompanied by a Digitized Sky Survey image which illustrates our rectangular spectroscopic aperture as a \emph{solid} outline and the $25$~mag~arcsec$^{-2}$ isophotal size of the galaxy as a \emph{dashed} ellipse.  The image legend gives the galaxy name, the unique identification number in parenthesis, and the morphological type as listed in Table~\ref{table:general_properties}.  The horizontal solid line in the lower-left corner of each image represents $30\arcsec$.}
\figsetgrpend
 
\figsetgrpstart
\figsetgrpnum{8.92}
\figsetgrptitle{Visualization of UGCA 130.}
\figsetplot{\includegraphics[scale=0.7,angle=90]{f8_92.eps}}
\figsetgrpnote{Presentation of our integrated spectral atlas.  We plot the spectrum as $f_{\lambda}(\lambda)$ versus rest wavelength, normalized in a way that attempts to achieve a balance between showing the full range in flux and illustrating the finer details of the continuum.  Our integrated spectrum is accompanied by a Digitized Sky Survey image which illustrates our rectangular spectroscopic aperture as a \emph{solid} outline and the $25$~mag~arcsec$^{-2}$ isophotal size of the galaxy as a \emph{dashed} ellipse.  The image legend gives the galaxy name, the unique identification number in parenthesis, and the morphological type as listed in Table~\ref{table:general_properties}.  The horizontal solid line in the lower-left corner of each image represents $30\arcsec$.}
\figsetgrpend
 
\figsetgrpstart
\figsetgrpnum{8.93}
\figsetgrptitle{Visualization of NGC 2337.}
\figsetplot{\includegraphics[scale=0.7,angle=90]{f8_93.eps}}
\figsetgrpnote{Presentation of our integrated spectral atlas.  We plot the spectrum as $f_{\lambda}(\lambda)$ versus rest wavelength, normalized in a way that attempts to achieve a balance between showing the full range in flux and illustrating the finer details of the continuum.  Our integrated spectrum is accompanied by a Digitized Sky Survey image which illustrates our rectangular spectroscopic aperture as a \emph{solid} outline and the $25$~mag~arcsec$^{-2}$ isophotal size of the galaxy as a \emph{dashed} ellipse.  The image legend gives the galaxy name, the unique identification number in parenthesis, and the morphological type as listed in Table~\ref{table:general_properties}.  The horizontal solid line in the lower-left corner of each image represents $30\arcsec$.}
\figsetgrpend
 
\figsetgrpstart
\figsetgrpnum{8.94}
\figsetgrptitle{Visualization of UGC 03838.}
\figsetplot{\includegraphics[scale=0.7,angle=90]{f8_94.eps}}
\figsetgrpnote{Presentation of our integrated spectral atlas.  We plot the spectrum as $f_{\lambda}(\lambda)$ versus rest wavelength, normalized in a way that attempts to achieve a balance between showing the full range in flux and illustrating the finer details of the continuum.  Our integrated spectrum is accompanied by a Digitized Sky Survey image which illustrates our rectangular spectroscopic aperture as a \emph{solid} outline and the $25$~mag~arcsec$^{-2}$ isophotal size of the galaxy as a \emph{dashed} ellipse.  The image legend gives the galaxy name, the unique identification number in parenthesis, and the morphological type as listed in Table~\ref{table:general_properties}.  The horizontal solid line in the lower-left corner of each image represents $30\arcsec$.}
\figsetgrpend
 
\figsetgrpstart
\figsetgrpnum{8.95}
\figsetgrptitle{Visualization of NGC 2363.}
\figsetplot{\includegraphics[scale=0.7,angle=90]{f8_95.eps}}
\figsetgrpnote{Presentation of our integrated spectral atlas.  We plot the spectrum as $f_{\lambda}(\lambda)$ versus rest wavelength, normalized in a way that attempts to achieve a balance between showing the full range in flux and illustrating the finer details of the continuum.  Our integrated spectrum is accompanied by a Digitized Sky Survey image which illustrates our rectangular spectroscopic aperture as a \emph{solid} outline and the $25$~mag~arcsec$^{-2}$ isophotal size of the galaxy as a \emph{dashed} ellipse.  The image legend gives the galaxy name, the unique identification number in parenthesis, and the morphological type as listed in Table~\ref{table:general_properties}.  The horizontal solid line in the lower-left corner of each image represents $30\arcsec$.}
\figsetgrpend
 
\figsetgrpstart
\figsetgrpnum{8.96}
\figsetgrptitle{Visualization of NGC 2363 A.}
\figsetplot{\includegraphics[scale=0.7,angle=90]{f8_96.eps}}
\figsetgrpnote{Presentation of our integrated spectral atlas.  We plot the spectrum as $f_{\lambda}(\lambda)$ versus rest wavelength, normalized in a way that attempts to achieve a balance between showing the full range in flux and illustrating the finer details of the continuum.  Our integrated spectrum is accompanied by a Digitized Sky Survey image which illustrates our rectangular spectroscopic aperture as a \emph{solid} outline and the $25$~mag~arcsec$^{-2}$ isophotal size of the galaxy as a \emph{dashed} ellipse.  The image legend gives the galaxy name, the unique identification number in parenthesis, and the morphological type as listed in Table~\ref{table:general_properties}.  The horizontal solid line in the lower-left corner of each image represents $30\arcsec$.}
\figsetgrpend
 
\figsetgrpstart
\figsetgrpnum{8.97}
\figsetgrptitle{Visualization of NGC 2388.}
\figsetplot{\includegraphics[scale=0.7,angle=90]{f8_97.eps}}
\figsetgrpnote{Presentation of our integrated spectral atlas.  We plot the spectrum as $f_{\lambda}(\lambda)$ versus rest wavelength, normalized in a way that attempts to achieve a balance between showing the full range in flux and illustrating the finer details of the continuum.  Our integrated spectrum is accompanied by a Digitized Sky Survey image which illustrates our rectangular spectroscopic aperture as a \emph{solid} outline and the $25$~mag~arcsec$^{-2}$ isophotal size of the galaxy as a \emph{dashed} ellipse.  The image legend gives the galaxy name, the unique identification number in parenthesis, and the morphological type as listed in Table~\ref{table:general_properties}.  The horizontal solid line in the lower-left corner of each image represents $30\arcsec$.}
\figsetgrpend
 
\figsetgrpstart
\figsetgrpnum{8.98}
\figsetgrptitle{Visualization of NGC 2415.}
\figsetplot{\includegraphics[scale=0.7,angle=90]{f8_98.eps}}
\figsetgrpnote{Presentation of our integrated spectral atlas.  We plot the spectrum as $f_{\lambda}(\lambda)$ versus rest wavelength, normalized in a way that attempts to achieve a balance between showing the full range in flux and illustrating the finer details of the continuum.  Our integrated spectrum is accompanied by a Digitized Sky Survey image which illustrates our rectangular spectroscopic aperture as a \emph{solid} outline and the $25$~mag~arcsec$^{-2}$ isophotal size of the galaxy as a \emph{dashed} ellipse.  The image legend gives the galaxy name, the unique identification number in parenthesis, and the morphological type as listed in Table~\ref{table:general_properties}.  The horizontal solid line in the lower-left corner of each image represents $30\arcsec$.}
\figsetgrpend
 
\figsetgrpstart
\figsetgrpnum{8.99}
\figsetgrptitle{Visualization of NGC 2500.}
\figsetplot{\includegraphics[scale=0.7,angle=90]{f8_99.eps}}
\figsetgrpnote{Presentation of our integrated spectral atlas.  We plot the spectrum as $f_{\lambda}(\lambda)$ versus rest wavelength, normalized in a way that attempts to achieve a balance between showing the full range in flux and illustrating the finer details of the continuum.  Our integrated spectrum is accompanied by a Digitized Sky Survey image which illustrates our rectangular spectroscopic aperture as a \emph{solid} outline and the $25$~mag~arcsec$^{-2}$ isophotal size of the galaxy as a \emph{dashed} ellipse.  The image legend gives the galaxy name, the unique identification number in parenthesis, and the morphological type as listed in Table~\ref{table:general_properties}.  The horizontal solid line in the lower-left corner of each image represents $30\arcsec$.}
\figsetgrpend
 
\figsetgrpstart
\figsetgrpnum{8.100}
\figsetgrptitle{Visualization of NGC 2537.}
\figsetplot{\includegraphics[scale=0.7,angle=90]{f8_100.eps}}
\figsetgrpnote{Presentation of our integrated spectral atlas.  We plot the spectrum as $f_{\lambda}(\lambda)$ versus rest wavelength, normalized in a way that attempts to achieve a balance between showing the full range in flux and illustrating the finer details of the continuum.  Our integrated spectrum is accompanied by a Digitized Sky Survey image which illustrates our rectangular spectroscopic aperture as a \emph{solid} outline and the $25$~mag~arcsec$^{-2}$ isophotal size of the galaxy as a \emph{dashed} ellipse.  The image legend gives the galaxy name, the unique identification number in parenthesis, and the morphological type as listed in Table~\ref{table:general_properties}.  The horizontal solid line in the lower-left corner of each image represents $30\arcsec$.}
\figsetgrpend
 
\figsetgrpstart
\figsetgrpnum{8.101}
\figsetgrptitle{Visualization of NGC 2541.}
\figsetplot{\includegraphics[scale=0.7,angle=90]{f8_101.eps}}
\figsetgrpnote{Presentation of our integrated spectral atlas.  We plot the spectrum as $f_{\lambda}(\lambda)$ versus rest wavelength, normalized in a way that attempts to achieve a balance between showing the full range in flux and illustrating the finer details of the continuum.  Our integrated spectrum is accompanied by a Digitized Sky Survey image which illustrates our rectangular spectroscopic aperture as a \emph{solid} outline and the $25$~mag~arcsec$^{-2}$ isophotal size of the galaxy as a \emph{dashed} ellipse.  The image legend gives the galaxy name, the unique identification number in parenthesis, and the morphological type as listed in Table~\ref{table:general_properties}.  The horizontal solid line in the lower-left corner of each image represents $30\arcsec$.}
\figsetgrpend
 
\figsetgrpstart
\figsetgrpnum{8.102}
\figsetgrptitle{Visualization of NGC 2552.}
\figsetplot{\includegraphics[scale=0.7,angle=90]{f8_102.eps}}
\figsetgrpnote{Presentation of our integrated spectral atlas.  We plot the spectrum as $f_{\lambda}(\lambda)$ versus rest wavelength, normalized in a way that attempts to achieve a balance between showing the full range in flux and illustrating the finer details of the continuum.  Our integrated spectrum is accompanied by a Digitized Sky Survey image which illustrates our rectangular spectroscopic aperture as a \emph{solid} outline and the $25$~mag~arcsec$^{-2}$ isophotal size of the galaxy as a \emph{dashed} ellipse.  The image legend gives the galaxy name, the unique identification number in parenthesis, and the morphological type as listed in Table~\ref{table:general_properties}.  The horizontal solid line in the lower-left corner of each image represents $30\arcsec$.}
\figsetgrpend
 
\figsetgrpstart
\figsetgrpnum{8.103}
\figsetgrptitle{Visualization of HS 0822+3542.}
\figsetplot{\includegraphics[scale=0.7,angle=90]{f8_103.eps}}
\figsetgrpnote{Presentation of our integrated spectral atlas.  We plot the spectrum as $f_{\lambda}(\lambda)$ versus rest wavelength, normalized in a way that attempts to achieve a balance between showing the full range in flux and illustrating the finer details of the continuum.  Our integrated spectrum is accompanied by a Digitized Sky Survey image which illustrates our rectangular spectroscopic aperture as a \emph{solid} outline and the $25$~mag~arcsec$^{-2}$ isophotal size of the galaxy as a \emph{dashed} ellipse.  The image legend gives the galaxy name, the unique identification number in parenthesis, and the morphological type as listed in Table~\ref{table:general_properties}.  The horizontal solid line in the lower-left corner of each image represents $30\arcsec$.}
\figsetgrpend
 
\figsetgrpstart
\figsetgrpnum{8.104}
\figsetgrptitle{Visualization of NGC 2599.}
\figsetplot{\includegraphics[scale=0.7,angle=90]{f8_104.eps}}
\figsetgrpnote{Presentation of our integrated spectral atlas.  We plot the spectrum as $f_{\lambda}(\lambda)$ versus rest wavelength, normalized in a way that attempts to achieve a balance between showing the full range in flux and illustrating the finer details of the continuum.  Our integrated spectrum is accompanied by a Digitized Sky Survey image which illustrates our rectangular spectroscopic aperture as a \emph{solid} outline and the $25$~mag~arcsec$^{-2}$ isophotal size of the galaxy as a \emph{dashed} ellipse.  The image legend gives the galaxy name, the unique identification number in parenthesis, and the morphological type as listed in Table~\ref{table:general_properties}.  The horizontal solid line in the lower-left corner of each image represents $30\arcsec$.}
\figsetgrpend
 
\figsetgrpstart
\figsetgrpnum{8.105}
\figsetgrptitle{Visualization of UGC 04459.}
\figsetplot{\includegraphics[scale=0.7,angle=90]{f8_105.eps}}
\figsetgrpnote{Presentation of our integrated spectral atlas.  We plot the spectrum as $f_{\lambda}(\lambda)$ versus rest wavelength, normalized in a way that attempts to achieve a balance between showing the full range in flux and illustrating the finer details of the continuum.  Our integrated spectrum is accompanied by a Digitized Sky Survey image which illustrates our rectangular spectroscopic aperture as a \emph{solid} outline and the $25$~mag~arcsec$^{-2}$ isophotal size of the galaxy as a \emph{dashed} ellipse.  The image legend gives the galaxy name, the unique identification number in parenthesis, and the morphological type as listed in Table~\ref{table:general_properties}.  The horizontal solid line in the lower-left corner of each image represents $30\arcsec$.}
\figsetgrpend
 
\figsetgrpstart
\figsetgrpnum{8.106}
\figsetgrptitle{Visualization of UGC 04483.}
\figsetplot{\includegraphics[scale=0.7,angle=90]{f8_106.eps}}
\figsetgrpnote{Presentation of our integrated spectral atlas.  We plot the spectrum as $f_{\lambda}(\lambda)$ versus rest wavelength, normalized in a way that attempts to achieve a balance between showing the full range in flux and illustrating the finer details of the continuum.  Our integrated spectrum is accompanied by a Digitized Sky Survey image which illustrates our rectangular spectroscopic aperture as a \emph{solid} outline and the $25$~mag~arcsec$^{-2}$ isophotal size of the galaxy as a \emph{dashed} ellipse.  The image legend gives the galaxy name, the unique identification number in parenthesis, and the morphological type as listed in Table~\ref{table:general_properties}.  The horizontal solid line in the lower-left corner of each image represents $30\arcsec$.}
\figsetgrpend
 
\figsetgrpstart
\figsetgrpnum{8.107}
\figsetgrptitle{Visualization of NGC 2623.}
\figsetplot{\includegraphics[scale=0.7,angle=90]{f8_107.eps}}
\figsetgrpnote{Presentation of our integrated spectral atlas.  We plot the spectrum as $f_{\lambda}(\lambda)$ versus rest wavelength, normalized in a way that attempts to achieve a balance between showing the full range in flux and illustrating the finer details of the continuum.  Our integrated spectrum is accompanied by a Digitized Sky Survey image which illustrates our rectangular spectroscopic aperture as a \emph{solid} outline and the $25$~mag~arcsec$^{-2}$ isophotal size of the galaxy as a \emph{dashed} ellipse.  The image legend gives the galaxy name, the unique identification number in parenthesis, and the morphological type as listed in Table~\ref{table:general_properties}.  The horizontal solid line in the lower-left corner of each image represents $30\arcsec$.}
\figsetgrpend
 
\figsetgrpstart
\figsetgrpnum{8.108}
\figsetgrptitle{Visualization of IRAS 08572+3915.}
\figsetplot{\includegraphics[scale=0.7,angle=90]{f8_108.eps}}
\figsetgrpnote{Presentation of our integrated spectral atlas.  We plot the spectrum as $f_{\lambda}(\lambda)$ versus rest wavelength, normalized in a way that attempts to achieve a balance between showing the full range in flux and illustrating the finer details of the continuum.  Our integrated spectrum is accompanied by a Digitized Sky Survey image which illustrates our rectangular spectroscopic aperture as a \emph{solid} outline and the $25$~mag~arcsec$^{-2}$ isophotal size of the galaxy as a \emph{dashed} ellipse.  The image legend gives the galaxy name, the unique identification number in parenthesis, and the morphological type as listed in Table~\ref{table:general_properties}.  The horizontal solid line in the lower-left corner of each image represents $30\arcsec$.}
\figsetgrpend
 
\figsetgrpstart
\figsetgrpnum{8.109}
\figsetgrptitle{Visualization of UGC 04787.}
\figsetplot{\includegraphics[scale=0.7,angle=90]{f8_109.eps}}
\figsetgrpnote{Presentation of our integrated spectral atlas.  We plot the spectrum as $f_{\lambda}(\lambda)$ versus rest wavelength, normalized in a way that attempts to achieve a balance between showing the full range in flux and illustrating the finer details of the continuum.  Our integrated spectrum is accompanied by a Digitized Sky Survey image which illustrates our rectangular spectroscopic aperture as a \emph{solid} outline and the $25$~mag~arcsec$^{-2}$ isophotal size of the galaxy as a \emph{dashed} ellipse.  The image legend gives the galaxy name, the unique identification number in parenthesis, and the morphological type as listed in Table~\ref{table:general_properties}.  The horizontal solid line in the lower-left corner of each image represents $30\arcsec$.}
\figsetgrpend
 
\figsetgrpstart
\figsetgrpnum{8.110}
\figsetgrptitle{Visualization of NGC 2782.}
\figsetplot{\includegraphics[scale=0.7,angle=90]{f8_110.eps}}
\figsetgrpnote{Presentation of our integrated spectral atlas.  We plot the spectrum as $f_{\lambda}(\lambda)$ versus rest wavelength, normalized in a way that attempts to achieve a balance between showing the full range in flux and illustrating the finer details of the continuum.  Our integrated spectrum is accompanied by a Digitized Sky Survey image which illustrates our rectangular spectroscopic aperture as a \emph{solid} outline and the $25$~mag~arcsec$^{-2}$ isophotal size of the galaxy as a \emph{dashed} ellipse.  The image legend gives the galaxy name, the unique identification number in parenthesis, and the morphological type as listed in Table~\ref{table:general_properties}.  The horizontal solid line in the lower-left corner of each image represents $30\arcsec$.}
\figsetgrpend
 
\figsetgrpstart
\figsetgrpnum{8.111}
\figsetgrptitle{Visualization of UGC 04881.}
\figsetplot{\includegraphics[scale=0.7,angle=90]{f8_111.eps}}
\figsetgrpnote{Presentation of our integrated spectral atlas.  We plot the spectrum as $f_{\lambda}(\lambda)$ versus rest wavelength, normalized in a way that attempts to achieve a balance between showing the full range in flux and illustrating the finer details of the continuum.  Our integrated spectrum is accompanied by a Digitized Sky Survey image which illustrates our rectangular spectroscopic aperture as a \emph{solid} outline and the $25$~mag~arcsec$^{-2}$ isophotal size of the galaxy as a \emph{dashed} ellipse.  The image legend gives the galaxy name, the unique identification number in parenthesis, and the morphological type as listed in Table~\ref{table:general_properties}.  The horizontal solid line in the lower-left corner of each image represents $30\arcsec$.}
\figsetgrpend
 
\figsetgrpstart
\figsetgrpnum{8.112}
\figsetgrptitle{Visualization of MRK 0019.}
\figsetplot{\includegraphics[scale=0.7,angle=90]{f8_112.eps}}
\figsetgrpnote{Presentation of our integrated spectral atlas.  We plot the spectrum as $f_{\lambda}(\lambda)$ versus rest wavelength, normalized in a way that attempts to achieve a balance between showing the full range in flux and illustrating the finer details of the continuum.  Our integrated spectrum is accompanied by a Digitized Sky Survey image which illustrates our rectangular spectroscopic aperture as a \emph{solid} outline and the $25$~mag~arcsec$^{-2}$ isophotal size of the galaxy as a \emph{dashed} ellipse.  The image legend gives the galaxy name, the unique identification number in parenthesis, and the morphological type as listed in Table~\ref{table:general_properties}.  The horizontal solid line in the lower-left corner of each image represents $30\arcsec$.}
\figsetgrpend
 
\figsetgrpstart
\figsetgrpnum{8.113}
\figsetgrptitle{Visualization of UGC 05028.}
\figsetplot{\includegraphics[scale=0.7,angle=90]{f8_113.eps}}
\figsetgrpnote{Presentation of our integrated spectral atlas.  We plot the spectrum as $f_{\lambda}(\lambda)$ versus rest wavelength, normalized in a way that attempts to achieve a balance between showing the full range in flux and illustrating the finer details of the continuum.  Our integrated spectrum is accompanied by a Digitized Sky Survey image which illustrates our rectangular spectroscopic aperture as a \emph{solid} outline and the $25$~mag~arcsec$^{-2}$ isophotal size of the galaxy as a \emph{dashed} ellipse.  The image legend gives the galaxy name, the unique identification number in parenthesis, and the morphological type as listed in Table~\ref{table:general_properties}.  The horizontal solid line in the lower-left corner of each image represents $30\arcsec$.}
\figsetgrpend
 
\figsetgrpstart
\figsetgrpnum{8.114}
\figsetgrptitle{Visualization of NGC 2893.}
\figsetplot{\includegraphics[scale=0.7,angle=90]{f8_114.eps}}
\figsetgrpnote{Presentation of our integrated spectral atlas.  We plot the spectrum as $f_{\lambda}(\lambda)$ versus rest wavelength, normalized in a way that attempts to achieve a balance between showing the full range in flux and illustrating the finer details of the continuum.  Our integrated spectrum is accompanied by a Digitized Sky Survey image which illustrates our rectangular spectroscopic aperture as a \emph{solid} outline and the $25$~mag~arcsec$^{-2}$ isophotal size of the galaxy as a \emph{dashed} ellipse.  The image legend gives the galaxy name, the unique identification number in parenthesis, and the morphological type as listed in Table~\ref{table:general_properties}.  The horizontal solid line in the lower-left corner of each image represents $30\arcsec$.}
\figsetgrpend
 
\figsetgrpstart
\figsetgrpnum{8.115}
\figsetgrptitle{Visualization of NGC 2903.}
\figsetplot{\includegraphics[scale=0.7,angle=90]{f8_115.eps}}
\figsetgrpnote{Presentation of our integrated spectral atlas.  We plot the spectrum as $f_{\lambda}(\lambda)$ versus rest wavelength, normalized in a way that attempts to achieve a balance between showing the full range in flux and illustrating the finer details of the continuum.  Our integrated spectrum is accompanied by a Digitized Sky Survey image which illustrates our rectangular spectroscopic aperture as a \emph{solid} outline and the $25$~mag~arcsec$^{-2}$ isophotal size of the galaxy as a \emph{dashed} ellipse.  The image legend gives the galaxy name, the unique identification number in parenthesis, and the morphological type as listed in Table~\ref{table:general_properties}.  The horizontal solid line in the lower-left corner of each image represents $30\arcsec$.}
\figsetgrpend
 
\figsetgrpstart
\figsetgrpnum{8.116}
\figsetgrptitle{Visualization of UGCA 166.}
\figsetplot{\includegraphics[scale=0.7,angle=90]{f8_116.eps}}
\figsetgrpnote{Presentation of our integrated spectral atlas.  We plot the spectrum as $f_{\lambda}(\lambda)$ versus rest wavelength, normalized in a way that attempts to achieve a balance between showing the full range in flux and illustrating the finer details of the continuum.  Our integrated spectrum is accompanied by a Digitized Sky Survey image which illustrates our rectangular spectroscopic aperture as a \emph{solid} outline and the $25$~mag~arcsec$^{-2}$ isophotal size of the galaxy as a \emph{dashed} ellipse.  The image legend gives the galaxy name, the unique identification number in parenthesis, and the morphological type as listed in Table~\ref{table:general_properties}.  The horizontal solid line in the lower-left corner of each image represents $30\arcsec$.}
\figsetgrpend
 
\figsetgrpstart
\figsetgrpnum{8.117}
\figsetgrptitle{Visualization of UGC 05101.}
\figsetplot{\includegraphics[scale=0.7,angle=90]{f8_117.eps}}
\figsetgrpnote{Presentation of our integrated spectral atlas.  We plot the spectrum as $f_{\lambda}(\lambda)$ versus rest wavelength, normalized in a way that attempts to achieve a balance between showing the full range in flux and illustrating the finer details of the continuum.  Our integrated spectrum is accompanied by a Digitized Sky Survey image which illustrates our rectangular spectroscopic aperture as a \emph{solid} outline and the $25$~mag~arcsec$^{-2}$ isophotal size of the galaxy as a \emph{dashed} ellipse.  The image legend gives the galaxy name, the unique identification number in parenthesis, and the morphological type as listed in Table~\ref{table:general_properties}.  The horizontal solid line in the lower-left corner of each image represents $30\arcsec$.}
\figsetgrpend
 
\figsetgrpstart
\figsetgrpnum{8.118}
\figsetgrptitle{Visualization of CGCG 239-011 W.}
\figsetplot{\includegraphics[scale=0.7,angle=90]{f8_118.eps}}
\figsetgrpnote{Presentation of our integrated spectral atlas.  We plot the spectrum as $f_{\lambda}(\lambda)$ versus rest wavelength, normalized in a way that attempts to achieve a balance between showing the full range in flux and illustrating the finer details of the continuum.  Our integrated spectrum is accompanied by a Digitized Sky Survey image which illustrates our rectangular spectroscopic aperture as a \emph{solid} outline and the $25$~mag~arcsec$^{-2}$ isophotal size of the galaxy as a \emph{dashed} ellipse.  The image legend gives the galaxy name, the unique identification number in parenthesis, and the morphological type as listed in Table~\ref{table:general_properties}.  The horizontal solid line in the lower-left corner of each image represents $30\arcsec$.}
\figsetgrpend
 
\figsetgrpstart
\figsetgrpnum{8.119}
\figsetgrptitle{Visualization of CGCG 239-011 E.}
\figsetplot{\includegraphics[scale=0.7,angle=90]{f8_119.eps}}
\figsetgrpnote{Presentation of our integrated spectral atlas.  We plot the spectrum as $f_{\lambda}(\lambda)$ versus rest wavelength, normalized in a way that attempts to achieve a balance between showing the full range in flux and illustrating the finer details of the continuum.  Our integrated spectrum is accompanied by a Digitized Sky Survey image which illustrates our rectangular spectroscopic aperture as a \emph{solid} outline and the $25$~mag~arcsec$^{-2}$ isophotal size of the galaxy as a \emph{dashed} ellipse.  The image legend gives the galaxy name, the unique identification number in parenthesis, and the morphological type as listed in Table~\ref{table:general_properties}.  The horizontal solid line in the lower-left corner of each image represents $30\arcsec$.}
\figsetgrpend
 
\figsetgrpstart
\figsetgrpnum{8.120}
\figsetgrptitle{Visualization of UGC 05151.}
\figsetplot{\includegraphics[scale=0.7,angle=90]{f8_120.eps}}
\figsetgrpnote{Presentation of our integrated spectral atlas.  We plot the spectrum as $f_{\lambda}(\lambda)$ versus rest wavelength, normalized in a way that attempts to achieve a balance between showing the full range in flux and illustrating the finer details of the continuum.  Our integrated spectrum is accompanied by a Digitized Sky Survey image which illustrates our rectangular spectroscopic aperture as a \emph{solid} outline and the $25$~mag~arcsec$^{-2}$ isophotal size of the galaxy as a \emph{dashed} ellipse.  The image legend gives the galaxy name, the unique identification number in parenthesis, and the morphological type as listed in Table~\ref{table:general_properties}.  The horizontal solid line in the lower-left corner of each image represents $30\arcsec$.}
\figsetgrpend
 
\figsetgrpstart
\figsetgrpnum{8.121}
\figsetgrptitle{Visualization of NGC 3049.}
\figsetplot{\includegraphics[scale=0.7,angle=90]{f8_121.eps}}
\figsetgrpnote{Presentation of our integrated spectral atlas.  We plot the spectrum as $f_{\lambda}(\lambda)$ versus rest wavelength, normalized in a way that attempts to achieve a balance between showing the full range in flux and illustrating the finer details of the continuum.  Our integrated spectrum is accompanied by a Digitized Sky Survey image which illustrates our rectangular spectroscopic aperture as a \emph{solid} outline and the $25$~mag~arcsec$^{-2}$ isophotal size of the galaxy as a \emph{dashed} ellipse.  The image legend gives the galaxy name, the unique identification number in parenthesis, and the morphological type as listed in Table~\ref{table:general_properties}.  The horizontal solid line in the lower-left corner of each image represents $30\arcsec$.}
\figsetgrpend
 
\figsetgrpstart
\figsetgrpnum{8.122}
\figsetgrptitle{Visualization of NGC 3079.}
\figsetplot{\includegraphics[scale=0.7,angle=90]{f8_122.eps}}
\figsetgrpnote{Presentation of our integrated spectral atlas.  We plot the spectrum as $f_{\lambda}(\lambda)$ versus rest wavelength, normalized in a way that attempts to achieve a balance between showing the full range in flux and illustrating the finer details of the continuum.  Our integrated spectrum is accompanied by a Digitized Sky Survey image which illustrates our rectangular spectroscopic aperture as a \emph{solid} outline and the $25$~mag~arcsec$^{-2}$ isophotal size of the galaxy as a \emph{dashed} ellipse.  The image legend gives the galaxy name, the unique identification number in parenthesis, and the morphological type as listed in Table~\ref{table:general_properties}.  The horizontal solid line in the lower-left corner of each image represents $30\arcsec$.}
\figsetgrpend
 
\figsetgrpstart
\figsetgrpnum{8.123}
\figsetgrptitle{Visualization of NGC 3077.}
\figsetplot{\includegraphics[scale=0.7,angle=90]{f8_123.eps}}
\figsetgrpnote{Presentation of our integrated spectral atlas.  We plot the spectrum as $f_{\lambda}(\lambda)$ versus rest wavelength, normalized in a way that attempts to achieve a balance between showing the full range in flux and illustrating the finer details of the continuum.  Our integrated spectrum is accompanied by a Digitized Sky Survey image which illustrates our rectangular spectroscopic aperture as a \emph{solid} outline and the $25$~mag~arcsec$^{-2}$ isophotal size of the galaxy as a \emph{dashed} ellipse.  The image legend gives the galaxy name, the unique identification number in parenthesis, and the morphological type as listed in Table~\ref{table:general_properties}.  The horizontal solid line in the lower-left corner of each image represents $30\arcsec$.}
\figsetgrpend
 
\figsetgrpstart
\figsetgrpnum{8.124}
\figsetgrptitle{Visualization of NGC 3104.}
\figsetplot{\includegraphics[scale=0.7,angle=90]{f8_124.eps}}
\figsetgrpnote{Presentation of our integrated spectral atlas.  We plot the spectrum as $f_{\lambda}(\lambda)$ versus rest wavelength, normalized in a way that attempts to achieve a balance between showing the full range in flux and illustrating the finer details of the continuum.  Our integrated spectrum is accompanied by a Digitized Sky Survey image which illustrates our rectangular spectroscopic aperture as a \emph{solid} outline and the $25$~mag~arcsec$^{-2}$ isophotal size of the galaxy as a \emph{dashed} ellipse.  The image legend gives the galaxy name, the unique identification number in parenthesis, and the morphological type as listed in Table~\ref{table:general_properties}.  The horizontal solid line in the lower-left corner of each image represents $30\arcsec$.}
\figsetgrpend
 
\figsetgrpstart
\figsetgrpnum{8.125}
\figsetgrptitle{Visualization of UGC 05423.}
\figsetplot{\includegraphics[scale=0.7,angle=90]{f8_125.eps}}
\figsetgrpnote{Presentation of our integrated spectral atlas.  We plot the spectrum as $f_{\lambda}(\lambda)$ versus rest wavelength, normalized in a way that attempts to achieve a balance between showing the full range in flux and illustrating the finer details of the continuum.  Our integrated spectrum is accompanied by a Digitized Sky Survey image which illustrates our rectangular spectroscopic aperture as a \emph{solid} outline and the $25$~mag~arcsec$^{-2}$ isophotal size of the galaxy as a \emph{dashed} ellipse.  The image legend gives the galaxy name, the unique identification number in parenthesis, and the morphological type as listed in Table~\ref{table:general_properties}.  The horizontal solid line in the lower-left corner of each image represents $30\arcsec$.}
\figsetgrpend
 
\figsetgrpstart
\figsetgrpnum{8.126}
\figsetgrptitle{Visualization of MRK 0026.}
\figsetplot{\includegraphics[scale=0.7,angle=90]{f8_126.eps}}
\figsetgrpnote{Presentation of our integrated spectral atlas.  We plot the spectrum as $f_{\lambda}(\lambda)$ versus rest wavelength, normalized in a way that attempts to achieve a balance between showing the full range in flux and illustrating the finer details of the continuum.  Our integrated spectrum is accompanied by a Digitized Sky Survey image which illustrates our rectangular spectroscopic aperture as a \emph{solid} outline and the $25$~mag~arcsec$^{-2}$ isophotal size of the galaxy as a \emph{dashed} ellipse.  The image legend gives the galaxy name, the unique identification number in parenthesis, and the morphological type as listed in Table~\ref{table:general_properties}.  The horizontal solid line in the lower-left corner of each image represents $30\arcsec$.}
\figsetgrpend
 
\figsetgrpstart
\figsetgrpnum{8.127}
\figsetgrptitle{Visualization of UGCA 208.}
\figsetplot{\includegraphics[scale=0.7,angle=90]{f8_127.eps}}
\figsetgrpnote{Presentation of our integrated spectral atlas.  We plot the spectrum as $f_{\lambda}(\lambda)$ versus rest wavelength, normalized in a way that attempts to achieve a balance between showing the full range in flux and illustrating the finer details of the continuum.  Our integrated spectrum is accompanied by a Digitized Sky Survey image which illustrates our rectangular spectroscopic aperture as a \emph{solid} outline and the $25$~mag~arcsec$^{-2}$ isophotal size of the galaxy as a \emph{dashed} ellipse.  The image legend gives the galaxy name, the unique identification number in parenthesis, and the morphological type as listed in Table~\ref{table:general_properties}.  The horizontal solid line in the lower-left corner of each image represents $30\arcsec$.}
\figsetgrpend
 
\figsetgrpstart
\figsetgrpnum{8.128}
\figsetgrptitle{Visualization of NGC 3198.}
\figsetplot{\includegraphics[scale=0.7,angle=90]{f8_128.eps}}
\figsetgrpnote{Presentation of our integrated spectral atlas.  We plot the spectrum as $f_{\lambda}(\lambda)$ versus rest wavelength, normalized in a way that attempts to achieve a balance between showing the full range in flux and illustrating the finer details of the continuum.  Our integrated spectrum is accompanied by a Digitized Sky Survey image which illustrates our rectangular spectroscopic aperture as a \emph{solid} outline and the $25$~mag~arcsec$^{-2}$ isophotal size of the galaxy as a \emph{dashed} ellipse.  The image legend gives the galaxy name, the unique identification number in parenthesis, and the morphological type as listed in Table~\ref{table:general_properties}.  The horizontal solid line in the lower-left corner of each image represents $30\arcsec$.}
\figsetgrpend
 
\figsetgrpstart
\figsetgrpnum{8.129}
\figsetgrptitle{Visualization of NGC 3239.}
\figsetplot{\includegraphics[scale=0.7,angle=90]{f8_129.eps}}
\figsetgrpnote{Presentation of our integrated spectral atlas.  We plot the spectrum as $f_{\lambda}(\lambda)$ versus rest wavelength, normalized in a way that attempts to achieve a balance between showing the full range in flux and illustrating the finer details of the continuum.  Our integrated spectrum is accompanied by a Digitized Sky Survey image which illustrates our rectangular spectroscopic aperture as a \emph{solid} outline and the $25$~mag~arcsec$^{-2}$ isophotal size of the galaxy as a \emph{dashed} ellipse.  The image legend gives the galaxy name, the unique identification number in parenthesis, and the morphological type as listed in Table~\ref{table:general_properties}.  The horizontal solid line in the lower-left corner of each image represents $30\arcsec$.}
\figsetgrpend
 
\figsetgrpstart
\figsetgrpnum{8.130}
\figsetgrptitle{Visualization of NGC 3265.}
\figsetplot{\includegraphics[scale=0.7,angle=90]{f8_130.eps}}
\figsetgrpnote{Presentation of our integrated spectral atlas.  We plot the spectrum as $f_{\lambda}(\lambda)$ versus rest wavelength, normalized in a way that attempts to achieve a balance between showing the full range in flux and illustrating the finer details of the continuum.  Our integrated spectrum is accompanied by a Digitized Sky Survey image which illustrates our rectangular spectroscopic aperture as a \emph{solid} outline and the $25$~mag~arcsec$^{-2}$ isophotal size of the galaxy as a \emph{dashed} ellipse.  The image legend gives the galaxy name, the unique identification number in parenthesis, and the morphological type as listed in Table~\ref{table:general_properties}.  The horizontal solid line in the lower-left corner of each image represents $30\arcsec$.}
\figsetgrpend
 
\figsetgrpstart
\figsetgrpnum{8.131}
\figsetgrptitle{Visualization of NGC 3274.}
\figsetplot{\includegraphics[scale=0.7,angle=90]{f8_131.eps}}
\figsetgrpnote{Presentation of our integrated spectral atlas.  We plot the spectrum as $f_{\lambda}(\lambda)$ versus rest wavelength, normalized in a way that attempts to achieve a balance between showing the full range in flux and illustrating the finer details of the continuum.  Our integrated spectrum is accompanied by a Digitized Sky Survey image which illustrates our rectangular spectroscopic aperture as a \emph{solid} outline and the $25$~mag~arcsec$^{-2}$ isophotal size of the galaxy as a \emph{dashed} ellipse.  The image legend gives the galaxy name, the unique identification number in parenthesis, and the morphological type as listed in Table~\ref{table:general_properties}.  The horizontal solid line in the lower-left corner of each image represents $30\arcsec$.}
\figsetgrpend
 
\figsetgrpstart
\figsetgrpnum{8.132}
\figsetgrptitle{Visualization of UGC 05720.}
\figsetplot{\includegraphics[scale=0.7,angle=90]{f8_132.eps}}
\figsetgrpnote{Presentation of our integrated spectral atlas.  We plot the spectrum as $f_{\lambda}(\lambda)$ versus rest wavelength, normalized in a way that attempts to achieve a balance between showing the full range in flux and illustrating the finer details of the continuum.  Our integrated spectrum is accompanied by a Digitized Sky Survey image which illustrates our rectangular spectroscopic aperture as a \emph{solid} outline and the $25$~mag~arcsec$^{-2}$ isophotal size of the galaxy as a \emph{dashed} ellipse.  The image legend gives the galaxy name, the unique identification number in parenthesis, and the morphological type as listed in Table~\ref{table:general_properties}.  The horizontal solid line in the lower-left corner of each image represents $30\arcsec$.}
\figsetgrpend
 
\figsetgrpstart
\figsetgrpnum{8.133}
\figsetgrptitle{Visualization of NGC 3303.}
\figsetplot{\includegraphics[scale=0.7,angle=90]{f8_133.eps}}
\figsetgrpnote{Presentation of our integrated spectral atlas.  We plot the spectrum as $f_{\lambda}(\lambda)$ versus rest wavelength, normalized in a way that attempts to achieve a balance between showing the full range in flux and illustrating the finer details of the continuum.  Our integrated spectrum is accompanied by a Digitized Sky Survey image which illustrates our rectangular spectroscopic aperture as a \emph{solid} outline and the $25$~mag~arcsec$^{-2}$ isophotal size of the galaxy as a \emph{dashed} ellipse.  The image legend gives the galaxy name, the unique identification number in parenthesis, and the morphological type as listed in Table~\ref{table:general_properties}.  The horizontal solid line in the lower-left corner of each image represents $30\arcsec$.}
\figsetgrpend
 
\figsetgrpstart
\figsetgrpnum{8.134}
\figsetgrptitle{Visualization of NGC 3310.}
\figsetplot{\includegraphics[scale=0.7,angle=90]{f8_134.eps}}
\figsetgrpnote{Presentation of our integrated spectral atlas.  We plot the spectrum as $f_{\lambda}(\lambda)$ versus rest wavelength, normalized in a way that attempts to achieve a balance between showing the full range in flux and illustrating the finer details of the continuum.  Our integrated spectrum is accompanied by a Digitized Sky Survey image which illustrates our rectangular spectroscopic aperture as a \emph{solid} outline and the $25$~mag~arcsec$^{-2}$ isophotal size of the galaxy as a \emph{dashed} ellipse.  The image legend gives the galaxy name, the unique identification number in parenthesis, and the morphological type as listed in Table~\ref{table:general_properties}.  The horizontal solid line in the lower-left corner of each image represents $30\arcsec$.}
\figsetgrpend
 
\figsetgrpstart
\figsetgrpnum{8.135}
\figsetgrptitle{Visualization of NGC 3344.}
\figsetplot{\includegraphics[scale=0.7,angle=90]{f8_135.eps}}
\figsetgrpnote{Presentation of our integrated spectral atlas.  We plot the spectrum as $f_{\lambda}(\lambda)$ versus rest wavelength, normalized in a way that attempts to achieve a balance between showing the full range in flux and illustrating the finer details of the continuum.  Our integrated spectrum is accompanied by a Digitized Sky Survey image which illustrates our rectangular spectroscopic aperture as a \emph{solid} outline and the $25$~mag~arcsec$^{-2}$ isophotal size of the galaxy as a \emph{dashed} ellipse.  The image legend gives the galaxy name, the unique identification number in parenthesis, and the morphological type as listed in Table~\ref{table:general_properties}.  The horizontal solid line in the lower-left corner of each image represents $30\arcsec$.}
\figsetgrpend
 
\figsetgrpstart
\figsetgrpnum{8.136}
\figsetgrptitle{Visualization of NGC 3351.}
\figsetplot{\includegraphics[scale=0.7,angle=90]{f8_136.eps}}
\figsetgrpnote{Presentation of our integrated spectral atlas.  We plot the spectrum as $f_{\lambda}(\lambda)$ versus rest wavelength, normalized in a way that attempts to achieve a balance between showing the full range in flux and illustrating the finer details of the continuum.  Our integrated spectrum is accompanied by a Digitized Sky Survey image which illustrates our rectangular spectroscopic aperture as a \emph{solid} outline and the $25$~mag~arcsec$^{-2}$ isophotal size of the galaxy as a \emph{dashed} ellipse.  The image legend gives the galaxy name, the unique identification number in parenthesis, and the morphological type as listed in Table~\ref{table:general_properties}.  The horizontal solid line in the lower-left corner of each image represents $30\arcsec$.}
\figsetgrpend
 
\figsetgrpstart
\figsetgrpnum{8.137}
\figsetgrptitle{Visualization of NGC 3353.}
\figsetplot{\includegraphics[scale=0.7,angle=90]{f8_137.eps}}
\figsetgrpnote{Presentation of our integrated spectral atlas.  We plot the spectrum as $f_{\lambda}(\lambda)$ versus rest wavelength, normalized in a way that attempts to achieve a balance between showing the full range in flux and illustrating the finer details of the continuum.  Our integrated spectrum is accompanied by a Digitized Sky Survey image which illustrates our rectangular spectroscopic aperture as a \emph{solid} outline and the $25$~mag~arcsec$^{-2}$ isophotal size of the galaxy as a \emph{dashed} ellipse.  The image legend gives the galaxy name, the unique identification number in parenthesis, and the morphological type as listed in Table~\ref{table:general_properties}.  The horizontal solid line in the lower-left corner of each image represents $30\arcsec$.}
\figsetgrpend
 
\figsetgrpstart
\figsetgrpnum{8.138}
\figsetgrptitle{Visualization of NGC 3365.}
\figsetplot{\includegraphics[scale=0.7,angle=90]{f8_138.eps}}
\figsetgrpnote{Presentation of our integrated spectral atlas.  We plot the spectrum as $f_{\lambda}(\lambda)$ versus rest wavelength, normalized in a way that attempts to achieve a balance between showing the full range in flux and illustrating the finer details of the continuum.  Our integrated spectrum is accompanied by a Digitized Sky Survey image which illustrates our rectangular spectroscopic aperture as a \emph{solid} outline and the $25$~mag~arcsec$^{-2}$ isophotal size of the galaxy as a \emph{dashed} ellipse.  The image legend gives the galaxy name, the unique identification number in parenthesis, and the morphological type as listed in Table~\ref{table:general_properties}.  The horizontal solid line in the lower-left corner of each image represents $30\arcsec$.}
\figsetgrpend
 
\figsetgrpstart
\figsetgrpnum{8.139}
\figsetgrptitle{Visualization of NGC 3367.}
\figsetplot{\includegraphics[scale=0.7,angle=90]{f8_139.eps}}
\figsetgrpnote{Presentation of our integrated spectral atlas.  We plot the spectrum as $f_{\lambda}(\lambda)$ versus rest wavelength, normalized in a way that attempts to achieve a balance between showing the full range in flux and illustrating the finer details of the continuum.  Our integrated spectrum is accompanied by a Digitized Sky Survey image which illustrates our rectangular spectroscopic aperture as a \emph{solid} outline and the $25$~mag~arcsec$^{-2}$ isophotal size of the galaxy as a \emph{dashed} ellipse.  The image legend gives the galaxy name, the unique identification number in parenthesis, and the morphological type as listed in Table~\ref{table:general_properties}.  The horizontal solid line in the lower-left corner of each image represents $30\arcsec$.}
\figsetgrpend
 
\figsetgrpstart
\figsetgrpnum{8.140}
\figsetgrptitle{Visualization of UGCA 219.}
\figsetplot{\includegraphics[scale=0.7,angle=90]{f8_140.eps}}
\figsetgrpnote{Presentation of our integrated spectral atlas.  We plot the spectrum as $f_{\lambda}(\lambda)$ versus rest wavelength, normalized in a way that attempts to achieve a balance between showing the full range in flux and illustrating the finer details of the continuum.  Our integrated spectrum is accompanied by a Digitized Sky Survey image which illustrates our rectangular spectroscopic aperture as a \emph{solid} outline and the $25$~mag~arcsec$^{-2}$ isophotal size of the galaxy as a \emph{dashed} ellipse.  The image legend gives the galaxy name, the unique identification number in parenthesis, and the morphological type as listed in Table~\ref{table:general_properties}.  The horizontal solid line in the lower-left corner of each image represents $30\arcsec$.}
\figsetgrpend
 
\figsetgrpstart
\figsetgrpnum{8.141}
\figsetgrptitle{Visualization of NGC 3395.}
\figsetplot{\includegraphics[scale=0.7,angle=90]{f8_141.eps}}
\figsetgrpnote{Presentation of our integrated spectral atlas.  We plot the spectrum as $f_{\lambda}(\lambda)$ versus rest wavelength, normalized in a way that attempts to achieve a balance between showing the full range in flux and illustrating the finer details of the continuum.  Our integrated spectrum is accompanied by a Digitized Sky Survey image which illustrates our rectangular spectroscopic aperture as a \emph{solid} outline and the $25$~mag~arcsec$^{-2}$ isophotal size of the galaxy as a \emph{dashed} ellipse.  The image legend gives the galaxy name, the unique identification number in parenthesis, and the morphological type as listed in Table~\ref{table:general_properties}.  The horizontal solid line in the lower-left corner of each image represents $30\arcsec$.}
\figsetgrpend
 
\figsetgrpstart
\figsetgrpnum{8.142}
\figsetgrptitle{Visualization of ARP 270.}
\figsetplot{\includegraphics[scale=0.7,angle=90]{f8_142.eps}}
\figsetgrpnote{Presentation of our integrated spectral atlas.  We plot the spectrum as $f_{\lambda}(\lambda)$ versus rest wavelength, normalized in a way that attempts to achieve a balance between showing the full range in flux and illustrating the finer details of the continuum.  Our integrated spectrum is accompanied by a Digitized Sky Survey image which illustrates our rectangular spectroscopic aperture as a \emph{solid} outline and the $25$~mag~arcsec$^{-2}$ isophotal size of the galaxy as a \emph{dashed} ellipse.  The image legend gives the galaxy name, the unique identification number in parenthesis, and the morphological type as listed in Table~\ref{table:general_properties}.  The horizontal solid line in the lower-left corner of each image represents $30\arcsec$.}
\figsetgrpend
 
\figsetgrpstart
\figsetgrpnum{8.143}
\figsetgrptitle{Visualization of NGC 3396.}
\figsetplot{\includegraphics[scale=0.7,angle=90]{f8_143.eps}}
\figsetgrpnote{Presentation of our integrated spectral atlas.  We plot the spectrum as $f_{\lambda}(\lambda)$ versus rest wavelength, normalized in a way that attempts to achieve a balance between showing the full range in flux and illustrating the finer details of the continuum.  Our integrated spectrum is accompanied by a Digitized Sky Survey image which illustrates our rectangular spectroscopic aperture as a \emph{solid} outline and the $25$~mag~arcsec$^{-2}$ isophotal size of the galaxy as a \emph{dashed} ellipse.  The image legend gives the galaxy name, the unique identification number in parenthesis, and the morphological type as listed in Table~\ref{table:general_properties}.  The horizontal solid line in the lower-left corner of each image represents $30\arcsec$.}
\figsetgrpend
 
\figsetgrpstart
\figsetgrpnum{8.144}
\figsetgrptitle{Visualization of NGC 3432.}
\figsetplot{\includegraphics[scale=0.7,angle=90]{f8_144.eps}}
\figsetgrpnote{Presentation of our integrated spectral atlas.  We plot the spectrum as $f_{\lambda}(\lambda)$ versus rest wavelength, normalized in a way that attempts to achieve a balance between showing the full range in flux and illustrating the finer details of the continuum.  Our integrated spectrum is accompanied by a Digitized Sky Survey image which illustrates our rectangular spectroscopic aperture as a \emph{solid} outline and the $25$~mag~arcsec$^{-2}$ isophotal size of the galaxy as a \emph{dashed} ellipse.  The image legend gives the galaxy name, the unique identification number in parenthesis, and the morphological type as listed in Table~\ref{table:general_properties}.  The horizontal solid line in the lower-left corner of each image represents $30\arcsec$.}
\figsetgrpend
 
\figsetgrpstart
\figsetgrpnum{8.145}
\figsetgrptitle{Visualization of NGC 3442.}
\figsetplot{\includegraphics[scale=0.7,angle=90]{f8_145.eps}}
\figsetgrpnote{Presentation of our integrated spectral atlas.  We plot the spectrum as $f_{\lambda}(\lambda)$ versus rest wavelength, normalized in a way that attempts to achieve a balance between showing the full range in flux and illustrating the finer details of the continuum.  Our integrated spectrum is accompanied by a Digitized Sky Survey image which illustrates our rectangular spectroscopic aperture as a \emph{solid} outline and the $25$~mag~arcsec$^{-2}$ isophotal size of the galaxy as a \emph{dashed} ellipse.  The image legend gives the galaxy name, the unique identification number in parenthesis, and the morphological type as listed in Table~\ref{table:general_properties}.  The horizontal solid line in the lower-left corner of each image represents $30\arcsec$.}
\figsetgrpend
 
\figsetgrpstart
\figsetgrpnum{8.146}
\figsetgrptitle{Visualization of UGC 05998 W.}
\figsetplot{\includegraphics[scale=0.7,angle=90]{f8_146.eps}}
\figsetgrpnote{Presentation of our integrated spectral atlas.  We plot the spectrum as $f_{\lambda}(\lambda)$ versus rest wavelength, normalized in a way that attempts to achieve a balance between showing the full range in flux and illustrating the finer details of the continuum.  Our integrated spectrum is accompanied by a Digitized Sky Survey image which illustrates our rectangular spectroscopic aperture as a \emph{solid} outline and the $25$~mag~arcsec$^{-2}$ isophotal size of the galaxy as a \emph{dashed} ellipse.  The image legend gives the galaxy name, the unique identification number in parenthesis, and the morphological type as listed in Table~\ref{table:general_properties}.  The horizontal solid line in the lower-left corner of each image represents $30\arcsec$.}
\figsetgrpend
 
\figsetgrpstart
\figsetgrpnum{8.147}
\figsetgrptitle{Visualization of UGC 05998.}
\figsetplot{\includegraphics[scale=0.7,angle=90]{f8_147.eps}}
\figsetgrpnote{Presentation of our integrated spectral atlas.  We plot the spectrum as $f_{\lambda}(\lambda)$ versus rest wavelength, normalized in a way that attempts to achieve a balance between showing the full range in flux and illustrating the finer details of the continuum.  Our integrated spectrum is accompanied by a Digitized Sky Survey image which illustrates our rectangular spectroscopic aperture as a \emph{solid} outline and the $25$~mag~arcsec$^{-2}$ isophotal size of the galaxy as a \emph{dashed} ellipse.  The image legend gives the galaxy name, the unique identification number in parenthesis, and the morphological type as listed in Table~\ref{table:general_properties}.  The horizontal solid line in the lower-left corner of each image represents $30\arcsec$.}
\figsetgrpend
 
\figsetgrpstart
\figsetgrpnum{8.148}
\figsetgrptitle{Visualization of UGC 05998 E.}
\figsetplot{\includegraphics[scale=0.7,angle=90]{f8_148.eps}}
\figsetgrpnote{Presentation of our integrated spectral atlas.  We plot the spectrum as $f_{\lambda}(\lambda)$ versus rest wavelength, normalized in a way that attempts to achieve a balance between showing the full range in flux and illustrating the finer details of the continuum.  Our integrated spectrum is accompanied by a Digitized Sky Survey image which illustrates our rectangular spectroscopic aperture as a \emph{solid} outline and the $25$~mag~arcsec$^{-2}$ isophotal size of the galaxy as a \emph{dashed} ellipse.  The image legend gives the galaxy name, the unique identification number in parenthesis, and the morphological type as listed in Table~\ref{table:general_properties}.  The horizontal solid line in the lower-left corner of each image represents $30\arcsec$.}
\figsetgrpend
 
\figsetgrpstart
\figsetgrpnum{8.149}
\figsetgrptitle{Visualization of NGC 3448.}
\figsetplot{\includegraphics[scale=0.7,angle=90]{f8_149.eps}}
\figsetgrpnote{Presentation of our integrated spectral atlas.  We plot the spectrum as $f_{\lambda}(\lambda)$ versus rest wavelength, normalized in a way that attempts to achieve a balance between showing the full range in flux and illustrating the finer details of the continuum.  Our integrated spectrum is accompanied by a Digitized Sky Survey image which illustrates our rectangular spectroscopic aperture as a \emph{solid} outline and the $25$~mag~arcsec$^{-2}$ isophotal size of the galaxy as a \emph{dashed} ellipse.  The image legend gives the galaxy name, the unique identification number in parenthesis, and the morphological type as listed in Table~\ref{table:general_properties}.  The horizontal solid line in the lower-left corner of each image represents $30\arcsec$.}
\figsetgrpend
 
\figsetgrpstart
\figsetgrpnum{8.150}
\figsetgrptitle{Visualization of UGC 06029.}
\figsetplot{\includegraphics[scale=0.7,angle=90]{f8_150.eps}}
\figsetgrpnote{Presentation of our integrated spectral atlas.  We plot the spectrum as $f_{\lambda}(\lambda)$ versus rest wavelength, normalized in a way that attempts to achieve a balance between showing the full range in flux and illustrating the finer details of the continuum.  Our integrated spectrum is accompanied by a Digitized Sky Survey image which illustrates our rectangular spectroscopic aperture as a \emph{solid} outline and the $25$~mag~arcsec$^{-2}$ isophotal size of the galaxy as a \emph{dashed} ellipse.  The image legend gives the galaxy name, the unique identification number in parenthesis, and the morphological type as listed in Table~\ref{table:general_properties}.  The horizontal solid line in the lower-left corner of each image represents $30\arcsec$.}
\figsetgrpend
 
\figsetgrpstart
\figsetgrpnum{8.151}
\figsetgrptitle{Visualization of UGC 06103.}
\figsetplot{\includegraphics[scale=0.7,angle=90]{f8_151.eps}}
\figsetgrpnote{Presentation of our integrated spectral atlas.  We plot the spectrum as $f_{\lambda}(\lambda)$ versus rest wavelength, normalized in a way that attempts to achieve a balance between showing the full range in flux and illustrating the finer details of the continuum.  Our integrated spectrum is accompanied by a Digitized Sky Survey image which illustrates our rectangular spectroscopic aperture as a \emph{solid} outline and the $25$~mag~arcsec$^{-2}$ isophotal size of the galaxy as a \emph{dashed} ellipse.  The image legend gives the galaxy name, the unique identification number in parenthesis, and the morphological type as listed in Table~\ref{table:general_properties}.  The horizontal solid line in the lower-left corner of each image represents $30\arcsec$.}
\figsetgrpend
 
\figsetgrpstart
\figsetgrpnum{8.152}
\figsetgrptitle{Visualization of NGC 3504.}
\figsetplot{\includegraphics[scale=0.7,angle=90]{f8_152.eps}}
\figsetgrpnote{Presentation of our integrated spectral atlas.  We plot the spectrum as $f_{\lambda}(\lambda)$ versus rest wavelength, normalized in a way that attempts to achieve a balance between showing the full range in flux and illustrating the finer details of the continuum.  Our integrated spectrum is accompanied by a Digitized Sky Survey image which illustrates our rectangular spectroscopic aperture as a \emph{solid} outline and the $25$~mag~arcsec$^{-2}$ isophotal size of the galaxy as a \emph{dashed} ellipse.  The image legend gives the galaxy name, the unique identification number in parenthesis, and the morphological type as listed in Table~\ref{table:general_properties}.  The horizontal solid line in the lower-left corner of each image represents $30\arcsec$.}
\figsetgrpend
 
\figsetgrpstart
\figsetgrpnum{8.153}
\figsetgrptitle{Visualization of NGC 3510.}
\figsetplot{\includegraphics[scale=0.7,angle=90]{f8_153.eps}}
\figsetgrpnote{Presentation of our integrated spectral atlas.  We plot the spectrum as $f_{\lambda}(\lambda)$ versus rest wavelength, normalized in a way that attempts to achieve a balance between showing the full range in flux and illustrating the finer details of the continuum.  Our integrated spectrum is accompanied by a Digitized Sky Survey image which illustrates our rectangular spectroscopic aperture as a \emph{solid} outline and the $25$~mag~arcsec$^{-2}$ isophotal size of the galaxy as a \emph{dashed} ellipse.  The image legend gives the galaxy name, the unique identification number in parenthesis, and the morphological type as listed in Table~\ref{table:general_properties}.  The horizontal solid line in the lower-left corner of each image represents $30\arcsec$.}
\figsetgrpend
 
\figsetgrpstart
\figsetgrpnum{8.154}
\figsetgrptitle{Visualization of UGCA 225.}
\figsetplot{\includegraphics[scale=0.7,angle=90]{f8_154.eps}}
\figsetgrpnote{Presentation of our integrated spectral atlas.  We plot the spectrum as $f_{\lambda}(\lambda)$ versus rest wavelength, normalized in a way that attempts to achieve a balance between showing the full range in flux and illustrating the finer details of the continuum.  Our integrated spectrum is accompanied by a Digitized Sky Survey image which illustrates our rectangular spectroscopic aperture as a \emph{solid} outline and the $25$~mag~arcsec$^{-2}$ isophotal size of the galaxy as a \emph{dashed} ellipse.  The image legend gives the galaxy name, the unique identification number in parenthesis, and the morphological type as listed in Table~\ref{table:general_properties}.  The horizontal solid line in the lower-left corner of each image represents $30\arcsec$.}
\figsetgrpend
 
\figsetgrpstart
\figsetgrpnum{8.155}
\figsetgrptitle{Visualization of NGC 3521.}
\figsetplot{\includegraphics[scale=0.7,angle=90]{f8_155.eps}}
\figsetgrpnote{Presentation of our integrated spectral atlas.  We plot the spectrum as $f_{\lambda}(\lambda)$ versus rest wavelength, normalized in a way that attempts to achieve a balance between showing the full range in flux and illustrating the finer details of the continuum.  Our integrated spectrum is accompanied by a Digitized Sky Survey image which illustrates our rectangular spectroscopic aperture as a \emph{solid} outline and the $25$~mag~arcsec$^{-2}$ isophotal size of the galaxy as a \emph{dashed} ellipse.  The image legend gives the galaxy name, the unique identification number in parenthesis, and the morphological type as listed in Table~\ref{table:general_properties}.  The horizontal solid line in the lower-left corner of each image represents $30\arcsec$.}
\figsetgrpend
 
\figsetgrpstart
\figsetgrpnum{8.156}
\figsetgrptitle{Visualization of NGC 3600.}
\figsetplot{\includegraphics[scale=0.7,angle=90]{f8_156.eps}}
\figsetgrpnote{Presentation of our integrated spectral atlas.  We plot the spectrum as $f_{\lambda}(\lambda)$ versus rest wavelength, normalized in a way that attempts to achieve a balance between showing the full range in flux and illustrating the finer details of the continuum.  Our integrated spectrum is accompanied by a Digitized Sky Survey image which illustrates our rectangular spectroscopic aperture as a \emph{solid} outline and the $25$~mag~arcsec$^{-2}$ isophotal size of the galaxy as a \emph{dashed} ellipse.  The image legend gives the galaxy name, the unique identification number in parenthesis, and the morphological type as listed in Table~\ref{table:general_properties}.  The horizontal solid line in the lower-left corner of each image represents $30\arcsec$.}
\figsetgrpend
 
\figsetgrpstart
\figsetgrpnum{8.157}
\figsetgrptitle{Visualization of NGC 3628.}
\figsetplot{\includegraphics[scale=0.7,angle=90]{f8_157.eps}}
\figsetgrpnote{Presentation of our integrated spectral atlas.  We plot the spectrum as $f_{\lambda}(\lambda)$ versus rest wavelength, normalized in a way that attempts to achieve a balance between showing the full range in flux and illustrating the finer details of the continuum.  Our integrated spectrum is accompanied by a Digitized Sky Survey image which illustrates our rectangular spectroscopic aperture as a \emph{solid} outline and the $25$~mag~arcsec$^{-2}$ isophotal size of the galaxy as a \emph{dashed} ellipse.  The image legend gives the galaxy name, the unique identification number in parenthesis, and the morphological type as listed in Table~\ref{table:general_properties}.  The horizontal solid line in the lower-left corner of each image represents $30\arcsec$.}
\figsetgrpend
 
\figsetgrpstart
\figsetgrpnum{8.158}
\figsetgrptitle{Visualization of UGC 06399.}
\figsetplot{\includegraphics[scale=0.7,angle=90]{f8_158.eps}}
\figsetgrpnote{Presentation of our integrated spectral atlas.  We plot the spectrum as $f_{\lambda}(\lambda)$ versus rest wavelength, normalized in a way that attempts to achieve a balance between showing the full range in flux and illustrating the finer details of the continuum.  Our integrated spectrum is accompanied by a Digitized Sky Survey image which illustrates our rectangular spectroscopic aperture as a \emph{solid} outline and the $25$~mag~arcsec$^{-2}$ isophotal size of the galaxy as a \emph{dashed} ellipse.  The image legend gives the galaxy name, the unique identification number in parenthesis, and the morphological type as listed in Table~\ref{table:general_properties}.  The horizontal solid line in the lower-left corner of each image represents $30\arcsec$.}
\figsetgrpend
 
\figsetgrpstart
\figsetgrpnum{8.159}
\figsetgrptitle{Visualization of NGC 3656.}
\figsetplot{\includegraphics[scale=0.7,angle=90]{f8_159.eps}}
\figsetgrpnote{Presentation of our integrated spectral atlas.  We plot the spectrum as $f_{\lambda}(\lambda)$ versus rest wavelength, normalized in a way that attempts to achieve a balance between showing the full range in flux and illustrating the finer details of the continuum.  Our integrated spectrum is accompanied by a Digitized Sky Survey image which illustrates our rectangular spectroscopic aperture as a \emph{solid} outline and the $25$~mag~arcsec$^{-2}$ isophotal size of the galaxy as a \emph{dashed} ellipse.  The image legend gives the galaxy name, the unique identification number in parenthesis, and the morphological type as listed in Table~\ref{table:general_properties}.  The horizontal solid line in the lower-left corner of each image represents $30\arcsec$.}
\figsetgrpend
 
\figsetgrpstart
\figsetgrpnum{8.160}
\figsetgrptitle{Visualization of IC 2810.}
\figsetplot{\includegraphics[scale=0.7,angle=90]{f8_160.eps}}
\figsetgrpnote{Presentation of our integrated spectral atlas.  We plot the spectrum as $f_{\lambda}(\lambda)$ versus rest wavelength, normalized in a way that attempts to achieve a balance between showing the full range in flux and illustrating the finer details of the continuum.  Our integrated spectrum is accompanied by a Digitized Sky Survey image which illustrates our rectangular spectroscopic aperture as a \emph{solid} outline and the $25$~mag~arcsec$^{-2}$ isophotal size of the galaxy as a \emph{dashed} ellipse.  The image legend gives the galaxy name, the unique identification number in parenthesis, and the morphological type as listed in Table~\ref{table:general_properties}.  The horizontal solid line in the lower-left corner of each image represents $30\arcsec$.}
\figsetgrpend
 
\figsetgrpstart
\figsetgrpnum{8.161}
\figsetgrptitle{Visualization of UGC 06446.}
\figsetplot{\includegraphics[scale=0.7,angle=90]{f8_161.eps}}
\figsetgrpnote{Presentation of our integrated spectral atlas.  We plot the spectrum as $f_{\lambda}(\lambda)$ versus rest wavelength, normalized in a way that attempts to achieve a balance between showing the full range in flux and illustrating the finer details of the continuum.  Our integrated spectrum is accompanied by a Digitized Sky Survey image which illustrates our rectangular spectroscopic aperture as a \emph{solid} outline and the $25$~mag~arcsec$^{-2}$ isophotal size of the galaxy as a \emph{dashed} ellipse.  The image legend gives the galaxy name, the unique identification number in parenthesis, and the morphological type as listed in Table~\ref{table:general_properties}.  The horizontal solid line in the lower-left corner of each image represents $30\arcsec$.}
\figsetgrpend
 
\figsetgrpstart
\figsetgrpnum{8.162}
\figsetgrptitle{Visualization of IC 0691.}
\figsetplot{\includegraphics[scale=0.7,angle=90]{f8_162.eps}}
\figsetgrpnote{Presentation of our integrated spectral atlas.  We plot the spectrum as $f_{\lambda}(\lambda)$ versus rest wavelength, normalized in a way that attempts to achieve a balance between showing the full range in flux and illustrating the finer details of the continuum.  Our integrated spectrum is accompanied by a Digitized Sky Survey image which illustrates our rectangular spectroscopic aperture as a \emph{solid} outline and the $25$~mag~arcsec$^{-2}$ isophotal size of the galaxy as a \emph{dashed} ellipse.  The image legend gives the galaxy name, the unique identification number in parenthesis, and the morphological type as listed in Table~\ref{table:general_properties}.  The horizontal solid line in the lower-left corner of each image represents $30\arcsec$.}
\figsetgrpend
 
\figsetgrpstart
\figsetgrpnum{8.163}
\figsetgrptitle{Visualization of UGC 06448.}
\figsetplot{\includegraphics[scale=0.7,angle=90]{f8_163.eps}}
\figsetgrpnote{Presentation of our integrated spectral atlas.  We plot the spectrum as $f_{\lambda}(\lambda)$ versus rest wavelength, normalized in a way that attempts to achieve a balance between showing the full range in flux and illustrating the finer details of the continuum.  Our integrated spectrum is accompanied by a Digitized Sky Survey image which illustrates our rectangular spectroscopic aperture as a \emph{solid} outline and the $25$~mag~arcsec$^{-2}$ isophotal size of the galaxy as a \emph{dashed} ellipse.  The image legend gives the galaxy name, the unique identification number in parenthesis, and the morphological type as listed in Table~\ref{table:general_properties}.  The horizontal solid line in the lower-left corner of each image represents $30\arcsec$.}
\figsetgrpend
 
\figsetgrpstart
\figsetgrpnum{8.164}
\figsetgrptitle{Visualization of UGC 06456.}
\figsetplot{\includegraphics[scale=0.7,angle=90]{f8_164.eps}}
\figsetgrpnote{Presentation of our integrated spectral atlas.  We plot the spectrum as $f_{\lambda}(\lambda)$ versus rest wavelength, normalized in a way that attempts to achieve a balance between showing the full range in flux and illustrating the finer details of the continuum.  Our integrated spectrum is accompanied by a Digitized Sky Survey image which illustrates our rectangular spectroscopic aperture as a \emph{solid} outline and the $25$~mag~arcsec$^{-2}$ isophotal size of the galaxy as a \emph{dashed} ellipse.  The image legend gives the galaxy name, the unique identification number in parenthesis, and the morphological type as listed in Table~\ref{table:general_properties}.  The horizontal solid line in the lower-left corner of each image represents $30\arcsec$.}
\figsetgrpend
 
\figsetgrpstart
\figsetgrpnum{8.165}
\figsetgrptitle{Visualization of NGC 3690 SW.}
\figsetplot{\includegraphics[scale=0.7,angle=90]{f8_165.eps}}
\figsetgrpnote{Presentation of our integrated spectral atlas.  We plot the spectrum as $f_{\lambda}(\lambda)$ versus rest wavelength, normalized in a way that attempts to achieve a balance between showing the full range in flux and illustrating the finer details of the continuum.  Our integrated spectrum is accompanied by a Digitized Sky Survey image which illustrates our rectangular spectroscopic aperture as a \emph{solid} outline and the $25$~mag~arcsec$^{-2}$ isophotal size of the galaxy as a \emph{dashed} ellipse.  The image legend gives the galaxy name, the unique identification number in parenthesis, and the morphological type as listed in Table~\ref{table:general_properties}.  The horizontal solid line in the lower-left corner of each image represents $30\arcsec$.}
\figsetgrpend
 
\figsetgrpstart
\figsetgrpnum{8.166}
\figsetgrptitle{Visualization of NGC 3690.}
\figsetplot{\includegraphics[scale=0.7,angle=90]{f8_166.eps}}
\figsetgrpnote{Presentation of our integrated spectral atlas.  We plot the spectrum as $f_{\lambda}(\lambda)$ versus rest wavelength, normalized in a way that attempts to achieve a balance between showing the full range in flux and illustrating the finer details of the continuum.  Our integrated spectrum is accompanied by a Digitized Sky Survey image which illustrates our rectangular spectroscopic aperture as a \emph{solid} outline and the $25$~mag~arcsec$^{-2}$ isophotal size of the galaxy as a \emph{dashed} ellipse.  The image legend gives the galaxy name, the unique identification number in parenthesis, and the morphological type as listed in Table~\ref{table:general_properties}.  The horizontal solid line in the lower-left corner of each image represents $30\arcsec$.}
\figsetgrpend
 
\figsetgrpstart
\figsetgrpnum{8.167}
\figsetgrptitle{Visualization of NGC 3690 NE.}
\figsetplot{\includegraphics[scale=0.7,angle=90]{f8_167.eps}}
\figsetgrpnote{Presentation of our integrated spectral atlas.  We plot the spectrum as $f_{\lambda}(\lambda)$ versus rest wavelength, normalized in a way that attempts to achieve a balance between showing the full range in flux and illustrating the finer details of the continuum.  Our integrated spectrum is accompanied by a Digitized Sky Survey image which illustrates our rectangular spectroscopic aperture as a \emph{solid} outline and the $25$~mag~arcsec$^{-2}$ isophotal size of the galaxy as a \emph{dashed} ellipse.  The image legend gives the galaxy name, the unique identification number in parenthesis, and the morphological type as listed in Table~\ref{table:general_properties}.  The horizontal solid line in the lower-left corner of each image represents $30\arcsec$.}
\figsetgrpend
 
\figsetgrpstart
\figsetgrpnum{8.168}
\figsetgrptitle{Visualization of UGC 06520.}
\figsetplot{\includegraphics[scale=0.7,angle=90]{f8_168.eps}}
\figsetgrpnote{Presentation of our integrated spectral atlas.  We plot the spectrum as $f_{\lambda}(\lambda)$ versus rest wavelength, normalized in a way that attempts to achieve a balance between showing the full range in flux and illustrating the finer details of the continuum.  Our integrated spectrum is accompanied by a Digitized Sky Survey image which illustrates our rectangular spectroscopic aperture as a \emph{solid} outline and the $25$~mag~arcsec$^{-2}$ isophotal size of the galaxy as a \emph{dashed} ellipse.  The image legend gives the galaxy name, the unique identification number in parenthesis, and the morphological type as listed in Table~\ref{table:general_properties}.  The horizontal solid line in the lower-left corner of each image represents $30\arcsec$.}
\figsetgrpend
 
\figsetgrpstart
\figsetgrpnum{8.169}
\figsetgrptitle{Visualization of NGC 3718.}
\figsetplot{\includegraphics[scale=0.7,angle=90]{f8_169.eps}}
\figsetgrpnote{Presentation of our integrated spectral atlas.  We plot the spectrum as $f_{\lambda}(\lambda)$ versus rest wavelength, normalized in a way that attempts to achieve a balance between showing the full range in flux and illustrating the finer details of the continuum.  Our integrated spectrum is accompanied by a Digitized Sky Survey image which illustrates our rectangular spectroscopic aperture as a \emph{solid} outline and the $25$~mag~arcsec$^{-2}$ isophotal size of the galaxy as a \emph{dashed} ellipse.  The image legend gives the galaxy name, the unique identification number in parenthesis, and the morphological type as listed in Table~\ref{table:general_properties}.  The horizontal solid line in the lower-left corner of each image represents $30\arcsec$.}
\figsetgrpend
 
\figsetgrpstart
\figsetgrpnum{8.170}
\figsetgrptitle{Visualization of NGC 3726.}
\figsetplot{\includegraphics[scale=0.7,angle=90]{f8_170.eps}}
\figsetgrpnote{Presentation of our integrated spectral atlas.  We plot the spectrum as $f_{\lambda}(\lambda)$ versus rest wavelength, normalized in a way that attempts to achieve a balance between showing the full range in flux and illustrating the finer details of the continuum.  Our integrated spectrum is accompanied by a Digitized Sky Survey image which illustrates our rectangular spectroscopic aperture as a \emph{solid} outline and the $25$~mag~arcsec$^{-2}$ isophotal size of the galaxy as a \emph{dashed} ellipse.  The image legend gives the galaxy name, the unique identification number in parenthesis, and the morphological type as listed in Table~\ref{table:general_properties}.  The horizontal solid line in the lower-left corner of each image represents $30\arcsec$.}
\figsetgrpend
 
\figsetgrpstart
\figsetgrpnum{8.171}
\figsetgrptitle{Visualization of UGC 06541.}
\figsetplot{\includegraphics[scale=0.7,angle=90]{f8_171.eps}}
\figsetgrpnote{Presentation of our integrated spectral atlas.  We plot the spectrum as $f_{\lambda}(\lambda)$ versus rest wavelength, normalized in a way that attempts to achieve a balance between showing the full range in flux and illustrating the finer details of the continuum.  Our integrated spectrum is accompanied by a Digitized Sky Survey image which illustrates our rectangular spectroscopic aperture as a \emph{solid} outline and the $25$~mag~arcsec$^{-2}$ isophotal size of the galaxy as a \emph{dashed} ellipse.  The image legend gives the galaxy name, the unique identification number in parenthesis, and the morphological type as listed in Table~\ref{table:general_properties}.  The horizontal solid line in the lower-left corner of each image represents $30\arcsec$.}
\figsetgrpend
 
\figsetgrpstart
\figsetgrpnum{8.172}
\figsetgrptitle{Visualization of NGC 3729.}
\figsetplot{\includegraphics[scale=0.7,angle=90]{f8_172.eps}}
\figsetgrpnote{Presentation of our integrated spectral atlas.  We plot the spectrum as $f_{\lambda}(\lambda)$ versus rest wavelength, normalized in a way that attempts to achieve a balance between showing the full range in flux and illustrating the finer details of the continuum.  Our integrated spectrum is accompanied by a Digitized Sky Survey image which illustrates our rectangular spectroscopic aperture as a \emph{solid} outline and the $25$~mag~arcsec$^{-2}$ isophotal size of the galaxy as a \emph{dashed} ellipse.  The image legend gives the galaxy name, the unique identification number in parenthesis, and the morphological type as listed in Table~\ref{table:general_properties}.  The horizontal solid line in the lower-left corner of each image represents $30\arcsec$.}
\figsetgrpend
 
\figsetgrpstart
\figsetgrpnum{8.173}
\figsetgrptitle{Visualization of NGC 3738.}
\figsetplot{\includegraphics[scale=0.7,angle=90]{f8_173.eps}}
\figsetgrpnote{Presentation of our integrated spectral atlas.  We plot the spectrum as $f_{\lambda}(\lambda)$ versus rest wavelength, normalized in a way that attempts to achieve a balance between showing the full range in flux and illustrating the finer details of the continuum.  Our integrated spectrum is accompanied by a Digitized Sky Survey image which illustrates our rectangular spectroscopic aperture as a \emph{solid} outline and the $25$~mag~arcsec$^{-2}$ isophotal size of the galaxy as a \emph{dashed} ellipse.  The image legend gives the galaxy name, the unique identification number in parenthesis, and the morphological type as listed in Table~\ref{table:general_properties}.  The horizontal solid line in the lower-left corner of each image represents $30\arcsec$.}
\figsetgrpend
 
\figsetgrpstart
\figsetgrpnum{8.174}
\figsetgrptitle{Visualization of NGC 3741.}
\figsetplot{\includegraphics[scale=0.7,angle=90]{f8_174.eps}}
\figsetgrpnote{Presentation of our integrated spectral atlas.  We plot the spectrum as $f_{\lambda}(\lambda)$ versus rest wavelength, normalized in a way that attempts to achieve a balance between showing the full range in flux and illustrating the finer details of the continuum.  Our integrated spectrum is accompanied by a Digitized Sky Survey image which illustrates our rectangular spectroscopic aperture as a \emph{solid} outline and the $25$~mag~arcsec$^{-2}$ isophotal size of the galaxy as a \emph{dashed} ellipse.  The image legend gives the galaxy name, the unique identification number in parenthesis, and the morphological type as listed in Table~\ref{table:general_properties}.  The horizontal solid line in the lower-left corner of each image represents $30\arcsec$.}
\figsetgrpend
 
\figsetgrpstart
\figsetgrpnum{8.175}
\figsetgrptitle{Visualization of NGC 3769.}
\figsetplot{\includegraphics[scale=0.7,angle=90]{f8_175.eps}}
\figsetgrpnote{Presentation of our integrated spectral atlas.  We plot the spectrum as $f_{\lambda}(\lambda)$ versus rest wavelength, normalized in a way that attempts to achieve a balance between showing the full range in flux and illustrating the finer details of the continuum.  Our integrated spectrum is accompanied by a Digitized Sky Survey image which illustrates our rectangular spectroscopic aperture as a \emph{solid} outline and the $25$~mag~arcsec$^{-2}$ isophotal size of the galaxy as a \emph{dashed} ellipse.  The image legend gives the galaxy name, the unique identification number in parenthesis, and the morphological type as listed in Table~\ref{table:general_properties}.  The horizontal solid line in the lower-left corner of each image represents $30\arcsec$.}
\figsetgrpend
 
\figsetgrpstart
\figsetgrpnum{8.176}
\figsetgrptitle{Visualization of ARP 280.}
\figsetplot{\includegraphics[scale=0.7,angle=90]{f8_176.eps}}
\figsetgrpnote{Presentation of our integrated spectral atlas.  We plot the spectrum as $f_{\lambda}(\lambda)$ versus rest wavelength, normalized in a way that attempts to achieve a balance between showing the full range in flux and illustrating the finer details of the continuum.  Our integrated spectrum is accompanied by a Digitized Sky Survey image which illustrates our rectangular spectroscopic aperture as a \emph{solid} outline and the $25$~mag~arcsec$^{-2}$ isophotal size of the galaxy as a \emph{dashed} ellipse.  The image legend gives the galaxy name, the unique identification number in parenthesis, and the morphological type as listed in Table~\ref{table:general_properties}.  The horizontal solid line in the lower-left corner of each image represents $30\arcsec$.}
\figsetgrpend
 
\figsetgrpstart
\figsetgrpnum{8.177}
\figsetgrptitle{Visualization of NGC 3769 A.}
\figsetplot{\includegraphics[scale=0.7,angle=90]{f8_177.eps}}
\figsetgrpnote{Presentation of our integrated spectral atlas.  We plot the spectrum as $f_{\lambda}(\lambda)$ versus rest wavelength, normalized in a way that attempts to achieve a balance between showing the full range in flux and illustrating the finer details of the continuum.  Our integrated spectrum is accompanied by a Digitized Sky Survey image which illustrates our rectangular spectroscopic aperture as a \emph{solid} outline and the $25$~mag~arcsec$^{-2}$ isophotal size of the galaxy as a \emph{dashed} ellipse.  The image legend gives the galaxy name, the unique identification number in parenthesis, and the morphological type as listed in Table~\ref{table:general_properties}.  The horizontal solid line in the lower-left corner of each image represents $30\arcsec$.}
\figsetgrpend
 
\figsetgrpstart
\figsetgrpnum{8.178}
\figsetgrptitle{Visualization of NGC 3773.}
\figsetplot{\includegraphics[scale=0.7,angle=90]{f8_178.eps}}
\figsetgrpnote{Presentation of our integrated spectral atlas.  We plot the spectrum as $f_{\lambda}(\lambda)$ versus rest wavelength, normalized in a way that attempts to achieve a balance between showing the full range in flux and illustrating the finer details of the continuum.  Our integrated spectrum is accompanied by a Digitized Sky Survey image which illustrates our rectangular spectroscopic aperture as a \emph{solid} outline and the $25$~mag~arcsec$^{-2}$ isophotal size of the galaxy as a \emph{dashed} ellipse.  The image legend gives the galaxy name, the unique identification number in parenthesis, and the morphological type as listed in Table~\ref{table:general_properties}.  The horizontal solid line in the lower-left corner of each image represents $30\arcsec$.}
\figsetgrpend
 
\figsetgrpstart
\figsetgrpnum{8.179}
\figsetgrptitle{Visualization of MRK 1450.}
\figsetplot{\includegraphics[scale=0.7,angle=90]{f8_179.eps}}
\figsetgrpnote{Presentation of our integrated spectral atlas.  We plot the spectrum as $f_{\lambda}(\lambda)$ versus rest wavelength, normalized in a way that attempts to achieve a balance between showing the full range in flux and illustrating the finer details of the continuum.  Our integrated spectrum is accompanied by a Digitized Sky Survey image which illustrates our rectangular spectroscopic aperture as a \emph{solid} outline and the $25$~mag~arcsec$^{-2}$ isophotal size of the galaxy as a \emph{dashed} ellipse.  The image legend gives the galaxy name, the unique identification number in parenthesis, and the morphological type as listed in Table~\ref{table:general_properties}.  The horizontal solid line in the lower-left corner of each image represents $30\arcsec$.}
\figsetgrpend
 
\figsetgrpstart
\figsetgrpnum{8.180}
\figsetgrptitle{Visualization of NGC 3782.}
\figsetplot{\includegraphics[scale=0.7,angle=90]{f8_180.eps}}
\figsetgrpnote{Presentation of our integrated spectral atlas.  We plot the spectrum as $f_{\lambda}(\lambda)$ versus rest wavelength, normalized in a way that attempts to achieve a balance between showing the full range in flux and illustrating the finer details of the continuum.  Our integrated spectrum is accompanied by a Digitized Sky Survey image which illustrates our rectangular spectroscopic aperture as a \emph{solid} outline and the $25$~mag~arcsec$^{-2}$ isophotal size of the galaxy as a \emph{dashed} ellipse.  The image legend gives the galaxy name, the unique identification number in parenthesis, and the morphological type as listed in Table~\ref{table:general_properties}.  The horizontal solid line in the lower-left corner of each image represents $30\arcsec$.}
\figsetgrpend
 
\figsetgrpstart
\figsetgrpnum{8.181}
\figsetgrptitle{Visualization of UGC 06628.}
\figsetplot{\includegraphics[scale=0.7,angle=90]{f8_181.eps}}
\figsetgrpnote{Presentation of our integrated spectral atlas.  We plot the spectrum as $f_{\lambda}(\lambda)$ versus rest wavelength, normalized in a way that attempts to achieve a balance between showing the full range in flux and illustrating the finer details of the continuum.  Our integrated spectrum is accompanied by a Digitized Sky Survey image which illustrates our rectangular spectroscopic aperture as a \emph{solid} outline and the $25$~mag~arcsec$^{-2}$ isophotal size of the galaxy as a \emph{dashed} ellipse.  The image legend gives the galaxy name, the unique identification number in parenthesis, and the morphological type as listed in Table~\ref{table:general_properties}.  The horizontal solid line in the lower-left corner of each image represents $30\arcsec$.}
\figsetgrpend
 
\figsetgrpstart
\figsetgrpnum{8.182}
\figsetgrptitle{Visualization of UGC 06665.}
\figsetplot{\includegraphics[scale=0.7,angle=90]{f8_182.eps}}
\figsetgrpnote{Presentation of our integrated spectral atlas.  We plot the spectrum as $f_{\lambda}(\lambda)$ versus rest wavelength, normalized in a way that attempts to achieve a balance between showing the full range in flux and illustrating the finer details of the continuum.  Our integrated spectrum is accompanied by a Digitized Sky Survey image which illustrates our rectangular spectroscopic aperture as a \emph{solid} outline and the $25$~mag~arcsec$^{-2}$ isophotal size of the galaxy as a \emph{dashed} ellipse.  The image legend gives the galaxy name, the unique identification number in parenthesis, and the morphological type as listed in Table~\ref{table:general_properties}.  The horizontal solid line in the lower-left corner of each image represents $30\arcsec$.}
\figsetgrpend
 
\figsetgrpstart
\figsetgrpnum{8.183}
\figsetgrptitle{Visualization of UGC 06667.}
\figsetplot{\includegraphics[scale=0.7,angle=90]{f8_183.eps}}
\figsetgrpnote{Presentation of our integrated spectral atlas.  We plot the spectrum as $f_{\lambda}(\lambda)$ versus rest wavelength, normalized in a way that attempts to achieve a balance between showing the full range in flux and illustrating the finer details of the continuum.  Our integrated spectrum is accompanied by a Digitized Sky Survey image which illustrates our rectangular spectroscopic aperture as a \emph{solid} outline and the $25$~mag~arcsec$^{-2}$ isophotal size of the galaxy as a \emph{dashed} ellipse.  The image legend gives the galaxy name, the unique identification number in parenthesis, and the morphological type as listed in Table~\ref{table:general_properties}.  The horizontal solid line in the lower-left corner of each image represents $30\arcsec$.}
\figsetgrpend
 
\figsetgrpstart
\figsetgrpnum{8.184}
\figsetgrptitle{Visualization of NGC 3870.}
\figsetplot{\includegraphics[scale=0.7,angle=90]{f8_184.eps}}
\figsetgrpnote{Presentation of our integrated spectral atlas.  We plot the spectrum as $f_{\lambda}(\lambda)$ versus rest wavelength, normalized in a way that attempts to achieve a balance between showing the full range in flux and illustrating the finer details of the continuum.  Our integrated spectrum is accompanied by a Digitized Sky Survey image which illustrates our rectangular spectroscopic aperture as a \emph{solid} outline and the $25$~mag~arcsec$^{-2}$ isophotal size of the galaxy as a \emph{dashed} ellipse.  The image legend gives the galaxy name, the unique identification number in parenthesis, and the morphological type as listed in Table~\ref{table:general_properties}.  The horizontal solid line in the lower-left corner of each image represents $30\arcsec$.}
\figsetgrpend
 
\figsetgrpstart
\figsetgrpnum{8.185}
\figsetgrptitle{Visualization of NGC 3877.}
\figsetplot{\includegraphics[scale=0.7,angle=90]{f8_185.eps}}
\figsetgrpnote{Presentation of our integrated spectral atlas.  We plot the spectrum as $f_{\lambda}(\lambda)$ versus rest wavelength, normalized in a way that attempts to achieve a balance between showing the full range in flux and illustrating the finer details of the continuum.  Our integrated spectrum is accompanied by a Digitized Sky Survey image which illustrates our rectangular spectroscopic aperture as a \emph{solid} outline and the $25$~mag~arcsec$^{-2}$ isophotal size of the galaxy as a \emph{dashed} ellipse.  The image legend gives the galaxy name, the unique identification number in parenthesis, and the morphological type as listed in Table~\ref{table:general_properties}.  The horizontal solid line in the lower-left corner of each image represents $30\arcsec$.}
\figsetgrpend
 
\figsetgrpstart
\figsetgrpnum{8.186}
\figsetgrptitle{Visualization of UGC 06773.}
\figsetplot{\includegraphics[scale=0.7,angle=90]{f8_186.eps}}
\figsetgrpnote{Presentation of our integrated spectral atlas.  We plot the spectrum as $f_{\lambda}(\lambda)$ versus rest wavelength, normalized in a way that attempts to achieve a balance between showing the full range in flux and illustrating the finer details of the continuum.  Our integrated spectrum is accompanied by a Digitized Sky Survey image which illustrates our rectangular spectroscopic aperture as a \emph{solid} outline and the $25$~mag~arcsec$^{-2}$ isophotal size of the galaxy as a \emph{dashed} ellipse.  The image legend gives the galaxy name, the unique identification number in parenthesis, and the morphological type as listed in Table~\ref{table:general_properties}.  The horizontal solid line in the lower-left corner of each image represents $30\arcsec$.}
\figsetgrpend
 
\figsetgrpstart
\figsetgrpnum{8.187}
\figsetgrptitle{Visualization of NGC 3893.}
\figsetplot{\includegraphics[scale=0.7,angle=90]{f8_187.eps}}
\figsetgrpnote{Presentation of our integrated spectral atlas.  We plot the spectrum as $f_{\lambda}(\lambda)$ versus rest wavelength, normalized in a way that attempts to achieve a balance between showing the full range in flux and illustrating the finer details of the continuum.  Our integrated spectrum is accompanied by a Digitized Sky Survey image which illustrates our rectangular spectroscopic aperture as a \emph{solid} outline and the $25$~mag~arcsec$^{-2}$ isophotal size of the galaxy as a \emph{dashed} ellipse.  The image legend gives the galaxy name, the unique identification number in parenthesis, and the morphological type as listed in Table~\ref{table:general_properties}.  The horizontal solid line in the lower-left corner of each image represents $30\arcsec$.}
\figsetgrpend
 
\figsetgrpstart
\figsetgrpnum{8.188}
\figsetgrptitle{Visualization of NGC 3896.}
\figsetplot{\includegraphics[scale=0.7,angle=90]{f8_188.eps}}
\figsetgrpnote{Presentation of our integrated spectral atlas.  We plot the spectrum as $f_{\lambda}(\lambda)$ versus rest wavelength, normalized in a way that attempts to achieve a balance between showing the full range in flux and illustrating the finer details of the continuum.  Our integrated spectrum is accompanied by a Digitized Sky Survey image which illustrates our rectangular spectroscopic aperture as a \emph{solid} outline and the $25$~mag~arcsec$^{-2}$ isophotal size of the galaxy as a \emph{dashed} ellipse.  The image legend gives the galaxy name, the unique identification number in parenthesis, and the morphological type as listed in Table~\ref{table:general_properties}.  The horizontal solid line in the lower-left corner of each image represents $30\arcsec$.}
\figsetgrpend
 
\figsetgrpstart
\figsetgrpnum{8.189}
\figsetgrptitle{Visualization of NGC 3906.}
\figsetplot{\includegraphics[scale=0.7,angle=90]{f8_189.eps}}
\figsetgrpnote{Presentation of our integrated spectral atlas.  We plot the spectrum as $f_{\lambda}(\lambda)$ versus rest wavelength, normalized in a way that attempts to achieve a balance between showing the full range in flux and illustrating the finer details of the continuum.  Our integrated spectrum is accompanied by a Digitized Sky Survey image which illustrates our rectangular spectroscopic aperture as a \emph{solid} outline and the $25$~mag~arcsec$^{-2}$ isophotal size of the galaxy as a \emph{dashed} ellipse.  The image legend gives the galaxy name, the unique identification number in parenthesis, and the morphological type as listed in Table~\ref{table:general_properties}.  The horizontal solid line in the lower-left corner of each image represents $30\arcsec$.}
\figsetgrpend
 
\figsetgrpstart
\figsetgrpnum{8.190}
\figsetgrptitle{Visualization of NGC 3913.}
\figsetplot{\includegraphics[scale=0.7,angle=90]{f8_190.eps}}
\figsetgrpnote{Presentation of our integrated spectral atlas.  We plot the spectrum as $f_{\lambda}(\lambda)$ versus rest wavelength, normalized in a way that attempts to achieve a balance between showing the full range in flux and illustrating the finer details of the continuum.  Our integrated spectrum is accompanied by a Digitized Sky Survey image which illustrates our rectangular spectroscopic aperture as a \emph{solid} outline and the $25$~mag~arcsec$^{-2}$ isophotal size of the galaxy as a \emph{dashed} ellipse.  The image legend gives the galaxy name, the unique identification number in parenthesis, and the morphological type as listed in Table~\ref{table:general_properties}.  The horizontal solid line in the lower-left corner of each image represents $30\arcsec$.}
\figsetgrpend
 
\figsetgrpstart
\figsetgrpnum{8.191}
\figsetgrptitle{Visualization of NGC 3917.}
\figsetplot{\includegraphics[scale=0.7,angle=90]{f8_191.eps}}
\figsetgrpnote{Presentation of our integrated spectral atlas.  We plot the spectrum as $f_{\lambda}(\lambda)$ versus rest wavelength, normalized in a way that attempts to achieve a balance between showing the full range in flux and illustrating the finer details of the continuum.  Our integrated spectrum is accompanied by a Digitized Sky Survey image which illustrates our rectangular spectroscopic aperture as a \emph{solid} outline and the $25$~mag~arcsec$^{-2}$ isophotal size of the galaxy as a \emph{dashed} ellipse.  The image legend gives the galaxy name, the unique identification number in parenthesis, and the morphological type as listed in Table~\ref{table:general_properties}.  The horizontal solid line in the lower-left corner of each image represents $30\arcsec$.}
\figsetgrpend
 
\figsetgrpstart
\figsetgrpnum{8.192}
\figsetgrptitle{Visualization of UGC 06818.}
\figsetplot{\includegraphics[scale=0.7,angle=90]{f8_192.eps}}
\figsetgrpnote{Presentation of our integrated spectral atlas.  We plot the spectrum as $f_{\lambda}(\lambda)$ versus rest wavelength, normalized in a way that attempts to achieve a balance between showing the full range in flux and illustrating the finer details of the continuum.  Our integrated spectrum is accompanied by a Digitized Sky Survey image which illustrates our rectangular spectroscopic aperture as a \emph{solid} outline and the $25$~mag~arcsec$^{-2}$ isophotal size of the galaxy as a \emph{dashed} ellipse.  The image legend gives the galaxy name, the unique identification number in parenthesis, and the morphological type as listed in Table~\ref{table:general_properties}.  The horizontal solid line in the lower-left corner of each image represents $30\arcsec$.}
\figsetgrpend
 
\figsetgrpstart
\figsetgrpnum{8.193}
\figsetgrptitle{Visualization of UGC 06816.}
\figsetplot{\includegraphics[scale=0.7,angle=90]{f8_193.eps}}
\figsetgrpnote{Presentation of our integrated spectral atlas.  We plot the spectrum as $f_{\lambda}(\lambda)$ versus rest wavelength, normalized in a way that attempts to achieve a balance between showing the full range in flux and illustrating the finer details of the continuum.  Our integrated spectrum is accompanied by a Digitized Sky Survey image which illustrates our rectangular spectroscopic aperture as a \emph{solid} outline and the $25$~mag~arcsec$^{-2}$ isophotal size of the galaxy as a \emph{dashed} ellipse.  The image legend gives the galaxy name, the unique identification number in parenthesis, and the morphological type as listed in Table~\ref{table:general_properties}.  The horizontal solid line in the lower-left corner of each image represents $30\arcsec$.}
\figsetgrpend
 
\figsetgrpstart
\figsetgrpnum{8.194}
\figsetgrptitle{Visualization of MRK 1460.}
\figsetplot{\includegraphics[scale=0.7,angle=90]{f8_194.eps}}
\figsetgrpnote{Presentation of our integrated spectral atlas.  We plot the spectrum as $f_{\lambda}(\lambda)$ versus rest wavelength, normalized in a way that attempts to achieve a balance between showing the full range in flux and illustrating the finer details of the continuum.  Our integrated spectrum is accompanied by a Digitized Sky Survey image which illustrates our rectangular spectroscopic aperture as a \emph{solid} outline and the $25$~mag~arcsec$^{-2}$ isophotal size of the galaxy as a \emph{dashed} ellipse.  The image legend gives the galaxy name, the unique identification number in parenthesis, and the morphological type as listed in Table~\ref{table:general_properties}.  The horizontal solid line in the lower-left corner of each image represents $30\arcsec$.}
\figsetgrpend
 
\figsetgrpstart
\figsetgrpnum{8.195}
\figsetgrptitle{Visualization of NGC 3921.}
\figsetplot{\includegraphics[scale=0.7,angle=90]{f8_195.eps}}
\figsetgrpnote{Presentation of our integrated spectral atlas.  We plot the spectrum as $f_{\lambda}(\lambda)$ versus rest wavelength, normalized in a way that attempts to achieve a balance between showing the full range in flux and illustrating the finer details of the continuum.  Our integrated spectrum is accompanied by a Digitized Sky Survey image which illustrates our rectangular spectroscopic aperture as a \emph{solid} outline and the $25$~mag~arcsec$^{-2}$ isophotal size of the galaxy as a \emph{dashed} ellipse.  The image legend gives the galaxy name, the unique identification number in parenthesis, and the morphological type as listed in Table~\ref{table:general_properties}.  The horizontal solid line in the lower-left corner of each image represents $30\arcsec$.}
\figsetgrpend
 
\figsetgrpstart
\figsetgrpnum{8.196}
\figsetgrptitle{Visualization of UM 461.}
\figsetplot{\includegraphics[scale=0.7,angle=90]{f8_196.eps}}
\figsetgrpnote{Presentation of our integrated spectral atlas.  We plot the spectrum as $f_{\lambda}(\lambda)$ versus rest wavelength, normalized in a way that attempts to achieve a balance between showing the full range in flux and illustrating the finer details of the continuum.  Our integrated spectrum is accompanied by a Digitized Sky Survey image which illustrates our rectangular spectroscopic aperture as a \emph{solid} outline and the $25$~mag~arcsec$^{-2}$ isophotal size of the galaxy as a \emph{dashed} ellipse.  The image legend gives the galaxy name, the unique identification number in parenthesis, and the morphological type as listed in Table~\ref{table:general_properties}.  The horizontal solid line in the lower-left corner of each image represents $30\arcsec$.}
\figsetgrpend
 
\figsetgrpstart
\figsetgrpnum{8.197}
\figsetgrptitle{Visualization of NGC 3928.}
\figsetplot{\includegraphics[scale=0.7,angle=90]{f8_197.eps}}
\figsetgrpnote{Presentation of our integrated spectral atlas.  We plot the spectrum as $f_{\lambda}(\lambda)$ versus rest wavelength, normalized in a way that attempts to achieve a balance between showing the full range in flux and illustrating the finer details of the continuum.  Our integrated spectrum is accompanied by a Digitized Sky Survey image which illustrates our rectangular spectroscopic aperture as a \emph{solid} outline and the $25$~mag~arcsec$^{-2}$ isophotal size of the galaxy as a \emph{dashed} ellipse.  The image legend gives the galaxy name, the unique identification number in parenthesis, and the morphological type as listed in Table~\ref{table:general_properties}.  The horizontal solid line in the lower-left corner of each image represents $30\arcsec$.}
\figsetgrpend
 
\figsetgrpstart
\figsetgrpnum{8.198}
\figsetgrptitle{Visualization of UGC 06850.}
\figsetplot{\includegraphics[scale=0.7,angle=90]{f8_198.eps}}
\figsetgrpnote{Presentation of our integrated spectral atlas.  We plot the spectrum as $f_{\lambda}(\lambda)$ versus rest wavelength, normalized in a way that attempts to achieve a balance between showing the full range in flux and illustrating the finer details of the continuum.  Our integrated spectrum is accompanied by a Digitized Sky Survey image which illustrates our rectangular spectroscopic aperture as a \emph{solid} outline and the $25$~mag~arcsec$^{-2}$ isophotal size of the galaxy as a \emph{dashed} ellipse.  The image legend gives the galaxy name, the unique identification number in parenthesis, and the morphological type as listed in Table~\ref{table:general_properties}.  The horizontal solid line in the lower-left corner of each image represents $30\arcsec$.}
\figsetgrpend
 
\figsetgrpstart
\figsetgrpnum{8.199}
\figsetgrptitle{Visualization of NGC 3949.}
\figsetplot{\includegraphics[scale=0.7,angle=90]{f8_199.eps}}
\figsetgrpnote{Presentation of our integrated spectral atlas.  We plot the spectrum as $f_{\lambda}(\lambda)$ versus rest wavelength, normalized in a way that attempts to achieve a balance between showing the full range in flux and illustrating the finer details of the continuum.  Our integrated spectrum is accompanied by a Digitized Sky Survey image which illustrates our rectangular spectroscopic aperture as a \emph{solid} outline and the $25$~mag~arcsec$^{-2}$ isophotal size of the galaxy as a \emph{dashed} ellipse.  The image legend gives the galaxy name, the unique identification number in parenthesis, and the morphological type as listed in Table~\ref{table:general_properties}.  The horizontal solid line in the lower-left corner of each image represents $30\arcsec$.}
\figsetgrpend
 
\figsetgrpstart
\figsetgrpnum{8.200}
\figsetgrptitle{Visualization of NGC 3953.}
\figsetplot{\includegraphics[scale=0.7,angle=90]{f8_200.eps}}
\figsetgrpnote{Presentation of our integrated spectral atlas.  We plot the spectrum as $f_{\lambda}(\lambda)$ versus rest wavelength, normalized in a way that attempts to achieve a balance between showing the full range in flux and illustrating the finer details of the continuum.  Our integrated spectrum is accompanied by a Digitized Sky Survey image which illustrates our rectangular spectroscopic aperture as a \emph{solid} outline and the $25$~mag~arcsec$^{-2}$ isophotal size of the galaxy as a \emph{dashed} ellipse.  The image legend gives the galaxy name, the unique identification number in parenthesis, and the morphological type as listed in Table~\ref{table:general_properties}.  The horizontal solid line in the lower-left corner of each image represents $30\arcsec$.}
\figsetgrpend
 
\figsetgrpstart
\figsetgrpnum{8.201}
\figsetgrptitle{Visualization of UGC 06894.}
\figsetplot{\includegraphics[scale=0.7,angle=90]{f8_201.eps}}
\figsetgrpnote{Presentation of our integrated spectral atlas.  We plot the spectrum as $f_{\lambda}(\lambda)$ versus rest wavelength, normalized in a way that attempts to achieve a balance between showing the full range in flux and illustrating the finer details of the continuum.  Our integrated spectrum is accompanied by a Digitized Sky Survey image which illustrates our rectangular spectroscopic aperture as a \emph{solid} outline and the $25$~mag~arcsec$^{-2}$ isophotal size of the galaxy as a \emph{dashed} ellipse.  The image legend gives the galaxy name, the unique identification number in parenthesis, and the morphological type as listed in Table~\ref{table:general_properties}.  The horizontal solid line in the lower-left corner of each image represents $30\arcsec$.}
\figsetgrpend
 
\figsetgrpstart
\figsetgrpnum{8.202}
\figsetgrptitle{Visualization of NGC 3972.}
\figsetplot{\includegraphics[scale=0.7,angle=90]{f8_202.eps}}
\figsetgrpnote{Presentation of our integrated spectral atlas.  We plot the spectrum as $f_{\lambda}(\lambda)$ versus rest wavelength, normalized in a way that attempts to achieve a balance between showing the full range in flux and illustrating the finer details of the continuum.  Our integrated spectrum is accompanied by a Digitized Sky Survey image which illustrates our rectangular spectroscopic aperture as a \emph{solid} outline and the $25$~mag~arcsec$^{-2}$ isophotal size of the galaxy as a \emph{dashed} ellipse.  The image legend gives the galaxy name, the unique identification number in parenthesis, and the morphological type as listed in Table~\ref{table:general_properties}.  The horizontal solid line in the lower-left corner of each image represents $30\arcsec$.}
\figsetgrpend
 
\figsetgrpstart
\figsetgrpnum{8.203}
\figsetgrptitle{Visualization of NGC 3982.}
\figsetplot{\includegraphics[scale=0.7,angle=90]{f8_203.eps}}
\figsetgrpnote{Presentation of our integrated spectral atlas.  We plot the spectrum as $f_{\lambda}(\lambda)$ versus rest wavelength, normalized in a way that attempts to achieve a balance between showing the full range in flux and illustrating the finer details of the continuum.  Our integrated spectrum is accompanied by a Digitized Sky Survey image which illustrates our rectangular spectroscopic aperture as a \emph{solid} outline and the $25$~mag~arcsec$^{-2}$ isophotal size of the galaxy as a \emph{dashed} ellipse.  The image legend gives the galaxy name, the unique identification number in parenthesis, and the morphological type as listed in Table~\ref{table:general_properties}.  The horizontal solid line in the lower-left corner of each image represents $30\arcsec$.}
\figsetgrpend
 
\figsetgrpstart
\figsetgrpnum{8.204}
\figsetgrptitle{Visualization of UGC 06917.}
\figsetplot{\includegraphics[scale=0.7,angle=90]{f8_204.eps}}
\figsetgrpnote{Presentation of our integrated spectral atlas.  We plot the spectrum as $f_{\lambda}(\lambda)$ versus rest wavelength, normalized in a way that attempts to achieve a balance between showing the full range in flux and illustrating the finer details of the continuum.  Our integrated spectrum is accompanied by a Digitized Sky Survey image which illustrates our rectangular spectroscopic aperture as a \emph{solid} outline and the $25$~mag~arcsec$^{-2}$ isophotal size of the galaxy as a \emph{dashed} ellipse.  The image legend gives the galaxy name, the unique identification number in parenthesis, and the morphological type as listed in Table~\ref{table:general_properties}.  The horizontal solid line in the lower-left corner of each image represents $30\arcsec$.}
\figsetgrpend
 
\figsetgrpstart
\figsetgrpnum{8.205}
\figsetgrptitle{Visualization of NGC 3985.}
\figsetplot{\includegraphics[scale=0.7,angle=90]{f8_205.eps}}
\figsetgrpnote{Presentation of our integrated spectral atlas.  We plot the spectrum as $f_{\lambda}(\lambda)$ versus rest wavelength, normalized in a way that attempts to achieve a balance between showing the full range in flux and illustrating the finer details of the continuum.  Our integrated spectrum is accompanied by a Digitized Sky Survey image which illustrates our rectangular spectroscopic aperture as a \emph{solid} outline and the $25$~mag~arcsec$^{-2}$ isophotal size of the galaxy as a \emph{dashed} ellipse.  The image legend gives the galaxy name, the unique identification number in parenthesis, and the morphological type as listed in Table~\ref{table:general_properties}.  The horizontal solid line in the lower-left corner of each image represents $30\arcsec$.}
\figsetgrpend
 
\figsetgrpstart
\figsetgrpnum{8.206}
\figsetgrptitle{Visualization of UGC 06923.}
\figsetplot{\includegraphics[scale=0.7,angle=90]{f8_206.eps}}
\figsetgrpnote{Presentation of our integrated spectral atlas.  We plot the spectrum as $f_{\lambda}(\lambda)$ versus rest wavelength, normalized in a way that attempts to achieve a balance between showing the full range in flux and illustrating the finer details of the continuum.  Our integrated spectrum is accompanied by a Digitized Sky Survey image which illustrates our rectangular spectroscopic aperture as a \emph{solid} outline and the $25$~mag~arcsec$^{-2}$ isophotal size of the galaxy as a \emph{dashed} ellipse.  The image legend gives the galaxy name, the unique identification number in parenthesis, and the morphological type as listed in Table~\ref{table:general_properties}.  The horizontal solid line in the lower-left corner of each image represents $30\arcsec$.}
\figsetgrpend
 
\figsetgrpstart
\figsetgrpnum{8.207}
\figsetgrptitle{Visualization of UGC 06922.}
\figsetplot{\includegraphics[scale=0.7,angle=90]{f8_207.eps}}
\figsetgrpnote{Presentation of our integrated spectral atlas.  We plot the spectrum as $f_{\lambda}(\lambda)$ versus rest wavelength, normalized in a way that attempts to achieve a balance between showing the full range in flux and illustrating the finer details of the continuum.  Our integrated spectrum is accompanied by a Digitized Sky Survey image which illustrates our rectangular spectroscopic aperture as a \emph{solid} outline and the $25$~mag~arcsec$^{-2}$ isophotal size of the galaxy as a \emph{dashed} ellipse.  The image legend gives the galaxy name, the unique identification number in parenthesis, and the morphological type as listed in Table~\ref{table:general_properties}.  The horizontal solid line in the lower-left corner of each image represents $30\arcsec$.}
\figsetgrpend
 
\figsetgrpstart
\figsetgrpnum{8.208}
\figsetgrptitle{Visualization of UGC 06930.}
\figsetplot{\includegraphics[scale=0.7,angle=90]{f8_208.eps}}
\figsetgrpnote{Presentation of our integrated spectral atlas.  We plot the spectrum as $f_{\lambda}(\lambda)$ versus rest wavelength, normalized in a way that attempts to achieve a balance between showing the full range in flux and illustrating the finer details of the continuum.  Our integrated spectrum is accompanied by a Digitized Sky Survey image which illustrates our rectangular spectroscopic aperture as a \emph{solid} outline and the $25$~mag~arcsec$^{-2}$ isophotal size of the galaxy as a \emph{dashed} ellipse.  The image legend gives the galaxy name, the unique identification number in parenthesis, and the morphological type as listed in Table~\ref{table:general_properties}.  The horizontal solid line in the lower-left corner of each image represents $30\arcsec$.}
\figsetgrpend
 
\figsetgrpstart
\figsetgrpnum{8.209}
\figsetgrptitle{Visualization of NGC 3991 S.}
\figsetplot{\includegraphics[scale=0.7,angle=90]{f8_209.eps}}
\figsetgrpnote{Presentation of our integrated spectral atlas.  We plot the spectrum as $f_{\lambda}(\lambda)$ versus rest wavelength, normalized in a way that attempts to achieve a balance between showing the full range in flux and illustrating the finer details of the continuum.  Our integrated spectrum is accompanied by a Digitized Sky Survey image which illustrates our rectangular spectroscopic aperture as a \emph{solid} outline and the $25$~mag~arcsec$^{-2}$ isophotal size of the galaxy as a \emph{dashed} ellipse.  The image legend gives the galaxy name, the unique identification number in parenthesis, and the morphological type as listed in Table~\ref{table:general_properties}.  The horizontal solid line in the lower-left corner of each image represents $30\arcsec$.}
\figsetgrpend
 
\figsetgrpstart
\figsetgrpnum{8.210}
\figsetgrptitle{Visualization of NGC 3991.}
\figsetplot{\includegraphics[scale=0.7,angle=90]{f8_210.eps}}
\figsetgrpnote{Presentation of our integrated spectral atlas.  We plot the spectrum as $f_{\lambda}(\lambda)$ versus rest wavelength, normalized in a way that attempts to achieve a balance between showing the full range in flux and illustrating the finer details of the continuum.  Our integrated spectrum is accompanied by a Digitized Sky Survey image which illustrates our rectangular spectroscopic aperture as a \emph{solid} outline and the $25$~mag~arcsec$^{-2}$ isophotal size of the galaxy as a \emph{dashed} ellipse.  The image legend gives the galaxy name, the unique identification number in parenthesis, and the morphological type as listed in Table~\ref{table:general_properties}.  The horizontal solid line in the lower-left corner of each image represents $30\arcsec$.}
\figsetgrpend
 
\figsetgrpstart
\figsetgrpnum{8.211}
\figsetgrptitle{Visualization of NGC 3991 N.}
\figsetplot{\includegraphics[scale=0.7,angle=90]{f8_211.eps}}
\figsetgrpnote{Presentation of our integrated spectral atlas.  We plot the spectrum as $f_{\lambda}(\lambda)$ versus rest wavelength, normalized in a way that attempts to achieve a balance between showing the full range in flux and illustrating the finer details of the continuum.  Our integrated spectrum is accompanied by a Digitized Sky Survey image which illustrates our rectangular spectroscopic aperture as a \emph{solid} outline and the $25$~mag~arcsec$^{-2}$ isophotal size of the galaxy as a \emph{dashed} ellipse.  The image legend gives the galaxy name, the unique identification number in parenthesis, and the morphological type as listed in Table~\ref{table:general_properties}.  The horizontal solid line in the lower-left corner of each image represents $30\arcsec$.}
\figsetgrpend
 
\figsetgrpstart
\figsetgrpnum{8.212}
\figsetgrptitle{Visualization of NGC 3994.}
\figsetplot{\includegraphics[scale=0.7,angle=90]{f8_212.eps}}
\figsetgrpnote{Presentation of our integrated spectral atlas.  We plot the spectrum as $f_{\lambda}(\lambda)$ versus rest wavelength, normalized in a way that attempts to achieve a balance between showing the full range in flux and illustrating the finer details of the continuum.  Our integrated spectrum is accompanied by a Digitized Sky Survey image which illustrates our rectangular spectroscopic aperture as a \emph{solid} outline and the $25$~mag~arcsec$^{-2}$ isophotal size of the galaxy as a \emph{dashed} ellipse.  The image legend gives the galaxy name, the unique identification number in parenthesis, and the morphological type as listed in Table~\ref{table:general_properties}.  The horizontal solid line in the lower-left corner of each image represents $30\arcsec$.}
\figsetgrpend
 
\figsetgrpstart
\figsetgrpnum{8.213}
\figsetgrptitle{Visualization of NGC 3995.}
\figsetplot{\includegraphics[scale=0.7,angle=90]{f8_213.eps}}
\figsetgrpnote{Presentation of our integrated spectral atlas.  We plot the spectrum as $f_{\lambda}(\lambda)$ versus rest wavelength, normalized in a way that attempts to achieve a balance between showing the full range in flux and illustrating the finer details of the continuum.  Our integrated spectrum is accompanied by a Digitized Sky Survey image which illustrates our rectangular spectroscopic aperture as a \emph{solid} outline and the $25$~mag~arcsec$^{-2}$ isophotal size of the galaxy as a \emph{dashed} ellipse.  The image legend gives the galaxy name, the unique identification number in parenthesis, and the morphological type as listed in Table~\ref{table:general_properties}.  The horizontal solid line in the lower-left corner of each image represents $30\arcsec$.}
\figsetgrpend
 
\figsetgrpstart
\figsetgrpnum{8.214}
\figsetgrptitle{Visualization of NGC 3998.}
\figsetplot{\includegraphics[scale=0.7,angle=90]{f8_214.eps}}
\figsetgrpnote{Presentation of our integrated spectral atlas.  We plot the spectrum as $f_{\lambda}(\lambda)$ versus rest wavelength, normalized in a way that attempts to achieve a balance between showing the full range in flux and illustrating the finer details of the continuum.  Our integrated spectrum is accompanied by a Digitized Sky Survey image which illustrates our rectangular spectroscopic aperture as a \emph{solid} outline and the $25$~mag~arcsec$^{-2}$ isophotal size of the galaxy as a \emph{dashed} ellipse.  The image legend gives the galaxy name, the unique identification number in parenthesis, and the morphological type as listed in Table~\ref{table:general_properties}.  The horizontal solid line in the lower-left corner of each image represents $30\arcsec$.}
\figsetgrpend
 
\figsetgrpstart
\figsetgrpnum{8.215}
\figsetgrptitle{Visualization of NGC 4004.}
\figsetplot{\includegraphics[scale=0.7,angle=90]{f8_215.eps}}
\figsetgrpnote{Presentation of our integrated spectral atlas.  We plot the spectrum as $f_{\lambda}(\lambda)$ versus rest wavelength, normalized in a way that attempts to achieve a balance between showing the full range in flux and illustrating the finer details of the continuum.  Our integrated spectrum is accompanied by a Digitized Sky Survey image which illustrates our rectangular spectroscopic aperture as a \emph{solid} outline and the $25$~mag~arcsec$^{-2}$ isophotal size of the galaxy as a \emph{dashed} ellipse.  The image legend gives the galaxy name, the unique identification number in parenthesis, and the morphological type as listed in Table~\ref{table:general_properties}.  The horizontal solid line in the lower-left corner of each image represents $30\arcsec$.}
\figsetgrpend
 
\figsetgrpstart
\figsetgrpnum{8.216}
\figsetgrptitle{Visualization of IC 0749.}
\figsetplot{\includegraphics[scale=0.7,angle=90]{f8_216.eps}}
\figsetgrpnote{Presentation of our integrated spectral atlas.  We plot the spectrum as $f_{\lambda}(\lambda)$ versus rest wavelength, normalized in a way that attempts to achieve a balance between showing the full range in flux and illustrating the finer details of the continuum.  Our integrated spectrum is accompanied by a Digitized Sky Survey image which illustrates our rectangular spectroscopic aperture as a \emph{solid} outline and the $25$~mag~arcsec$^{-2}$ isophotal size of the galaxy as a \emph{dashed} ellipse.  The image legend gives the galaxy name, the unique identification number in parenthesis, and the morphological type as listed in Table~\ref{table:general_properties}.  The horizontal solid line in the lower-left corner of each image represents $30\arcsec$.}
\figsetgrpend
 
\figsetgrpstart
\figsetgrpnum{8.217}
\figsetgrptitle{Visualization of NGC 4010.}
\figsetplot{\includegraphics[scale=0.7,angle=90]{f8_217.eps}}
\figsetgrpnote{Presentation of our integrated spectral atlas.  We plot the spectrum as $f_{\lambda}(\lambda)$ versus rest wavelength, normalized in a way that attempts to achieve a balance between showing the full range in flux and illustrating the finer details of the continuum.  Our integrated spectrum is accompanied by a Digitized Sky Survey image which illustrates our rectangular spectroscopic aperture as a \emph{solid} outline and the $25$~mag~arcsec$^{-2}$ isophotal size of the galaxy as a \emph{dashed} ellipse.  The image legend gives the galaxy name, the unique identification number in parenthesis, and the morphological type as listed in Table~\ref{table:general_properties}.  The horizontal solid line in the lower-left corner of each image represents $30\arcsec$.}
\figsetgrpend
 
\figsetgrpstart
\figsetgrpnum{8.218}
\figsetgrptitle{Visualization of UGC 06969.}
\figsetplot{\includegraphics[scale=0.7,angle=90]{f8_218.eps}}
\figsetgrpnote{Presentation of our integrated spectral atlas.  We plot the spectrum as $f_{\lambda}(\lambda)$ versus rest wavelength, normalized in a way that attempts to achieve a balance between showing the full range in flux and illustrating the finer details of the continuum.  Our integrated spectrum is accompanied by a Digitized Sky Survey image which illustrates our rectangular spectroscopic aperture as a \emph{solid} outline and the $25$~mag~arcsec$^{-2}$ isophotal size of the galaxy as a \emph{dashed} ellipse.  The image legend gives the galaxy name, the unique identification number in parenthesis, and the morphological type as listed in Table~\ref{table:general_properties}.  The horizontal solid line in the lower-left corner of each image represents $30\arcsec$.}
\figsetgrpend
 
\figsetgrpstart
\figsetgrpnum{8.219}
\figsetgrptitle{Visualization of IC 0750.}
\figsetplot{\includegraphics[scale=0.7,angle=90]{f8_219.eps}}
\figsetgrpnote{Presentation of our integrated spectral atlas.  We plot the spectrum as $f_{\lambda}(\lambda)$ versus rest wavelength, normalized in a way that attempts to achieve a balance between showing the full range in flux and illustrating the finer details of the continuum.  Our integrated spectrum is accompanied by a Digitized Sky Survey image which illustrates our rectangular spectroscopic aperture as a \emph{solid} outline and the $25$~mag~arcsec$^{-2}$ isophotal size of the galaxy as a \emph{dashed} ellipse.  The image legend gives the galaxy name, the unique identification number in parenthesis, and the morphological type as listed in Table~\ref{table:general_properties}.  The horizontal solid line in the lower-left corner of each image represents $30\arcsec$.}
\figsetgrpend
 
\figsetgrpstart
\figsetgrpnum{8.220}
\figsetgrptitle{Visualization of NGC 4020.}
\figsetplot{\includegraphics[scale=0.7,angle=90]{f8_220.eps}}
\figsetgrpnote{Presentation of our integrated spectral atlas.  We plot the spectrum as $f_{\lambda}(\lambda)$ versus rest wavelength, normalized in a way that attempts to achieve a balance between showing the full range in flux and illustrating the finer details of the continuum.  Our integrated spectrum is accompanied by a Digitized Sky Survey image which illustrates our rectangular spectroscopic aperture as a \emph{solid} outline and the $25$~mag~arcsec$^{-2}$ isophotal size of the galaxy as a \emph{dashed} ellipse.  The image legend gives the galaxy name, the unique identification number in parenthesis, and the morphological type as listed in Table~\ref{table:general_properties}.  The horizontal solid line in the lower-left corner of each image represents $30\arcsec$.}
\figsetgrpend
 
\figsetgrpstart
\figsetgrpnum{8.221}
\figsetgrptitle{Visualization of UGC 06983.}
\figsetplot{\includegraphics[scale=0.7,angle=90]{f8_221.eps}}
\figsetgrpnote{Presentation of our integrated spectral atlas.  We plot the spectrum as $f_{\lambda}(\lambda)$ versus rest wavelength, normalized in a way that attempts to achieve a balance between showing the full range in flux and illustrating the finer details of the continuum.  Our integrated spectrum is accompanied by a Digitized Sky Survey image which illustrates our rectangular spectroscopic aperture as a \emph{solid} outline and the $25$~mag~arcsec$^{-2}$ isophotal size of the galaxy as a \emph{dashed} ellipse.  The image legend gives the galaxy name, the unique identification number in parenthesis, and the morphological type as listed in Table~\ref{table:general_properties}.  The horizontal solid line in the lower-left corner of each image represents $30\arcsec$.}
\figsetgrpend
 
\figsetgrpstart
\figsetgrpnum{8.222}
\figsetgrptitle{Visualization of ARP 244.}
\figsetplot{\includegraphics[scale=0.7,angle=90]{f8_222.eps}}
\figsetgrpnote{Presentation of our integrated spectral atlas.  We plot the spectrum as $f_{\lambda}(\lambda)$ versus rest wavelength, normalized in a way that attempts to achieve a balance between showing the full range in flux and illustrating the finer details of the continuum.  Our integrated spectrum is accompanied by a Digitized Sky Survey image which illustrates our rectangular spectroscopic aperture as a \emph{solid} outline and the $25$~mag~arcsec$^{-2}$ isophotal size of the galaxy as a \emph{dashed} ellipse.  The image legend gives the galaxy name, the unique identification number in parenthesis, and the morphological type as listed in Table~\ref{table:general_properties}.  The horizontal solid line in the lower-left corner of each image represents $30\arcsec$.}
\figsetgrpend
 
\figsetgrpstart
\figsetgrpnum{8.223}
\figsetgrptitle{Visualization of NGC 4051.}
\figsetplot{\includegraphics[scale=0.7,angle=90]{f8_223.eps}}
\figsetgrpnote{Presentation of our integrated spectral atlas.  We plot the spectrum as $f_{\lambda}(\lambda)$ versus rest wavelength, normalized in a way that attempts to achieve a balance between showing the full range in flux and illustrating the finer details of the continuum.  Our integrated spectrum is accompanied by a Digitized Sky Survey image which illustrates our rectangular spectroscopic aperture as a \emph{solid} outline and the $25$~mag~arcsec$^{-2}$ isophotal size of the galaxy as a \emph{dashed} ellipse.  The image legend gives the galaxy name, the unique identification number in parenthesis, and the morphological type as listed in Table~\ref{table:general_properties}.  The horizontal solid line in the lower-left corner of each image represents $30\arcsec$.}
\figsetgrpend
 
\figsetgrpstart
\figsetgrpnum{8.224}
\figsetgrptitle{Visualization of NGC 4068.}
\figsetplot{\includegraphics[scale=0.7,angle=90]{f8_224.eps}}
\figsetgrpnote{Presentation of our integrated spectral atlas.  We plot the spectrum as $f_{\lambda}(\lambda)$ versus rest wavelength, normalized in a way that attempts to achieve a balance between showing the full range in flux and illustrating the finer details of the continuum.  Our integrated spectrum is accompanied by a Digitized Sky Survey image which illustrates our rectangular spectroscopic aperture as a \emph{solid} outline and the $25$~mag~arcsec$^{-2}$ isophotal size of the galaxy as a \emph{dashed} ellipse.  The image legend gives the galaxy name, the unique identification number in parenthesis, and the morphological type as listed in Table~\ref{table:general_properties}.  The horizontal solid line in the lower-left corner of each image represents $30\arcsec$.}
\figsetgrpend
 
\figsetgrpstart
\figsetgrpnum{8.225}
\figsetgrptitle{Visualization of NGC 4062.}
\figsetplot{\includegraphics[scale=0.7,angle=90]{f8_225.eps}}
\figsetgrpnote{Presentation of our integrated spectral atlas.  We plot the spectrum as $f_{\lambda}(\lambda)$ versus rest wavelength, normalized in a way that attempts to achieve a balance between showing the full range in flux and illustrating the finer details of the continuum.  Our integrated spectrum is accompanied by a Digitized Sky Survey image which illustrates our rectangular spectroscopic aperture as a \emph{solid} outline and the $25$~mag~arcsec$^{-2}$ isophotal size of the galaxy as a \emph{dashed} ellipse.  The image legend gives the galaxy name, the unique identification number in parenthesis, and the morphological type as listed in Table~\ref{table:general_properties}.  The horizontal solid line in the lower-left corner of each image represents $30\arcsec$.}
\figsetgrpend
 
\figsetgrpstart
\figsetgrpnum{8.226}
\figsetgrptitle{Visualization of NGC 4085.}
\figsetplot{\includegraphics[scale=0.7,angle=90]{f8_226.eps}}
\figsetgrpnote{Presentation of our integrated spectral atlas.  We plot the spectrum as $f_{\lambda}(\lambda)$ versus rest wavelength, normalized in a way that attempts to achieve a balance between showing the full range in flux and illustrating the finer details of the continuum.  Our integrated spectrum is accompanied by a Digitized Sky Survey image which illustrates our rectangular spectroscopic aperture as a \emph{solid} outline and the $25$~mag~arcsec$^{-2}$ isophotal size of the galaxy as a \emph{dashed} ellipse.  The image legend gives the galaxy name, the unique identification number in parenthesis, and the morphological type as listed in Table~\ref{table:general_properties}.  The horizontal solid line in the lower-left corner of each image represents $30\arcsec$.}
\figsetgrpend
 
\figsetgrpstart
\figsetgrpnum{8.227}
\figsetgrptitle{Visualization of NGC 4088.}
\figsetplot{\includegraphics[scale=0.7,angle=90]{f8_227.eps}}
\figsetgrpnote{Presentation of our integrated spectral atlas.  We plot the spectrum as $f_{\lambda}(\lambda)$ versus rest wavelength, normalized in a way that attempts to achieve a balance between showing the full range in flux and illustrating the finer details of the continuum.  Our integrated spectrum is accompanied by a Digitized Sky Survey image which illustrates our rectangular spectroscopic aperture as a \emph{solid} outline and the $25$~mag~arcsec$^{-2}$ isophotal size of the galaxy as a \emph{dashed} ellipse.  The image legend gives the galaxy name, the unique identification number in parenthesis, and the morphological type as listed in Table~\ref{table:general_properties}.  The horizontal solid line in the lower-left corner of each image represents $30\arcsec$.}
\figsetgrpend
 
\figsetgrpstart
\figsetgrpnum{8.228}
\figsetgrptitle{Visualization of UGC 07089.}
\figsetplot{\includegraphics[scale=0.7,angle=90]{f8_228.eps}}
\figsetgrpnote{Presentation of our integrated spectral atlas.  We plot the spectrum as $f_{\lambda}(\lambda)$ versus rest wavelength, normalized in a way that attempts to achieve a balance between showing the full range in flux and illustrating the finer details of the continuum.  Our integrated spectrum is accompanied by a Digitized Sky Survey image which illustrates our rectangular spectroscopic aperture as a \emph{solid} outline and the $25$~mag~arcsec$^{-2}$ isophotal size of the galaxy as a \emph{dashed} ellipse.  The image legend gives the galaxy name, the unique identification number in parenthesis, and the morphological type as listed in Table~\ref{table:general_properties}.  The horizontal solid line in the lower-left corner of each image represents $30\arcsec$.}
\figsetgrpend
 
\figsetgrpstart
\figsetgrpnum{8.229}
\figsetgrptitle{Visualization of NGC 4096.}
\figsetplot{\includegraphics[scale=0.7,angle=90]{f8_229.eps}}
\figsetgrpnote{Presentation of our integrated spectral atlas.  We plot the spectrum as $f_{\lambda}(\lambda)$ versus rest wavelength, normalized in a way that attempts to achieve a balance between showing the full range in flux and illustrating the finer details of the continuum.  Our integrated spectrum is accompanied by a Digitized Sky Survey image which illustrates our rectangular spectroscopic aperture as a \emph{solid} outline and the $25$~mag~arcsec$^{-2}$ isophotal size of the galaxy as a \emph{dashed} ellipse.  The image legend gives the galaxy name, the unique identification number in parenthesis, and the morphological type as listed in Table~\ref{table:general_properties}.  The horizontal solid line in the lower-left corner of each image represents $30\arcsec$.}
\figsetgrpend
 
\figsetgrpstart
\figsetgrpnum{8.230}
\figsetgrptitle{Visualization of NGC 4100.}
\figsetplot{\includegraphics[scale=0.7,angle=90]{f8_230.eps}}
\figsetgrpnote{Presentation of our integrated spectral atlas.  We plot the spectrum as $f_{\lambda}(\lambda)$ versus rest wavelength, normalized in a way that attempts to achieve a balance between showing the full range in flux and illustrating the finer details of the continuum.  Our integrated spectrum is accompanied by a Digitized Sky Survey image which illustrates our rectangular spectroscopic aperture as a \emph{solid} outline and the $25$~mag~arcsec$^{-2}$ isophotal size of the galaxy as a \emph{dashed} ellipse.  The image legend gives the galaxy name, the unique identification number in parenthesis, and the morphological type as listed in Table~\ref{table:general_properties}.  The horizontal solid line in the lower-left corner of each image represents $30\arcsec$.}
\figsetgrpend
 
\figsetgrpstart
\figsetgrpnum{8.231}
\figsetgrptitle{Visualization of UGC 07094.}
\figsetplot{\includegraphics[scale=0.7,angle=90]{f8_231.eps}}
\figsetgrpnote{Presentation of our integrated spectral atlas.  We plot the spectrum as $f_{\lambda}(\lambda)$ versus rest wavelength, normalized in a way that attempts to achieve a balance between showing the full range in flux and illustrating the finer details of the continuum.  Our integrated spectrum is accompanied by a Digitized Sky Survey image which illustrates our rectangular spectroscopic aperture as a \emph{solid} outline and the $25$~mag~arcsec$^{-2}$ isophotal size of the galaxy as a \emph{dashed} ellipse.  The image legend gives the galaxy name, the unique identification number in parenthesis, and the morphological type as listed in Table~\ref{table:general_properties}.  The horizontal solid line in the lower-left corner of each image represents $30\arcsec$.}
\figsetgrpend
 
\figsetgrpstart
\figsetgrpnum{8.232}
\figsetgrptitle{Visualization of NGC 4102.}
\figsetplot{\includegraphics[scale=0.7,angle=90]{f8_232.eps}}
\figsetgrpnote{Presentation of our integrated spectral atlas.  We plot the spectrum as $f_{\lambda}(\lambda)$ versus rest wavelength, normalized in a way that attempts to achieve a balance between showing the full range in flux and illustrating the finer details of the continuum.  Our integrated spectrum is accompanied by a Digitized Sky Survey image which illustrates our rectangular spectroscopic aperture as a \emph{solid} outline and the $25$~mag~arcsec$^{-2}$ isophotal size of the galaxy as a \emph{dashed} ellipse.  The image legend gives the galaxy name, the unique identification number in parenthesis, and the morphological type as listed in Table~\ref{table:general_properties}.  The horizontal solid line in the lower-left corner of each image represents $30\arcsec$.}
\figsetgrpend
 
\figsetgrpstart
\figsetgrpnum{8.233}
\figsetgrptitle{Visualization of NGC 4111.}
\figsetplot{\includegraphics[scale=0.7,angle=90]{f8_233.eps}}
\figsetgrpnote{Presentation of our integrated spectral atlas.  We plot the spectrum as $f_{\lambda}(\lambda)$ versus rest wavelength, normalized in a way that attempts to achieve a balance between showing the full range in flux and illustrating the finer details of the continuum.  Our integrated spectrum is accompanied by a Digitized Sky Survey image which illustrates our rectangular spectroscopic aperture as a \emph{solid} outline and the $25$~mag~arcsec$^{-2}$ isophotal size of the galaxy as a \emph{dashed} ellipse.  The image legend gives the galaxy name, the unique identification number in parenthesis, and the morphological type as listed in Table~\ref{table:general_properties}.  The horizontal solid line in the lower-left corner of each image represents $30\arcsec$.}
\figsetgrpend
 
\figsetgrpstart
\figsetgrpnum{8.234}
\figsetgrptitle{Visualization of NGC 4136.}
\figsetplot{\includegraphics[scale=0.7,angle=90]{f8_234.eps}}
\figsetgrpnote{Presentation of our integrated spectral atlas.  We plot the spectrum as $f_{\lambda}(\lambda)$ versus rest wavelength, normalized in a way that attempts to achieve a balance between showing the full range in flux and illustrating the finer details of the continuum.  Our integrated spectrum is accompanied by a Digitized Sky Survey image which illustrates our rectangular spectroscopic aperture as a \emph{solid} outline and the $25$~mag~arcsec$^{-2}$ isophotal size of the galaxy as a \emph{dashed} ellipse.  The image legend gives the galaxy name, the unique identification number in parenthesis, and the morphological type as listed in Table~\ref{table:general_properties}.  The horizontal solid line in the lower-left corner of each image represents $30\arcsec$.}
\figsetgrpend
 
\figsetgrpstart
\figsetgrpnum{8.235}
\figsetgrptitle{Visualization of NGC 4138.}
\figsetplot{\includegraphics[scale=0.7,angle=90]{f8_235.eps}}
\figsetgrpnote{Presentation of our integrated spectral atlas.  We plot the spectrum as $f_{\lambda}(\lambda)$ versus rest wavelength, normalized in a way that attempts to achieve a balance between showing the full range in flux and illustrating the finer details of the continuum.  Our integrated spectrum is accompanied by a Digitized Sky Survey image which illustrates our rectangular spectroscopic aperture as a \emph{solid} outline and the $25$~mag~arcsec$^{-2}$ isophotal size of the galaxy as a \emph{dashed} ellipse.  The image legend gives the galaxy name, the unique identification number in parenthesis, and the morphological type as listed in Table~\ref{table:general_properties}.  The horizontal solid line in the lower-left corner of each image represents $30\arcsec$.}
\figsetgrpend
 
\figsetgrpstart
\figsetgrpnum{8.236}
\figsetgrptitle{Visualization of NGC 4144.}
\figsetplot{\includegraphics[scale=0.7,angle=90]{f8_236.eps}}
\figsetgrpnote{Presentation of our integrated spectral atlas.  We plot the spectrum as $f_{\lambda}(\lambda)$ versus rest wavelength, normalized in a way that attempts to achieve a balance between showing the full range in flux and illustrating the finer details of the continuum.  Our integrated spectrum is accompanied by a Digitized Sky Survey image which illustrates our rectangular spectroscopic aperture as a \emph{solid} outline and the $25$~mag~arcsec$^{-2}$ isophotal size of the galaxy as a \emph{dashed} ellipse.  The image legend gives the galaxy name, the unique identification number in parenthesis, and the morphological type as listed in Table~\ref{table:general_properties}.  The horizontal solid line in the lower-left corner of each image represents $30\arcsec$.}
\figsetgrpend
 
\figsetgrpstart
\figsetgrpnum{8.237}
\figsetgrptitle{Visualization of NGC 4150.}
\figsetplot{\includegraphics[scale=0.7,angle=90]{f8_237.eps}}
\figsetgrpnote{Presentation of our integrated spectral atlas.  We plot the spectrum as $f_{\lambda}(\lambda)$ versus rest wavelength, normalized in a way that attempts to achieve a balance between showing the full range in flux and illustrating the finer details of the continuum.  Our integrated spectrum is accompanied by a Digitized Sky Survey image which illustrates our rectangular spectroscopic aperture as a \emph{solid} outline and the $25$~mag~arcsec$^{-2}$ isophotal size of the galaxy as a \emph{dashed} ellipse.  The image legend gives the galaxy name, the unique identification number in parenthesis, and the morphological type as listed in Table~\ref{table:general_properties}.  The horizontal solid line in the lower-left corner of each image represents $30\arcsec$.}
\figsetgrpend
 
\figsetgrpstart
\figsetgrpnum{8.238}
\figsetgrptitle{Visualization of NGC 4157.}
\figsetplot{\includegraphics[scale=0.7,angle=90]{f8_238.eps}}
\figsetgrpnote{Presentation of our integrated spectral atlas.  We plot the spectrum as $f_{\lambda}(\lambda)$ versus rest wavelength, normalized in a way that attempts to achieve a balance between showing the full range in flux and illustrating the finer details of the continuum.  Our integrated spectrum is accompanied by a Digitized Sky Survey image which illustrates our rectangular spectroscopic aperture as a \emph{solid} outline and the $25$~mag~arcsec$^{-2}$ isophotal size of the galaxy as a \emph{dashed} ellipse.  The image legend gives the galaxy name, the unique identification number in parenthesis, and the morphological type as listed in Table~\ref{table:general_properties}.  The horizontal solid line in the lower-left corner of each image represents $30\arcsec$.}
\figsetgrpend
 
\figsetgrpstart
\figsetgrpnum{8.239}
\figsetgrptitle{Visualization of NGC 4163.}
\figsetplot{\includegraphics[scale=0.7,angle=90]{f8_239.eps}}
\figsetgrpnote{Presentation of our integrated spectral atlas.  We plot the spectrum as $f_{\lambda}(\lambda)$ versus rest wavelength, normalized in a way that attempts to achieve a balance between showing the full range in flux and illustrating the finer details of the continuum.  Our integrated spectrum is accompanied by a Digitized Sky Survey image which illustrates our rectangular spectroscopic aperture as a \emph{solid} outline and the $25$~mag~arcsec$^{-2}$ isophotal size of the galaxy as a \emph{dashed} ellipse.  The image legend gives the galaxy name, the unique identification number in parenthesis, and the morphological type as listed in Table~\ref{table:general_properties}.  The horizontal solid line in the lower-left corner of each image represents $30\arcsec$.}
\figsetgrpend
 
\figsetgrpstart
\figsetgrpnum{8.240}
\figsetgrptitle{Visualization of NGC 4183.}
\figsetplot{\includegraphics[scale=0.7,angle=90]{f8_240.eps}}
\figsetgrpnote{Presentation of our integrated spectral atlas.  We plot the spectrum as $f_{\lambda}(\lambda)$ versus rest wavelength, normalized in a way that attempts to achieve a balance between showing the full range in flux and illustrating the finer details of the continuum.  Our integrated spectrum is accompanied by a Digitized Sky Survey image which illustrates our rectangular spectroscopic aperture as a \emph{solid} outline and the $25$~mag~arcsec$^{-2}$ isophotal size of the galaxy as a \emph{dashed} ellipse.  The image legend gives the galaxy name, the unique identification number in parenthesis, and the morphological type as listed in Table~\ref{table:general_properties}.  The horizontal solid line in the lower-left corner of each image represents $30\arcsec$.}
\figsetgrpend
 
\figsetgrpstart
\figsetgrpnum{8.241}
\figsetgrptitle{Visualization of NGC 4190.}
\figsetplot{\includegraphics[scale=0.7,angle=90]{f8_241.eps}}
\figsetgrpnote{Presentation of our integrated spectral atlas.  We plot the spectrum as $f_{\lambda}(\lambda)$ versus rest wavelength, normalized in a way that attempts to achieve a balance between showing the full range in flux and illustrating the finer details of the continuum.  Our integrated spectrum is accompanied by a Digitized Sky Survey image which illustrates our rectangular spectroscopic aperture as a \emph{solid} outline and the $25$~mag~arcsec$^{-2}$ isophotal size of the galaxy as a \emph{dashed} ellipse.  The image legend gives the galaxy name, the unique identification number in parenthesis, and the morphological type as listed in Table~\ref{table:general_properties}.  The horizontal solid line in the lower-left corner of each image represents $30\arcsec$.}
\figsetgrpend
 
\figsetgrpstart
\figsetgrpnum{8.242}
\figsetgrptitle{Visualization of NGC 4194.}
\figsetplot{\includegraphics[scale=0.7,angle=90]{f8_242.eps}}
\figsetgrpnote{Presentation of our integrated spectral atlas.  We plot the spectrum as $f_{\lambda}(\lambda)$ versus rest wavelength, normalized in a way that attempts to achieve a balance between showing the full range in flux and illustrating the finer details of the continuum.  Our integrated spectrum is accompanied by a Digitized Sky Survey image which illustrates our rectangular spectroscopic aperture as a \emph{solid} outline and the $25$~mag~arcsec$^{-2}$ isophotal size of the galaxy as a \emph{dashed} ellipse.  The image legend gives the galaxy name, the unique identification number in parenthesis, and the morphological type as listed in Table~\ref{table:general_properties}.  The horizontal solid line in the lower-left corner of each image represents $30\arcsec$.}
\figsetgrpend
 
\figsetgrpstart
\figsetgrpnum{8.243}
\figsetgrptitle{Visualization of HARO 06.}
\figsetplot{\includegraphics[scale=0.7,angle=90]{f8_243.eps}}
\figsetgrpnote{Presentation of our integrated spectral atlas.  We plot the spectrum as $f_{\lambda}(\lambda)$ versus rest wavelength, normalized in a way that attempts to achieve a balance between showing the full range in flux and illustrating the finer details of the continuum.  Our integrated spectrum is accompanied by a Digitized Sky Survey image which illustrates our rectangular spectroscopic aperture as a \emph{solid} outline and the $25$~mag~arcsec$^{-2}$ isophotal size of the galaxy as a \emph{dashed} ellipse.  The image legend gives the galaxy name, the unique identification number in parenthesis, and the morphological type as listed in Table~\ref{table:general_properties}.  The horizontal solid line in the lower-left corner of each image represents $30\arcsec$.}
\figsetgrpend
 
\figsetgrpstart
\figsetgrpnum{8.244}
\figsetgrptitle{Visualization of NGC 4214.}
\figsetplot{\includegraphics[scale=0.7,angle=90]{f8_244.eps}}
\figsetgrpnote{Presentation of our integrated spectral atlas.  We plot the spectrum as $f_{\lambda}(\lambda)$ versus rest wavelength, normalized in a way that attempts to achieve a balance between showing the full range in flux and illustrating the finer details of the continuum.  Our integrated spectrum is accompanied by a Digitized Sky Survey image which illustrates our rectangular spectroscopic aperture as a \emph{solid} outline and the $25$~mag~arcsec$^{-2}$ isophotal size of the galaxy as a \emph{dashed} ellipse.  The image legend gives the galaxy name, the unique identification number in parenthesis, and the morphological type as listed in Table~\ref{table:general_properties}.  The horizontal solid line in the lower-left corner of each image represents $30\arcsec$.}
\figsetgrpend
 
\figsetgrpstart
\figsetgrpnum{8.245}
\figsetgrptitle{Visualization of NGC 4218.}
\figsetplot{\includegraphics[scale=0.7,angle=90]{f8_245.eps}}
\figsetgrpnote{Presentation of our integrated spectral atlas.  We plot the spectrum as $f_{\lambda}(\lambda)$ versus rest wavelength, normalized in a way that attempts to achieve a balance between showing the full range in flux and illustrating the finer details of the continuum.  Our integrated spectrum is accompanied by a Digitized Sky Survey image which illustrates our rectangular spectroscopic aperture as a \emph{solid} outline and the $25$~mag~arcsec$^{-2}$ isophotal size of the galaxy as a \emph{dashed} ellipse.  The image legend gives the galaxy name, the unique identification number in parenthesis, and the morphological type as listed in Table~\ref{table:general_properties}.  The horizontal solid line in the lower-left corner of each image represents $30\arcsec$.}
\figsetgrpend
 
\figsetgrpstart
\figsetgrpnum{8.246}
\figsetgrptitle{Visualization of NGC 4217.}
\figsetplot{\includegraphics[scale=0.7,angle=90]{f8_246.eps}}
\figsetgrpnote{Presentation of our integrated spectral atlas.  We plot the spectrum as $f_{\lambda}(\lambda)$ versus rest wavelength, normalized in a way that attempts to achieve a balance between showing the full range in flux and illustrating the finer details of the continuum.  Our integrated spectrum is accompanied by a Digitized Sky Survey image which illustrates our rectangular spectroscopic aperture as a \emph{solid} outline and the $25$~mag~arcsec$^{-2}$ isophotal size of the galaxy as a \emph{dashed} ellipse.  The image legend gives the galaxy name, the unique identification number in parenthesis, and the morphological type as listed in Table~\ref{table:general_properties}.  The horizontal solid line in the lower-left corner of each image represents $30\arcsec$.}
\figsetgrpend
 
\figsetgrpstart
\figsetgrpnum{8.247}
\figsetgrptitle{Visualization of NGC 4220.}
\figsetplot{\includegraphics[scale=0.7,angle=90]{f8_247.eps}}
\figsetgrpnote{Presentation of our integrated spectral atlas.  We plot the spectrum as $f_{\lambda}(\lambda)$ versus rest wavelength, normalized in a way that attempts to achieve a balance between showing the full range in flux and illustrating the finer details of the continuum.  Our integrated spectrum is accompanied by a Digitized Sky Survey image which illustrates our rectangular spectroscopic aperture as a \emph{solid} outline and the $25$~mag~arcsec$^{-2}$ isophotal size of the galaxy as a \emph{dashed} ellipse.  The image legend gives the galaxy name, the unique identification number in parenthesis, and the morphological type as listed in Table~\ref{table:general_properties}.  The horizontal solid line in the lower-left corner of each image represents $30\arcsec$.}
\figsetgrpend
 
\figsetgrpstart
\figsetgrpnum{8.248}
\figsetgrptitle{Visualization of NGC 4244.}
\figsetplot{\includegraphics[scale=0.7,angle=90]{f8_248.eps}}
\figsetgrpnote{Presentation of our integrated spectral atlas.  We plot the spectrum as $f_{\lambda}(\lambda)$ versus rest wavelength, normalized in a way that attempts to achieve a balance between showing the full range in flux and illustrating the finer details of the continuum.  Our integrated spectrum is accompanied by a Digitized Sky Survey image which illustrates our rectangular spectroscopic aperture as a \emph{solid} outline and the $25$~mag~arcsec$^{-2}$ isophotal size of the galaxy as a \emph{dashed} ellipse.  The image legend gives the galaxy name, the unique identification number in parenthesis, and the morphological type as listed in Table~\ref{table:general_properties}.  The horizontal solid line in the lower-left corner of each image represents $30\arcsec$.}
\figsetgrpend
 
\figsetgrpstart
\figsetgrpnum{8.249}
\figsetgrptitle{Visualization of NGC 4248.}
\figsetplot{\includegraphics[scale=0.7,angle=90]{f8_249.eps}}
\figsetgrpnote{Presentation of our integrated spectral atlas.  We plot the spectrum as $f_{\lambda}(\lambda)$ versus rest wavelength, normalized in a way that attempts to achieve a balance between showing the full range in flux and illustrating the finer details of the continuum.  Our integrated spectrum is accompanied by a Digitized Sky Survey image which illustrates our rectangular spectroscopic aperture as a \emph{solid} outline and the $25$~mag~arcsec$^{-2}$ isophotal size of the galaxy as a \emph{dashed} ellipse.  The image legend gives the galaxy name, the unique identification number in parenthesis, and the morphological type as listed in Table~\ref{table:general_properties}.  The horizontal solid line in the lower-left corner of each image represents $30\arcsec$.}
\figsetgrpend
 
\figsetgrpstart
\figsetgrpnum{8.250}
\figsetgrptitle{Visualization of NGC 4254.}
\figsetplot{\includegraphics[scale=0.7,angle=90]{f8_250.eps}}
\figsetgrpnote{Presentation of our integrated spectral atlas.  We plot the spectrum as $f_{\lambda}(\lambda)$ versus rest wavelength, normalized in a way that attempts to achieve a balance between showing the full range in flux and illustrating the finer details of the continuum.  Our integrated spectrum is accompanied by a Digitized Sky Survey image which illustrates our rectangular spectroscopic aperture as a \emph{solid} outline and the $25$~mag~arcsec$^{-2}$ isophotal size of the galaxy as a \emph{dashed} ellipse.  The image legend gives the galaxy name, the unique identification number in parenthesis, and the morphological type as listed in Table~\ref{table:general_properties}.  The horizontal solid line in the lower-left corner of each image represents $30\arcsec$.}
\figsetgrpend
 
\figsetgrpstart
\figsetgrpnum{8.251}
\figsetgrptitle{Visualization of UGC 07354.}
\figsetplot{\includegraphics[scale=0.7,angle=90]{f8_251.eps}}
\figsetgrpnote{Presentation of our integrated spectral atlas.  We plot the spectrum as $f_{\lambda}(\lambda)$ versus rest wavelength, normalized in a way that attempts to achieve a balance between showing the full range in flux and illustrating the finer details of the continuum.  Our integrated spectrum is accompanied by a Digitized Sky Survey image which illustrates our rectangular spectroscopic aperture as a \emph{solid} outline and the $25$~mag~arcsec$^{-2}$ isophotal size of the galaxy as a \emph{dashed} ellipse.  The image legend gives the galaxy name, the unique identification number in parenthesis, and the morphological type as listed in Table~\ref{table:general_properties}.  The horizontal solid line in the lower-left corner of each image represents $30\arcsec$.}
\figsetgrpend
 
\figsetgrpstart
\figsetgrpnum{8.252}
\figsetgrptitle{Visualization of NGC 4288.}
\figsetplot{\includegraphics[scale=0.7,angle=90]{f8_252.eps}}
\figsetgrpnote{Presentation of our integrated spectral atlas.  We plot the spectrum as $f_{\lambda}(\lambda)$ versus rest wavelength, normalized in a way that attempts to achieve a balance between showing the full range in flux and illustrating the finer details of the continuum.  Our integrated spectrum is accompanied by a Digitized Sky Survey image which illustrates our rectangular spectroscopic aperture as a \emph{solid} outline and the $25$~mag~arcsec$^{-2}$ isophotal size of the galaxy as a \emph{dashed} ellipse.  The image legend gives the galaxy name, the unique identification number in parenthesis, and the morphological type as listed in Table~\ref{table:general_properties}.  The horizontal solid line in the lower-left corner of each image represents $30\arcsec$.}
\figsetgrpend
 
\figsetgrpstart
\figsetgrpnum{8.253}
\figsetgrptitle{Visualization of NGC 4303.}
\figsetplot{\includegraphics[scale=0.7,angle=90]{f8_253.eps}}
\figsetgrpnote{Presentation of our integrated spectral atlas.  We plot the spectrum as $f_{\lambda}(\lambda)$ versus rest wavelength, normalized in a way that attempts to achieve a balance between showing the full range in flux and illustrating the finer details of the continuum.  Our integrated spectrum is accompanied by a Digitized Sky Survey image which illustrates our rectangular spectroscopic aperture as a \emph{solid} outline and the $25$~mag~arcsec$^{-2}$ isophotal size of the galaxy as a \emph{dashed} ellipse.  The image legend gives the galaxy name, the unique identification number in parenthesis, and the morphological type as listed in Table~\ref{table:general_properties}.  The horizontal solid line in the lower-left corner of each image represents $30\arcsec$.}
\figsetgrpend
 
\figsetgrpstart
\figsetgrpnum{8.254}
\figsetgrptitle{Visualization of NGC 4321.}
\figsetplot{\includegraphics[scale=0.7,angle=90]{f8_254.eps}}
\figsetgrpnote{Presentation of our integrated spectral atlas.  We plot the spectrum as $f_{\lambda}(\lambda)$ versus rest wavelength, normalized in a way that attempts to achieve a balance between showing the full range in flux and illustrating the finer details of the continuum.  Our integrated spectrum is accompanied by a Digitized Sky Survey image which illustrates our rectangular spectroscopic aperture as a \emph{solid} outline and the $25$~mag~arcsec$^{-2}$ isophotal size of the galaxy as a \emph{dashed} ellipse.  The image legend gives the galaxy name, the unique identification number in parenthesis, and the morphological type as listed in Table~\ref{table:general_properties}.  The horizontal solid line in the lower-left corner of each image represents $30\arcsec$.}
\figsetgrpend
 
\figsetgrpstart
\figsetgrpnum{8.255}
\figsetgrptitle{Visualization of NGC 4384.}
\figsetplot{\includegraphics[scale=0.7,angle=90]{f8_255.eps}}
\figsetgrpnote{Presentation of our integrated spectral atlas.  We plot the spectrum as $f_{\lambda}(\lambda)$ versus rest wavelength, normalized in a way that attempts to achieve a balance between showing the full range in flux and illustrating the finer details of the continuum.  Our integrated spectrum is accompanied by a Digitized Sky Survey image which illustrates our rectangular spectroscopic aperture as a \emph{solid} outline and the $25$~mag~arcsec$^{-2}$ isophotal size of the galaxy as a \emph{dashed} ellipse.  The image legend gives the galaxy name, the unique identification number in parenthesis, and the morphological type as listed in Table~\ref{table:general_properties}.  The horizontal solid line in the lower-left corner of each image represents $30\arcsec$.}
\figsetgrpend
 
\figsetgrpstart
\figsetgrpnum{8.256}
\figsetgrptitle{Visualization of NGC 4389.}
\figsetplot{\includegraphics[scale=0.7,angle=90]{f8_256.eps}}
\figsetgrpnote{Presentation of our integrated spectral atlas.  We plot the spectrum as $f_{\lambda}(\lambda)$ versus rest wavelength, normalized in a way that attempts to achieve a balance between showing the full range in flux and illustrating the finer details of the continuum.  Our integrated spectrum is accompanied by a Digitized Sky Survey image which illustrates our rectangular spectroscopic aperture as a \emph{solid} outline and the $25$~mag~arcsec$^{-2}$ isophotal size of the galaxy as a \emph{dashed} ellipse.  The image legend gives the galaxy name, the unique identification number in parenthesis, and the morphological type as listed in Table~\ref{table:general_properties}.  The horizontal solid line in the lower-left corner of each image represents $30\arcsec$.}
\figsetgrpend
 
\figsetgrpstart
\figsetgrpnum{8.257}
\figsetgrptitle{Visualization of NGC 4385.}
\figsetplot{\includegraphics[scale=0.7,angle=90]{f8_257.eps}}
\figsetgrpnote{Presentation of our integrated spectral atlas.  We plot the spectrum as $f_{\lambda}(\lambda)$ versus rest wavelength, normalized in a way that attempts to achieve a balance between showing the full range in flux and illustrating the finer details of the continuum.  Our integrated spectrum is accompanied by a Digitized Sky Survey image which illustrates our rectangular spectroscopic aperture as a \emph{solid} outline and the $25$~mag~arcsec$^{-2}$ isophotal size of the galaxy as a \emph{dashed} ellipse.  The image legend gives the galaxy name, the unique identification number in parenthesis, and the morphological type as listed in Table~\ref{table:general_properties}.  The horizontal solid line in the lower-left corner of each image represents $30\arcsec$.}
\figsetgrpend
 
\figsetgrpstart
\figsetgrpnum{8.258}
\figsetgrptitle{Visualization of UGCA 281.}
\figsetplot{\includegraphics[scale=0.7,angle=90]{f8_258.eps}}
\figsetgrpnote{Presentation of our integrated spectral atlas.  We plot the spectrum as $f_{\lambda}(\lambda)$ versus rest wavelength, normalized in a way that attempts to achieve a balance between showing the full range in flux and illustrating the finer details of the continuum.  Our integrated spectrum is accompanied by a Digitized Sky Survey image which illustrates our rectangular spectroscopic aperture as a \emph{solid} outline and the $25$~mag~arcsec$^{-2}$ isophotal size of the galaxy as a \emph{dashed} ellipse.  The image legend gives the galaxy name, the unique identification number in parenthesis, and the morphological type as listed in Table~\ref{table:general_properties}.  The horizontal solid line in the lower-left corner of each image represents $30\arcsec$.}
\figsetgrpend
 
\figsetgrpstart
\figsetgrpnum{8.259}
\figsetgrptitle{Visualization of NGC 4414.}
\figsetplot{\includegraphics[scale=0.7,angle=90]{f8_259.eps}}
\figsetgrpnote{Presentation of our integrated spectral atlas.  We plot the spectrum as $f_{\lambda}(\lambda)$ versus rest wavelength, normalized in a way that attempts to achieve a balance between showing the full range in flux and illustrating the finer details of the continuum.  Our integrated spectrum is accompanied by a Digitized Sky Survey image which illustrates our rectangular spectroscopic aperture as a \emph{solid} outline and the $25$~mag~arcsec$^{-2}$ isophotal size of the galaxy as a \emph{dashed} ellipse.  The image legend gives the galaxy name, the unique identification number in parenthesis, and the morphological type as listed in Table~\ref{table:general_properties}.  The horizontal solid line in the lower-left corner of each image represents $30\arcsec$.}
\figsetgrpend
 
\figsetgrpstart
\figsetgrpnum{8.260}
\figsetgrptitle{Visualization of NGC 4455.}
\figsetplot{\includegraphics[scale=0.7,angle=90]{f8_260.eps}}
\figsetgrpnote{Presentation of our integrated spectral atlas.  We plot the spectrum as $f_{\lambda}(\lambda)$ versus rest wavelength, normalized in a way that attempts to achieve a balance between showing the full range in flux and illustrating the finer details of the continuum.  Our integrated spectrum is accompanied by a Digitized Sky Survey image which illustrates our rectangular spectroscopic aperture as a \emph{solid} outline and the $25$~mag~arcsec$^{-2}$ isophotal size of the galaxy as a \emph{dashed} ellipse.  The image legend gives the galaxy name, the unique identification number in parenthesis, and the morphological type as listed in Table~\ref{table:general_properties}.  The horizontal solid line in the lower-left corner of each image represents $30\arcsec$.}
\figsetgrpend
 
\figsetgrpstart
\figsetgrpnum{8.261}
\figsetgrptitle{Visualization of NGC 4500.}
\figsetplot{\includegraphics[scale=0.7,angle=90]{f8_261.eps}}
\figsetgrpnote{Presentation of our integrated spectral atlas.  We plot the spectrum as $f_{\lambda}(\lambda)$ versus rest wavelength, normalized in a way that attempts to achieve a balance between showing the full range in flux and illustrating the finer details of the continuum.  Our integrated spectrum is accompanied by a Digitized Sky Survey image which illustrates our rectangular spectroscopic aperture as a \emph{solid} outline and the $25$~mag~arcsec$^{-2}$ isophotal size of the galaxy as a \emph{dashed} ellipse.  The image legend gives the galaxy name, the unique identification number in parenthesis, and the morphological type as listed in Table~\ref{table:general_properties}.  The horizontal solid line in the lower-left corner of each image represents $30\arcsec$.}
\figsetgrpend
 
\figsetgrpstart
\figsetgrpnum{8.262}
\figsetgrptitle{Visualization of UGC 07690.}
\figsetplot{\includegraphics[scale=0.7,angle=90]{f8_262.eps}}
\figsetgrpnote{Presentation of our integrated spectral atlas.  We plot the spectrum as $f_{\lambda}(\lambda)$ versus rest wavelength, normalized in a way that attempts to achieve a balance between showing the full range in flux and illustrating the finer details of the continuum.  Our integrated spectrum is accompanied by a Digitized Sky Survey image which illustrates our rectangular spectroscopic aperture as a \emph{solid} outline and the $25$~mag~arcsec$^{-2}$ isophotal size of the galaxy as a \emph{dashed} ellipse.  The image legend gives the galaxy name, the unique identification number in parenthesis, and the morphological type as listed in Table~\ref{table:general_properties}.  The horizontal solid line in the lower-left corner of each image represents $30\arcsec$.}
\figsetgrpend
 
\figsetgrpstart
\figsetgrpnum{8.263}
\figsetgrptitle{Visualization of NGC 4534.}
\figsetplot{\includegraphics[scale=0.7,angle=90]{f8_263.eps}}
\figsetgrpnote{Presentation of our integrated spectral atlas.  We plot the spectrum as $f_{\lambda}(\lambda)$ versus rest wavelength, normalized in a way that attempts to achieve a balance between showing the full range in flux and illustrating the finer details of the continuum.  Our integrated spectrum is accompanied by a Digitized Sky Survey image which illustrates our rectangular spectroscopic aperture as a \emph{solid} outline and the $25$~mag~arcsec$^{-2}$ isophotal size of the galaxy as a \emph{dashed} ellipse.  The image legend gives the galaxy name, the unique identification number in parenthesis, and the morphological type as listed in Table~\ref{table:general_properties}.  The horizontal solid line in the lower-left corner of each image represents $30\arcsec$.}
\figsetgrpend
 
\figsetgrpstart
\figsetgrpnum{8.264}
\figsetgrptitle{Visualization of NGC 4569.}
\figsetplot{\includegraphics[scale=0.7,angle=90]{f8_264.eps}}
\figsetgrpnote{Presentation of our integrated spectral atlas.  We plot the spectrum as $f_{\lambda}(\lambda)$ versus rest wavelength, normalized in a way that attempts to achieve a balance between showing the full range in flux and illustrating the finer details of the continuum.  Our integrated spectrum is accompanied by a Digitized Sky Survey image which illustrates our rectangular spectroscopic aperture as a \emph{solid} outline and the $25$~mag~arcsec$^{-2}$ isophotal size of the galaxy as a \emph{dashed} ellipse.  The image legend gives the galaxy name, the unique identification number in parenthesis, and the morphological type as listed in Table~\ref{table:general_properties}.  The horizontal solid line in the lower-left corner of each image represents $30\arcsec$.}
\figsetgrpend
 
\figsetgrpstart
\figsetgrpnum{8.265}
\figsetgrptitle{Visualization of NGC 4605.}
\figsetplot{\includegraphics[scale=0.7,angle=90]{f8_265.eps}}
\figsetgrpnote{Presentation of our integrated spectral atlas.  We plot the spectrum as $f_{\lambda}(\lambda)$ versus rest wavelength, normalized in a way that attempts to achieve a balance between showing the full range in flux and illustrating the finer details of the continuum.  Our integrated spectrum is accompanied by a Digitized Sky Survey image which illustrates our rectangular spectroscopic aperture as a \emph{solid} outline and the $25$~mag~arcsec$^{-2}$ isophotal size of the galaxy as a \emph{dashed} ellipse.  The image legend gives the galaxy name, the unique identification number in parenthesis, and the morphological type as listed in Table~\ref{table:general_properties}.  The horizontal solid line in the lower-left corner of each image represents $30\arcsec$.}
\figsetgrpend
 
\figsetgrpstart
\figsetgrpnum{8.266}
\figsetgrptitle{Visualization of NGC 4618.}
\figsetplot{\includegraphics[scale=0.7,angle=90]{f8_266.eps}}
\figsetgrpnote{Presentation of our integrated spectral atlas.  We plot the spectrum as $f_{\lambda}(\lambda)$ versus rest wavelength, normalized in a way that attempts to achieve a balance between showing the full range in flux and illustrating the finer details of the continuum.  Our integrated spectrum is accompanied by a Digitized Sky Survey image which illustrates our rectangular spectroscopic aperture as a \emph{solid} outline and the $25$~mag~arcsec$^{-2}$ isophotal size of the galaxy as a \emph{dashed} ellipse.  The image legend gives the galaxy name, the unique identification number in parenthesis, and the morphological type as listed in Table~\ref{table:general_properties}.  The horizontal solid line in the lower-left corner of each image represents $30\arcsec$.}
\figsetgrpend
 
\figsetgrpstart
\figsetgrpnum{8.267}
\figsetgrptitle{Visualization of NGC 4625.}
\figsetplot{\includegraphics[scale=0.7,angle=90]{f8_267.eps}}
\figsetgrpnote{Presentation of our integrated spectral atlas.  We plot the spectrum as $f_{\lambda}(\lambda)$ versus rest wavelength, normalized in a way that attempts to achieve a balance between showing the full range in flux and illustrating the finer details of the continuum.  Our integrated spectrum is accompanied by a Digitized Sky Survey image which illustrates our rectangular spectroscopic aperture as a \emph{solid} outline and the $25$~mag~arcsec$^{-2}$ isophotal size of the galaxy as a \emph{dashed} ellipse.  The image legend gives the galaxy name, the unique identification number in parenthesis, and the morphological type as listed in Table~\ref{table:general_properties}.  The horizontal solid line in the lower-left corner of each image represents $30\arcsec$.}
\figsetgrpend
 
\figsetgrpstart
\figsetgrpnum{8.268}
\figsetgrptitle{Visualization of NGC 4651.}
\figsetplot{\includegraphics[scale=0.7,angle=90]{f8_268.eps}}
\figsetgrpnote{Presentation of our integrated spectral atlas.  We plot the spectrum as $f_{\lambda}(\lambda)$ versus rest wavelength, normalized in a way that attempts to achieve a balance between showing the full range in flux and illustrating the finer details of the continuum.  Our integrated spectrum is accompanied by a Digitized Sky Survey image which illustrates our rectangular spectroscopic aperture as a \emph{solid} outline and the $25$~mag~arcsec$^{-2}$ isophotal size of the galaxy as a \emph{dashed} ellipse.  The image legend gives the galaxy name, the unique identification number in parenthesis, and the morphological type as listed in Table~\ref{table:general_properties}.  The horizontal solid line in the lower-left corner of each image represents $30\arcsec$.}
\figsetgrpend
 
\figsetgrpstart
\figsetgrpnum{8.269}
\figsetgrptitle{Visualization of NGC 4656.}
\figsetplot{\includegraphics[scale=0.7,angle=90]{f8_269.eps}}
\figsetgrpnote{Presentation of our integrated spectral atlas.  We plot the spectrum as $f_{\lambda}(\lambda)$ versus rest wavelength, normalized in a way that attempts to achieve a balance between showing the full range in flux and illustrating the finer details of the continuum.  Our integrated spectrum is accompanied by a Digitized Sky Survey image which illustrates our rectangular spectroscopic aperture as a \emph{solid} outline and the $25$~mag~arcsec$^{-2}$ isophotal size of the galaxy as a \emph{dashed} ellipse.  The image legend gives the galaxy name, the unique identification number in parenthesis, and the morphological type as listed in Table~\ref{table:general_properties}.  The horizontal solid line in the lower-left corner of each image represents $30\arcsec$.}
\figsetgrpend
 
\figsetgrpstart
\figsetgrpnum{8.270}
\figsetgrptitle{Visualization of NGC 4666.}
\figsetplot{\includegraphics[scale=0.7,angle=90]{f8_270.eps}}
\figsetgrpnote{Presentation of our integrated spectral atlas.  We plot the spectrum as $f_{\lambda}(\lambda)$ versus rest wavelength, normalized in a way that attempts to achieve a balance between showing the full range in flux and illustrating the finer details of the continuum.  Our integrated spectrum is accompanied by a Digitized Sky Survey image which illustrates our rectangular spectroscopic aperture as a \emph{solid} outline and the $25$~mag~arcsec$^{-2}$ isophotal size of the galaxy as a \emph{dashed} ellipse.  The image legend gives the galaxy name, the unique identification number in parenthesis, and the morphological type as listed in Table~\ref{table:general_properties}.  The horizontal solid line in the lower-left corner of each image represents $30\arcsec$.}
\figsetgrpend
 
\figsetgrpstart
\figsetgrpnum{8.271}
\figsetgrptitle{Visualization of NGC 4670.}
\figsetplot{\includegraphics[scale=0.7,angle=90]{f8_271.eps}}
\figsetgrpnote{Presentation of our integrated spectral atlas.  We plot the spectrum as $f_{\lambda}(\lambda)$ versus rest wavelength, normalized in a way that attempts to achieve a balance between showing the full range in flux and illustrating the finer details of the continuum.  Our integrated spectrum is accompanied by a Digitized Sky Survey image which illustrates our rectangular spectroscopic aperture as a \emph{solid} outline and the $25$~mag~arcsec$^{-2}$ isophotal size of the galaxy as a \emph{dashed} ellipse.  The image legend gives the galaxy name, the unique identification number in parenthesis, and the morphological type as listed in Table~\ref{table:general_properties}.  The horizontal solid line in the lower-left corner of each image represents $30\arcsec$.}
\figsetgrpend
 
\figsetgrpstart
\figsetgrpnum{8.272}
\figsetgrptitle{Visualization of NGC 4676.}
\figsetplot{\includegraphics[scale=0.7,angle=90]{f8_272.eps}}
\figsetgrpnote{Presentation of our integrated spectral atlas.  We plot the spectrum as $f_{\lambda}(\lambda)$ versus rest wavelength, normalized in a way that attempts to achieve a balance between showing the full range in flux and illustrating the finer details of the continuum.  Our integrated spectrum is accompanied by a Digitized Sky Survey image which illustrates our rectangular spectroscopic aperture as a \emph{solid} outline and the $25$~mag~arcsec$^{-2}$ isophotal size of the galaxy as a \emph{dashed} ellipse.  The image legend gives the galaxy name, the unique identification number in parenthesis, and the morphological type as listed in Table~\ref{table:general_properties}.  The horizontal solid line in the lower-left corner of each image represents $30\arcsec$.}
\figsetgrpend
 
\figsetgrpstart
\figsetgrpnum{8.273}
\figsetgrptitle{Visualization of NGC 4676 A.}
\figsetplot{\includegraphics[scale=0.7,angle=90]{f8_273.eps}}
\figsetgrpnote{Presentation of our integrated spectral atlas.  We plot the spectrum as $f_{\lambda}(\lambda)$ versus rest wavelength, normalized in a way that attempts to achieve a balance between showing the full range in flux and illustrating the finer details of the continuum.  Our integrated spectrum is accompanied by a Digitized Sky Survey image which illustrates our rectangular spectroscopic aperture as a \emph{solid} outline and the $25$~mag~arcsec$^{-2}$ isophotal size of the galaxy as a \emph{dashed} ellipse.  The image legend gives the galaxy name, the unique identification number in parenthesis, and the morphological type as listed in Table~\ref{table:general_properties}.  The horizontal solid line in the lower-left corner of each image represents $30\arcsec$.}
\figsetgrpend
 
\figsetgrpstart
\figsetgrpnum{8.274}
\figsetgrptitle{Visualization of NGC 4676 B.}
\figsetplot{\includegraphics[scale=0.7,angle=90]{f8_274.eps}}
\figsetgrpnote{Presentation of our integrated spectral atlas.  We plot the spectrum as $f_{\lambda}(\lambda)$ versus rest wavelength, normalized in a way that attempts to achieve a balance between showing the full range in flux and illustrating the finer details of the continuum.  Our integrated spectrum is accompanied by a Digitized Sky Survey image which illustrates our rectangular spectroscopic aperture as a \emph{solid} outline and the $25$~mag~arcsec$^{-2}$ isophotal size of the galaxy as a \emph{dashed} ellipse.  The image legend gives the galaxy name, the unique identification number in parenthesis, and the morphological type as listed in Table~\ref{table:general_properties}.  The horizontal solid line in the lower-left corner of each image represents $30\arcsec$.}
\figsetgrpend
 
\figsetgrpstart
\figsetgrpnum{8.275}
\figsetgrptitle{Visualization of UGC 07950.}
\figsetplot{\includegraphics[scale=0.7,angle=90]{f8_275.eps}}
\figsetgrpnote{Presentation of our integrated spectral atlas.  We plot the spectrum as $f_{\lambda}(\lambda)$ versus rest wavelength, normalized in a way that attempts to achieve a balance between showing the full range in flux and illustrating the finer details of the continuum.  Our integrated spectrum is accompanied by a Digitized Sky Survey image which illustrates our rectangular spectroscopic aperture as a \emph{solid} outline and the $25$~mag~arcsec$^{-2}$ isophotal size of the galaxy as a \emph{dashed} ellipse.  The image legend gives the galaxy name, the unique identification number in parenthesis, and the morphological type as listed in Table~\ref{table:general_properties}.  The horizontal solid line in the lower-left corner of each image represents $30\arcsec$.}
\figsetgrpend
 
\figsetgrpstart
\figsetgrpnum{8.276}
\figsetgrptitle{Visualization of NGC 4713.}
\figsetplot{\includegraphics[scale=0.7,angle=90]{f8_276.eps}}
\figsetgrpnote{Presentation of our integrated spectral atlas.  We plot the spectrum as $f_{\lambda}(\lambda)$ versus rest wavelength, normalized in a way that attempts to achieve a balance between showing the full range in flux and illustrating the finer details of the continuum.  Our integrated spectrum is accompanied by a Digitized Sky Survey image which illustrates our rectangular spectroscopic aperture as a \emph{solid} outline and the $25$~mag~arcsec$^{-2}$ isophotal size of the galaxy as a \emph{dashed} ellipse.  The image legend gives the galaxy name, the unique identification number in parenthesis, and the morphological type as listed in Table~\ref{table:general_properties}.  The horizontal solid line in the lower-left corner of each image represents $30\arcsec$.}
\figsetgrpend
 
\figsetgrpstart
\figsetgrpnum{8.277}
\figsetgrptitle{Visualization of NGC 4736.}
\figsetplot{\includegraphics[scale=0.7,angle=90]{f8_277.eps}}
\figsetgrpnote{Presentation of our integrated spectral atlas.  We plot the spectrum as $f_{\lambda}(\lambda)$ versus rest wavelength, normalized in a way that attempts to achieve a balance between showing the full range in flux and illustrating the finer details of the continuum.  Our integrated spectrum is accompanied by a Digitized Sky Survey image which illustrates our rectangular spectroscopic aperture as a \emph{solid} outline and the $25$~mag~arcsec$^{-2}$ isophotal size of the galaxy as a \emph{dashed} ellipse.  The image legend gives the galaxy name, the unique identification number in parenthesis, and the morphological type as listed in Table~\ref{table:general_properties}.  The horizontal solid line in the lower-left corner of each image represents $30\arcsec$.}
\figsetgrpend
 
\figsetgrpstart
\figsetgrpnum{8.278}
\figsetgrptitle{Visualization of MRK 0053.}
\figsetplot{\includegraphics[scale=0.7,angle=90]{f8_278.eps}}
\figsetgrpnote{Presentation of our integrated spectral atlas.  We plot the spectrum as $f_{\lambda}(\lambda)$ versus rest wavelength, normalized in a way that attempts to achieve a balance between showing the full range in flux and illustrating the finer details of the continuum.  Our integrated spectrum is accompanied by a Digitized Sky Survey image which illustrates our rectangular spectroscopic aperture as a \emph{solid} outline and the $25$~mag~arcsec$^{-2}$ isophotal size of the galaxy as a \emph{dashed} ellipse.  The image legend gives the galaxy name, the unique identification number in parenthesis, and the morphological type as listed in Table~\ref{table:general_properties}.  The horizontal solid line in the lower-left corner of each image represents $30\arcsec$.}
\figsetgrpend
 
\figsetgrpstart
\figsetgrpnum{8.279}
\figsetgrptitle{Visualization of UGC 08058.}
\figsetplot{\includegraphics[scale=0.7,angle=90]{f8_279.eps}}
\figsetgrpnote{Presentation of our integrated spectral atlas.  We plot the spectrum as $f_{\lambda}(\lambda)$ versus rest wavelength, normalized in a way that attempts to achieve a balance between showing the full range in flux and illustrating the finer details of the continuum.  Our integrated spectrum is accompanied by a Digitized Sky Survey image which illustrates our rectangular spectroscopic aperture as a \emph{solid} outline and the $25$~mag~arcsec$^{-2}$ isophotal size of the galaxy as a \emph{dashed} ellipse.  The image legend gives the galaxy name, the unique identification number in parenthesis, and the morphological type as listed in Table~\ref{table:general_properties}.  The horizontal solid line in the lower-left corner of each image represents $30\arcsec$.}
\figsetgrpend
 
\figsetgrpstart
\figsetgrpnum{8.280}
\figsetgrptitle{Visualization of MRK 0055.}
\figsetplot{\includegraphics[scale=0.7,angle=90]{f8_280.eps}}
\figsetgrpnote{Presentation of our integrated spectral atlas.  We plot the spectrum as $f_{\lambda}(\lambda)$ versus rest wavelength, normalized in a way that attempts to achieve a balance between showing the full range in flux and illustrating the finer details of the continuum.  Our integrated spectrum is accompanied by a Digitized Sky Survey image which illustrates our rectangular spectroscopic aperture as a \emph{solid} outline and the $25$~mag~arcsec$^{-2}$ isophotal size of the galaxy as a \emph{dashed} ellipse.  The image legend gives the galaxy name, the unique identification number in parenthesis, and the morphological type as listed in Table~\ref{table:general_properties}.  The horizontal solid line in the lower-left corner of each image represents $30\arcsec$.}
\figsetgrpend
 
\figsetgrpstart
\figsetgrpnum{8.281}
\figsetgrptitle{Visualization of NGC 4861.}
\figsetplot{\includegraphics[scale=0.7,angle=90]{f8_281.eps}}
\figsetgrpnote{Presentation of our integrated spectral atlas.  We plot the spectrum as $f_{\lambda}(\lambda)$ versus rest wavelength, normalized in a way that attempts to achieve a balance between showing the full range in flux and illustrating the finer details of the continuum.  Our integrated spectrum is accompanied by a Digitized Sky Survey image which illustrates our rectangular spectroscopic aperture as a \emph{solid} outline and the $25$~mag~arcsec$^{-2}$ isophotal size of the galaxy as a \emph{dashed} ellipse.  The image legend gives the galaxy name, the unique identification number in parenthesis, and the morphological type as listed in Table~\ref{table:general_properties}.  The horizontal solid line in the lower-left corner of each image represents $30\arcsec$.}
\figsetgrpend
 
\figsetgrpstart
\figsetgrpnum{8.282}
\figsetgrptitle{Visualization of NGC 4900.}
\figsetplot{\includegraphics[scale=0.7,angle=90]{f8_282.eps}}
\figsetgrpnote{Presentation of our integrated spectral atlas.  We plot the spectrum as $f_{\lambda}(\lambda)$ versus rest wavelength, normalized in a way that attempts to achieve a balance between showing the full range in flux and illustrating the finer details of the continuum.  Our integrated spectrum is accompanied by a Digitized Sky Survey image which illustrates our rectangular spectroscopic aperture as a \emph{solid} outline and the $25$~mag~arcsec$^{-2}$ isophotal size of the galaxy as a \emph{dashed} ellipse.  The image legend gives the galaxy name, the unique identification number in parenthesis, and the morphological type as listed in Table~\ref{table:general_properties}.  The horizontal solid line in the lower-left corner of each image represents $30\arcsec$.}
\figsetgrpend
 
\figsetgrpstart
\figsetgrpnum{8.283}
\figsetgrptitle{Visualization of NGC 4922 S.}
\figsetplot{\includegraphics[scale=0.7,angle=90]{f8_283.eps}}
\figsetgrpnote{Presentation of our integrated spectral atlas.  We plot the spectrum as $f_{\lambda}(\lambda)$ versus rest wavelength, normalized in a way that attempts to achieve a balance between showing the full range in flux and illustrating the finer details of the continuum.  Our integrated spectrum is accompanied by a Digitized Sky Survey image which illustrates our rectangular spectroscopic aperture as a \emph{solid} outline and the $25$~mag~arcsec$^{-2}$ isophotal size of the galaxy as a \emph{dashed} ellipse.  The image legend gives the galaxy name, the unique identification number in parenthesis, and the morphological type as listed in Table~\ref{table:general_properties}.  The horizontal solid line in the lower-left corner of each image represents $30\arcsec$.}
\figsetgrpend
 
\figsetgrpstart
\figsetgrpnum{8.284}
\figsetgrptitle{Visualization of NGC 4922.}
\figsetplot{\includegraphics[scale=0.7,angle=90]{f8_284.eps}}
\figsetgrpnote{Presentation of our integrated spectral atlas.  We plot the spectrum as $f_{\lambda}(\lambda)$ versus rest wavelength, normalized in a way that attempts to achieve a balance between showing the full range in flux and illustrating the finer details of the continuum.  Our integrated spectrum is accompanied by a Digitized Sky Survey image which illustrates our rectangular spectroscopic aperture as a \emph{solid} outline and the $25$~mag~arcsec$^{-2}$ isophotal size of the galaxy as a \emph{dashed} ellipse.  The image legend gives the galaxy name, the unique identification number in parenthesis, and the morphological type as listed in Table~\ref{table:general_properties}.  The horizontal solid line in the lower-left corner of each image represents $30\arcsec$.}
\figsetgrpend
 
\figsetgrpstart
\figsetgrpnum{8.285}
\figsetgrptitle{Visualization of NGC 4922 N.}
\figsetplot{\includegraphics[scale=0.7,angle=90]{f8_285.eps}}
\figsetgrpnote{Presentation of our integrated spectral atlas.  We plot the spectrum as $f_{\lambda}(\lambda)$ versus rest wavelength, normalized in a way that attempts to achieve a balance between showing the full range in flux and illustrating the finer details of the continuum.  Our integrated spectrum is accompanied by a Digitized Sky Survey image which illustrates our rectangular spectroscopic aperture as a \emph{solid} outline and the $25$~mag~arcsec$^{-2}$ isophotal size of the galaxy as a \emph{dashed} ellipse.  The image legend gives the galaxy name, the unique identification number in parenthesis, and the morphological type as listed in Table~\ref{table:general_properties}.  The horizontal solid line in the lower-left corner of each image represents $30\arcsec$.}
\figsetgrpend
 
\figsetgrpstart
\figsetgrpnum{8.286}
\figsetgrptitle{Visualization of MCG -02-33-098.}
\figsetplot{\includegraphics[scale=0.7,angle=90]{f8_286.eps}}
\figsetgrpnote{Presentation of our integrated spectral atlas.  We plot the spectrum as $f_{\lambda}(\lambda)$ versus rest wavelength, normalized in a way that attempts to achieve a balance between showing the full range in flux and illustrating the finer details of the continuum.  Our integrated spectrum is accompanied by a Digitized Sky Survey image which illustrates our rectangular spectroscopic aperture as a \emph{solid} outline and the $25$~mag~arcsec$^{-2}$ isophotal size of the galaxy as a \emph{dashed} ellipse.  The image legend gives the galaxy name, the unique identification number in parenthesis, and the morphological type as listed in Table~\ref{table:general_properties}.  The horizontal solid line in the lower-left corner of each image represents $30\arcsec$.}
\figsetgrpend
 
\figsetgrpstart
\figsetgrpnum{8.287}
\figsetgrptitle{Visualization of UGCA 320.}
\figsetplot{\includegraphics[scale=0.7,angle=90]{f8_287.eps}}
\figsetgrpnote{Presentation of our integrated spectral atlas.  We plot the spectrum as $f_{\lambda}(\lambda)$ versus rest wavelength, normalized in a way that attempts to achieve a balance between showing the full range in flux and illustrating the finer details of the continuum.  Our integrated spectrum is accompanied by a Digitized Sky Survey image which illustrates our rectangular spectroscopic aperture as a \emph{solid} outline and the $25$~mag~arcsec$^{-2}$ isophotal size of the galaxy as a \emph{dashed} ellipse.  The image legend gives the galaxy name, the unique identification number in parenthesis, and the morphological type as listed in Table~\ref{table:general_properties}.  The horizontal solid line in the lower-left corner of each image represents $30\arcsec$.}
\figsetgrpend
 
\figsetgrpstart
\figsetgrpnum{8.288}
\figsetgrptitle{Visualization of UGC 08201.}
\figsetplot{\includegraphics[scale=0.7,angle=90]{f8_288.eps}}
\figsetgrpnote{Presentation of our integrated spectral atlas.  We plot the spectrum as $f_{\lambda}(\lambda)$ versus rest wavelength, normalized in a way that attempts to achieve a balance between showing the full range in flux and illustrating the finer details of the continuum.  Our integrated spectrum is accompanied by a Digitized Sky Survey image which illustrates our rectangular spectroscopic aperture as a \emph{solid} outline and the $25$~mag~arcsec$^{-2}$ isophotal size of the galaxy as a \emph{dashed} ellipse.  The image legend gives the galaxy name, the unique identification number in parenthesis, and the morphological type as listed in Table~\ref{table:general_properties}.  The horizontal solid line in the lower-left corner of each image represents $30\arcsec$.}
\figsetgrpend
 
\figsetgrpstart
\figsetgrpnum{8.289}
\figsetgrptitle{Visualization of NGC 5014.}
\figsetplot{\includegraphics[scale=0.7,angle=90]{f8_289.eps}}
\figsetgrpnote{Presentation of our integrated spectral atlas.  We plot the spectrum as $f_{\lambda}(\lambda)$ versus rest wavelength, normalized in a way that attempts to achieve a balance between showing the full range in flux and illustrating the finer details of the continuum.  Our integrated spectrum is accompanied by a Digitized Sky Survey image which illustrates our rectangular spectroscopic aperture as a \emph{solid} outline and the $25$~mag~arcsec$^{-2}$ isophotal size of the galaxy as a \emph{dashed} ellipse.  The image legend gives the galaxy name, the unique identification number in parenthesis, and the morphological type as listed in Table~\ref{table:general_properties}.  The horizontal solid line in the lower-left corner of each image represents $30\arcsec$.}
\figsetgrpend
 
\figsetgrpstart
\figsetgrpnum{8.290}
\figsetgrptitle{Visualization of MRK 0247.}
\figsetplot{\includegraphics[scale=0.7,angle=90]{f8_290.eps}}
\figsetgrpnote{Presentation of our integrated spectral atlas.  We plot the spectrum as $f_{\lambda}(\lambda)$ versus rest wavelength, normalized in a way that attempts to achieve a balance between showing the full range in flux and illustrating the finer details of the continuum.  Our integrated spectrum is accompanied by a Digitized Sky Survey image which illustrates our rectangular spectroscopic aperture as a \emph{solid} outline and the $25$~mag~arcsec$^{-2}$ isophotal size of the galaxy as a \emph{dashed} ellipse.  The image legend gives the galaxy name, the unique identification number in parenthesis, and the morphological type as listed in Table~\ref{table:general_properties}.  The horizontal solid line in the lower-left corner of each image represents $30\arcsec$.}
\figsetgrpend
 
\figsetgrpstart
\figsetgrpnum{8.291}
\figsetgrptitle{Visualization of UGC 08323.}
\figsetplot{\includegraphics[scale=0.7,angle=90]{f8_291.eps}}
\figsetgrpnote{Presentation of our integrated spectral atlas.  We plot the spectrum as $f_{\lambda}(\lambda)$ versus rest wavelength, normalized in a way that attempts to achieve a balance between showing the full range in flux and illustrating the finer details of the continuum.  Our integrated spectrum is accompanied by a Digitized Sky Survey image which illustrates our rectangular spectroscopic aperture as a \emph{solid} outline and the $25$~mag~arcsec$^{-2}$ isophotal size of the galaxy as a \emph{dashed} ellipse.  The image legend gives the galaxy name, the unique identification number in parenthesis, and the morphological type as listed in Table~\ref{table:general_properties}.  The horizontal solid line in the lower-left corner of each image represents $30\arcsec$.}
\figsetgrpend
 
\figsetgrpstart
\figsetgrpnum{8.292}
\figsetgrptitle{Visualization of IC 0860.}
\figsetplot{\includegraphics[scale=0.7,angle=90]{f8_292.eps}}
\figsetgrpnote{Presentation of our integrated spectral atlas.  We plot the spectrum as $f_{\lambda}(\lambda)$ versus rest wavelength, normalized in a way that attempts to achieve a balance between showing the full range in flux and illustrating the finer details of the continuum.  Our integrated spectrum is accompanied by a Digitized Sky Survey image which illustrates our rectangular spectroscopic aperture as a \emph{solid} outline and the $25$~mag~arcsec$^{-2}$ isophotal size of the galaxy as a \emph{dashed} ellipse.  The image legend gives the galaxy name, the unique identification number in parenthesis, and the morphological type as listed in Table~\ref{table:general_properties}.  The horizontal solid line in the lower-left corner of each image represents $30\arcsec$.}
\figsetgrpend
 
\figsetgrpstart
\figsetgrpnum{8.293}
\figsetgrptitle{Visualization of UGC 08335 NW.}
\figsetplot{\includegraphics[scale=0.7,angle=90]{f8_293.eps}}
\figsetgrpnote{Presentation of our integrated spectral atlas.  We plot the spectrum as $f_{\lambda}(\lambda)$ versus rest wavelength, normalized in a way that attempts to achieve a balance between showing the full range in flux and illustrating the finer details of the continuum.  Our integrated spectrum is accompanied by a Digitized Sky Survey image which illustrates our rectangular spectroscopic aperture as a \emph{solid} outline and the $25$~mag~arcsec$^{-2}$ isophotal size of the galaxy as a \emph{dashed} ellipse.  The image legend gives the galaxy name, the unique identification number in parenthesis, and the morphological type as listed in Table~\ref{table:general_properties}.  The horizontal solid line in the lower-left corner of each image represents $30\arcsec$.}
\figsetgrpend
 
\figsetgrpstart
\figsetgrpnum{8.294}
\figsetgrptitle{Visualization of UGC 08335.}
\figsetplot{\includegraphics[scale=0.7,angle=90]{f8_294.eps}}
\figsetgrpnote{Presentation of our integrated spectral atlas.  We plot the spectrum as $f_{\lambda}(\lambda)$ versus rest wavelength, normalized in a way that attempts to achieve a balance between showing the full range in flux and illustrating the finer details of the continuum.  Our integrated spectrum is accompanied by a Digitized Sky Survey image which illustrates our rectangular spectroscopic aperture as a \emph{solid} outline and the $25$~mag~arcsec$^{-2}$ isophotal size of the galaxy as a \emph{dashed} ellipse.  The image legend gives the galaxy name, the unique identification number in parenthesis, and the morphological type as listed in Table~\ref{table:general_properties}.  The horizontal solid line in the lower-left corner of each image represents $30\arcsec$.}
\figsetgrpend
 
\figsetgrpstart
\figsetgrpnum{8.295}
\figsetgrptitle{Visualization of UGC 08335 SE.}
\figsetplot{\includegraphics[scale=0.7,angle=90]{f8_295.eps}}
\figsetgrpnote{Presentation of our integrated spectral atlas.  We plot the spectrum as $f_{\lambda}(\lambda)$ versus rest wavelength, normalized in a way that attempts to achieve a balance between showing the full range in flux and illustrating the finer details of the continuum.  Our integrated spectrum is accompanied by a Digitized Sky Survey image which illustrates our rectangular spectroscopic aperture as a \emph{solid} outline and the $25$~mag~arcsec$^{-2}$ isophotal size of the galaxy as a \emph{dashed} ellipse.  The image legend gives the galaxy name, the unique identification number in parenthesis, and the morphological type as listed in Table~\ref{table:general_properties}.  The horizontal solid line in the lower-left corner of each image represents $30\arcsec$.}
\figsetgrpend
 
\figsetgrpstart
\figsetgrpnum{8.296}
\figsetgrptitle{Visualization of IC 0883.}
\figsetplot{\includegraphics[scale=0.7,angle=90]{f8_296.eps}}
\figsetgrpnote{Presentation of our integrated spectral atlas.  We plot the spectrum as $f_{\lambda}(\lambda)$ versus rest wavelength, normalized in a way that attempts to achieve a balance between showing the full range in flux and illustrating the finer details of the continuum.  Our integrated spectrum is accompanied by a Digitized Sky Survey image which illustrates our rectangular spectroscopic aperture as a \emph{solid} outline and the $25$~mag~arcsec$^{-2}$ isophotal size of the galaxy as a \emph{dashed} ellipse.  The image legend gives the galaxy name, the unique identification number in parenthesis, and the morphological type as listed in Table~\ref{table:general_properties}.  The horizontal solid line in the lower-left corner of each image represents $30\arcsec$.}
\figsetgrpend
 
\figsetgrpstart
\figsetgrpnum{8.297}
\figsetgrptitle{Visualization of NGC 5104.}
\figsetplot{\includegraphics[scale=0.7,angle=90]{f8_297.eps}}
\figsetgrpnote{Presentation of our integrated spectral atlas.  We plot the spectrum as $f_{\lambda}(\lambda)$ versus rest wavelength, normalized in a way that attempts to achieve a balance between showing the full range in flux and illustrating the finer details of the continuum.  Our integrated spectrum is accompanied by a Digitized Sky Survey image which illustrates our rectangular spectroscopic aperture as a \emph{solid} outline and the $25$~mag~arcsec$^{-2}$ isophotal size of the galaxy as a \emph{dashed} ellipse.  The image legend gives the galaxy name, the unique identification number in parenthesis, and the morphological type as listed in Table~\ref{table:general_properties}.  The horizontal solid line in the lower-left corner of each image represents $30\arcsec$.}
\figsetgrpend
 
\figsetgrpstart
\figsetgrpnum{8.298}
\figsetgrptitle{Visualization of NGC 5144.}
\figsetplot{\includegraphics[scale=0.7,angle=90]{f8_298.eps}}
\figsetgrpnote{Presentation of our integrated spectral atlas.  We plot the spectrum as $f_{\lambda}(\lambda)$ versus rest wavelength, normalized in a way that attempts to achieve a balance between showing the full range in flux and illustrating the finer details of the continuum.  Our integrated spectrum is accompanied by a Digitized Sky Survey image which illustrates our rectangular spectroscopic aperture as a \emph{solid} outline and the $25$~mag~arcsec$^{-2}$ isophotal size of the galaxy as a \emph{dashed} ellipse.  The image legend gives the galaxy name, the unique identification number in parenthesis, and the morphological type as listed in Table~\ref{table:general_properties}.  The horizontal solid line in the lower-left corner of each image represents $30\arcsec$.}
\figsetgrpend
 
\figsetgrpstart
\figsetgrpnum{8.299}
\figsetgrptitle{Visualization of MRK 0066.}
\figsetplot{\includegraphics[scale=0.7,angle=90]{f8_299.eps}}
\figsetgrpnote{Presentation of our integrated spectral atlas.  We plot the spectrum as $f_{\lambda}(\lambda)$ versus rest wavelength, normalized in a way that attempts to achieve a balance between showing the full range in flux and illustrating the finer details of the continuum.  Our integrated spectrum is accompanied by a Digitized Sky Survey image which illustrates our rectangular spectroscopic aperture as a \emph{solid} outline and the $25$~mag~arcsec$^{-2}$ isophotal size of the galaxy as a \emph{dashed} ellipse.  The image legend gives the galaxy name, the unique identification number in parenthesis, and the morphological type as listed in Table~\ref{table:general_properties}.  The horizontal solid line in the lower-left corner of each image represents $30\arcsec$.}
\figsetgrpend
 
\figsetgrpstart
\figsetgrpnum{8.300}
\figsetgrptitle{Visualization of NGC 5194.}
\figsetplot{\includegraphics[scale=0.7,angle=90]{f8_300.eps}}
\figsetgrpnote{Presentation of our integrated spectral atlas.  We plot the spectrum as $f_{\lambda}(\lambda)$ versus rest wavelength, normalized in a way that attempts to achieve a balance between showing the full range in flux and illustrating the finer details of the continuum.  Our integrated spectrum is accompanied by a Digitized Sky Survey image which illustrates our rectangular spectroscopic aperture as a \emph{solid} outline and the $25$~mag~arcsec$^{-2}$ isophotal size of the galaxy as a \emph{dashed} ellipse.  The image legend gives the galaxy name, the unique identification number in parenthesis, and the morphological type as listed in Table~\ref{table:general_properties}.  The horizontal solid line in the lower-left corner of each image represents $30\arcsec$.}
\figsetgrpend
 
\figsetgrpstart
\figsetgrpnum{8.301}
\figsetgrptitle{Visualization of UGC 08508.}
\figsetplot{\includegraphics[scale=0.7,angle=90]{f8_301.eps}}
\figsetgrpnote{Presentation of our integrated spectral atlas.  We plot the spectrum as $f_{\lambda}(\lambda)$ versus rest wavelength, normalized in a way that attempts to achieve a balance between showing the full range in flux and illustrating the finer details of the continuum.  Our integrated spectrum is accompanied by a Digitized Sky Survey image which illustrates our rectangular spectroscopic aperture as a \emph{solid} outline and the $25$~mag~arcsec$^{-2}$ isophotal size of the galaxy as a \emph{dashed} ellipse.  The image legend gives the galaxy name, the unique identification number in parenthesis, and the morphological type as listed in Table~\ref{table:general_properties}.  The horizontal solid line in the lower-left corner of each image represents $30\arcsec$.}
\figsetgrpend
 
\figsetgrpstart
\figsetgrpnum{8.302}
\figsetgrptitle{Visualization of NGC 5238.}
\figsetplot{\includegraphics[scale=0.7,angle=90]{f8_302.eps}}
\figsetgrpnote{Presentation of our integrated spectral atlas.  We plot the spectrum as $f_{\lambda}(\lambda)$ versus rest wavelength, normalized in a way that attempts to achieve a balance between showing the full range in flux and illustrating the finer details of the continuum.  Our integrated spectrum is accompanied by a Digitized Sky Survey image which illustrates our rectangular spectroscopic aperture as a \emph{solid} outline and the $25$~mag~arcsec$^{-2}$ isophotal size of the galaxy as a \emph{dashed} ellipse.  The image legend gives the galaxy name, the unique identification number in parenthesis, and the morphological type as listed in Table~\ref{table:general_properties}.  The horizontal solid line in the lower-left corner of each image represents $30\arcsec$.}
\figsetgrpend
 
\figsetgrpstart
\figsetgrpnum{8.303}
\figsetgrptitle{Visualization of NGC 5256.}
\figsetplot{\includegraphics[scale=0.7,angle=90]{f8_303.eps}}
\figsetgrpnote{Presentation of our integrated spectral atlas.  We plot the spectrum as $f_{\lambda}(\lambda)$ versus rest wavelength, normalized in a way that attempts to achieve a balance between showing the full range in flux and illustrating the finer details of the continuum.  Our integrated spectrum is accompanied by a Digitized Sky Survey image which illustrates our rectangular spectroscopic aperture as a \emph{solid} outline and the $25$~mag~arcsec$^{-2}$ isophotal size of the galaxy as a \emph{dashed} ellipse.  The image legend gives the galaxy name, the unique identification number in parenthesis, and the morphological type as listed in Table~\ref{table:general_properties}.  The horizontal solid line in the lower-left corner of each image represents $30\arcsec$.}
\figsetgrpend
 
\figsetgrpstart
\figsetgrpnum{8.304}
\figsetgrptitle{Visualization of NGC 5257.}
\figsetplot{\includegraphics[scale=0.7,angle=90]{f8_304.eps}}
\figsetgrpnote{Presentation of our integrated spectral atlas.  We plot the spectrum as $f_{\lambda}(\lambda)$ versus rest wavelength, normalized in a way that attempts to achieve a balance between showing the full range in flux and illustrating the finer details of the continuum.  Our integrated spectrum is accompanied by a Digitized Sky Survey image which illustrates our rectangular spectroscopic aperture as a \emph{solid} outline and the $25$~mag~arcsec$^{-2}$ isophotal size of the galaxy as a \emph{dashed} ellipse.  The image legend gives the galaxy name, the unique identification number in parenthesis, and the morphological type as listed in Table~\ref{table:general_properties}.  The horizontal solid line in the lower-left corner of each image represents $30\arcsec$.}
\figsetgrpend
 
\figsetgrpstart
\figsetgrpnum{8.305}
\figsetgrptitle{Visualization of ARP 240.}
\figsetplot{\includegraphics[scale=0.7,angle=90]{f8_305.eps}}
\figsetgrpnote{Presentation of our integrated spectral atlas.  We plot the spectrum as $f_{\lambda}(\lambda)$ versus rest wavelength, normalized in a way that attempts to achieve a balance between showing the full range in flux and illustrating the finer details of the continuum.  Our integrated spectrum is accompanied by a Digitized Sky Survey image which illustrates our rectangular spectroscopic aperture as a \emph{solid} outline and the $25$~mag~arcsec$^{-2}$ isophotal size of the galaxy as a \emph{dashed} ellipse.  The image legend gives the galaxy name, the unique identification number in parenthesis, and the morphological type as listed in Table~\ref{table:general_properties}.  The horizontal solid line in the lower-left corner of each image represents $30\arcsec$.}
\figsetgrpend
 
\figsetgrpstart
\figsetgrpnum{8.306}
\figsetgrptitle{Visualization of NGC 5253.}
\figsetplot{\includegraphics[scale=0.7,angle=90]{f8_306.eps}}
\figsetgrpnote{Presentation of our integrated spectral atlas.  We plot the spectrum as $f_{\lambda}(\lambda)$ versus rest wavelength, normalized in a way that attempts to achieve a balance between showing the full range in flux and illustrating the finer details of the continuum.  Our integrated spectrum is accompanied by a Digitized Sky Survey image which illustrates our rectangular spectroscopic aperture as a \emph{solid} outline and the $25$~mag~arcsec$^{-2}$ isophotal size of the galaxy as a \emph{dashed} ellipse.  The image legend gives the galaxy name, the unique identification number in parenthesis, and the morphological type as listed in Table~\ref{table:general_properties}.  The horizontal solid line in the lower-left corner of each image represents $30\arcsec$.}
\figsetgrpend
 
\figsetgrpstart
\figsetgrpnum{8.307}
\figsetgrptitle{Visualization of NGC 5258.}
\figsetplot{\includegraphics[scale=0.7,angle=90]{f8_307.eps}}
\figsetgrpnote{Presentation of our integrated spectral atlas.  We plot the spectrum as $f_{\lambda}(\lambda)$ versus rest wavelength, normalized in a way that attempts to achieve a balance between showing the full range in flux and illustrating the finer details of the continuum.  Our integrated spectrum is accompanied by a Digitized Sky Survey image which illustrates our rectangular spectroscopic aperture as a \emph{solid} outline and the $25$~mag~arcsec$^{-2}$ isophotal size of the galaxy as a \emph{dashed} ellipse.  The image legend gives the galaxy name, the unique identification number in parenthesis, and the morphological type as listed in Table~\ref{table:general_properties}.  The horizontal solid line in the lower-left corner of each image represents $30\arcsec$.}
\figsetgrpend
 
\figsetgrpstart
\figsetgrpnum{8.308}
\figsetgrptitle{Visualization of NGC 5264.}
\figsetplot{\includegraphics[scale=0.7,angle=90]{f8_308.eps}}
\figsetgrpnote{Presentation of our integrated spectral atlas.  We plot the spectrum as $f_{\lambda}(\lambda)$ versus rest wavelength, normalized in a way that attempts to achieve a balance between showing the full range in flux and illustrating the finer details of the continuum.  Our integrated spectrum is accompanied by a Digitized Sky Survey image which illustrates our rectangular spectroscopic aperture as a \emph{solid} outline and the $25$~mag~arcsec$^{-2}$ isophotal size of the galaxy as a \emph{dashed} ellipse.  The image legend gives the galaxy name, the unique identification number in parenthesis, and the morphological type as listed in Table~\ref{table:general_properties}.  The horizontal solid line in the lower-left corner of each image represents $30\arcsec$.}
\figsetgrpend
 
\figsetgrpstart
\figsetgrpnum{8.309}
\figsetgrptitle{Visualization of NGC 5278.}
\figsetplot{\includegraphics[scale=0.7,angle=90]{f8_309.eps}}
\figsetgrpnote{Presentation of our integrated spectral atlas.  We plot the spectrum as $f_{\lambda}(\lambda)$ versus rest wavelength, normalized in a way that attempts to achieve a balance between showing the full range in flux and illustrating the finer details of the continuum.  Our integrated spectrum is accompanied by a Digitized Sky Survey image which illustrates our rectangular spectroscopic aperture as a \emph{solid} outline and the $25$~mag~arcsec$^{-2}$ isophotal size of the galaxy as a \emph{dashed} ellipse.  The image legend gives the galaxy name, the unique identification number in parenthesis, and the morphological type as listed in Table~\ref{table:general_properties}.  The horizontal solid line in the lower-left corner of each image represents $30\arcsec$.}
\figsetgrpend
 
\figsetgrpstart
\figsetgrpnum{8.310}
\figsetgrptitle{Visualization of ARP 239.}
\figsetplot{\includegraphics[scale=0.7,angle=90]{f8_310.eps}}
\figsetgrpnote{Presentation of our integrated spectral atlas.  We plot the spectrum as $f_{\lambda}(\lambda)$ versus rest wavelength, normalized in a way that attempts to achieve a balance between showing the full range in flux and illustrating the finer details of the continuum.  Our integrated spectrum is accompanied by a Digitized Sky Survey image which illustrates our rectangular spectroscopic aperture as a \emph{solid} outline and the $25$~mag~arcsec$^{-2}$ isophotal size of the galaxy as a \emph{dashed} ellipse.  The image legend gives the galaxy name, the unique identification number in parenthesis, and the morphological type as listed in Table~\ref{table:general_properties}.  The horizontal solid line in the lower-left corner of each image represents $30\arcsec$.}
\figsetgrpend
 
\figsetgrpstart
\figsetgrpnum{8.311}
\figsetgrptitle{Visualization of NGC 5279.}
\figsetplot{\includegraphics[scale=0.7,angle=90]{f8_311.eps}}
\figsetgrpnote{Presentation of our integrated spectral atlas.  We plot the spectrum as $f_{\lambda}(\lambda)$ versus rest wavelength, normalized in a way that attempts to achieve a balance between showing the full range in flux and illustrating the finer details of the continuum.  Our integrated spectrum is accompanied by a Digitized Sky Survey image which illustrates our rectangular spectroscopic aperture as a \emph{solid} outline and the $25$~mag~arcsec$^{-2}$ isophotal size of the galaxy as a \emph{dashed} ellipse.  The image legend gives the galaxy name, the unique identification number in parenthesis, and the morphological type as listed in Table~\ref{table:general_properties}.  The horizontal solid line in the lower-left corner of each image represents $30\arcsec$.}
\figsetgrpend
 
\figsetgrpstart
\figsetgrpnum{8.312}
\figsetgrptitle{Visualization of UGCA 372.}
\figsetplot{\includegraphics[scale=0.7,angle=90]{f8_312.eps}}
\figsetgrpnote{Presentation of our integrated spectral atlas.  We plot the spectrum as $f_{\lambda}(\lambda)$ versus rest wavelength, normalized in a way that attempts to achieve a balance between showing the full range in flux and illustrating the finer details of the continuum.  Our integrated spectrum is accompanied by a Digitized Sky Survey image which illustrates our rectangular spectroscopic aperture as a \emph{solid} outline and the $25$~mag~arcsec$^{-2}$ isophotal size of the galaxy as a \emph{dashed} ellipse.  The image legend gives the galaxy name, the unique identification number in parenthesis, and the morphological type as listed in Table~\ref{table:general_properties}.  The horizontal solid line in the lower-left corner of each image represents $30\arcsec$.}
\figsetgrpend
 
\figsetgrpstart
\figsetgrpnum{8.313}
\figsetgrptitle{Visualization of UGC 08696.}
\figsetplot{\includegraphics[scale=0.7,angle=90]{f8_313.eps}}
\figsetgrpnote{Presentation of our integrated spectral atlas.  We plot the spectrum as $f_{\lambda}(\lambda)$ versus rest wavelength, normalized in a way that attempts to achieve a balance between showing the full range in flux and illustrating the finer details of the continuum.  Our integrated spectrum is accompanied by a Digitized Sky Survey image which illustrates our rectangular spectroscopic aperture as a \emph{solid} outline and the $25$~mag~arcsec$^{-2}$ isophotal size of the galaxy as a \emph{dashed} ellipse.  The image legend gives the galaxy name, the unique identification number in parenthesis, and the morphological type as listed in Table~\ref{table:general_properties}.  The horizontal solid line in the lower-left corner of each image represents $30\arcsec$.}
\figsetgrpend
 
\figsetgrpstart
\figsetgrpnum{8.314}
\figsetgrptitle{Visualization of NGC 5430.}
\figsetplot{\includegraphics[scale=0.7,angle=90]{f8_314.eps}}
\figsetgrpnote{Presentation of our integrated spectral atlas.  We plot the spectrum as $f_{\lambda}(\lambda)$ versus rest wavelength, normalized in a way that attempts to achieve a balance between showing the full range in flux and illustrating the finer details of the continuum.  Our integrated spectrum is accompanied by a Digitized Sky Survey image which illustrates our rectangular spectroscopic aperture as a \emph{solid} outline and the $25$~mag~arcsec$^{-2}$ isophotal size of the galaxy as a \emph{dashed} ellipse.  The image legend gives the galaxy name, the unique identification number in parenthesis, and the morphological type as listed in Table~\ref{table:general_properties}.  The horizontal solid line in the lower-left corner of each image represents $30\arcsec$.}
\figsetgrpend
 
\figsetgrpstart
\figsetgrpnum{8.315}
\figsetgrptitle{Visualization of UGC 09081.}
\figsetplot{\includegraphics[scale=0.7,angle=90]{f8_315.eps}}
\figsetgrpnote{Presentation of our integrated spectral atlas.  We plot the spectrum as $f_{\lambda}(\lambda)$ versus rest wavelength, normalized in a way that attempts to achieve a balance between showing the full range in flux and illustrating the finer details of the continuum.  Our integrated spectrum is accompanied by a Digitized Sky Survey image which illustrates our rectangular spectroscopic aperture as a \emph{solid} outline and the $25$~mag~arcsec$^{-2}$ isophotal size of the galaxy as a \emph{dashed} ellipse.  The image legend gives the galaxy name, the unique identification number in parenthesis, and the morphological type as listed in Table~\ref{table:general_properties}.  The horizontal solid line in the lower-left corner of each image represents $30\arcsec$.}
\figsetgrpend
 
\figsetgrpstart
\figsetgrpnum{8.316}
\figsetgrptitle{Visualization of UGC 09128.}
\figsetplot{\includegraphics[scale=0.7,angle=90]{f8_316.eps}}
\figsetgrpnote{Presentation of our integrated spectral atlas.  We plot the spectrum as $f_{\lambda}(\lambda)$ versus rest wavelength, normalized in a way that attempts to achieve a balance between showing the full range in flux and illustrating the finer details of the continuum.  Our integrated spectrum is accompanied by a Digitized Sky Survey image which illustrates our rectangular spectroscopic aperture as a \emph{solid} outline and the $25$~mag~arcsec$^{-2}$ isophotal size of the galaxy as a \emph{dashed} ellipse.  The image legend gives the galaxy name, the unique identification number in parenthesis, and the morphological type as listed in Table~\ref{table:general_properties}.  The horizontal solid line in the lower-left corner of each image represents $30\arcsec$.}
\figsetgrpend
 
\figsetgrpstart
\figsetgrpnum{8.317}
\figsetgrptitle{Visualization of NGC 5607.}
\figsetplot{\includegraphics[scale=0.7,angle=90]{f8_317.eps}}
\figsetgrpnote{Presentation of our integrated spectral atlas.  We plot the spectrum as $f_{\lambda}(\lambda)$ versus rest wavelength, normalized in a way that attempts to achieve a balance between showing the full range in flux and illustrating the finer details of the continuum.  Our integrated spectrum is accompanied by a Digitized Sky Survey image which illustrates our rectangular spectroscopic aperture as a \emph{solid} outline and the $25$~mag~arcsec$^{-2}$ isophotal size of the galaxy as a \emph{dashed} ellipse.  The image legend gives the galaxy name, the unique identification number in parenthesis, and the morphological type as listed in Table~\ref{table:general_properties}.  The horizontal solid line in the lower-left corner of each image represents $30\arcsec$.}
\figsetgrpend
 
\figsetgrpstart
\figsetgrpnum{8.318}
\figsetgrptitle{Visualization of MRK 1490.}
\figsetplot{\includegraphics[scale=0.7,angle=90]{f8_318.eps}}
\figsetgrpnote{Presentation of our integrated spectral atlas.  We plot the spectrum as $f_{\lambda}(\lambda)$ versus rest wavelength, normalized in a way that attempts to achieve a balance between showing the full range in flux and illustrating the finer details of the continuum.  Our integrated spectrum is accompanied by a Digitized Sky Survey image which illustrates our rectangular spectroscopic aperture as a \emph{solid} outline and the $25$~mag~arcsec$^{-2}$ isophotal size of the galaxy as a \emph{dashed} ellipse.  The image legend gives the galaxy name, the unique identification number in parenthesis, and the morphological type as listed in Table~\ref{table:general_properties}.  The horizontal solid line in the lower-left corner of each image represents $30\arcsec$.}
\figsetgrpend
 
\figsetgrpstart
\figsetgrpnum{8.319}
\figsetgrptitle{Visualization of NGC 5591 E.}
\figsetplot{\includegraphics[scale=0.7,angle=90]{f8_319.eps}}
\figsetgrpnote{Presentation of our integrated spectral atlas.  We plot the spectrum as $f_{\lambda}(\lambda)$ versus rest wavelength, normalized in a way that attempts to achieve a balance between showing the full range in flux and illustrating the finer details of the continuum.  Our integrated spectrum is accompanied by a Digitized Sky Survey image which illustrates our rectangular spectroscopic aperture as a \emph{solid} outline and the $25$~mag~arcsec$^{-2}$ isophotal size of the galaxy as a \emph{dashed} ellipse.  The image legend gives the galaxy name, the unique identification number in parenthesis, and the morphological type as listed in Table~\ref{table:general_properties}.  The horizontal solid line in the lower-left corner of each image represents $30\arcsec$.}
\figsetgrpend
 
\figsetgrpstart
\figsetgrpnum{8.320}
\figsetgrptitle{Visualization of NGC 5591.}
\figsetplot{\includegraphics[scale=0.7,angle=90]{f8_320.eps}}
\figsetgrpnote{Presentation of our integrated spectral atlas.  We plot the spectrum as $f_{\lambda}(\lambda)$ versus rest wavelength, normalized in a way that attempts to achieve a balance between showing the full range in flux and illustrating the finer details of the continuum.  Our integrated spectrum is accompanied by a Digitized Sky Survey image which illustrates our rectangular spectroscopic aperture as a \emph{solid} outline and the $25$~mag~arcsec$^{-2}$ isophotal size of the galaxy as a \emph{dashed} ellipse.  The image legend gives the galaxy name, the unique identification number in parenthesis, and the morphological type as listed in Table~\ref{table:general_properties}.  The horizontal solid line in the lower-left corner of each image represents $30\arcsec$.}
\figsetgrpend
 
\figsetgrpstart
\figsetgrpnum{8.321}
\figsetgrptitle{Visualization of NGC 5591 W.}
\figsetplot{\includegraphics[scale=0.7,angle=90]{f8_321.eps}}
\figsetgrpnote{Presentation of our integrated spectral atlas.  We plot the spectrum as $f_{\lambda}(\lambda)$ versus rest wavelength, normalized in a way that attempts to achieve a balance between showing the full range in flux and illustrating the finer details of the continuum.  Our integrated spectrum is accompanied by a Digitized Sky Survey image which illustrates our rectangular spectroscopic aperture as a \emph{solid} outline and the $25$~mag~arcsec$^{-2}$ isophotal size of the galaxy as a \emph{dashed} ellipse.  The image legend gives the galaxy name, the unique identification number in parenthesis, and the morphological type as listed in Table~\ref{table:general_properties}.  The horizontal solid line in the lower-left corner of each image represents $30\arcsec$.}
\figsetgrpend
 
\figsetgrpstart
\figsetgrpnum{8.322}
\figsetgrptitle{Visualization of UGC 09240.}
\figsetplot{\includegraphics[scale=0.7,angle=90]{f8_322.eps}}
\figsetgrpnote{Presentation of our integrated spectral atlas.  We plot the spectrum as $f_{\lambda}(\lambda)$ versus rest wavelength, normalized in a way that attempts to achieve a balance between showing the full range in flux and illustrating the finer details of the continuum.  Our integrated spectrum is accompanied by a Digitized Sky Survey image which illustrates our rectangular spectroscopic aperture as a \emph{solid} outline and the $25$~mag~arcsec$^{-2}$ isophotal size of the galaxy as a \emph{dashed} ellipse.  The image legend gives the galaxy name, the unique identification number in parenthesis, and the morphological type as listed in Table~\ref{table:general_properties}.  The horizontal solid line in the lower-left corner of each image represents $30\arcsec$.}
\figsetgrpend
 
\figsetgrpstart
\figsetgrpnum{8.323}
\figsetgrptitle{Visualization of NGC 5653.}
\figsetplot{\includegraphics[scale=0.7,angle=90]{f8_323.eps}}
\figsetgrpnote{Presentation of our integrated spectral atlas.  We plot the spectrum as $f_{\lambda}(\lambda)$ versus rest wavelength, normalized in a way that attempts to achieve a balance between showing the full range in flux and illustrating the finer details of the continuum.  Our integrated spectrum is accompanied by a Digitized Sky Survey image which illustrates our rectangular spectroscopic aperture as a \emph{solid} outline and the $25$~mag~arcsec$^{-2}$ isophotal size of the galaxy as a \emph{dashed} ellipse.  The image legend gives the galaxy name, the unique identification number in parenthesis, and the morphological type as listed in Table~\ref{table:general_properties}.  The horizontal solid line in the lower-left corner of each image represents $30\arcsec$.}
\figsetgrpend
 
\figsetgrpstart
\figsetgrpnum{8.324}
\figsetgrptitle{Visualization of MRK 0685.}
\figsetplot{\includegraphics[scale=0.7,angle=90]{f8_324.eps}}
\figsetgrpnote{Presentation of our integrated spectral atlas.  We plot the spectrum as $f_{\lambda}(\lambda)$ versus rest wavelength, normalized in a way that attempts to achieve a balance between showing the full range in flux and illustrating the finer details of the continuum.  Our integrated spectrum is accompanied by a Digitized Sky Survey image which illustrates our rectangular spectroscopic aperture as a \emph{solid} outline and the $25$~mag~arcsec$^{-2}$ isophotal size of the galaxy as a \emph{dashed} ellipse.  The image legend gives the galaxy name, the unique identification number in parenthesis, and the morphological type as listed in Table~\ref{table:general_properties}.  The horizontal solid line in the lower-left corner of each image represents $30\arcsec$.}
\figsetgrpend
 
\figsetgrpstart
\figsetgrpnum{8.325}
\figsetgrptitle{Visualization of NGC 5676.}
\figsetplot{\includegraphics[scale=0.7,angle=90]{f8_325.eps}}
\figsetgrpnote{Presentation of our integrated spectral atlas.  We plot the spectrum as $f_{\lambda}(\lambda)$ versus rest wavelength, normalized in a way that attempts to achieve a balance between showing the full range in flux and illustrating the finer details of the continuum.  Our integrated spectrum is accompanied by a Digitized Sky Survey image which illustrates our rectangular spectroscopic aperture as a \emph{solid} outline and the $25$~mag~arcsec$^{-2}$ isophotal size of the galaxy as a \emph{dashed} ellipse.  The image legend gives the galaxy name, the unique identification number in parenthesis, and the morphological type as listed in Table~\ref{table:general_properties}.  The horizontal solid line in the lower-left corner of each image represents $30\arcsec$.}
\figsetgrpend
 
\figsetgrpstart
\figsetgrpnum{8.326}
\figsetgrptitle{Visualization of UGC 09425 NW.}
\figsetplot{\includegraphics[scale=0.7,angle=90]{f8_326.eps}}
\figsetgrpnote{Presentation of our integrated spectral atlas.  We plot the spectrum as $f_{\lambda}(\lambda)$ versus rest wavelength, normalized in a way that attempts to achieve a balance between showing the full range in flux and illustrating the finer details of the continuum.  Our integrated spectrum is accompanied by a Digitized Sky Survey image which illustrates our rectangular spectroscopic aperture as a \emph{solid} outline and the $25$~mag~arcsec$^{-2}$ isophotal size of the galaxy as a \emph{dashed} ellipse.  The image legend gives the galaxy name, the unique identification number in parenthesis, and the morphological type as listed in Table~\ref{table:general_properties}.  The horizontal solid line in the lower-left corner of each image represents $30\arcsec$.}
\figsetgrpend
 
\figsetgrpstart
\figsetgrpnum{8.327}
\figsetgrptitle{Visualization of UGC 09425.}
\figsetplot{\includegraphics[scale=0.7,angle=90]{f8_327.eps}}
\figsetgrpnote{Presentation of our integrated spectral atlas.  We plot the spectrum as $f_{\lambda}(\lambda)$ versus rest wavelength, normalized in a way that attempts to achieve a balance between showing the full range in flux and illustrating the finer details of the continuum.  Our integrated spectrum is accompanied by a Digitized Sky Survey image which illustrates our rectangular spectroscopic aperture as a \emph{solid} outline and the $25$~mag~arcsec$^{-2}$ isophotal size of the galaxy as a \emph{dashed} ellipse.  The image legend gives the galaxy name, the unique identification number in parenthesis, and the morphological type as listed in Table~\ref{table:general_properties}.  The horizontal solid line in the lower-left corner of each image represents $30\arcsec$.}
\figsetgrpend
 
\figsetgrpstart
\figsetgrpnum{8.328}
\figsetgrptitle{Visualization of UGC 09425 SE.}
\figsetplot{\includegraphics[scale=0.7,angle=90]{f8_328.eps}}
\figsetgrpnote{Presentation of our integrated spectral atlas.  We plot the spectrum as $f_{\lambda}(\lambda)$ versus rest wavelength, normalized in a way that attempts to achieve a balance between showing the full range in flux and illustrating the finer details of the continuum.  Our integrated spectrum is accompanied by a Digitized Sky Survey image which illustrates our rectangular spectroscopic aperture as a \emph{solid} outline and the $25$~mag~arcsec$^{-2}$ isophotal size of the galaxy as a \emph{dashed} ellipse.  The image legend gives the galaxy name, the unique identification number in parenthesis, and the morphological type as listed in Table~\ref{table:general_properties}.  The horizontal solid line in the lower-left corner of each image represents $30\arcsec$.}
\figsetgrpend
 
\figsetgrpstart
\figsetgrpnum{8.329}
\figsetgrptitle{Visualization of MRK 0475.}
\figsetplot{\includegraphics[scale=0.7,angle=90]{f8_329.eps}}
\figsetgrpnote{Presentation of our integrated spectral atlas.  We plot the spectrum as $f_{\lambda}(\lambda)$ versus rest wavelength, normalized in a way that attempts to achieve a balance between showing the full range in flux and illustrating the finer details of the continuum.  Our integrated spectrum is accompanied by a Digitized Sky Survey image which illustrates our rectangular spectroscopic aperture as a \emph{solid} outline and the $25$~mag~arcsec$^{-2}$ isophotal size of the galaxy as a \emph{dashed} ellipse.  The image legend gives the galaxy name, the unique identification number in parenthesis, and the morphological type as listed in Table~\ref{table:general_properties}.  The horizontal solid line in the lower-left corner of each image represents $30\arcsec$.}
\figsetgrpend
 
\figsetgrpstart
\figsetgrpnum{8.330}
\figsetgrptitle{Visualization of UGC 09560.}
\figsetplot{\includegraphics[scale=0.7,angle=90]{f8_330.eps}}
\figsetgrpnote{Presentation of our integrated spectral atlas.  We plot the spectrum as $f_{\lambda}(\lambda)$ versus rest wavelength, normalized in a way that attempts to achieve a balance between showing the full range in flux and illustrating the finer details of the continuum.  Our integrated spectrum is accompanied by a Digitized Sky Survey image which illustrates our rectangular spectroscopic aperture as a \emph{solid} outline and the $25$~mag~arcsec$^{-2}$ isophotal size of the galaxy as a \emph{dashed} ellipse.  The image legend gives the galaxy name, the unique identification number in parenthesis, and the morphological type as listed in Table~\ref{table:general_properties}.  The horizontal solid line in the lower-left corner of each image represents $30\arcsec$.}
\figsetgrpend
 
\figsetgrpstart
\figsetgrpnum{8.331}
\figsetgrptitle{Visualization of IC 1076.}
\figsetplot{\includegraphics[scale=0.7,angle=90]{f8_331.eps}}
\figsetgrpnote{Presentation of our integrated spectral atlas.  We plot the spectrum as $f_{\lambda}(\lambda)$ versus rest wavelength, normalized in a way that attempts to achieve a balance between showing the full range in flux and illustrating the finer details of the continuum.  Our integrated spectrum is accompanied by a Digitized Sky Survey image which illustrates our rectangular spectroscopic aperture as a \emph{solid} outline and the $25$~mag~arcsec$^{-2}$ isophotal size of the galaxy as a \emph{dashed} ellipse.  The image legend gives the galaxy name, the unique identification number in parenthesis, and the morphological type as listed in Table~\ref{table:general_properties}.  The horizontal solid line in the lower-left corner of each image represents $30\arcsec$.}
\figsetgrpend
 
\figsetgrpstart
\figsetgrpnum{8.332}
\figsetgrptitle{Visualization of UGC 09618 S.}
\figsetplot{\includegraphics[scale=0.7,angle=90]{f8_332.eps}}
\figsetgrpnote{Presentation of our integrated spectral atlas.  We plot the spectrum as $f_{\lambda}(\lambda)$ versus rest wavelength, normalized in a way that attempts to achieve a balance between showing the full range in flux and illustrating the finer details of the continuum.  Our integrated spectrum is accompanied by a Digitized Sky Survey image which illustrates our rectangular spectroscopic aperture as a \emph{solid} outline and the $25$~mag~arcsec$^{-2}$ isophotal size of the galaxy as a \emph{dashed} ellipse.  The image legend gives the galaxy name, the unique identification number in parenthesis, and the morphological type as listed in Table~\ref{table:general_properties}.  The horizontal solid line in the lower-left corner of each image represents $30\arcsec$.}
\figsetgrpend
 
\figsetgrpstart
\figsetgrpnum{8.333}
\figsetgrptitle{Visualization of UGC 09618.}
\figsetplot{\includegraphics[scale=0.7,angle=90]{f8_333.eps}}
\figsetgrpnote{Presentation of our integrated spectral atlas.  We plot the spectrum as $f_{\lambda}(\lambda)$ versus rest wavelength, normalized in a way that attempts to achieve a balance between showing the full range in flux and illustrating the finer details of the continuum.  Our integrated spectrum is accompanied by a Digitized Sky Survey image which illustrates our rectangular spectroscopic aperture as a \emph{solid} outline and the $25$~mag~arcsec$^{-2}$ isophotal size of the galaxy as a \emph{dashed} ellipse.  The image legend gives the galaxy name, the unique identification number in parenthesis, and the morphological type as listed in Table~\ref{table:general_properties}.  The horizontal solid line in the lower-left corner of each image represents $30\arcsec$.}
\figsetgrpend
 
\figsetgrpstart
\figsetgrpnum{8.334}
\figsetgrptitle{Visualization of UGC 09618 N.}
\figsetplot{\includegraphics[scale=0.7,angle=90]{f8_334.eps}}
\figsetgrpnote{Presentation of our integrated spectral atlas.  We plot the spectrum as $f_{\lambda}(\lambda)$ versus rest wavelength, normalized in a way that attempts to achieve a balance between showing the full range in flux and illustrating the finer details of the continuum.  Our integrated spectrum is accompanied by a Digitized Sky Survey image which illustrates our rectangular spectroscopic aperture as a \emph{solid} outline and the $25$~mag~arcsec$^{-2}$ isophotal size of the galaxy as a \emph{dashed} ellipse.  The image legend gives the galaxy name, the unique identification number in parenthesis, and the morphological type as listed in Table~\ref{table:general_properties}.  The horizontal solid line in the lower-left corner of each image represents $30\arcsec$.}
\figsetgrpend
 
\figsetgrpstart
\figsetgrpnum{8.335}
\figsetgrptitle{Visualization of CGCG 049-057.}
\figsetplot{\includegraphics[scale=0.7,angle=90]{f8_335.eps}}
\figsetgrpnote{Presentation of our integrated spectral atlas.  We plot the spectrum as $f_{\lambda}(\lambda)$ versus rest wavelength, normalized in a way that attempts to achieve a balance between showing the full range in flux and illustrating the finer details of the continuum.  Our integrated spectrum is accompanied by a Digitized Sky Survey image which illustrates our rectangular spectroscopic aperture as a \emph{solid} outline and the $25$~mag~arcsec$^{-2}$ isophotal size of the galaxy as a \emph{dashed} ellipse.  The image legend gives the galaxy name, the unique identification number in parenthesis, and the morphological type as listed in Table~\ref{table:general_properties}.  The horizontal solid line in the lower-left corner of each image represents $30\arcsec$.}
\figsetgrpend
 
\figsetgrpstart
\figsetgrpnum{8.336}
\figsetgrptitle{Visualization of MRK 0848.}
\figsetplot{\includegraphics[scale=0.7,angle=90]{f8_336.eps}}
\figsetgrpnote{Presentation of our integrated spectral atlas.  We plot the spectrum as $f_{\lambda}(\lambda)$ versus rest wavelength, normalized in a way that attempts to achieve a balance between showing the full range in flux and illustrating the finer details of the continuum.  Our integrated spectrum is accompanied by a Digitized Sky Survey image which illustrates our rectangular spectroscopic aperture as a \emph{solid} outline and the $25$~mag~arcsec$^{-2}$ isophotal size of the galaxy as a \emph{dashed} ellipse.  The image legend gives the galaxy name, the unique identification number in parenthesis, and the morphological type as listed in Table~\ref{table:general_properties}.  The horizontal solid line in the lower-left corner of each image represents $30\arcsec$.}
\figsetgrpend
 
\figsetgrpstart
\figsetgrpnum{8.337}
\figsetgrptitle{Visualization of NGC 5953.}
\figsetplot{\includegraphics[scale=0.7,angle=90]{f8_337.eps}}
\figsetgrpnote{Presentation of our integrated spectral atlas.  We plot the spectrum as $f_{\lambda}(\lambda)$ versus rest wavelength, normalized in a way that attempts to achieve a balance between showing the full range in flux and illustrating the finer details of the continuum.  Our integrated spectrum is accompanied by a Digitized Sky Survey image which illustrates our rectangular spectroscopic aperture as a \emph{solid} outline and the $25$~mag~arcsec$^{-2}$ isophotal size of the galaxy as a \emph{dashed} ellipse.  The image legend gives the galaxy name, the unique identification number in parenthesis, and the morphological type as listed in Table~\ref{table:general_properties}.  The horizontal solid line in the lower-left corner of each image represents $30\arcsec$.}
\figsetgrpend
 
\figsetgrpstart
\figsetgrpnum{8.338}
\figsetgrptitle{Visualization of ARP 091.}
\figsetplot{\includegraphics[scale=0.7,angle=90]{f8_338.eps}}
\figsetgrpnote{Presentation of our integrated spectral atlas.  We plot the spectrum as $f_{\lambda}(\lambda)$ versus rest wavelength, normalized in a way that attempts to achieve a balance between showing the full range in flux and illustrating the finer details of the continuum.  Our integrated spectrum is accompanied by a Digitized Sky Survey image which illustrates our rectangular spectroscopic aperture as a \emph{solid} outline and the $25$~mag~arcsec$^{-2}$ isophotal size of the galaxy as a \emph{dashed} ellipse.  The image legend gives the galaxy name, the unique identification number in parenthesis, and the morphological type as listed in Table~\ref{table:general_properties}.  The horizontal solid line in the lower-left corner of each image represents $30\arcsec$.}
\figsetgrpend
 
\figsetgrpstart
\figsetgrpnum{8.339}
\figsetgrptitle{Visualization of IC 4553.}
\figsetplot{\includegraphics[scale=0.7,angle=90]{f8_339.eps}}
\figsetgrpnote{Presentation of our integrated spectral atlas.  We plot the spectrum as $f_{\lambda}(\lambda)$ versus rest wavelength, normalized in a way that attempts to achieve a balance between showing the full range in flux and illustrating the finer details of the continuum.  Our integrated spectrum is accompanied by a Digitized Sky Survey image which illustrates our rectangular spectroscopic aperture as a \emph{solid} outline and the $25$~mag~arcsec$^{-2}$ isophotal size of the galaxy as a \emph{dashed} ellipse.  The image legend gives the galaxy name, the unique identification number in parenthesis, and the morphological type as listed in Table~\ref{table:general_properties}.  The horizontal solid line in the lower-left corner of each image represents $30\arcsec$.}
\figsetgrpend
 
\figsetgrpstart
\figsetgrpnum{8.340}
\figsetgrptitle{Visualization of UGCA 410.}
\figsetplot{\includegraphics[scale=0.7,angle=90]{f8_340.eps}}
\figsetgrpnote{Presentation of our integrated spectral atlas.  We plot the spectrum as $f_{\lambda}(\lambda)$ versus rest wavelength, normalized in a way that attempts to achieve a balance between showing the full range in flux and illustrating the finer details of the continuum.  Our integrated spectrum is accompanied by a Digitized Sky Survey image which illustrates our rectangular spectroscopic aperture as a \emph{solid} outline and the $25$~mag~arcsec$^{-2}$ isophotal size of the galaxy as a \emph{dashed} ellipse.  The image legend gives the galaxy name, the unique identification number in parenthesis, and the morphological type as listed in Table~\ref{table:general_properties}.  The horizontal solid line in the lower-left corner of each image represents $30\arcsec$.}
\figsetgrpend
 
\figsetgrpstart
\figsetgrpnum{8.341}
\figsetgrptitle{Visualization of NGC 5992.}
\figsetplot{\includegraphics[scale=0.7,angle=90]{f8_341.eps}}
\figsetgrpnote{Presentation of our integrated spectral atlas.  We plot the spectrum as $f_{\lambda}(\lambda)$ versus rest wavelength, normalized in a way that attempts to achieve a balance between showing the full range in flux and illustrating the finer details of the continuum.  Our integrated spectrum is accompanied by a Digitized Sky Survey image which illustrates our rectangular spectroscopic aperture as a \emph{solid} outline and the $25$~mag~arcsec$^{-2}$ isophotal size of the galaxy as a \emph{dashed} ellipse.  The image legend gives the galaxy name, the unique identification number in parenthesis, and the morphological type as listed in Table~\ref{table:general_properties}.  The horizontal solid line in the lower-left corner of each image represents $30\arcsec$.}
\figsetgrpend
 
\figsetgrpstart
\figsetgrpnum{8.342}
\figsetgrptitle{Visualization of NGC 5996.}
\figsetplot{\includegraphics[scale=0.7,angle=90]{f8_342.eps}}
\figsetgrpnote{Presentation of our integrated spectral atlas.  We plot the spectrum as $f_{\lambda}(\lambda)$ versus rest wavelength, normalized in a way that attempts to achieve a balance between showing the full range in flux and illustrating the finer details of the continuum.  Our integrated spectrum is accompanied by a Digitized Sky Survey image which illustrates our rectangular spectroscopic aperture as a \emph{solid} outline and the $25$~mag~arcsec$^{-2}$ isophotal size of the galaxy as a \emph{dashed} ellipse.  The image legend gives the galaxy name, the unique identification number in parenthesis, and the morphological type as listed in Table~\ref{table:general_properties}.  The horizontal solid line in the lower-left corner of each image represents $30\arcsec$.}
\figsetgrpend
 
\figsetgrpstart
\figsetgrpnum{8.343}
\figsetgrptitle{Visualization of MRK 0863.}
\figsetplot{\includegraphics[scale=0.7,angle=90]{f8_343.eps}}
\figsetgrpnote{Presentation of our integrated spectral atlas.  We plot the spectrum as $f_{\lambda}(\lambda)$ versus rest wavelength, normalized in a way that attempts to achieve a balance between showing the full range in flux and illustrating the finer details of the continuum.  Our integrated spectrum is accompanied by a Digitized Sky Survey image which illustrates our rectangular spectroscopic aperture as a \emph{solid} outline and the $25$~mag~arcsec$^{-2}$ isophotal size of the galaxy as a \emph{dashed} ellipse.  The image legend gives the galaxy name, the unique identification number in parenthesis, and the morphological type as listed in Table~\ref{table:general_properties}.  The horizontal solid line in the lower-left corner of each image represents $30\arcsec$.}
\figsetgrpend
 
\figsetgrpstart
\figsetgrpnum{8.344}
\figsetgrptitle{Visualization of MRK 0296.}
\figsetplot{\includegraphics[scale=0.7,angle=90]{f8_344.eps}}
\figsetgrpnote{Presentation of our integrated spectral atlas.  We plot the spectrum as $f_{\lambda}(\lambda)$ versus rest wavelength, normalized in a way that attempts to achieve a balance between showing the full range in flux and illustrating the finer details of the continuum.  Our integrated spectrum is accompanied by a Digitized Sky Survey image which illustrates our rectangular spectroscopic aperture as a \emph{solid} outline and the $25$~mag~arcsec$^{-2}$ isophotal size of the galaxy as a \emph{dashed} ellipse.  The image legend gives the galaxy name, the unique identification number in parenthesis, and the morphological type as listed in Table~\ref{table:general_properties}.  The horizontal solid line in the lower-left corner of each image represents $30\arcsec$.}
\figsetgrpend
 
\figsetgrpstart
\figsetgrpnum{8.345}
\figsetgrptitle{Visualization of NGC 6052.}
\figsetplot{\includegraphics[scale=0.7,angle=90]{f8_345.eps}}
\figsetgrpnote{Presentation of our integrated spectral atlas.  We plot the spectrum as $f_{\lambda}(\lambda)$ versus rest wavelength, normalized in a way that attempts to achieve a balance between showing the full range in flux and illustrating the finer details of the continuum.  Our integrated spectrum is accompanied by a Digitized Sky Survey image which illustrates our rectangular spectroscopic aperture as a \emph{solid} outline and the $25$~mag~arcsec$^{-2}$ isophotal size of the galaxy as a \emph{dashed} ellipse.  The image legend gives the galaxy name, the unique identification number in parenthesis, and the morphological type as listed in Table~\ref{table:general_properties}.  The horizontal solid line in the lower-left corner of each image represents $30\arcsec$.}
\figsetgrpend
 
\figsetgrpstart
\figsetgrpnum{8.346}
\figsetgrptitle{Visualization of NGC 6090.}
\figsetplot{\includegraphics[scale=0.7,angle=90]{f8_346.eps}}
\figsetgrpnote{Presentation of our integrated spectral atlas.  We plot the spectrum as $f_{\lambda}(\lambda)$ versus rest wavelength, normalized in a way that attempts to achieve a balance between showing the full range in flux and illustrating the finer details of the continuum.  Our integrated spectrum is accompanied by a Digitized Sky Survey image which illustrates our rectangular spectroscopic aperture as a \emph{solid} outline and the $25$~mag~arcsec$^{-2}$ isophotal size of the galaxy as a \emph{dashed} ellipse.  The image legend gives the galaxy name, the unique identification number in parenthesis, and the morphological type as listed in Table~\ref{table:general_properties}.  The horizontal solid line in the lower-left corner of each image represents $30\arcsec$.}
\figsetgrpend
 
\figsetgrpstart
\figsetgrpnum{8.347}
\figsetgrptitle{Visualization of NGC 6207.}
\figsetplot{\includegraphics[scale=0.7,angle=90]{f8_347.eps}}
\figsetgrpnote{Presentation of our integrated spectral atlas.  We plot the spectrum as $f_{\lambda}(\lambda)$ versus rest wavelength, normalized in a way that attempts to achieve a balance between showing the full range in flux and illustrating the finer details of the continuum.  Our integrated spectrum is accompanied by a Digitized Sky Survey image which illustrates our rectangular spectroscopic aperture as a \emph{solid} outline and the $25$~mag~arcsec$^{-2}$ isophotal size of the galaxy as a \emph{dashed} ellipse.  The image legend gives the galaxy name, the unique identification number in parenthesis, and the morphological type as listed in Table~\ref{table:general_properties}.  The horizontal solid line in the lower-left corner of each image represents $30\arcsec$.}
\figsetgrpend
 
\figsetgrpstart
\figsetgrpnum{8.348}
\figsetgrptitle{Visualization of MRK 0499.}
\figsetplot{\includegraphics[scale=0.7,angle=90]{f8_348.eps}}
\figsetgrpnote{Presentation of our integrated spectral atlas.  We plot the spectrum as $f_{\lambda}(\lambda)$ versus rest wavelength, normalized in a way that attempts to achieve a balance between showing the full range in flux and illustrating the finer details of the continuum.  Our integrated spectrum is accompanied by a Digitized Sky Survey image which illustrates our rectangular spectroscopic aperture as a \emph{solid} outline and the $25$~mag~arcsec$^{-2}$ isophotal size of the galaxy as a \emph{dashed} ellipse.  The image legend gives the galaxy name, the unique identification number in parenthesis, and the morphological type as listed in Table~\ref{table:general_properties}.  The horizontal solid line in the lower-left corner of each image represents $30\arcsec$.}
\figsetgrpend
 
\figsetgrpstart
\figsetgrpnum{8.349}
\figsetgrptitle{Visualization of NGC 6240.}
\figsetplot{\includegraphics[scale=0.7,angle=90]{f8_349.eps}}
\figsetgrpnote{Presentation of our integrated spectral atlas.  We plot the spectrum as $f_{\lambda}(\lambda)$ versus rest wavelength, normalized in a way that attempts to achieve a balance between showing the full range in flux and illustrating the finer details of the continuum.  Our integrated spectrum is accompanied by a Digitized Sky Survey image which illustrates our rectangular spectroscopic aperture as a \emph{solid} outline and the $25$~mag~arcsec$^{-2}$ isophotal size of the galaxy as a \emph{dashed} ellipse.  The image legend gives the galaxy name, the unique identification number in parenthesis, and the morphological type as listed in Table~\ref{table:general_properties}.  The horizontal solid line in the lower-left corner of each image represents $30\arcsec$.}
\figsetgrpend
 
\figsetgrpstart
\figsetgrpnum{8.350}
\figsetgrptitle{Visualization of NGC 6285.}
\figsetplot{\includegraphics[scale=0.7,angle=90]{f8_350.eps}}
\figsetgrpnote{Presentation of our integrated spectral atlas.  We plot the spectrum as $f_{\lambda}(\lambda)$ versus rest wavelength, normalized in a way that attempts to achieve a balance between showing the full range in flux and illustrating the finer details of the continuum.  Our integrated spectrum is accompanied by a Digitized Sky Survey image which illustrates our rectangular spectroscopic aperture as a \emph{solid} outline and the $25$~mag~arcsec$^{-2}$ isophotal size of the galaxy as a \emph{dashed} ellipse.  The image legend gives the galaxy name, the unique identification number in parenthesis, and the morphological type as listed in Table~\ref{table:general_properties}.  The horizontal solid line in the lower-left corner of each image represents $30\arcsec$.}
\figsetgrpend
 
\figsetgrpstart
\figsetgrpnum{8.351}
\figsetgrptitle{Visualization of NGC 6286.}
\figsetplot{\includegraphics[scale=0.7,angle=90]{f8_351.eps}}
\figsetgrpnote{Presentation of our integrated spectral atlas.  We plot the spectrum as $f_{\lambda}(\lambda)$ versus rest wavelength, normalized in a way that attempts to achieve a balance between showing the full range in flux and illustrating the finer details of the continuum.  Our integrated spectrum is accompanied by a Digitized Sky Survey image which illustrates our rectangular spectroscopic aperture as a \emph{solid} outline and the $25$~mag~arcsec$^{-2}$ isophotal size of the galaxy as a \emph{dashed} ellipse.  The image legend gives the galaxy name, the unique identification number in parenthesis, and the morphological type as listed in Table~\ref{table:general_properties}.  The horizontal solid line in the lower-left corner of each image represents $30\arcsec$.}
\figsetgrpend
 
\figsetgrpstart
\figsetgrpnum{8.352}
\figsetgrptitle{Visualization of IRAS 17132+5313.}
\figsetplot{\includegraphics[scale=0.7,angle=90]{f8_352.eps}}
\figsetgrpnote{Presentation of our integrated spectral atlas.  We plot the spectrum as $f_{\lambda}(\lambda)$ versus rest wavelength, normalized in a way that attempts to achieve a balance between showing the full range in flux and illustrating the finer details of the continuum.  Our integrated spectrum is accompanied by a Digitized Sky Survey image which illustrates our rectangular spectroscopic aperture as a \emph{solid} outline and the $25$~mag~arcsec$^{-2}$ isophotal size of the galaxy as a \emph{dashed} ellipse.  The image legend gives the galaxy name, the unique identification number in parenthesis, and the morphological type as listed in Table~\ref{table:general_properties}.  The horizontal solid line in the lower-left corner of each image represents $30\arcsec$.}
\figsetgrpend
 
\figsetgrpstart
\figsetgrpnum{8.353}
\figsetgrptitle{Visualization of IRAS 17208-0014.}
\figsetplot{\includegraphics[scale=0.7,angle=90]{f8_353.eps}}
\figsetgrpnote{Presentation of our integrated spectral atlas.  We plot the spectrum as $f_{\lambda}(\lambda)$ versus rest wavelength, normalized in a way that attempts to achieve a balance between showing the full range in flux and illustrating the finer details of the continuum.  Our integrated spectrum is accompanied by a Digitized Sky Survey image which illustrates our rectangular spectroscopic aperture as a \emph{solid} outline and the $25$~mag~arcsec$^{-2}$ isophotal size of the galaxy as a \emph{dashed} ellipse.  The image legend gives the galaxy name, the unique identification number in parenthesis, and the morphological type as listed in Table~\ref{table:general_properties}.  The horizontal solid line in the lower-left corner of each image represents $30\arcsec$.}
\figsetgrpend
 
\figsetgrpstart
\figsetgrpnum{8.354}
\figsetgrptitle{Visualization of NGC 6621.}
\figsetplot{\includegraphics[scale=0.7,angle=90]{f8_354.eps}}
\figsetgrpnote{Presentation of our integrated spectral atlas.  We plot the spectrum as $f_{\lambda}(\lambda)$ versus rest wavelength, normalized in a way that attempts to achieve a balance between showing the full range in flux and illustrating the finer details of the continuum.  Our integrated spectrum is accompanied by a Digitized Sky Survey image which illustrates our rectangular spectroscopic aperture as a \emph{solid} outline and the $25$~mag~arcsec$^{-2}$ isophotal size of the galaxy as a \emph{dashed} ellipse.  The image legend gives the galaxy name, the unique identification number in parenthesis, and the morphological type as listed in Table~\ref{table:general_properties}.  The horizontal solid line in the lower-left corner of each image represents $30\arcsec$.}
\figsetgrpend
 
\figsetgrpstart
\figsetgrpnum{8.355}
\figsetgrptitle{Visualization of UGC 11175.}
\figsetplot{\includegraphics[scale=0.7,angle=90]{f8_355.eps}}
\figsetgrpnote{Presentation of our integrated spectral atlas.  We plot the spectrum as $f_{\lambda}(\lambda)$ versus rest wavelength, normalized in a way that attempts to achieve a balance between showing the full range in flux and illustrating the finer details of the continuum.  Our integrated spectrum is accompanied by a Digitized Sky Survey image which illustrates our rectangular spectroscopic aperture as a \emph{solid} outline and the $25$~mag~arcsec$^{-2}$ isophotal size of the galaxy as a \emph{dashed} ellipse.  The image legend gives the galaxy name, the unique identification number in parenthesis, and the morphological type as listed in Table~\ref{table:general_properties}.  The horizontal solid line in the lower-left corner of each image represents $30\arcsec$.}
\figsetgrpend
 
\figsetgrpstart
\figsetgrpnum{8.356}
\figsetgrptitle{Visualization of NGC 6622.}
\figsetplot{\includegraphics[scale=0.7,angle=90]{f8_356.eps}}
\figsetgrpnote{Presentation of our integrated spectral atlas.  We plot the spectrum as $f_{\lambda}(\lambda)$ versus rest wavelength, normalized in a way that attempts to achieve a balance between showing the full range in flux and illustrating the finer details of the continuum.  Our integrated spectrum is accompanied by a Digitized Sky Survey image which illustrates our rectangular spectroscopic aperture as a \emph{solid} outline and the $25$~mag~arcsec$^{-2}$ isophotal size of the galaxy as a \emph{dashed} ellipse.  The image legend gives the galaxy name, the unique identification number in parenthesis, and the morphological type as listed in Table~\ref{table:general_properties}.  The horizontal solid line in the lower-left corner of each image represents $30\arcsec$.}
\figsetgrpend
 
\figsetgrpstart
\figsetgrpnum{8.357}
\figsetgrptitle{Visualization of NGC 6670 B.}
\figsetplot{\includegraphics[scale=0.7,angle=90]{f8_357.eps}}
\figsetgrpnote{Presentation of our integrated spectral atlas.  We plot the spectrum as $f_{\lambda}(\lambda)$ versus rest wavelength, normalized in a way that attempts to achieve a balance between showing the full range in flux and illustrating the finer details of the continuum.  Our integrated spectrum is accompanied by a Digitized Sky Survey image which illustrates our rectangular spectroscopic aperture as a \emph{solid} outline and the $25$~mag~arcsec$^{-2}$ isophotal size of the galaxy as a \emph{dashed} ellipse.  The image legend gives the galaxy name, the unique identification number in parenthesis, and the morphological type as listed in Table~\ref{table:general_properties}.  The horizontal solid line in the lower-left corner of each image represents $30\arcsec$.}
\figsetgrpend
 
\figsetgrpstart
\figsetgrpnum{8.358}
\figsetgrptitle{Visualization of NGC 6670.}
\figsetplot{\includegraphics[scale=0.7,angle=90]{f8_358.eps}}
\figsetgrpnote{Presentation of our integrated spectral atlas.  We plot the spectrum as $f_{\lambda}(\lambda)$ versus rest wavelength, normalized in a way that attempts to achieve a balance between showing the full range in flux and illustrating the finer details of the continuum.  Our integrated spectrum is accompanied by a Digitized Sky Survey image which illustrates our rectangular spectroscopic aperture as a \emph{solid} outline and the $25$~mag~arcsec$^{-2}$ isophotal size of the galaxy as a \emph{dashed} ellipse.  The image legend gives the galaxy name, the unique identification number in parenthesis, and the morphological type as listed in Table~\ref{table:general_properties}.  The horizontal solid line in the lower-left corner of each image represents $30\arcsec$.}
\figsetgrpend
 
\figsetgrpstart
\figsetgrpnum{8.359}
\figsetgrptitle{Visualization of NGC 6670 A.}
\figsetplot{\includegraphics[scale=0.7,angle=90]{f8_359.eps}}
\figsetgrpnote{Presentation of our integrated spectral atlas.  We plot the spectrum as $f_{\lambda}(\lambda)$ versus rest wavelength, normalized in a way that attempts to achieve a balance between showing the full range in flux and illustrating the finer details of the continuum.  Our integrated spectrum is accompanied by a Digitized Sky Survey image which illustrates our rectangular spectroscopic aperture as a \emph{solid} outline and the $25$~mag~arcsec$^{-2}$ isophotal size of the galaxy as a \emph{dashed} ellipse.  The image legend gives the galaxy name, the unique identification number in parenthesis, and the morphological type as listed in Table~\ref{table:general_properties}.  The horizontal solid line in the lower-left corner of each image represents $30\arcsec$.}
\figsetgrpend
 
\figsetgrpstart
\figsetgrpnum{8.360}
\figsetgrptitle{Visualization of NGC 6701.}
\figsetplot{\includegraphics[scale=0.7,angle=90]{f8_360.eps}}
\figsetgrpnote{Presentation of our integrated spectral atlas.  We plot the spectrum as $f_{\lambda}(\lambda)$ versus rest wavelength, normalized in a way that attempts to achieve a balance between showing the full range in flux and illustrating the finer details of the continuum.  Our integrated spectrum is accompanied by a Digitized Sky Survey image which illustrates our rectangular spectroscopic aperture as a \emph{solid} outline and the $25$~mag~arcsec$^{-2}$ isophotal size of the galaxy as a \emph{dashed} ellipse.  The image legend gives the galaxy name, the unique identification number in parenthesis, and the morphological type as listed in Table~\ref{table:general_properties}.  The horizontal solid line in the lower-left corner of each image represents $30\arcsec$.}
\figsetgrpend
 
\figsetgrpstart
\figsetgrpnum{8.361}
\figsetgrptitle{Visualization of NGC 6926.}
\figsetplot{\includegraphics[scale=0.7,angle=90]{f8_361.eps}}
\figsetgrpnote{Presentation of our integrated spectral atlas.  We plot the spectrum as $f_{\lambda}(\lambda)$ versus rest wavelength, normalized in a way that attempts to achieve a balance between showing the full range in flux and illustrating the finer details of the continuum.  Our integrated spectrum is accompanied by a Digitized Sky Survey image which illustrates our rectangular spectroscopic aperture as a \emph{solid} outline and the $25$~mag~arcsec$^{-2}$ isophotal size of the galaxy as a \emph{dashed} ellipse.  The image legend gives the galaxy name, the unique identification number in parenthesis, and the morphological type as listed in Table~\ref{table:general_properties}.  The horizontal solid line in the lower-left corner of each image represents $30\arcsec$.}
\figsetgrpend
 
\figsetgrpstart
\figsetgrpnum{8.362}
\figsetgrptitle{Visualization of II Zw 096.}
\figsetplot{\includegraphics[scale=0.7,angle=90]{f8_362.eps}}
\figsetgrpnote{Presentation of our integrated spectral atlas.  We plot the spectrum as $f_{\lambda}(\lambda)$ versus rest wavelength, normalized in a way that attempts to achieve a balance between showing the full range in flux and illustrating the finer details of the continuum.  Our integrated spectrum is accompanied by a Digitized Sky Survey image which illustrates our rectangular spectroscopic aperture as a \emph{solid} outline and the $25$~mag~arcsec$^{-2}$ isophotal size of the galaxy as a \emph{dashed} ellipse.  The image legend gives the galaxy name, the unique identification number in parenthesis, and the morphological type as listed in Table~\ref{table:general_properties}.  The horizontal solid line in the lower-left corner of each image represents $30\arcsec$.}
\figsetgrpend
 
\figsetgrpstart
\figsetgrpnum{8.363}
\figsetgrptitle{Visualization of II Zw 096 NW.}
\figsetplot{\includegraphics[scale=0.7,angle=90]{f8_363.eps}}
\figsetgrpnote{Presentation of our integrated spectral atlas.  We plot the spectrum as $f_{\lambda}(\lambda)$ versus rest wavelength, normalized in a way that attempts to achieve a balance between showing the full range in flux and illustrating the finer details of the continuum.  Our integrated spectrum is accompanied by a Digitized Sky Survey image which illustrates our rectangular spectroscopic aperture as a \emph{solid} outline and the $25$~mag~arcsec$^{-2}$ isophotal size of the galaxy as a \emph{dashed} ellipse.  The image legend gives the galaxy name, the unique identification number in parenthesis, and the morphological type as listed in Table~\ref{table:general_properties}.  The horizontal solid line in the lower-left corner of each image represents $30\arcsec$.}
\figsetgrpend
 
\figsetgrpstart
\figsetgrpnum{8.364}
\figsetgrptitle{Visualization of II Zw 096 SE.}
\figsetplot{\includegraphics[scale=0.7,angle=90]{f8_364.eps}}
\figsetgrpnote{Presentation of our integrated spectral atlas.  We plot the spectrum as $f_{\lambda}(\lambda)$ versus rest wavelength, normalized in a way that attempts to achieve a balance between showing the full range in flux and illustrating the finer details of the continuum.  Our integrated spectrum is accompanied by a Digitized Sky Survey image which illustrates our rectangular spectroscopic aperture as a \emph{solid} outline and the $25$~mag~arcsec$^{-2}$ isophotal size of the galaxy as a \emph{dashed} ellipse.  The image legend gives the galaxy name, the unique identification number in parenthesis, and the morphological type as listed in Table~\ref{table:general_properties}.  The horizontal solid line in the lower-left corner of each image represents $30\arcsec$.}
\figsetgrpend
 
\figsetgrpstart
\figsetgrpnum{8.365}
\figsetgrptitle{Visualization of UGC 11680 W.}
\figsetplot{\includegraphics[scale=0.7,angle=90]{f8_365.eps}}
\figsetgrpnote{Presentation of our integrated spectral atlas.  We plot the spectrum as $f_{\lambda}(\lambda)$ versus rest wavelength, normalized in a way that attempts to achieve a balance between showing the full range in flux and illustrating the finer details of the continuum.  Our integrated spectrum is accompanied by a Digitized Sky Survey image which illustrates our rectangular spectroscopic aperture as a \emph{solid} outline and the $25$~mag~arcsec$^{-2}$ isophotal size of the galaxy as a \emph{dashed} ellipse.  The image legend gives the galaxy name, the unique identification number in parenthesis, and the morphological type as listed in Table~\ref{table:general_properties}.  The horizontal solid line in the lower-left corner of each image represents $30\arcsec$.}
\figsetgrpend
 
\figsetgrpstart
\figsetgrpnum{8.366}
\figsetgrptitle{Visualization of NGC 7137.}
\figsetplot{\includegraphics[scale=0.7,angle=90]{f8_366.eps}}
\figsetgrpnote{Presentation of our integrated spectral atlas.  We plot the spectrum as $f_{\lambda}(\lambda)$ versus rest wavelength, normalized in a way that attempts to achieve a balance between showing the full range in flux and illustrating the finer details of the continuum.  Our integrated spectrum is accompanied by a Digitized Sky Survey image which illustrates our rectangular spectroscopic aperture as a \emph{solid} outline and the $25$~mag~arcsec$^{-2}$ isophotal size of the galaxy as a \emph{dashed} ellipse.  The image legend gives the galaxy name, the unique identification number in parenthesis, and the morphological type as listed in Table~\ref{table:general_properties}.  The horizontal solid line in the lower-left corner of each image represents $30\arcsec$.}
\figsetgrpend
 
\figsetgrpstart
\figsetgrpnum{8.367}
\figsetgrptitle{Visualization of NGC 7130.}
\figsetplot{\includegraphics[scale=0.7,angle=90]{f8_367.eps}}
\figsetgrpnote{Presentation of our integrated spectral atlas.  We plot the spectrum as $f_{\lambda}(\lambda)$ versus rest wavelength, normalized in a way that attempts to achieve a balance between showing the full range in flux and illustrating the finer details of the continuum.  Our integrated spectrum is accompanied by a Digitized Sky Survey image which illustrates our rectangular spectroscopic aperture as a \emph{solid} outline and the $25$~mag~arcsec$^{-2}$ isophotal size of the galaxy as a \emph{dashed} ellipse.  The image legend gives the galaxy name, the unique identification number in parenthesis, and the morphological type as listed in Table~\ref{table:general_properties}.  The horizontal solid line in the lower-left corner of each image represents $30\arcsec$.}
\figsetgrpend
 
\figsetgrpstart
\figsetgrpnum{8.368}
\figsetgrptitle{Visualization of IC 5179.}
\figsetplot{\includegraphics[scale=0.7,angle=90]{f8_368.eps}}
\figsetgrpnote{Presentation of our integrated spectral atlas.  We plot the spectrum as $f_{\lambda}(\lambda)$ versus rest wavelength, normalized in a way that attempts to achieve a balance between showing the full range in flux and illustrating the finer details of the continuum.  Our integrated spectrum is accompanied by a Digitized Sky Survey image which illustrates our rectangular spectroscopic aperture as a \emph{solid} outline and the $25$~mag~arcsec$^{-2}$ isophotal size of the galaxy as a \emph{dashed} ellipse.  The image legend gives the galaxy name, the unique identification number in parenthesis, and the morphological type as listed in Table~\ref{table:general_properties}.  The horizontal solid line in the lower-left corner of each image represents $30\arcsec$.}
\figsetgrpend
 
\figsetgrpstart
\figsetgrpnum{8.369}
\figsetgrptitle{Visualization of NGC 7244.}
\figsetplot{\includegraphics[scale=0.7,angle=90]{f8_369.eps}}
\figsetgrpnote{Presentation of our integrated spectral atlas.  We plot the spectrum as $f_{\lambda}(\lambda)$ versus rest wavelength, normalized in a way that attempts to achieve a balance between showing the full range in flux and illustrating the finer details of the continuum.  Our integrated spectrum is accompanied by a Digitized Sky Survey image which illustrates our rectangular spectroscopic aperture as a \emph{solid} outline and the $25$~mag~arcsec$^{-2}$ isophotal size of the galaxy as a \emph{dashed} ellipse.  The image legend gives the galaxy name, the unique identification number in parenthesis, and the morphological type as listed in Table~\ref{table:general_properties}.  The horizontal solid line in the lower-left corner of each image represents $30\arcsec$.}
\figsetgrpend
 
\figsetgrpstart
\figsetgrpnum{8.370}
\figsetgrptitle{Visualization of NGC 7284.}
\figsetplot{\includegraphics[scale=0.7,angle=90]{f8_370.eps}}
\figsetgrpnote{Presentation of our integrated spectral atlas.  We plot the spectrum as $f_{\lambda}(\lambda)$ versus rest wavelength, normalized in a way that attempts to achieve a balance between showing the full range in flux and illustrating the finer details of the continuum.  Our integrated spectrum is accompanied by a Digitized Sky Survey image which illustrates our rectangular spectroscopic aperture as a \emph{solid} outline and the $25$~mag~arcsec$^{-2}$ isophotal size of the galaxy as a \emph{dashed} ellipse.  The image legend gives the galaxy name, the unique identification number in parenthesis, and the morphological type as listed in Table~\ref{table:general_properties}.  The horizontal solid line in the lower-left corner of each image represents $30\arcsec$.}
\figsetgrpend
 
\figsetgrpstart
\figsetgrpnum{8.371}
\figsetgrptitle{Visualization of ARP 093.}
\figsetplot{\includegraphics[scale=0.7,angle=90]{f8_371.eps}}
\figsetgrpnote{Presentation of our integrated spectral atlas.  We plot the spectrum as $f_{\lambda}(\lambda)$ versus rest wavelength, normalized in a way that attempts to achieve a balance between showing the full range in flux and illustrating the finer details of the continuum.  Our integrated spectrum is accompanied by a Digitized Sky Survey image which illustrates our rectangular spectroscopic aperture as a \emph{solid} outline and the $25$~mag~arcsec$^{-2}$ isophotal size of the galaxy as a \emph{dashed} ellipse.  The image legend gives the galaxy name, the unique identification number in parenthesis, and the morphological type as listed in Table~\ref{table:general_properties}.  The horizontal solid line in the lower-left corner of each image represents $30\arcsec$.}
\figsetgrpend
 
\figsetgrpstart
\figsetgrpnum{8.372}
\figsetgrptitle{Visualization of NGC 7285.}
\figsetplot{\includegraphics[scale=0.7,angle=90]{f8_372.eps}}
\figsetgrpnote{Presentation of our integrated spectral atlas.  We plot the spectrum as $f_{\lambda}(\lambda)$ versus rest wavelength, normalized in a way that attempts to achieve a balance between showing the full range in flux and illustrating the finer details of the continuum.  Our integrated spectrum is accompanied by a Digitized Sky Survey image which illustrates our rectangular spectroscopic aperture as a \emph{solid} outline and the $25$~mag~arcsec$^{-2}$ isophotal size of the galaxy as a \emph{dashed} ellipse.  The image legend gives the galaxy name, the unique identification number in parenthesis, and the morphological type as listed in Table~\ref{table:general_properties}.  The horizontal solid line in the lower-left corner of each image represents $30\arcsec$.}
\figsetgrpend
 
\figsetgrpstart
\figsetgrpnum{8.373}
\figsetgrptitle{Visualization of ESO 602-G025.}
\figsetplot{\includegraphics[scale=0.7,angle=90]{f8_373.eps}}
\figsetgrpnote{Presentation of our integrated spectral atlas.  We plot the spectrum as $f_{\lambda}(\lambda)$ versus rest wavelength, normalized in a way that attempts to achieve a balance between showing the full range in flux and illustrating the finer details of the continuum.  Our integrated spectrum is accompanied by a Digitized Sky Survey image which illustrates our rectangular spectroscopic aperture as a \emph{solid} outline and the $25$~mag~arcsec$^{-2}$ isophotal size of the galaxy as a \emph{dashed} ellipse.  The image legend gives the galaxy name, the unique identification number in parenthesis, and the morphological type as listed in Table~\ref{table:general_properties}.  The horizontal solid line in the lower-left corner of each image represents $30\arcsec$.}
\figsetgrpend
 
\figsetgrpstart
\figsetgrpnum{8.374}
\figsetgrptitle{Visualization of NGC 7316.}
\figsetplot{\includegraphics[scale=0.7,angle=90]{f8_374.eps}}
\figsetgrpnote{Presentation of our integrated spectral atlas.  We plot the spectrum as $f_{\lambda}(\lambda)$ versus rest wavelength, normalized in a way that attempts to achieve a balance between showing the full range in flux and illustrating the finer details of the continuum.  Our integrated spectrum is accompanied by a Digitized Sky Survey image which illustrates our rectangular spectroscopic aperture as a \emph{solid} outline and the $25$~mag~arcsec$^{-2}$ isophotal size of the galaxy as a \emph{dashed} ellipse.  The image legend gives the galaxy name, the unique identification number in parenthesis, and the morphological type as listed in Table~\ref{table:general_properties}.  The horizontal solid line in the lower-left corner of each image represents $30\arcsec$.}
\figsetgrpend
 
\figsetgrpstart
\figsetgrpnum{8.375}
\figsetgrptitle{Visualization of UGC 12150.}
\figsetplot{\includegraphics[scale=0.7,angle=90]{f8_375.eps}}
\figsetgrpnote{Presentation of our integrated spectral atlas.  We plot the spectrum as $f_{\lambda}(\lambda)$ versus rest wavelength, normalized in a way that attempts to achieve a balance between showing the full range in flux and illustrating the finer details of the continuum.  Our integrated spectrum is accompanied by a Digitized Sky Survey image which illustrates our rectangular spectroscopic aperture as a \emph{solid} outline and the $25$~mag~arcsec$^{-2}$ isophotal size of the galaxy as a \emph{dashed} ellipse.  The image legend gives the galaxy name, the unique identification number in parenthesis, and the morphological type as listed in Table~\ref{table:general_properties}.  The horizontal solid line in the lower-left corner of each image represents $30\arcsec$.}
\figsetgrpend
 
\figsetgrpstart
\figsetgrpnum{8.376}
\figsetgrptitle{Visualization of NGC 7448.}
\figsetplot{\includegraphics[scale=0.7,angle=90]{f8_376.eps}}
\figsetgrpnote{Presentation of our integrated spectral atlas.  We plot the spectrum as $f_{\lambda}(\lambda)$ versus rest wavelength, normalized in a way that attempts to achieve a balance between showing the full range in flux and illustrating the finer details of the continuum.  Our integrated spectrum is accompanied by a Digitized Sky Survey image which illustrates our rectangular spectroscopic aperture as a \emph{solid} outline and the $25$~mag~arcsec$^{-2}$ isophotal size of the galaxy as a \emph{dashed} ellipse.  The image legend gives the galaxy name, the unique identification number in parenthesis, and the morphological type as listed in Table~\ref{table:general_properties}.  The horizontal solid line in the lower-left corner of each image represents $30\arcsec$.}
\figsetgrpend
 
\figsetgrpstart
\figsetgrpnum{8.377}
\figsetgrptitle{Visualization of MRK 0312.}
\figsetplot{\includegraphics[scale=0.7,angle=90]{f8_377.eps}}
\figsetgrpnote{Presentation of our integrated spectral atlas.  We plot the spectrum as $f_{\lambda}(\lambda)$ versus rest wavelength, normalized in a way that attempts to achieve a balance between showing the full range in flux and illustrating the finer details of the continuum.  Our integrated spectrum is accompanied by a Digitized Sky Survey image which illustrates our rectangular spectroscopic aperture as a \emph{solid} outline and the $25$~mag~arcsec$^{-2}$ isophotal size of the galaxy as a \emph{dashed} ellipse.  The image legend gives the galaxy name, the unique identification number in parenthesis, and the morphological type as listed in Table~\ref{table:general_properties}.  The horizontal solid line in the lower-left corner of each image represents $30\arcsec$.}
\figsetgrpend
 
\figsetgrpstart
\figsetgrpnum{8.378}
\figsetgrptitle{Visualization of NGC 7465.}
\figsetplot{\includegraphics[scale=0.7,angle=90]{f8_378.eps}}
\figsetgrpnote{Presentation of our integrated spectral atlas.  We plot the spectrum as $f_{\lambda}(\lambda)$ versus rest wavelength, normalized in a way that attempts to achieve a balance between showing the full range in flux and illustrating the finer details of the continuum.  Our integrated spectrum is accompanied by a Digitized Sky Survey image which illustrates our rectangular spectroscopic aperture as a \emph{solid} outline and the $25$~mag~arcsec$^{-2}$ isophotal size of the galaxy as a \emph{dashed} ellipse.  The image legend gives the galaxy name, the unique identification number in parenthesis, and the morphological type as listed in Table~\ref{table:general_properties}.  The horizontal solid line in the lower-left corner of each image represents $30\arcsec$.}
\figsetgrpend
 
\figsetgrpstart
\figsetgrpnum{8.379}
\figsetgrptitle{Visualization of NGC 7468.}
\figsetplot{\includegraphics[scale=0.7,angle=90]{f8_379.eps}}
\figsetgrpnote{Presentation of our integrated spectral atlas.  We plot the spectrum as $f_{\lambda}(\lambda)$ versus rest wavelength, normalized in a way that attempts to achieve a balance between showing the full range in flux and illustrating the finer details of the continuum.  Our integrated spectrum is accompanied by a Digitized Sky Survey image which illustrates our rectangular spectroscopic aperture as a \emph{solid} outline and the $25$~mag~arcsec$^{-2}$ isophotal size of the galaxy as a \emph{dashed} ellipse.  The image legend gives the galaxy name, the unique identification number in parenthesis, and the morphological type as listed in Table~\ref{table:general_properties}.  The horizontal solid line in the lower-left corner of each image represents $30\arcsec$.}
\figsetgrpend
 
\figsetgrpstart
\figsetgrpnum{8.380}
\figsetgrptitle{Visualization of NGC 7469.}
\figsetplot{\includegraphics[scale=0.7,angle=90]{f8_380.eps}}
\figsetgrpnote{Presentation of our integrated spectral atlas.  We plot the spectrum as $f_{\lambda}(\lambda)$ versus rest wavelength, normalized in a way that attempts to achieve a balance between showing the full range in flux and illustrating the finer details of the continuum.  Our integrated spectrum is accompanied by a Digitized Sky Survey image which illustrates our rectangular spectroscopic aperture as a \emph{solid} outline and the $25$~mag~arcsec$^{-2}$ isophotal size of the galaxy as a \emph{dashed} ellipse.  The image legend gives the galaxy name, the unique identification number in parenthesis, and the morphological type as listed in Table~\ref{table:general_properties}.  The horizontal solid line in the lower-left corner of each image represents $30\arcsec$.}
\figsetgrpend
 
\figsetgrpstart
\figsetgrpnum{8.381}
\figsetgrptitle{Visualization of MRK 0315.}
\figsetplot{\includegraphics[scale=0.7,angle=90]{f8_381.eps}}
\figsetgrpnote{Presentation of our integrated spectral atlas.  We plot the spectrum as $f_{\lambda}(\lambda)$ versus rest wavelength, normalized in a way that attempts to achieve a balance between showing the full range in flux and illustrating the finer details of the continuum.  Our integrated spectrum is accompanied by a Digitized Sky Survey image which illustrates our rectangular spectroscopic aperture as a \emph{solid} outline and the $25$~mag~arcsec$^{-2}$ isophotal size of the galaxy as a \emph{dashed} ellipse.  The image legend gives the galaxy name, the unique identification number in parenthesis, and the morphological type as listed in Table~\ref{table:general_properties}.  The horizontal solid line in the lower-left corner of each image represents $30\arcsec$.}
\figsetgrpend
 
\figsetgrpstart
\figsetgrpnum{8.382}
\figsetgrptitle{Visualization of CGCG 453-062.}
\figsetplot{\includegraphics[scale=0.7,angle=90]{f8_382.eps}}
\figsetgrpnote{Presentation of our integrated spectral atlas.  We plot the spectrum as $f_{\lambda}(\lambda)$ versus rest wavelength, normalized in a way that attempts to achieve a balance between showing the full range in flux and illustrating the finer details of the continuum.  Our integrated spectrum is accompanied by a Digitized Sky Survey image which illustrates our rectangular spectroscopic aperture as a \emph{solid} outline and the $25$~mag~arcsec$^{-2}$ isophotal size of the galaxy as a \emph{dashed} ellipse.  The image legend gives the galaxy name, the unique identification number in parenthesis, and the morphological type as listed in Table~\ref{table:general_properties}.  The horizontal solid line in the lower-left corner of each image represents $30\arcsec$.}
\figsetgrpend
 
\figsetgrpstart
\figsetgrpnum{8.383}
\figsetgrptitle{Visualization of NGC 7518.}
\figsetplot{\includegraphics[scale=0.7,angle=90]{f8_383.eps}}
\figsetgrpnote{Presentation of our integrated spectral atlas.  We plot the spectrum as $f_{\lambda}(\lambda)$ versus rest wavelength, normalized in a way that attempts to achieve a balance between showing the full range in flux and illustrating the finer details of the continuum.  Our integrated spectrum is accompanied by a Digitized Sky Survey image which illustrates our rectangular spectroscopic aperture as a \emph{solid} outline and the $25$~mag~arcsec$^{-2}$ isophotal size of the galaxy as a \emph{dashed} ellipse.  The image legend gives the galaxy name, the unique identification number in parenthesis, and the morphological type as listed in Table~\ref{table:general_properties}.  The horizontal solid line in the lower-left corner of each image represents $30\arcsec$.}
\figsetgrpend
 
\figsetgrpstart
\figsetgrpnum{8.384}
\figsetgrptitle{Visualization of IC 5298.}
\figsetplot{\includegraphics[scale=0.7,angle=90]{f8_384.eps}}
\figsetgrpnote{Presentation of our integrated spectral atlas.  We plot the spectrum as $f_{\lambda}(\lambda)$ versus rest wavelength, normalized in a way that attempts to achieve a balance between showing the full range in flux and illustrating the finer details of the continuum.  Our integrated spectrum is accompanied by a Digitized Sky Survey image which illustrates our rectangular spectroscopic aperture as a \emph{solid} outline and the $25$~mag~arcsec$^{-2}$ isophotal size of the galaxy as a \emph{dashed} ellipse.  The image legend gives the galaxy name, the unique identification number in parenthesis, and the morphological type as listed in Table~\ref{table:general_properties}.  The horizontal solid line in the lower-left corner of each image represents $30\arcsec$.}
\figsetgrpend
 
\figsetgrpstart
\figsetgrpnum{8.385}
\figsetgrptitle{Visualization of NGC 7580.}
\figsetplot{\includegraphics[scale=0.7,angle=90]{f8_385.eps}}
\figsetgrpnote{Presentation of our integrated spectral atlas.  We plot the spectrum as $f_{\lambda}(\lambda)$ versus rest wavelength, normalized in a way that attempts to achieve a balance between showing the full range in flux and illustrating the finer details of the continuum.  Our integrated spectrum is accompanied by a Digitized Sky Survey image which illustrates our rectangular spectroscopic aperture as a \emph{solid} outline and the $25$~mag~arcsec$^{-2}$ isophotal size of the galaxy as a \emph{dashed} ellipse.  The image legend gives the galaxy name, the unique identification number in parenthesis, and the morphological type as listed in Table~\ref{table:general_properties}.  The horizontal solid line in the lower-left corner of each image represents $30\arcsec$.}
\figsetgrpend
 
\figsetgrpstart
\figsetgrpnum{8.386}
\figsetgrptitle{Visualization of NGC 5954.}
\figsetplot{\includegraphics[scale=0.7,angle=90]{f8_386.eps}}
\figsetgrpnote{Presentation of our integrated spectral atlas.  We plot the spectrum as $f_{\lambda}(\lambda)$ versus rest wavelength, normalized in a way that attempts to achieve a balance between showing the full range in flux and illustrating the finer details of the continuum.  Our integrated spectrum is accompanied by a Digitized Sky Survey image which illustrates our rectangular spectroscopic aperture as a \emph{solid} outline and the $25$~mag~arcsec$^{-2}$ isophotal size of the galaxy as a \emph{dashed} ellipse.  The image legend gives the galaxy name, the unique identification number in parenthesis, and the morphological type as listed in Table~\ref{table:general_properties}.  The horizontal solid line in the lower-left corner of each image represents $30\arcsec$.}
\figsetgrpend
 
\figsetgrpstart
\figsetgrpnum{8.387}
\figsetgrptitle{Visualization of NGC 7585.}
\figsetplot{\includegraphics[scale=0.7,angle=90]{f8_387.eps}}
\figsetgrpnote{Presentation of our integrated spectral atlas.  We plot the spectrum as $f_{\lambda}(\lambda)$ versus rest wavelength, normalized in a way that attempts to achieve a balance between showing the full range in flux and illustrating the finer details of the continuum.  Our integrated spectrum is accompanied by a Digitized Sky Survey image which illustrates our rectangular spectroscopic aperture as a \emph{solid} outline and the $25$~mag~arcsec$^{-2}$ isophotal size of the galaxy as a \emph{dashed} ellipse.  The image legend gives the galaxy name, the unique identification number in parenthesis, and the morphological type as listed in Table~\ref{table:general_properties}.  The horizontal solid line in the lower-left corner of each image represents $30\arcsec$.}
\figsetgrpend
 
\figsetgrpstart
\figsetgrpnum{8.388}
\figsetgrptitle{Visualization of NGC 7591.}
\figsetplot{\includegraphics[scale=0.7,angle=90]{f8_388.eps}}
\figsetgrpnote{Presentation of our integrated spectral atlas.  We plot the spectrum as $f_{\lambda}(\lambda)$ versus rest wavelength, normalized in a way that attempts to achieve a balance between showing the full range in flux and illustrating the finer details of the continuum.  Our integrated spectrum is accompanied by a Digitized Sky Survey image which illustrates our rectangular spectroscopic aperture as a \emph{solid} outline and the $25$~mag~arcsec$^{-2}$ isophotal size of the galaxy as a \emph{dashed} ellipse.  The image legend gives the galaxy name, the unique identification number in parenthesis, and the morphological type as listed in Table~\ref{table:general_properties}.  The horizontal solid line in the lower-left corner of each image represents $30\arcsec$.}
\figsetgrpend
 
\figsetgrpstart
\figsetgrpnum{8.389}
\figsetgrptitle{Visualization of NGC 7592 A.}
\figsetplot{\includegraphics[scale=0.7,angle=90]{f8_389.eps}}
\figsetgrpnote{Presentation of our integrated spectral atlas.  We plot the spectrum as $f_{\lambda}(\lambda)$ versus rest wavelength, normalized in a way that attempts to achieve a balance between showing the full range in flux and illustrating the finer details of the continuum.  Our integrated spectrum is accompanied by a Digitized Sky Survey image which illustrates our rectangular spectroscopic aperture as a \emph{solid} outline and the $25$~mag~arcsec$^{-2}$ isophotal size of the galaxy as a \emph{dashed} ellipse.  The image legend gives the galaxy name, the unique identification number in parenthesis, and the morphological type as listed in Table~\ref{table:general_properties}.  The horizontal solid line in the lower-left corner of each image represents $30\arcsec$.}
\figsetgrpend
 
\figsetgrpstart
\figsetgrpnum{8.390}
\figsetgrptitle{Visualization of NGC 7592.}
\figsetplot{\includegraphics[scale=0.7,angle=90]{f8_390.eps}}
\figsetgrpnote{Presentation of our integrated spectral atlas.  We plot the spectrum as $f_{\lambda}(\lambda)$ versus rest wavelength, normalized in a way that attempts to achieve a balance between showing the full range in flux and illustrating the finer details of the continuum.  Our integrated spectrum is accompanied by a Digitized Sky Survey image which illustrates our rectangular spectroscopic aperture as a \emph{solid} outline and the $25$~mag~arcsec$^{-2}$ isophotal size of the galaxy as a \emph{dashed} ellipse.  The image legend gives the galaxy name, the unique identification number in parenthesis, and the morphological type as listed in Table~\ref{table:general_properties}.  The horizontal solid line in the lower-left corner of each image represents $30\arcsec$.}
\figsetgrpend
 
\figsetgrpstart
\figsetgrpnum{8.391}
\figsetgrptitle{Visualization of NGC 7592 B.}
\figsetplot{\includegraphics[scale=0.7,angle=90]{f8_391.eps}}
\figsetgrpnote{Presentation of our integrated spectral atlas.  We plot the spectrum as $f_{\lambda}(\lambda)$ versus rest wavelength, normalized in a way that attempts to achieve a balance between showing the full range in flux and illustrating the finer details of the continuum.  Our integrated spectrum is accompanied by a Digitized Sky Survey image which illustrates our rectangular spectroscopic aperture as a \emph{solid} outline and the $25$~mag~arcsec$^{-2}$ isophotal size of the galaxy as a \emph{dashed} ellipse.  The image legend gives the galaxy name, the unique identification number in parenthesis, and the morphological type as listed in Table~\ref{table:general_properties}.  The horizontal solid line in the lower-left corner of each image represents $30\arcsec$.}
\figsetgrpend
 
\figsetgrpstart
\figsetgrpnum{8.392}
\figsetgrptitle{Visualization of UGC 12490.}
\figsetplot{\includegraphics[scale=0.7,angle=90]{f8_392.eps}}
\figsetgrpnote{Presentation of our integrated spectral atlas.  We plot the spectrum as $f_{\lambda}(\lambda)$ versus rest wavelength, normalized in a way that attempts to achieve a balance between showing the full range in flux and illustrating the finer details of the continuum.  Our integrated spectrum is accompanied by a Digitized Sky Survey image which illustrates our rectangular spectroscopic aperture as a \emph{solid} outline and the $25$~mag~arcsec$^{-2}$ isophotal size of the galaxy as a \emph{dashed} ellipse.  The image legend gives the galaxy name, the unique identification number in parenthesis, and the morphological type as listed in Table~\ref{table:general_properties}.  The horizontal solid line in the lower-left corner of each image represents $30\arcsec$.}
\figsetgrpend
 
\figsetgrpstart
\figsetgrpnum{8.393}
\figsetgrptitle{Visualization of NGC 7620.}
\figsetplot{\includegraphics[scale=0.7,angle=90]{f8_393.eps}}
\figsetgrpnote{Presentation of our integrated spectral atlas.  We plot the spectrum as $f_{\lambda}(\lambda)$ versus rest wavelength, normalized in a way that attempts to achieve a balance between showing the full range in flux and illustrating the finer details of the continuum.  Our integrated spectrum is accompanied by a Digitized Sky Survey image which illustrates our rectangular spectroscopic aperture as a \emph{solid} outline and the $25$~mag~arcsec$^{-2}$ isophotal size of the galaxy as a \emph{dashed} ellipse.  The image legend gives the galaxy name, the unique identification number in parenthesis, and the morphological type as listed in Table~\ref{table:general_properties}.  The horizontal solid line in the lower-left corner of each image represents $30\arcsec$.}
\figsetgrpend
 
\figsetgrpstart
\figsetgrpnum{8.394}
\figsetgrptitle{Visualization of NGC 7624.}
\figsetplot{\includegraphics[scale=0.7,angle=90]{f8_394.eps}}
\figsetgrpnote{Presentation of our integrated spectral atlas.  We plot the spectrum as $f_{\lambda}(\lambda)$ versus rest wavelength, normalized in a way that attempts to achieve a balance between showing the full range in flux and illustrating the finer details of the continuum.  Our integrated spectrum is accompanied by a Digitized Sky Survey image which illustrates our rectangular spectroscopic aperture as a \emph{solid} outline and the $25$~mag~arcsec$^{-2}$ isophotal size of the galaxy as a \emph{dashed} ellipse.  The image legend gives the galaxy name, the unique identification number in parenthesis, and the morphological type as listed in Table~\ref{table:general_properties}.  The horizontal solid line in the lower-left corner of each image represents $30\arcsec$.}
\figsetgrpend
 
\figsetgrpstart
\figsetgrpnum{8.395}
\figsetgrptitle{Visualization of NGC 7625.}
\figsetplot{\includegraphics[scale=0.7,angle=90]{f8_395.eps}}
\figsetgrpnote{Presentation of our integrated spectral atlas.  We plot the spectrum as $f_{\lambda}(\lambda)$ versus rest wavelength, normalized in a way that attempts to achieve a balance between showing the full range in flux and illustrating the finer details of the continuum.  Our integrated spectrum is accompanied by a Digitized Sky Survey image which illustrates our rectangular spectroscopic aperture as a \emph{solid} outline and the $25$~mag~arcsec$^{-2}$ isophotal size of the galaxy as a \emph{dashed} ellipse.  The image legend gives the galaxy name, the unique identification number in parenthesis, and the morphological type as listed in Table~\ref{table:general_properties}.  The horizontal solid line in the lower-left corner of each image represents $30\arcsec$.}
\figsetgrpend
 
\figsetgrpstart
\figsetgrpnum{8.396}
\figsetgrptitle{Visualization of NGC 7640.}
\figsetplot{\includegraphics[scale=0.7,angle=90]{f8_396.eps}}
\figsetgrpnote{Presentation of our integrated spectral atlas.  We plot the spectrum as $f_{\lambda}(\lambda)$ versus rest wavelength, normalized in a way that attempts to achieve a balance between showing the full range in flux and illustrating the finer details of the continuum.  Our integrated spectrum is accompanied by a Digitized Sky Survey image which illustrates our rectangular spectroscopic aperture as a \emph{solid} outline and the $25$~mag~arcsec$^{-2}$ isophotal size of the galaxy as a \emph{dashed} ellipse.  The image legend gives the galaxy name, the unique identification number in parenthesis, and the morphological type as listed in Table~\ref{table:general_properties}.  The horizontal solid line in the lower-left corner of each image represents $30\arcsec$.}
\figsetgrpend
 
\figsetgrpstart
\figsetgrpnum{8.397}
\figsetgrptitle{Visualization of UGC 12588.}
\figsetplot{\includegraphics[scale=0.7,angle=90]{f8_397.eps}}
\figsetgrpnote{Presentation of our integrated spectral atlas.  We plot the spectrum as $f_{\lambda}(\lambda)$ versus rest wavelength, normalized in a way that attempts to achieve a balance between showing the full range in flux and illustrating the finer details of the continuum.  Our integrated spectrum is accompanied by a Digitized Sky Survey image which illustrates our rectangular spectroscopic aperture as a \emph{solid} outline and the $25$~mag~arcsec$^{-2}$ isophotal size of the galaxy as a \emph{dashed} ellipse.  The image legend gives the galaxy name, the unique identification number in parenthesis, and the morphological type as listed in Table~\ref{table:general_properties}.  The horizontal solid line in the lower-left corner of each image represents $30\arcsec$.}
\figsetgrpend
 
\figsetgrpstart
\figsetgrpnum{8.398}
\figsetgrptitle{Visualization of UGCA 439.}
\figsetplot{\includegraphics[scale=0.7,angle=90]{f8_398.eps}}
\figsetgrpnote{Presentation of our integrated spectral atlas.  We plot the spectrum as $f_{\lambda}(\lambda)$ versus rest wavelength, normalized in a way that attempts to achieve a balance between showing the full range in flux and illustrating the finer details of the continuum.  Our integrated spectrum is accompanied by a Digitized Sky Survey image which illustrates our rectangular spectroscopic aperture as a \emph{solid} outline and the $25$~mag~arcsec$^{-2}$ isophotal size of the galaxy as a \emph{dashed} ellipse.  The image legend gives the galaxy name, the unique identification number in parenthesis, and the morphological type as listed in Table~\ref{table:general_properties}.  The horizontal solid line in the lower-left corner of each image represents $30\arcsec$.}
\figsetgrpend
 
\figsetgrpstart
\figsetgrpnum{8.399}
\figsetgrptitle{Visualization of NGC 7673.}
\figsetplot{\includegraphics[scale=0.7,angle=90]{f8_399.eps}}
\figsetgrpnote{Presentation of our integrated spectral atlas.  We plot the spectrum as $f_{\lambda}(\lambda)$ versus rest wavelength, normalized in a way that attempts to achieve a balance between showing the full range in flux and illustrating the finer details of the continuum.  Our integrated spectrum is accompanied by a Digitized Sky Survey image which illustrates our rectangular spectroscopic aperture as a \emph{solid} outline and the $25$~mag~arcsec$^{-2}$ isophotal size of the galaxy as a \emph{dashed} ellipse.  The image legend gives the galaxy name, the unique identification number in parenthesis, and the morphological type as listed in Table~\ref{table:general_properties}.  The horizontal solid line in the lower-left corner of each image represents $30\arcsec$.}
\figsetgrpend
 
\figsetgrpstart
\figsetgrpnum{8.400}
\figsetgrptitle{Visualization of NGC 7674.}
\figsetplot{\includegraphics[scale=0.7,angle=90]{f8_400.eps}}
\figsetgrpnote{Presentation of our integrated spectral atlas.  We plot the spectrum as $f_{\lambda}(\lambda)$ versus rest wavelength, normalized in a way that attempts to achieve a balance between showing the full range in flux and illustrating the finer details of the continuum.  Our integrated spectrum is accompanied by a Digitized Sky Survey image which illustrates our rectangular spectroscopic aperture as a \emph{solid} outline and the $25$~mag~arcsec$^{-2}$ isophotal size of the galaxy as a \emph{dashed} ellipse.  The image legend gives the galaxy name, the unique identification number in parenthesis, and the morphological type as listed in Table~\ref{table:general_properties}.  The horizontal solid line in the lower-left corner of each image represents $30\arcsec$.}
\figsetgrpend
 
\figsetgrpstart
\figsetgrpnum{8.401}
\figsetgrptitle{Visualization of ARP 182.}
\figsetplot{\includegraphics[scale=0.7,angle=90]{f8_401.eps}}
\figsetgrpnote{Presentation of our integrated spectral atlas.  We plot the spectrum as $f_{\lambda}(\lambda)$ versus rest wavelength, normalized in a way that attempts to achieve a balance between showing the full range in flux and illustrating the finer details of the continuum.  Our integrated spectrum is accompanied by a Digitized Sky Survey image which illustrates our rectangular spectroscopic aperture as a \emph{solid} outline and the $25$~mag~arcsec$^{-2}$ isophotal size of the galaxy as a \emph{dashed} ellipse.  The image legend gives the galaxy name, the unique identification number in parenthesis, and the morphological type as listed in Table~\ref{table:general_properties}.  The horizontal solid line in the lower-left corner of each image represents $30\arcsec$.}
\figsetgrpend
 
\figsetgrpstart
\figsetgrpnum{8.402}
\figsetgrptitle{Visualization of NGC 7674 A.}
\figsetplot{\includegraphics[scale=0.7,angle=90]{f8_402.eps}}
\figsetgrpnote{Presentation of our integrated spectral atlas.  We plot the spectrum as $f_{\lambda}(\lambda)$ versus rest wavelength, normalized in a way that attempts to achieve a balance between showing the full range in flux and illustrating the finer details of the continuum.  Our integrated spectrum is accompanied by a Digitized Sky Survey image which illustrates our rectangular spectroscopic aperture as a \emph{solid} outline and the $25$~mag~arcsec$^{-2}$ isophotal size of the galaxy as a \emph{dashed} ellipse.  The image legend gives the galaxy name, the unique identification number in parenthesis, and the morphological type as listed in Table~\ref{table:general_properties}.  The horizontal solid line in the lower-left corner of each image represents $30\arcsec$.}
\figsetgrpend
 
\figsetgrpstart
\figsetgrpnum{8.403}
\figsetgrptitle{Visualization of NGC 7677.}
\figsetplot{\includegraphics[scale=0.7,angle=90]{f8_403.eps}}
\figsetgrpnote{Presentation of our integrated spectral atlas.  We plot the spectrum as $f_{\lambda}(\lambda)$ versus rest wavelength, normalized in a way that attempts to achieve a balance between showing the full range in flux and illustrating the finer details of the continuum.  Our integrated spectrum is accompanied by a Digitized Sky Survey image which illustrates our rectangular spectroscopic aperture as a \emph{solid} outline and the $25$~mag~arcsec$^{-2}$ isophotal size of the galaxy as a \emph{dashed} ellipse.  The image legend gives the galaxy name, the unique identification number in parenthesis, and the morphological type as listed in Table~\ref{table:general_properties}.  The horizontal solid line in the lower-left corner of each image represents $30\arcsec$.}
\figsetgrpend
 
\figsetgrpstart
\figsetgrpnum{8.404}
\figsetgrptitle{Visualization of NGC 7678.}
\figsetplot{\includegraphics[scale=0.7,angle=90]{f8_404.eps}}
\figsetgrpnote{Presentation of our integrated spectral atlas.  We plot the spectrum as $f_{\lambda}(\lambda)$ versus rest wavelength, normalized in a way that attempts to achieve a balance between showing the full range in flux and illustrating the finer details of the continuum.  Our integrated spectrum is accompanied by a Digitized Sky Survey image which illustrates our rectangular spectroscopic aperture as a \emph{solid} outline and the $25$~mag~arcsec$^{-2}$ isophotal size of the galaxy as a \emph{dashed} ellipse.  The image legend gives the galaxy name, the unique identification number in parenthesis, and the morphological type as listed in Table~\ref{table:general_properties}.  The horizontal solid line in the lower-left corner of each image represents $30\arcsec$.}
\figsetgrpend
 
\figsetgrpstart
\figsetgrpnum{8.405}
\figsetgrptitle{Visualization of NGC 7679.}
\figsetplot{\includegraphics[scale=0.7,angle=90]{f8_405.eps}}
\figsetgrpnote{Presentation of our integrated spectral atlas.  We plot the spectrum as $f_{\lambda}(\lambda)$ versus rest wavelength, normalized in a way that attempts to achieve a balance between showing the full range in flux and illustrating the finer details of the continuum.  Our integrated spectrum is accompanied by a Digitized Sky Survey image which illustrates our rectangular spectroscopic aperture as a \emph{solid} outline and the $25$~mag~arcsec$^{-2}$ isophotal size of the galaxy as a \emph{dashed} ellipse.  The image legend gives the galaxy name, the unique identification number in parenthesis, and the morphological type as listed in Table~\ref{table:general_properties}.  The horizontal solid line in the lower-left corner of each image represents $30\arcsec$.}
\figsetgrpend
 
\figsetgrpstart
\figsetgrpnum{8.406}
\figsetgrptitle{Visualization of MRK 0930.}
\figsetplot{\includegraphics[scale=0.7,angle=90]{f8_406.eps}}
\figsetgrpnote{Presentation of our integrated spectral atlas.  We plot the spectrum as $f_{\lambda}(\lambda)$ versus rest wavelength, normalized in a way that attempts to achieve a balance between showing the full range in flux and illustrating the finer details of the continuum.  Our integrated spectrum is accompanied by a Digitized Sky Survey image which illustrates our rectangular spectroscopic aperture as a \emph{solid} outline and the $25$~mag~arcsec$^{-2}$ isophotal size of the galaxy as a \emph{dashed} ellipse.  The image legend gives the galaxy name, the unique identification number in parenthesis, and the morphological type as listed in Table~\ref{table:general_properties}.  The horizontal solid line in the lower-left corner of each image represents $30\arcsec$.}
\figsetgrpend
 
\figsetgrpstart
\figsetgrpnum{8.407}
\figsetgrptitle{Visualization of NGC 7714.}
\figsetplot{\includegraphics[scale=0.7,angle=90]{f8_407.eps}}
\figsetgrpnote{Presentation of our integrated spectral atlas.  We plot the spectrum as $f_{\lambda}(\lambda)$ versus rest wavelength, normalized in a way that attempts to achieve a balance between showing the full range in flux and illustrating the finer details of the continuum.  Our integrated spectrum is accompanied by a Digitized Sky Survey image which illustrates our rectangular spectroscopic aperture as a \emph{solid} outline and the $25$~mag~arcsec$^{-2}$ isophotal size of the galaxy as a \emph{dashed} ellipse.  The image legend gives the galaxy name, the unique identification number in parenthesis, and the morphological type as listed in Table~\ref{table:general_properties}.  The horizontal solid line in the lower-left corner of each image represents $30\arcsec$.}
\figsetgrpend
 
\figsetgrpstart
\figsetgrpnum{8.408}
\figsetgrptitle{Visualization of NGC 7713.}
\figsetplot{\includegraphics[scale=0.7,angle=90]{f8_408.eps}}
\figsetgrpnote{Presentation of our integrated spectral atlas.  We plot the spectrum as $f_{\lambda}(\lambda)$ versus rest wavelength, normalized in a way that attempts to achieve a balance between showing the full range in flux and illustrating the finer details of the continuum.  Our integrated spectrum is accompanied by a Digitized Sky Survey image which illustrates our rectangular spectroscopic aperture as a \emph{solid} outline and the $25$~mag~arcsec$^{-2}$ isophotal size of the galaxy as a \emph{dashed} ellipse.  The image legend gives the galaxy name, the unique identification number in parenthesis, and the morphological type as listed in Table~\ref{table:general_properties}.  The horizontal solid line in the lower-left corner of each image represents $30\arcsec$.}
\figsetgrpend
 
\figsetgrpstart
\figsetgrpnum{8.409}
\figsetgrptitle{Visualization of IRAS 23365+3604.}
\figsetplot{\includegraphics[scale=0.7,angle=90]{f8_409.eps}}
\figsetgrpnote{Presentation of our integrated spectral atlas.  We plot the spectrum as $f_{\lambda}(\lambda)$ versus rest wavelength, normalized in a way that attempts to achieve a balance between showing the full range in flux and illustrating the finer details of the continuum.  Our integrated spectrum is accompanied by a Digitized Sky Survey image which illustrates our rectangular spectroscopic aperture as a \emph{solid} outline and the $25$~mag~arcsec$^{-2}$ isophotal size of the galaxy as a \emph{dashed} ellipse.  The image legend gives the galaxy name, the unique identification number in parenthesis, and the morphological type as listed in Table~\ref{table:general_properties}.  The horizontal solid line in the lower-left corner of each image represents $30\arcsec$.}
\figsetgrpend
 
\figsetgrpstart
\figsetgrpnum{8.410}
\figsetgrptitle{Visualization of NGC 7727.}
\figsetplot{\includegraphics[scale=0.7,angle=90]{f8_410.eps}}
\figsetgrpnote{Presentation of our integrated spectral atlas.  We plot the spectrum as $f_{\lambda}(\lambda)$ versus rest wavelength, normalized in a way that attempts to achieve a balance between showing the full range in flux and illustrating the finer details of the continuum.  Our integrated spectrum is accompanied by a Digitized Sky Survey image which illustrates our rectangular spectroscopic aperture as a \emph{solid} outline and the $25$~mag~arcsec$^{-2}$ isophotal size of the galaxy as a \emph{dashed} ellipse.  The image legend gives the galaxy name, the unique identification number in parenthesis, and the morphological type as listed in Table~\ref{table:general_properties}.  The horizontal solid line in the lower-left corner of each image represents $30\arcsec$.}
\figsetgrpend
 
\figsetgrpstart
\figsetgrpnum{8.411}
\figsetgrptitle{Visualization of UGC 12747.}
\figsetplot{\includegraphics[scale=0.7,angle=90]{f8_411.eps}}
\figsetgrpnote{Presentation of our integrated spectral atlas.  We plot the spectrum as $f_{\lambda}(\lambda)$ versus rest wavelength, normalized in a way that attempts to achieve a balance between showing the full range in flux and illustrating the finer details of the continuum.  Our integrated spectrum is accompanied by a Digitized Sky Survey image which illustrates our rectangular spectroscopic aperture as a \emph{solid} outline and the $25$~mag~arcsec$^{-2}$ isophotal size of the galaxy as a \emph{dashed} ellipse.  The image legend gives the galaxy name, the unique identification number in parenthesis, and the morphological type as listed in Table~\ref{table:general_properties}.  The horizontal solid line in the lower-left corner of each image represents $30\arcsec$.}
\figsetgrpend
 
\figsetgrpstart
\figsetgrpnum{8.412}
\figsetgrptitle{Visualization of NGC 7742.}
\figsetplot{\includegraphics[scale=0.7,angle=90]{f8_412.eps}}
\figsetgrpnote{Presentation of our integrated spectral atlas.  We plot the spectrum as $f_{\lambda}(\lambda)$ versus rest wavelength, normalized in a way that attempts to achieve a balance between showing the full range in flux and illustrating the finer details of the continuum.  Our integrated spectrum is accompanied by a Digitized Sky Survey image which illustrates our rectangular spectroscopic aperture as a \emph{solid} outline and the $25$~mag~arcsec$^{-2}$ isophotal size of the galaxy as a \emph{dashed} ellipse.  The image legend gives the galaxy name, the unique identification number in parenthesis, and the morphological type as listed in Table~\ref{table:general_properties}.  The horizontal solid line in the lower-left corner of each image represents $30\arcsec$.}
\figsetgrpend
 
\figsetgrpstart
\figsetgrpnum{8.413}
\figsetgrptitle{Visualization of NGC 7771.}
\figsetplot{\includegraphics[scale=0.7,angle=90]{f8_413.eps}}
\figsetgrpnote{Presentation of our integrated spectral atlas.  We plot the spectrum as $f_{\lambda}(\lambda)$ versus rest wavelength, normalized in a way that attempts to achieve a balance between showing the full range in flux and illustrating the finer details of the continuum.  Our integrated spectrum is accompanied by a Digitized Sky Survey image which illustrates our rectangular spectroscopic aperture as a \emph{solid} outline and the $25$~mag~arcsec$^{-2}$ isophotal size of the galaxy as a \emph{dashed} ellipse.  The image legend gives the galaxy name, the unique identification number in parenthesis, and the morphological type as listed in Table~\ref{table:general_properties}.  The horizontal solid line in the lower-left corner of each image represents $30\arcsec$.}
\figsetgrpend
 
\figsetgrpstart
\figsetgrpnum{8.414}
\figsetgrptitle{Visualization of MRK 0331.}
\figsetplot{\includegraphics[scale=0.7,angle=90]{f8_414.eps}}
\figsetgrpnote{Presentation of our integrated spectral atlas.  We plot the spectrum as $f_{\lambda}(\lambda)$ versus rest wavelength, normalized in a way that attempts to achieve a balance between showing the full range in flux and illustrating the finer details of the continuum.  Our integrated spectrum is accompanied by a Digitized Sky Survey image which illustrates our rectangular spectroscopic aperture as a \emph{solid} outline and the $25$~mag~arcsec$^{-2}$ isophotal size of the galaxy as a \emph{dashed} ellipse.  The image legend gives the galaxy name, the unique identification number in parenthesis, and the morphological type as listed in Table~\ref{table:general_properties}.  The horizontal solid line in the lower-left corner of each image represents $30\arcsec$.}
\figsetgrpend
 
\figsetgrpstart
\figsetgrpnum{8.415}
\figsetgrptitle{Visualization of NGC 7782.}
\figsetplot{\includegraphics[scale=0.7,angle=90]{f8_415.eps}}
\figsetgrpnote{Presentation of our integrated spectral atlas.  We plot the spectrum as $f_{\lambda}(\lambda)$ versus rest wavelength, normalized in a way that attempts to achieve a balance between showing the full range in flux and illustrating the finer details of the continuum.  Our integrated spectrum is accompanied by a Digitized Sky Survey image which illustrates our rectangular spectroscopic aperture as a \emph{solid} outline and the $25$~mag~arcsec$^{-2}$ isophotal size of the galaxy as a \emph{dashed} ellipse.  The image legend gives the galaxy name, the unique identification number in parenthesis, and the morphological type as listed in Table~\ref{table:general_properties}.  The horizontal solid line in the lower-left corner of each image represents $30\arcsec$.}
\figsetgrpend
 
\figsetgrpstart
\figsetgrpnum{8.416}
\figsetgrptitle{Visualization of NGC 7798.}
\figsetplot{\includegraphics[scale=0.7,angle=90]{f8_416.eps}}
\figsetgrpnote{Presentation of our integrated spectral atlas.  We plot the spectrum as $f_{\lambda}(\lambda)$ versus rest wavelength, normalized in a way that attempts to achieve a balance between showing the full range in flux and illustrating the finer details of the continuum.  Our integrated spectrum is accompanied by a Digitized Sky Survey image which illustrates our rectangular spectroscopic aperture as a \emph{solid} outline and the $25$~mag~arcsec$^{-2}$ isophotal size of the galaxy as a \emph{dashed} ellipse.  The image legend gives the galaxy name, the unique identification number in parenthesis, and the morphological type as listed in Table~\ref{table:general_properties}.  The horizontal solid line in the lower-left corner of each image represents $30\arcsec$.}
\figsetgrpend
 
\figsetgrpstart
\figsetgrpnum{8.417}
\figsetgrptitle{Visualization of NGC 7800.}
\figsetplot{\includegraphics[scale=0.7,angle=90]{f8_417.eps}}
\figsetgrpnote{Presentation of our integrated spectral atlas.  We plot the spectrum as $f_{\lambda}(\lambda)$ versus rest wavelength, normalized in a way that attempts to achieve a balance between showing the full range in flux and illustrating the finer details of the continuum.  Our integrated spectrum is accompanied by a Digitized Sky Survey image which illustrates our rectangular spectroscopic aperture as a \emph{solid} outline and the $25$~mag~arcsec$^{-2}$ isophotal size of the galaxy as a \emph{dashed} ellipse.  The image legend gives the galaxy name, the unique identification number in parenthesis, and the morphological type as listed in Table~\ref{table:general_properties}.  The horizontal solid line in the lower-left corner of each image represents $30\arcsec$.}
\figsetgrpend
 
\figsetend


\begin{figure}
\figurenum{8.1}
\begin{center}
\includegraphics[scale=0.7,angle=90]{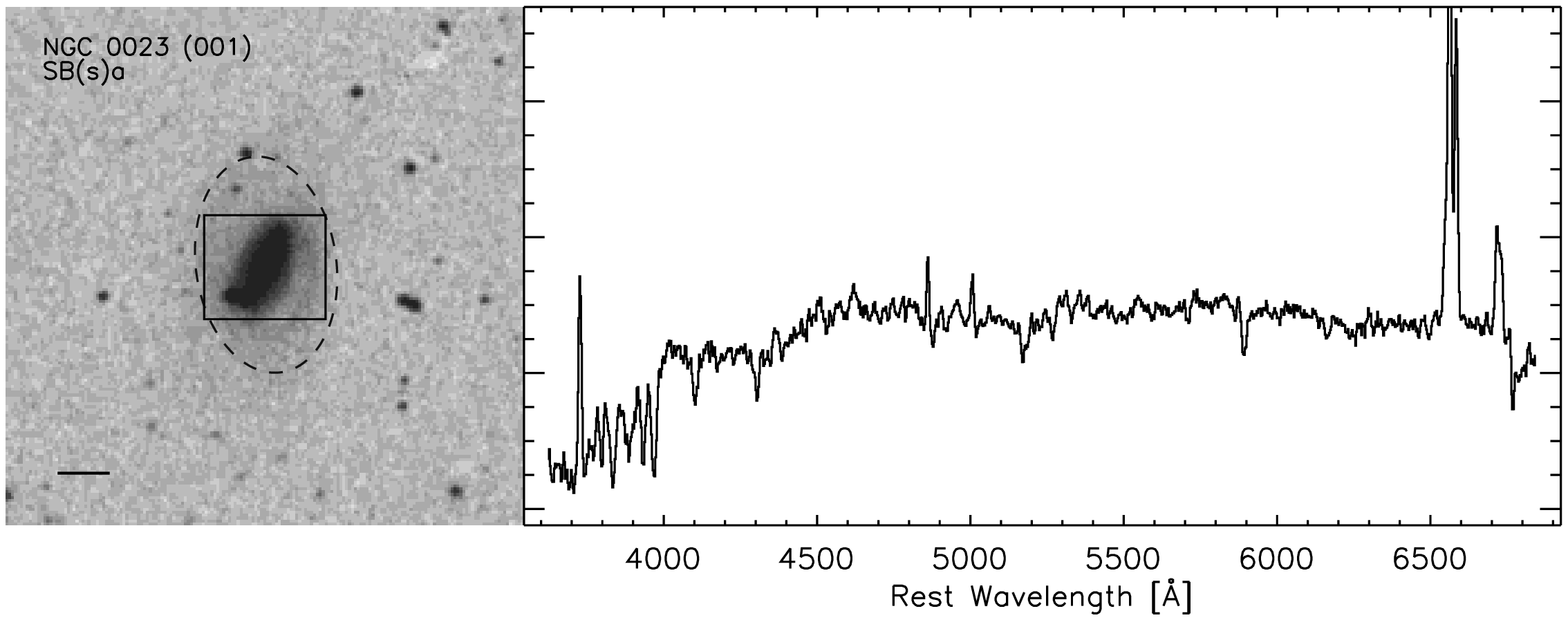}
\caption{Presentation of our integrated spectral atlas.  We plot the
spectrum as $f_{\lambda}(\lambda)$ versus rest wavelength, normalized
in a way that attempts to achieve a balance between showing the full
range in flux and illustrating the finer details of the continuum.
Our integrated spectrum is accompanied by a Digitized Sky Survey image
which illustrates our rectangular spectroscopic aperture as a
\emph{solid} outline and the $25$~mag~arcsec$^{-2}$ isophotal size of
the galaxy as a \emph{dashed} ellipse.  The image legend gives the
galaxy name, the unique identification number in parenthesis, and the
morphological type as listed in Table~\ref{table:general_properties}.
The horizontal solid line in the lower-left corner of each image
represents $30\arcsec$.  [{\em See the electronic edition of the
Astrophysical Journal for Figs. 8.2-8.417}.]
\label{fig:atlas_00}}
\end{center}
\end{figure}

\setcounter{figure}{8} 

\begin{figure}
\epsscale{1.0}
\plotone{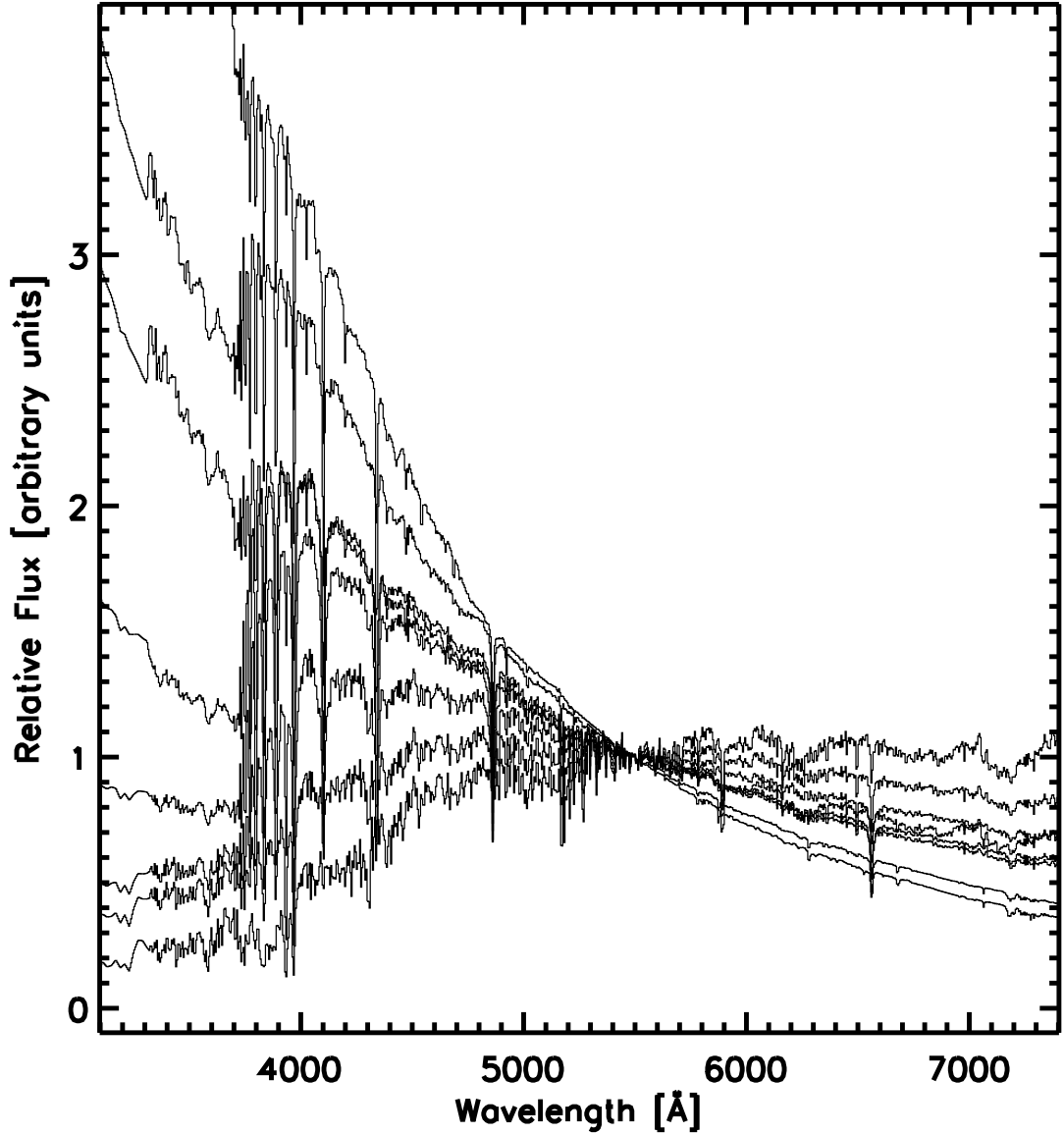}
\caption{The eight instantaneous-burst, solar metallicity population
synthesis models from \citet{bruzual03}, normalized at $5500$~\AA,
that we use to model the stellar continuum of each galaxy.  Each
spectrum corresponds to the following ages, plotted from top to bottom
on the left-hand side of the figure: $0$, $5$, $25$, $100$, $255$,
$640$, $1400$, and $12000$~Myr.  The FWHM spectral resolution of these
models is $3$~\AA. \label{fig:templates}}
\end{figure}

\clearpage

\begin{figure}
\epsscale{0.9}
\plotone{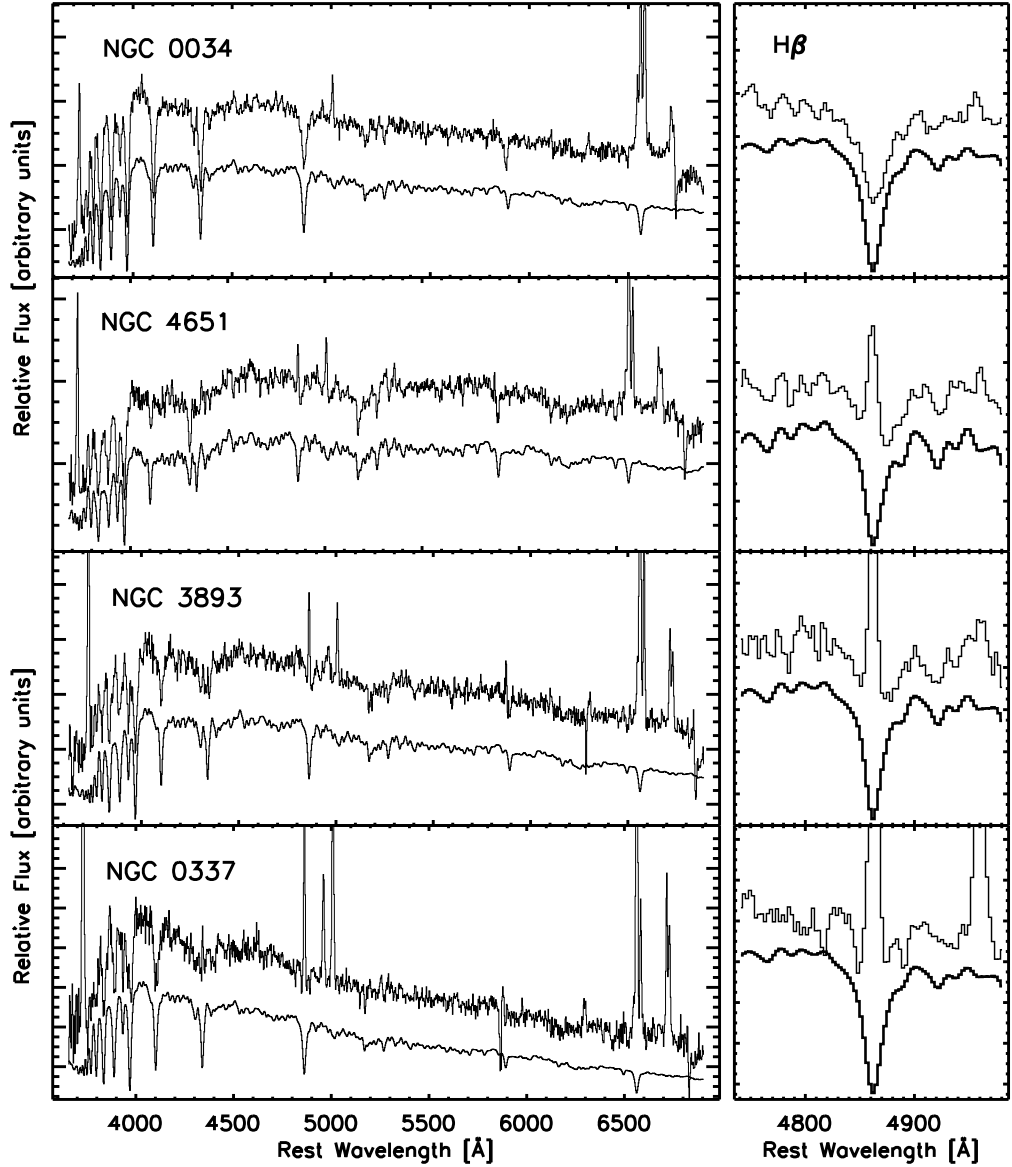}
\caption{Illustration of our population synthesis fitting technique
for the integrated spectra of NGC~0034, NGC~4651, NGC~3893, and
NGC~0337.  For each object we offset the best-fitting stellar
continuum spectrum downward from the data for clarity.  The left
panels show the full spectral range while the right panel zooms in on
the \hblam{} nebular emission line.  
\label{fig:example_specfit}}
\end{figure}

\clearpage

\begin{figure}
\epsscale{0.9}
\plotone{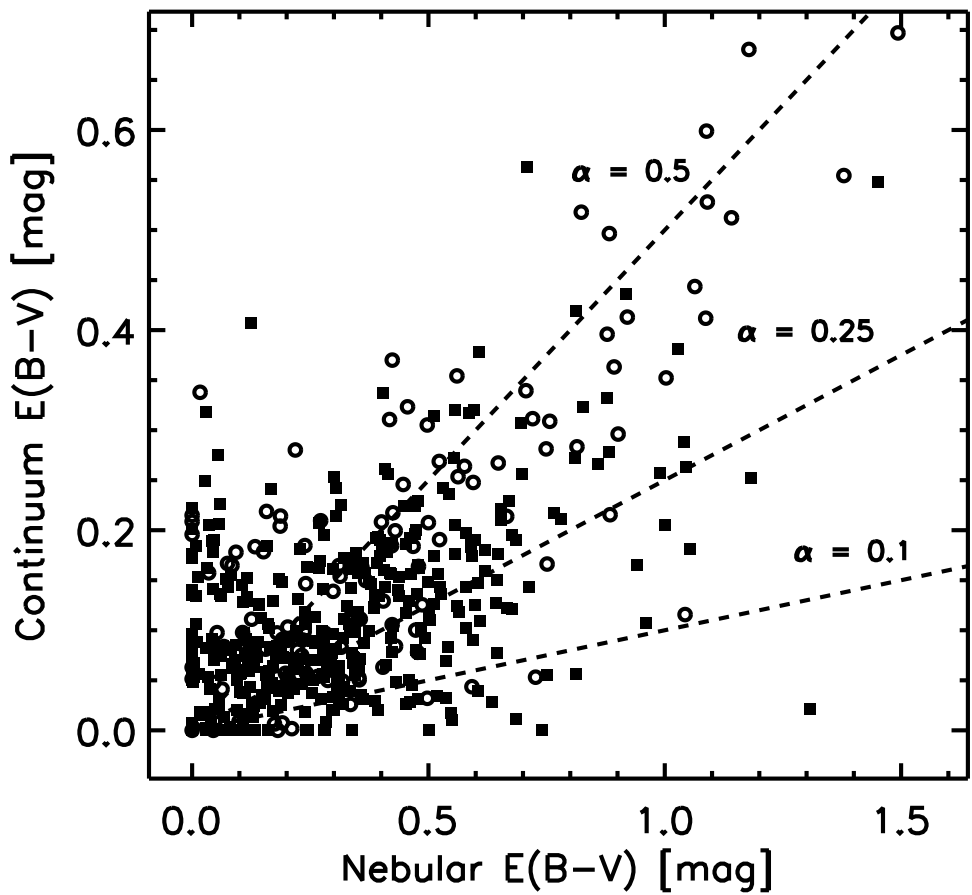}
\caption{Correlation between the best-fitting continuum reddening,
\ebv, and the nebular reddening as derived from the \ha/\hb{} Balmer
decrement for our nuclear spectra (open circles) and our integrated
spectra (filled squares).  The dashed lines show three values of the
$\alpha$ parameter, assuming that the nebular reddening traces the
reddening of the young ($<10$~Myr) stellar populations.  See
\S\ref{sec:ebvmodel} for more details.
\label{fig:ebv_correlation}}
\end{figure}

\clearpage


\voffset0.5in
\begin{deluxetable}{cllrrcllcccrc}
\tabletypesize{\tiny}
\rotate
\tablecaption{General Properties \label{table:general_properties}}
\tablewidth{0pt}
\tablehead{
\colhead{ID} & 
\colhead{Galaxy Name} & 
\colhead{Other Names} & 
\colhead{$\alpha_{\rm J2000.0}$} & 
\colhead{$\delta_{\rm J2000.0}$} & 
\colhead{cz} & 
\colhead{E(\bv)} & 
\colhead{Type} & 
\colhead{D$_{25}^{\rm maj}$} & 
\colhead{D$_{25}^{\rm min}$} & 
\colhead{$\theta$} & 
\colhead{D} & 
\colhead{Ref.} \\ 
\colhead{(1)} & 
\colhead{(2)} & 
\colhead{(3)} & 
\colhead{(4)} & 
\colhead{(5)} & 
\colhead{(6)} & 
\colhead{(7)} & 
\colhead{(8)} & 
\colhead{(9)} & 
\colhead{(10)} & 
\colhead{(11)} & 
\colhead{(12)} & 
\colhead{(13)} 
}
\startdata
001 &  \object[NGC0023]{NGC 0023} &  UGC 00089, MRK 0545 &  00:09:53.4 & 
+25:55:26 &  4556 &  0.038 &  SB(s)a &  2.09 &  1.35 &  8 &  64.2 &  1 \\ 
002 &  \object[NGC0017]{NGC 0034} &  NGC 0017, MRK 0938 &  00:11:06.5 & 
-12:06:26 &  5935 &  0.027 &  Sc &  2.19 &  0.81 &  31\tablenotemark{b} & 
83.2 &  1 \\ 
003 &  \object[ARP256NED02]{ARP 256 N} &  MCG -02-01-052 &  00:18:50.1 & 
-10:21:42 &  8184 &  0.036 &  SB(s)c &  1.12 &  0.76 &  50\tablenotemark{b} & 
114.3 &  1 \\ 
004 &  \object[ARP256]{ARP 256} &  VV 352 &  00:18:50.4 &  -10:22:08 &  8154 & 
0.036 &  Pec &  \nodata &  \nodata &  \nodata &  113.9 &  1 \\ 
005 &  \object[ARP256NED01]{ARP 256 S} &  MCG -02-01-051 &  00:18:50.9 & 
-10:22:37 &  8124 &  0.036 &  SB(s)b &  1.10 &  0.63 &  71\tablenotemark{b} & 
113.4 &  1 \\ 
006 &  \object[NGC0095]{NGC 0095} &  UGC 00214 &  00:22:13.5 &  +10:29:30 & 
5366 &  0.116 &  SAB(rs)c &  1.95 &  1.12 &  87\tablenotemark{b} &  74.9 & 
1 \\ 
007 &  \object[NGC0151]{NGC 0151} &  NGC 0153 &  00:34:02.8 &  -09:42:19 & 
3747 &  0.032 &  SB(r)bc &  3.71 &  1.70 &  75 &  52.5 &  1 \\ 
008 &  \object[NGC0157]{NGC 0157} &  MCG -02-02-056 &  00:34:46.7 & 
-08:23:47 &  1678 &  0.044 &  SAB(rs)bc &  4.17 &  2.69 &  30 &  24.0 &  1 \\ 
009 &  \object[NGC0178]{NGC 0178} &  IC 0039, VIII Zw 034 &  00:39:08.4 & 
-14:10:22 &  1439 &  0.021 &  SB(s)m &  2.04 &  0.98 &  175 &  20.7 &  1 \\ 
010 &  \object[NGC0232]{NGC 0232} &  MCG -04-02-040 &  00:42:45.8 & 
-23:33:41 &  6655 &  0.019 &  SB(r)a &  0.95 &  0.78 &  13\tablenotemark{b} & 
93.3 &  1 \\ 
\enddata
\tablecomments{Col. (1) ID number; Col. (2) Galaxy name; Col. (3) Other common galaxy name or names; Col. (4) Right ascension from NED, unless otherwise noted (J2000); Col. (5) Declination from NED, unless otherwise noted (J2000); Col. (6) Heliocentric redshift from NED (km~s$^{-1}$); Col. (7) Foreground Galactic reddening from \citet{schlegel98} (mag); Col. (8) Morphological type (see \S\ref{sec:properties}); Col. (9) Major axis diameter at the $25$~mag~arcsec$^{-2}$ isophote from the RC3, unless otherwise noted (arcmin); Col. (10) Minor axis diameter at the $25$~mag~arcsec$^{-2}$ isophote from the RC3, unless otherwise noted (arcmin); Col. (11) Position angle, measured positive from North to East from the RC3, unless otherwise noted (deg);  Col. (12) Distance (Mpc);  Col. (13) Distance reference. Table \ref{table:general_properties} is published in its entirety in the electronic edition of the {\it Astrophysical Journal}.  A portion is shown here for guidance regarding its form and content.}
\tablenotetext{a}{Coordinates taken from SIMBAD.}
\tablenotetext{b}{Coordinates, major/minor axis diameters, and/or position angle taken from LEDA.}
\tablenotetext{c}{Major/minor axis diameter from NED "Basic Data".}
\tablenotetext{d}{Major/minor axis diameter and/or position angle from the Uppsala General Catalog of Galaxies \citep[UGC;][]{nilson73}.}
\tablenotetext{e}{Major/minor axis diameter and/or position angle from the ESO/Uppsala Survey of the ESO(B) Atlas \citep{lauberts82}.}
\tablenotetext{f}{Major/minor axis diameter and/or position angle from the Morphological Catalogue of Galaxies \citep[MCG;][]{vorontsov62}.}
\tablenotetext{g}{Position angle from the 2MASS Large Galaxy Atlas \citep{jarrett03}.}
\tablenotetext{h}{Position angle from the 2MASS Extended Source Catalog \citep{jarrett00}.}
\tablerefs{(1) \citet{mould00}; (2) \citet{karachentsev04}; (3) \citet{shapley01b}; (4) \citet{freedman01}; (5) \citet{tully96}; (6) \citet{tonry01}}
\end{deluxetable}

\clearpage

\voffset0.5in
\begin{deluxetable}{lcccccccccc}
\tabletypesize{\tiny}
\rotate
\tablecaption{Photometric Properties \label{table:photometric_properties}}
\tablewidth{0pt}
\tablehead{
\colhead{ID} & 
\colhead{U} & 
\colhead{B} & 
\colhead{V} & 
\colhead{J} & 
\colhead{H} & 
\colhead{K$_{\rm s}$} & 
\colhead{$S_{\nu}(12~\micron)$} & 
\colhead{$S_{\nu}(25~\micron)$} & 
\colhead{$S_{\nu}(60~\micron)$} & 
\colhead{$S_{\nu}(100~\micron)$} \\ 
\colhead{(1)} & 
\colhead{(2)} & 
\colhead{(3)} & 
\colhead{(4)} & 
\colhead{(5)} & 
\colhead{(6)} & 
\colhead{(7)} & 
\colhead{(8)} & 
\colhead{(9)} & 
\colhead{(10)} & 
\colhead{(11)} 
}
\startdata
001 &  \nodata &  $       12.69\pm        0.17$ & 
$       11.91\pm        0.18$ &  $       9.877\pm       0.013$ & 
$       9.188\pm       0.014$ &  $       8.862\pm       0.020$ & 
$       0.590\pm       0.042$ &  $       1.240\pm       0.054$ & 
$       8.770\pm       0.047$ &  $      14.960\pm       0.096$ \\ 
002 &  \nodata &  $       15.03\pm        0.46$ &  \nodata & 
$      11.216\pm       0.025$ &  $      10.449\pm       0.029$ & 
$      10.064\pm       0.030$ &  $       0.360\pm       0.038$ & 
$       2.380\pm       0.066$ &  $      16.080\pm       0.058$ & 
$       16.97\pm        0.20$ \\ 
003 &  \nodata &  $       14.70\pm        0.15$ &  \nodata & 
$      12.656\pm       0.052$ &  $      12.329\pm       0.088$ & 
$      11.811\pm       0.092$ &  \nodata &  \nodata &  \nodata &  \nodata \\ 
004 &  \nodata &  \nodata &  \nodata &  \nodata &  \nodata &  \nodata & 
$       0.230\pm       0.040$ &  $       1.180\pm       0.066$ & 
$       7.350\pm       0.046$ &  $        9.48\pm        0.16$ \\ 
005 &  \nodata &  $       14.66\pm        0.13$ & 
$       14.22\pm        0.13$ &  $      12.404\pm       0.028$ & 
$      11.677\pm       0.037$ &  $      11.330\pm       0.046$ &  \nodata & 
\nodata &  \nodata &  \nodata \\ 
006 &  $       12.67\pm        0.14$ &  $       12.71\pm        0.14$ & 
$       12.16\pm        0.15$ &  $      10.246\pm       0.020$ & 
$       9.642\pm       0.025$ &  $       9.407\pm       0.033$ & 
$       0.127\pm       0.033$ &  $       0.190\pm       0.053$ & 
$        2.20\pm        0.15$ &  $        5.28\pm        0.48$ \\ 
007 &  $       12.28\pm        0.13$ &  $       12.18\pm        0.13$ & 
$       11.49\pm        0.13$ &  $       9.520\pm       0.015$ & 
$       8.929\pm       0.022$ &  $       8.741\pm       0.030$ & 
$       0.215\pm       0.039$ &  $       0.255\pm       0.051$ & 
$        1.53\pm        0.11$ &  $        5.58\pm        0.61$ \\ 
008 &  $       10.76\pm        0.12$ &  $       10.82\pm        0.12$ & 
$       10.27\pm        0.12$ &  $      8.5600\pm      0.0060$ & 
$      7.8970\pm      0.0080$ &  $       7.665\pm       0.012$ & 
$       1.570\pm       0.044$ &  $       2.080\pm       0.059$ & 
$      17.570\pm       0.045$ &  $       43.10\pm        0.12$ \\ 
009 &  $       12.65\pm        0.21$ &  $       13.01\pm        0.20$ & 
$       12.56\pm        0.20$ &  $      11.670\pm       0.036$ & 
$      11.085\pm       0.050$ &  $      10.794\pm       0.065$ &  \nodata & 
\nodata &  $        1.62\pm        0.13$ &  $        2.59\pm        0.21$ \\ 
010 &  \nodata &  $       14.31\pm        0.21$ &  \nodata & 
$      10.968\pm       0.025$ &  $      10.275\pm       0.028$ & 
$       9.889\pm       0.030$ &  $       0.360\pm       0.042$ & 
$       1.220\pm       0.040$ &  $      10.060\pm       0.036$ & 
$       16.83\pm        0.11$ \\ 
\enddata
\tablecomments{Col. (1) ID number; Col. (2) U magnitude; Col. (3) B magnitude; Col. (4) V magnitude; Col. (5) 2MASS total J magnitude; Col. (6) 2MASS total H magnitude; Col. (7) 2MASS total K$_{\mathrm s}$ magnitude; Col. (8) IRAS $12~\micron$ flux density (Jy); Col. (9) IRAS $25~\micron$ flux density (Jy); Col. (10) IRAS $60~\micron$ flux density (Jy); Col. (11) IRAS $100~\micron$ flux density (Jy). Table \ref{table:photometric_properties} is published in its entirety in the electronic edition of the {\it Astrophysical Journal}.  A portion is shown here for guidance regarding its form and content.}
\tablenotetext{a}{\citet{soifer89} do not provide measurement errors for the IRAS fluxes of NGC~5194, therefore we assume a fixed $25\%$ uncertainty at every wavelength.}
\end{deluxetable}

\clearpage

\voffset0.5in
\begin{deluxetable}{llccccccl}
\tabletypesize{\tiny}
\tablecaption{Summary of Integrated Spectrophotometric Observations\label{table:journal}}
\tablewidth{0pt}
\tablehead{
\colhead{ID} & 
\colhead{Galaxy name} & 
\colhead{$\Delta_{\rm scan}$} & 
\colhead{Ap.} & 
\colhead{$\theta_{\rm slit}$} & 
\colhead{$t$} & 
\colhead{Clear} & 
\colhead{$\Delta$m} & 
\colhead{Remarks\tablenotemark{a}} \\ 
\colhead{(1)} & 
\colhead{(2)} & 
\colhead{(3)} & 
\colhead{(4)} & 
\colhead{(5)} & 
\colhead{(6)} & 
\colhead{(7)} & 
\colhead{(8)} & 
\colhead{(9)} 
}
\startdata
001 &  NGC 0023 &  60 &  70 &  90 &  1800 &  Y &  0.054 & 
supernova 1955C subtracted \\ 
002 &  NGC 0034 &  40 &  45 &  90 &  1800 &  Y &  0.054 &   \\ 
003 &  ARP 256 N &  40 &  60 &  180 &  1800 &  Y &  0.054 & 
interacting with ARP 256 S \\ 
004 &  ARP 256 &  40 &  100 &  180 &  1800 &  Y &  0.054 &  interacting pair \\ 
005 &  ARP 256 S &  40 &  40 &  180 &  1800 &  Y &  0.054 & 
interacting with ARP 256 N \\ 
006 &  NGC 0095 &  60 &  100 &  90 &  2400 &  Y &  0.054 &   \\ 
007 &  NGC 0151 &  150 &  120 &  180 &  2400 &  N &  0.531 &   \\ 
008 &  NGC 0157 &  180 &  130 &  90 &  2400 &  Y &  0.054 & 
foreground star excluded \\ 
009 &  NGC 0178 &  90 &  60 &  90 &  2400 &  Y &  0.054 &   \\ 
010 &  NGC 0232 &  40 &  60 &  90 &  3600 &  Y &  0.078 &   \\ 
\enddata
\tablecomments{Col. (1) ID number; Col. (2) Galaxy name; Col. (3) Drift scan length perpendicular to the slit (arcsec); Col. (4) Extraction aperture along the slit (arcsec); Col. (5) Slit position angle measured positive from North to East (deg); Col. (6) Total exposure time (seconds); Col. (7) Flag indicating clear (Y) or non-photometric (N) observing conditions; Col. (8) Absolute spectrophotometric uncertainty ($1\sigma$) based only on the scatter in the observed sensitivity function (mag); Col. (9) Remarks regarding the object or the spectral extraction. Table \ref{table:journal} is published in its entirety in the electronic edition of the {\it Astrophysical Journal}.  A portion is shown here for guidance regarding its form and content.}
\tablenotetext{a}{In this footnote we clarify some of the remarks that appear in column (9): {\em stellar contamination}: indicates foreground stellar contamination that could not be subtracted; {\em foreground star(s) excluded}: a smaller extraction aperture was adopted to avoid one or more foreground stars; {\em scan avoids star(s)}: the drift scan center and length were selected to avoid nearby bright stars.}
\tablenotetext{b}{Variance-weighted average of three indistinguishable spectra extracted using $80\arcsec\times25\arcsec$ (PA=$180^{\circ}$), $80\arcsec\times60\arcsec$ (PA=$180^{\circ}$), and $80\arcsec\times60\arcsec$ (PA=$90^{\circ}$) apertures.}
\tablenotetext{c}{Variance-weighted average of two indistinguishable spectra obtained at a $90^{\circ}$ slit position angle and extracted using $110\arcsec\times60\arcsec$ and $110\arcsec\times100\arcsec$ apertures.}
\tablenotetext{d}{Variance-weighted average of two indistinguishable spectra obtained at a $90^{\circ}$ slit position angle and extracted using $90\arcsec\times360\arcsec$ and $100\arcsec\times360\arcsec$ apertures.}
\tablenotetext{e}{Variance-weighted average of two indistinguishable spectra extracted using $160\arcsec\times60\arcsec$ (PA=$70^{\circ}$) and $140\arcsec\times90\arcsec$ (PA=$90^{\circ}$) apertures.}
\tablenotetext{f}{Variance-weighted average of two indistinguishable spectra obtained at a $90^{\circ}$ slit position angle and extracted using $80\arcsec\times40\arcsec$ and $80\arcsec\times50\arcsec$ apertures.}
\tablenotetext{g}{Variance-weighted average of two indistinguishable spectra taken at $114^{\circ}$ and $108^{\circ}$.}
\tablenotetext{h}{Variance-weighted average of two indistinguishable spectra extracted using $125\arcsec\times30\arcsec$ (PA=$165^{\circ}$) and $60\arcsec\times60\arcsec$ (PA=$90^{\circ}$) apertures.}
\tablenotetext{i}{Variance-weighted average of three indistinguishable spectra obtained at a $90^{\circ}$ slit position angle and extracted using $140\arcsec\times60\arcsec$ and two $140\arcsec\times75\arcsec$ apertures.}
\tablenotetext{j}{Variance-weighted average of two indistinguishable spectra extracted using $75\arcsec\times150\arcsec$ (PA=$90^{\circ}$) and $95\arcsec\times155\arcsec$ (PA=$180^{\circ}$) apertures.}
\tablenotetext{k}{Variance-weighted average of two indistinguishable spectra extracted using $40\arcsec\times150\arcsec$ (PA=$90^{\circ}$) and $45\arcsec\times155\arcsec$ (PA=$180^{\circ}$) apertures.}
\tablenotetext{l}{Variance-weighted average of two indistinguishable spectra extracted using $35\arcsec\times150\arcsec$ (PA=$90^{\circ}$) and $50\arcsec\times155\arcsec$ (PA=$180^{\circ}$) apertures.}
\tablenotetext{m}{Variance-weighted average of three indistinguishable spectra extracted using $90\arcsec\times20\arcsec$ (PA=$130^{\circ}$), $90\arcsec\times20\arcsec$ (PA=$130^{\circ}$), and $50\arcsec\times30\arcsec$ (PA=$90^{\circ}$) apertures.}
\tablenotetext{n}{Variance-weighted average of two indistinguishable spectra obtained at a $90^{\circ}$ slit position angle and extracted using $45\arcsec\times10\arcsec$ and $45\arcsec\times20\arcsec$ apertures.}
\tablenotetext{o}{Variance-weighted average of two indistinguishable spectra obtained at a $90^{\circ}$ slit position angle and extracted using $50\arcsec\times30\arcsec$ and $50\arcsec\times40\arcsec$ apertures.}
\tablenotetext{p}{Variance-weighted average of two indistinguishable spectra taken at $90^{\circ}$ and $153^{\circ}$.}
\end{deluxetable}

\clearpage

\include{tab4}
\clearpage

\voffset0.5in
\begin{deluxetable}{lcccccccccc}
\tabletypesize{\tiny}
\rotate
\tablecaption{Integrated Emission-line Fluxes\tablenotemark{a} \label{table:int_lineflux}}
\tablewidth{0pt}
\tablehead{
\colhead{ID} & 
\colhead{[O~{\sc ii}]$~\lambda3727$} & 
\colhead{H$\delta~\lambda4101$} & 
\colhead{H$\gamma~\lambda4340$} & 
\colhead{H$\beta~\lambda4861$} & 
\colhead{[O~{\sc iii}]$~\lambda5007$} & 
\colhead{[O~{\sc i}]$~\lambda6300$} & 
\colhead{H$\alpha~\lambda6563$} & 
\colhead{[N~{\sc ii}]$~\lambda6584$} & 
\colhead{[S~{\sc ii}]$~\lambda6716$} & 
\colhead{[S~{\sc ii}]$~\lambda6731$}
}
\startdata
001 &  $         352\pm          30$ &  $        64.2\pm         6.1$ & 
$        95.5\pm         6.1$ &  $       223.0\pm         9.9$ & 
$        97.9\pm         5.8$ &  $        32.9\pm         5.0$ & 
$        1158\pm          47$ &  $         603\pm          25$ & 
$         219\pm          11$ &  $       149.8\pm         9.4$ \\ 
002 &  $         124\pm          17$ &  $         9.9\pm         3.0$ & 
$         7.3\pm         2.6$ &  $        23.5\pm         2.3$ & 
$        44.5\pm         2.9$ &  $        18.7\pm         2.7$ & 
$       203.7\pm         9.1$ &  $       220.5\pm         9.6$ & 
$        62.6\pm         4.1$ &  $        12.7\pm         3.6$ \\ 
003 &  $         233\pm          18$ &  $        20.5\pm         2.9$ & 
$        43.9\pm         3.0$ &  $       106.3\pm         4.8$ & 
$        92.8\pm         4.4$ &  $        17.8\pm         2.6$ & 
$         400\pm          16$ &  $       115.9\pm         5.4$ & 
$        60.1\pm         3.7$ &  $        34.9\pm         3.2$ \\ 
004 &  $         571\pm          32$ &  $        50.7\pm         4.3$ & 
$       102.8\pm         5.3$ &  $         260\pm          11$ & 
$       230.7\pm         9.8$ &  $        46.8\pm         3.9$ & 
$        1103\pm          44$ &  $         344\pm          14$ & 
$       167.8\pm         7.8$ &  $       110.8\pm         6.0$ \\ 
005 &  $         336\pm          20$ &  $        30.2\pm         2.8$ & 
$        57.4\pm         3.2$ &  $       152.0\pm         6.5$ & 
$       138.6\pm         6.0$ &  $        28.9\pm         2.6$ & 
$         701\pm          28$ &  $       227.2\pm         9.5$ & 
$       107.5\pm         5.1$ &  $        76.8\pm         4.1$ \\ 
006 &  $         559\pm          42$ &  $        72.4\pm         7.4$ & 
$       111.7\pm         7.3$ &  $         225\pm          10$ & 
$       169.1\pm         8.3$ &  $        41.3\pm         5.3$ & 
$         890\pm          36$ &  $         315\pm          14$ & 
$       155.7\pm         9.6$ &  $       112.6\pm         7.5$ \\ 
007 &  $         590\pm         110$ &  $         146\pm          20$ & 
$          73\pm          17$ &  $         309\pm          20$ & 
$         206\pm          18$ &  $<48$ &  $         971\pm          45$ & 
$         465\pm          29$ &  $         214\pm          22$ & 
$         126\pm          21$ \\ 
008 &  $        1350\pm         120$ &  $         431\pm          27$ & 
$         462\pm          25$ &  $        1065\pm          45$ & 
$         326\pm          19$ &  $         182\pm          18$ & 
$        4400\pm         180$ &  $        1602\pm          67$ & 
$         893\pm          40$ &  $         671\pm          32$ \\ 
009 &  $         880\pm          51$ &  $        61.3\pm         6.1$ & 
$       111.7\pm         6.7$ &  $         260\pm          11$ & 
$         641\pm          26$ &  $        42.8\pm         5.1$ & 
$         844\pm          34$ &  $        97.6\pm         5.8$ & 
$       181.1\pm         8.8$ &  $       111.5\pm         6.6$ \\ 
010 &  $          64\pm          11$ &  $        15.3\pm         2.2$ & 
$         8.7\pm         1.9$ &  $        28.9\pm         2.1$ & 
$        31.9\pm         2.1$ &  $        15.5\pm         2.2$ & 
$       235.1\pm         9.9$ &  $       165.5\pm         7.2$ &  $<16$ & 
$<16$ \\ 
\enddata
\tablecomments{Table \ref{table:int_lineflux} is published in its entirety in the electronic edition of the {\it Astrophysical Journal}.  A portion is shown here for guidance regarding its form and content.}
\tablenotetext{a}{Integrated emission-line fluxes in units of $10^{-15}$~erg~s$^{-1}$~cm$^{-2}$.  We give $1\sigma$ upper limits assuming two significant figures and identify them using a $<$ sign.  These flux measurements have been corrected for foreground Galactic extinction using the reddening values in Table~\ref{table:general_properties} and the \citet{odonnell94} Milky Way extinction curve assuming $R_{\rm V}=3.1$, as well as underlying stellar absorption as described in \S~\ref{sec:algorithm}.  The errors only include statistical measurement uncertainties.  We do not give fluxes for the following objects because their integrated spectra cannot be modeled reliably using the algorithm described in \S~\ref{sec:fitting}: NGC~1275 (070), IRAS~05189-2524 (087), UGC~08058 (279), NGC~7469 (381),  MRK~0315 (382), NGC~7674 (400), and ARP~182 (401).}
\end{deluxetable}

\clearpage

\voffset0.5in
\begin{deluxetable}{lcccccccccc}
\tabletypesize{\tiny}
\rotate
\tablecaption{Nuclear Emission-line Fluxes\tablenotemark{a} \label{table:nuc_lineflux}}
\tablewidth{0pt}
\tablehead{
\colhead{ID} & 
\colhead{[O~{\sc ii}]$~\lambda3727$} & 
\colhead{H$\delta~\lambda4101$} & 
\colhead{H$\gamma~\lambda4340$} & 
\colhead{H$\beta~\lambda4861$} & 
\colhead{[O~{\sc iii}]$~\lambda5007$} & 
\colhead{[O~{\sc i}]$~\lambda6300$} & 
\colhead{H$\alpha~\lambda6563$} & 
\colhead{[N~{\sc ii}]$~\lambda6584$} & 
\colhead{[S~{\sc ii}]$~\lambda6716$} & 
\colhead{[S~{\sc ii}]$~\lambda6731$}
}
\startdata
019 &  $         7.9\pm         1.1$ &  $        0.59\pm        0.16$ & 
$        1.23\pm        0.15$ &  $        2.56\pm        0.16$ & 
$        1.19\pm        0.13$ &  $        0.38\pm        0.12$ & 
$        7.47\pm        0.36$ &  $        1.99\pm        0.16$ & 
$        2.21\pm        0.18$ &  $        1.71\pm        0.16$ \\ 
020 &  $        0.88\pm        0.56$ &  $<0.08$ & 
$       0.166\pm       0.069$ &  $       0.283\pm       0.062$ & 
$       0.183\pm       0.063$ &  $<0.06$ &  $        1.03\pm        0.10$ & 
$<0.07$ &  $       0.233\pm       0.080$ &  $       0.190\pm       0.079$ \\ 
030 &  $        4.24\pm        0.99$ &  $        1.10\pm        0.36$ & 
$<2.8$ &  $<1.8$ &  $        1.63\pm        0.25$ & 
$        1.01\pm        0.29$ &  $        6.53\pm        0.65$ & 
$        7.78\pm        0.54$ &  $        1.99\pm        0.35$ & 
$        2.04\pm        0.36$ \\ 
037 &  $        9.72\pm        0.96$ &  $        1.16\pm        0.16$ & 
$        1.95\pm        0.16$ &  $        4.84\pm        0.24$ & 
$        4.10\pm        0.22$ &  $        0.53\pm        0.14$ & 
$       22.94\pm        0.96$ &  $        5.73\pm        0.29$ & 
$        3.90\pm        0.24$ &  $        2.71\pm        0.20$ \\ 
038 &  $        2.80\pm        0.80$ &  $        1.24\pm        0.19$ & 
$        1.66\pm        0.18$ &  $        3.93\pm        0.23$ & 
$        1.17\pm        0.17$ &  $        0.51\pm        0.19$ & 
$        28.4\pm         1.2$ &  $       11.44\pm        0.55$ &  \nodata & 
\nodata \\ 
042 &  $        0.46\pm        0.45$ &  $<0.09$ & 
$       0.107\pm       0.070$ &  $       0.186\pm       0.061$ & 
$       0.228\pm       0.060$ &  $<0.1$ &  $       0.518\pm       0.090$ & 
$       0.062\pm       0.058$ &  $       0.247\pm       0.072$ & 
$       0.134\pm       0.066$ \\ 
043 &  $<0.27$ &  $<0.2$ &  $        0.21\pm        0.17$ & 
$        0.34\pm        0.15$ &  $        0.45\pm        0.16$ &  $<0.19$ & 
$        4.30\pm        0.31$ &  $        2.22\pm        0.21$ & 
$        0.59\pm        0.20$ &  $        0.57\pm        0.20$ \\ 
048 &  $<0.81$ &  $        0.51\pm        0.27$ &  $<0.63$ &  $<0.45$ & 
$<0.63$ &  $<0.97$ &  $        2.83\pm        0.43$ &  $<0.84$ &  $<1.1$ & 
\nodata \\ 
050 &  $        2.33\pm        0.96$ &  $        0.60\pm        0.22$ & 
$        0.26\pm        0.20$ &  $        0.59\pm        0.19$ & 
$        1.26\pm        0.22$ &  $        0.42\pm        0.26$ & 
$        5.04\pm        0.42$ &  $        5.34\pm        0.43$ & 
$        1.58\pm        0.32$ &  $        1.27\pm        0.32$ \\ 
051 &  $        2.54\pm        0.86$ &  $        0.96\pm        0.19$ & 
$        1.19\pm        0.17$ &  $        3.22\pm        0.21$ & 
$        3.45\pm        0.23$ &  $        0.70\pm        0.20$ & 
$        30.6\pm         1.3$ &  $       13.10\pm        0.60$ & 
$        5.08\pm        0.34$ &  $        4.27\pm        0.31$ \\ 
\enddata
\tablecomments{Table \ref{table:nuc_lineflux} is published in its entirety in the electronic edition of the {\it Astrophysical Journal}.  A portion is shown here for guidance regarding its form and content.}
\tablenotetext{a}{Nuclear emission-line fluxes in units of $10^{-15}$~erg~s$^{-1}$~cm$^{-2}$.  We give $1\sigma$ upper limits assuming two significant figures and identify them using a $<$ sign.  These flux measurements have been corrected for foreground Galactic extinction using the reddening values in Table~\ref{table:general_properties} and the \citet{odonnell94} Milky Way extinction curve assuming $R_{\rm V}=3.1$, as well as underlying stellar absorption as described in \S~\ref{sec:algorithm}.  The errors only include statistical measurement uncertainties.  We do not give fluxes for the following objects because their nuclear spectra cannot be modeled reliably using the algorithm described in \S~\ref{sec:fitting}: NGC~1068 (054), NGC~1275 (070), NGC~3998 (214), NGC~4051 (223), and MRK~0315 (382).}
\end{deluxetable}

\clearpage

\voffset0.5in
\begin{deluxetable}{lcccccccccc}
\tabletypesize{\tiny}
\rotate
\tablecaption{Integrated Emission-line Equivalent Widths\tablenotemark{a} \label{table:int_lineEW}}
\tablewidth{0pt}
\tablehead{
\colhead{ID} & 
\colhead{[O~{\sc ii}]$~\lambda3727$} & 
\colhead{H$\delta~\lambda4101$} & 
\colhead{H$\gamma~\lambda4340$} & 
\colhead{H$\beta~\lambda4861$} & 
\colhead{[O~{\sc iii}]$~\lambda5007$} & 
\colhead{[O~{\sc i}]$~\lambda6300$} & 
\colhead{H$\alpha~\lambda6563$} & 
\colhead{[N~{\sc ii}]$~\lambda6584$} & 
\colhead{[S~{\sc ii}]$~\lambda6716$} & 
\colhead{[S~{\sc ii}]$~\lambda6731$}
}
\startdata
001 &  $        13.9\pm         1.2$ &  $        1.52\pm        0.14$ & 
$        2.16\pm        0.14$ &  $        4.44\pm        0.20$ & 
$        2.06\pm        0.12$ &  $        0.70\pm        0.11$ & 
$       24.40\pm        0.99$ &  $       12.66\pm        0.53$ & 
$        4.77\pm        0.24$ &  $        3.29\pm        0.21$ \\ 
002 &  $        12.0\pm         1.6$ &  $        0.50\pm        0.15$ & 
$        0.38\pm        0.13$ &  $        1.22\pm        0.12$ & 
$        2.41\pm        0.16$ &  $        1.17\pm        0.17$ & 
$       12.90\pm        0.58$ &  $       13.57\pm        0.60$ & 
$        4.01\pm        0.28$ &  $        0.86\pm        0.25$ \\ 
003 &  $        32.9\pm         2.7$ &  $        2.26\pm        0.32$ & 
$        5.08\pm        0.35$ &  $       12.66\pm        0.59$ & 
$       11.92\pm        0.58$ &  $        2.71\pm        0.40$ & 
$        61.4\pm         2.6$ &  $       17.38\pm        0.90$ & 
$       10.40\pm        0.80$ &  $        6.34\pm        0.64$ \\ 
004 &  $        39.1\pm         2.3$ &  $        2.67\pm        0.23$ & 
$        5.73\pm        0.30$ &  $       15.16\pm        0.64$ & 
$       14.23\pm        0.61$ &  $        3.54\pm        0.31$ & 
$        83.8\pm         3.5$ &  $        25.4\pm         1.2$ & 
$       13.80\pm        0.82$ &  $        9.78\pm        0.62$ \\ 
005 &  $        45.6\pm         3.0$ &  $        3.07\pm        0.28$ & 
$        6.20\pm        0.35$ &  $       17.47\pm        0.76$ & 
$       16.72\pm        0.73$ &  $        4.33\pm        0.39$ & 
$       105.2\pm         4.4$ &  $        33.8\pm         1.6$ & 
$        17.4\pm         1.0$ &  $       12.69\pm        0.77$ \\ 
006 &  $        20.5\pm         1.7$ &  $        1.58\pm        0.16$ & 
$        2.45\pm        0.16$ &  $        4.73\pm        0.21$ & 
$        3.78\pm        0.19$ &  $        1.06\pm        0.14$ & 
$       22.82\pm        0.94$ &  $        7.96\pm        0.36$ & 
$        4.18\pm        0.28$ &  $        3.31\pm        0.23$ \\ 
007 &  $        11.6\pm         2.2$ &  $        1.77\pm        0.24$ & 
$        0.84\pm        0.19$ &  $        3.12\pm        0.20$ & 
$        2.18\pm        0.20$ &  $<0.56$ &  $       11.23\pm        0.53$ & 
$        5.40\pm        0.34$ &  $        2.60\pm        0.27$ & 
$        1.55\pm        0.26$ \\ 
008 &  $        11.5\pm         1.1$ &  $        2.04\pm        0.13$ & 
$        2.19\pm        0.12$ &  $        4.88\pm        0.21$ & 
$       1.603\pm       0.096$ &  $       0.983\pm       0.099$ & 
$       23.72\pm        0.96$ &  $        8.57\pm        0.36$ & 
$        4.83\pm        0.22$ &  $        3.64\pm        0.18$ \\ 
009 &  $        34.0\pm         2.1$ &  $        1.66\pm        0.16$ & 
$        3.29\pm        0.20$ &  $        8.68\pm        0.38$ & 
$       22.76\pm        0.94$ &  $        2.07\pm        0.25$ & 
$        43.4\pm         1.8$ &  $        4.95\pm        0.30$ & 
$        9.38\pm        0.48$ &  $        5.80\pm        0.35$ \\ 
010 &  $        13.1\pm         2.3$ &  $        1.64\pm        0.24$ & 
$        0.84\pm        0.18$ &  $        2.22\pm        0.16$ & 
$        2.50\pm        0.16$ &  $        1.10\pm        0.15$ & 
$       15.97\pm        0.68$ &  $       11.22\pm        0.49$ &  $<1.1$ & 
$<1.2$ \\ 
\enddata
\tablecomments{Table \ref{table:int_lineEW} is published in its entirety in the electronic edition of the {\it Astrophysical Journal}.  A portion is shown here for guidance regarding its form and content.}
\tablenotetext{a}{Integrated rest-frame emission-line equivalent widths in Angstroms, corrected for underlying stellar absorption as described in \S~\ref{sec:algorithm}.  We give $1\sigma$ upper limits assuming two significant figures and identify them using a $<$ sign.  The errors only include statistical measurement uncertainties.  We do not give equivalent widths for the following objects because their integrated spectra cannot be modeled reliably using the algorithm described in \S~\ref{sec:fitting}: NGC~1275 (070), IRAS~05189-2524 (087), UGC~08058 (279), NGC~7469 (381),  MRK~0315 (382), NGC~7674 (400), and ARP~182 (401).}
\end{deluxetable}

\clearpage

\voffset0.5in
\begin{deluxetable}{lcccccccccc}
\tabletypesize{\tiny}
\rotate
\tablecaption{Nuclear Emission-line Equivalent Widths\tablenotemark{a} \label{table:nuc_lineEW}}
\tablewidth{0pt}
\tablehead{
\colhead{ID} & 
\colhead{[O~{\sc ii}]$~\lambda3727$} & 
\colhead{H$\delta~\lambda4101$} & 
\colhead{H$\gamma~\lambda4340$} & 
\colhead{H$\beta~\lambda4861$} & 
\colhead{[O~{\sc iii}]$~\lambda5007$} & 
\colhead{[O~{\sc i}]$~\lambda6300$} & 
\colhead{H$\alpha~\lambda6563$} & 
\colhead{[N~{\sc ii}]$~\lambda6584$} & 
\colhead{[S~{\sc ii}]$~\lambda6716$} & 
\colhead{[S~{\sc ii}]$~\lambda6731$}
}
\startdata
019 &  $        29.7\pm         4.2$ &  $        1.92\pm        0.54$ & 
$        4.18\pm        0.51$ &  $        8.80\pm        0.58$ & 
$        4.42\pm        0.47$ &  $        1.50\pm        0.46$ & 
$        27.6\pm         1.4$ &  $        7.63\pm        0.63$ & 
$        8.69\pm        0.71$ &  $        6.58\pm        0.63$ \\ 
020 &  $          30\pm          20$ &  $<1.1$ & 
$         2.5\pm         1.0$ &  $         4.8\pm         1.1$ & 
$         3.5\pm         1.2$ &  $<1.6$ &  $        27.0\pm         3.2$ & 
$<1.8$ &  $         6.1\pm         2.1$ &  $         4.7\pm         1.9$ \\ 
030 &  $         5.0\pm         1.2$ &  $        0.67\pm        0.22$ & 
$<1.5$ &  $<0.66$ &  $       0.643\pm       0.099$ & 
$        0.35\pm        0.10$ &  $        2.15\pm        0.21$ & 
$        2.58\pm        0.18$ &  $        0.69\pm        0.12$ & 
$        0.70\pm        0.12$ \\ 
037 &  $        25.2\pm         2.5$ &  $        1.97\pm        0.28$ & 
$        3.34\pm        0.28$ &  $        8.09\pm        0.41$ & 
$        7.28\pm        0.40$ &  $        1.03\pm        0.27$ & 
$        45.5\pm         2.0$ &  $       11.28\pm        0.58$ & 
$        7.89\pm        0.49$ &  $        5.46\pm        0.42$ \\ 
038 &  $         7.4\pm         2.1$ &  $        1.92\pm        0.29$ & 
$        2.50\pm        0.27$ &  $        5.39\pm        0.32$ & 
$        1.67\pm        0.24$ &  $        0.67\pm        0.26$ & 
$        36.4\pm         1.6$ &  $       14.70\pm        0.73$ &  \nodata & 
\nodata \\ 
042 &  $          12\pm          12$ &  $<1.7$ & 
$         2.2\pm         1.4$ &  $         4.1\pm         1.4$ & 
$         5.4\pm         1.4$ &  $<3$ &  $        16.7\pm         3.2$ & 
$         2.1\pm         2.0$ &  $         8.1\pm         2.4$ & 
$         4.6\pm         2.3$ \\ 
043 &  $<0.79$ &  $<0.34$ &  $        0.32\pm        0.26$ & 
$        0.39\pm        0.18$ &  $        0.53\pm        0.19$ &  $<0.21$ & 
$        4.42\pm        0.32$ &  $        2.25\pm        0.22$ & 
$        0.62\pm        0.21$ &  $        0.60\pm        0.22$ \\ 
048 &  $<4.5$ &  $        1.64\pm        0.88$ &  $<1.7$ &  $<0.83$ &  $<1.2$ & 
$<1.4$ &  $        3.97\pm        0.60$ &  $<1.2$ &  $<1.6$ &  \nodata \\ 
050 &  $         4.7\pm         1.9$ &  $        0.70\pm        0.26$ & 
$        0.27\pm        0.20$ &  $        0.44\pm        0.15$ & 
$        0.98\pm        0.17$ &  $        0.29\pm        0.18$ & 
$        3.26\pm        0.28$ &  $        3.41\pm        0.27$ & 
$        1.04\pm        0.21$ &  $        0.85\pm        0.22$ \\ 
051 &  $         7.1\pm         2.4$ &  $        1.42\pm        0.28$ & 
$        1.63\pm        0.24$ &  $        3.55\pm        0.24$ & 
$        3.92\pm        0.26$ &  $        0.61\pm        0.18$ & 
$        25.7\pm         1.1$ &  $       10.92\pm        0.52$ & 
$        4.23\pm        0.28$ &  $        3.59\pm        0.27$ \\ 
\enddata
\tablecomments{Table \ref{table:nuc_lineEW} is published in its entirety in the electronic edition of the {\it Astrophysical Journal}.  A portion is shown here for guidance regarding its form and content.}
\tablenotetext{a}{Nuclear rest-frame emission-line equivalent widths in Angstroms, corrected for underlying stellar absorption as described in \S~\ref{sec:algorithm}.  We give $1\sigma$ upper limits assuming two significant figures and identify them using a $<$ sign.  The errors only include statistical measurement uncertainties.  We do not give equivalent widths for the following objects because their nuclear spectra cannot be modeled reliably using the algorithm described in \S~\ref{sec:fitting}: NGC~1068 (054), NGC~1275 (070), NGC~3998 (214), NGC~4051 (223), and MRK~0315 (382).}
\end{deluxetable}

\clearpage

\include{tab9}


\clearpage

\begin{thebibliography}{86}
\expandafter\ifx\csname natexlab\endcsname\relax\def\natexlab#1{#1}\fi

\bibitem[{{Arp}(1966)}]{arp66}
{Arp}, H. 1966, {Atlas of Peculiar Galaxies} (Atlas of Peculiar Galaxies
  Publisher: California Institute of Technology, Pasaadena, CA, 1966)

\bibitem[{{Balzano}(1983)}]{balzano83}
{Balzano}, V.~A. 1983, \apj, 268, 602

\bibitem[{{Bell}(2003)}]{bell03}
{Bell}, E.~F. 2003, \apj, 586, 794

\bibitem[{{Bessell}(1990)}]{bessell90}
{Bessell}, M.~S. 1990, \pasp, 102, 1181

\bibitem[{{Blitz} \& {Shu}(1980)}]{blitz80}
{Blitz}, L., \& {Shu}, F.~H. 1980, \apj, 238, 148

\bibitem[{{Bohlin} {et~al.}(2001){Bohlin}, {Dickinson}, \&
  {Calzetti}}]{bohlin01}
{Bohlin}, R.~C., {Dickinson}, M.~E., \& {Calzetti}, D. 2001, \aj, 122, 2118

\bibitem[{{Brinchmann} {et~al.}(2004){Brinchmann}, {Charlot}, {White},
  {Tremonti}, {Kauffmann}, {Heckman}, \& {Brinkmann}}]{brinchmann04}
{Brinchmann}, J., {Charlot}, S., {White}, S.~D.~M., {Tremonti}, C.,
  {Kauffmann}, G., {Heckman}, T., \& {Brinkmann}, J. 2004, \mnras, 351, 1151

\bibitem[{{Bruzual} \& {Charlot}(2003)}]{bruzual03}
{Bruzual}, G., \& {Charlot}, S. 2003, \mnras, 344, 1000

\bibitem[{{Calzetti}(1997{\natexlab{a}})}]{calzetti97a}
{Calzetti}, D. 1997{\natexlab{a}}, \aj, 113, 162

\bibitem[{{Calzetti}(1997{\natexlab{b}})}]{calzetti97b}
{Calzetti}, D. 1997{\natexlab{b}}, in American Institute of Physics Conference
  Series, 403--+

\bibitem[{{Calzetti} {et~al.}(2000){Calzetti}, {Armus}, {Bohlin}, {Kinney},
  {Koornneef}, \& {Storchi-Bergmann}}]{calzetti00}
{Calzetti}, D., {Armus}, L., {Bohlin}, R.~C., {Kinney}, A.~L., {Koornneef}, J.,
  \& {Storchi-Bergmann}, T. 2000, \apj, 533, 682

\bibitem[{{Calzetti} {et~al.}(1994){Calzetti}, {Kinney}, \&
  {Storchi-Bergmann}}]{calzetti94}
{Calzetti}, D., {Kinney}, A.~L., \& {Storchi-Bergmann}, T. 1994, \apj, 429, 582

\bibitem[{{Charlot} \& {Fall}(2000)}]{charlot00}
{Charlot}, S.~., \& {Fall}, S.~M. 2000, \apj, 539, 718

\bibitem[{{Cid Fernandes} {et~al.}(2005){Cid Fernandes}, {Mateus}, {Laerte},
  {Stasinska}, \& {Gomes}}]{cid05}
{Cid Fernandes}, R., {Mateus}, A., {Laerte}, S.~J., {Stasinska}, G., \&
  {Gomes}, J.~M. 2005, \mnras, 358, 363

\bibitem[{{Colless} {et~al.}(2001){Colless}, {Dalton}, {Maddox}, {Sutherland},
  {Norberg}, {Cole}, {Bland-Hawthorn}, {Bridges}, {Cannon}, {Collins}, {Couch},
  {Cross}, {Deeley}, {De Propris}, {Driver}, {Efstathiou}, {Ellis}, {Frenk},
  {Glazebrook}, {Jackson}, {Lahav}, {Lewis}, {Lumsden}, {Madgwick}, {Peacock},
  {Peterson}, {Price}, {Seaborne}, \& {Taylor}}]{colless01}
{Colless}, M., {Dalton}, G., {Maddox}, S., {Sutherland}, W., {Norberg}, P.,
  {Cole}, S., {Bland-Hawthorn}, J., {Bridges}, T., {Cannon}, R., {Collins}, C.,
  {Couch}, W., {Cross}, N., {Deeley}, K., {De Propris}, R., {Driver}, S.~P.,
  {Efstathiou}, G., {Ellis}, R.~S., {Frenk}, C.~S., {Glazebrook}, K.,
  {Jackson}, C., {Lahav}, O., {Lewis}, I., {Lumsden}, S., {Madgwick}, D.,
  {Peacock}, J.~A., {Peterson}, B.~A., {Price}, I., {Seaborne}, M., \&
  {Taylor}, K. 2001, \mnras, 328, 1039

\bibitem[{{de Vaucouleurs} {et~al.}(1991){de Vaucouleurs}, {de Vaucouleurs},
  {Corwin}, {Buta}, {Paturel}, \& {Fouque}}]{devac91}
{de Vaucouleurs}, G., {de Vaucouleurs}, A., {Corwin}, H.~G., {Buta}, R.~J.,
  {Paturel}, G., \& {Fouque}, P. 1991, {Third Reference Catalogue of Bright
  Galaxies} (Volume 1-3, XII, 2069 pp.~7 figs..~ Springer-Verlag Berlin
  Heidelberg New York)

\bibitem[{{Filippenko}(1982)}]{filippenko82}
{Filippenko}, A.~V. 1982, \pasp, 94, 715

\bibitem[{{Freedman} {et~al.}(2001){Freedman}, {Madore}, {Gibson}, {Ferrarese},
  {Kelson}, {Sakai}, {Mould}, {Kennicutt}, {Ford}, {Graham}, {Huchra},
  {Hughes}, {Illingworth}, {Macri}, \& {Stetson}}]{freedman01}
{Freedman}, W.~L., {Madore}, B.~F., {Gibson}, B.~K., {Ferrarese}, L., {Kelson},
  D.~D., {Sakai}, S., {Mould}, J.~R., {Kennicutt}, R.~C., {Ford}, H.~C.,
  {Graham}, J.~A., {Huchra}, J.~P., {Hughes}, S.~M.~G., {Illingworth}, G.~D.,
  {Macri}, L.~M., \& {Stetson}, P.~B. 2001, \apj, 553, 47

\bibitem[{{Gavazzi} {et~al.}(2002){Gavazzi}, {Bonfanti}, {Sanvito}, {Boselli},
  \& {Scodeggio}}]{gavazzi02}
{Gavazzi}, G., {Bonfanti}, C., {Sanvito}, G., {Boselli}, A., \& {Scodeggio}, M.
  2002, \apj, 576, 135

\bibitem[{{Gavazzi} {et~al.}(2004){Gavazzi}, {Zaccardo}, {Sanvito}, {Boselli},
  \& {Bonfanti}}]{gavazzi04}
{Gavazzi}, G., {Zaccardo}, A., {Sanvito}, G., {Boselli}, A., \& {Bonfanti}, C.
  2004, \aap, 417, 499

\bibitem[{{Girardi} {et~al.}(1996){Girardi}, {Bressan}, {Chiosi}, {Bertelli},
  \& {Nasi}}]{girardi96}
{Girardi}, L., {Bressan}, A., {Chiosi}, C., {Bertelli}, G., \& {Nasi}, E. 1996,
  \aaps, 117, 113

\bibitem[{{Gordon} {et~al.}(2003){Gordon}, {Clayton}, {Misselt}, {Landolt}, \&
  {Wolff}}]{gordon03}
{Gordon}, K.~D., {Clayton}, G.~C., {Misselt}, K.~A., {Landolt}, A.~U., \&
  {Wolff}, M.~J. 2003, \apj, 594, 279

\bibitem[{{Hayes}(1985)}]{hayes85}
{Hayes}, D.~S. 1985, in IAU Symp. 111: Calibration of Fundamental Stellar
  Quantities, 225--249

\bibitem[{{Helou} {et~al.}(1988){Helou}, {Khan}, {Malek}, \&
  {Boehmer}}]{helou88}
{Helou}, G., {Khan}, I.~R., {Malek}, L., \& {Boehmer}, L. 1988, \apjs, 68, 151

\bibitem[{{Huchra}(1977)}]{huchra77}
{Huchra}, J.~P. 1977, \apjs, 35, 171

\bibitem[{{Jansen} {et~al.}(2000{\natexlab{a}}){Jansen}, {Fabricant}, {Franx},
  \& {Caldwell}}]{jansen00a}
{Jansen}, R.~A., {Fabricant}, D., {Franx}, M., \& {Caldwell}, N.
  2000{\natexlab{a}}, \apjs, 126, 331

\bibitem[{{Jansen} {et~al.}(2001){Jansen}, {Franx}, \& {Fabricant}}]{jansen01}
{Jansen}, R.~A., {Franx}, M., \& {Fabricant}, D. 2001, \apj, 551, 825

\bibitem[{{Jansen} {et~al.}(2000{\natexlab{b}}){Jansen}, {Franx}, {Fabricant},
  \& {Caldwell}}]{jansen00b}
{Jansen}, R.~A., {Franx}, M., {Fabricant}, D., \& {Caldwell}, N.
  2000{\natexlab{b}}, \apjs, 126, 271

\bibitem[{{Jarrett} {et~al.}(2000){Jarrett}, {Chester}, {Cutri}, {Schneider},
  {Skrutskie}, \& {Huchra}}]{jarrett00}
{Jarrett}, T.~H., {Chester}, T., {Cutri}, R., {Schneider}, S., {Skrutskie}, M.,
  \& {Huchra}, J.~P. 2000, \aj, 119, 2498

\bibitem[{{Jarrett} {et~al.}(2003){Jarrett}, {Chester}, {Cutri}, {Schneider},
  \& {Huchra}}]{jarrett03}
{Jarrett}, T.~H., {Chester}, T., {Cutri}, R., {Schneider}, S.~E., \& {Huchra},
  J.~P. 2003, \aj, 125, 525

\bibitem[{{Karachentsev} {et~al.}(2004){Karachentsev}, {Karachentseva},
  {Huchtmeier}, \& {Makarov}}]{karachentsev04}
{Karachentsev}, I.~D., {Karachentseva}, V.~E., {Huchtmeier}, W.~K., \&
  {Makarov}, D.~I. 2004, \aj, 127, 2031

\bibitem[{{Kauffmann} {et~al.}(2003){Kauffmann}, {Heckman}, {White}, {Charlot},
  {Tremonti}, {Brinchmann}, {Bruzual}, {Peng}, {et~al.}}]{kauffmann03a}
{Kauffmann}, G., {Heckman}, T.~M., {White}, S.~D.~M., {Charlot}, S.,
  {Tremonti}, C., {Brinchmann}, J., {Bruzual}, G., {Peng}, E.~W., {et~al.}
  2003, \mnras, 341, 33

\bibitem[{{Kelson}(2003)}]{kelson03}
{Kelson}, D.~D. 2003, \pasp, 115, 688

\bibitem[{{Kennicutt}(1992{\natexlab{a}})}]{kenn92a}
{Kennicutt}, R.~C. 1992{\natexlab{a}}, \apjs, 79, 255

\bibitem[{{Kennicutt}(1992{\natexlab{b}})}]{kenn92b}
---. 1992{\natexlab{b}}, \apj, 388, 310

\bibitem[{{Kennicutt} {et~al.}(1994){Kennicutt}, {Tamblyn}, \&
  {Congdon}}]{kenn94}
{Kennicutt}, R.~C., {Tamblyn}, P., \& {Congdon}, C.~E. 1994, \apj, 435, 22

\bibitem[{{Kewley} {et~al.}(2005){Kewley}, {Jansen}, \& {Geller}}]{kewley05}
{Kewley}, L.~J., {Jansen}, R.~A., \& {Geller}, M.~J. 2005, \pasp, 117, 227

\bibitem[{{Kim} {et~al.}(1995){Kim}, {Sanders}, {Veilleux}, {Mazzarella}, \&
  {Soifer}}]{kim95}
{Kim}, D.-C., {Sanders}, D.~B., {Veilleux}, S., {Mazzarella}, J.~M., \&
  {Soifer}, B.~T. 1995, \apjs, 98, 129

\bibitem[{{Kinney} {et~al.}(1993){Kinney}, {Bohlin}, {Calzetti}, {Panagia}, \&
  {Wyse}}]{kinney93}
{Kinney}, A.~L., {Bohlin}, R.~C., {Calzetti}, D., {Panagia}, N., \& {Wyse},
  R.~F.~G. 1993, \apjs, 86, 5

\bibitem[{{Kobulnicky} {et~al.}(1999){Kobulnicky}, {Kennicutt}, \&
  {Pizagno}}]{kobulnicky99a}
{Kobulnicky}, H.~A., {Kennicutt}, R.~C., \& {Pizagno}, J.~L. 1999, \apj, 514,
  544

\bibitem[{{Lauberts}(1982)}]{lauberts82}
{Lauberts}, A. 1982, {ESO/Uppsala survey of the ESO(B) atlas} (Garching:
  European Southern Observatory (ESO), 1982)

\bibitem[{{Le Borgne} {et~al.}(2004){Le Borgne}, {Rocca-Volmerange},
  {Prugniel}, {Lan{\c c}on}, {Fioc}, \& {Soubiran}}]{leborgne04}
{Le Borgne}, D., {Rocca-Volmerange}, B., {Prugniel}, P., {Lan{\c c}on}, A.,
  {Fioc}, M., \& {Soubiran}, C. 2004, \aap, 425, 881

\bibitem[{{Le Borgne} {et~al.}(2003){Le Borgne}, {Bruzual}, {Pell{\' o}},
  {Lan{\c c}on}, {Rocca-Volmerange}, {Sanahuja}, {Schaerer}, {Soubiran}, \&
  {V{\'{\i}}lchez-G{\' o}mez}}]{leborgne03}
{Le Borgne}, J.-F., {Bruzual}, G., {Pell{\' o}}, R., {Lan{\c c}on}, A.,
  {Rocca-Volmerange}, B., {Sanahuja}, B., {Schaerer}, D., {Soubiran}, C., \&
  {V{\'{\i}}lchez-G{\' o}mez}, R. 2003, \aap, 402, 433

\bibitem[{{Lejeune} {et~al.}(1997){Lejeune}, {Cuisinier}, \&
  {Buser}}]{lejeune97}
{Lejeune}, T., {Cuisinier}, F., \& {Buser}, R. 1997, \aaps, 125, 229

\bibitem[{{Markarian} {et~al.}(1989){Markarian}, {Lipovetsky}, {Stepanian},
  {Erastova}, \& {Shapovalova}}]{markarian89}
{Markarian}, B.~E., {Lipovetsky}, V.~A., {Stepanian}, J.~A., {Erastova}, L.~K.,
  \& {Shapovalova}, A.~I. 1989, Soobshcheniya Spetsial'noj Astrofizicheskoj
  Observatorii, 62, 5

\bibitem[{{Massey} {et~al.}(1988){Massey}, {Strobel}, {Barnes}, \&
  {Anderson}}]{massey88}
{Massey}, P., {Strobel}, K., {Barnes}, J.~V., \& {Anderson}, E. 1988, \apj,
  328, 315

\bibitem[{{Mayya} \& {Prabhu}(1996)}]{mayya96}
{Mayya}, Y.~D., \& {Prabhu}, T.~P. 1996, \aj, 111, 1252

\bibitem[{{McCall} {et~al.}(1985){McCall}, {Rybski}, \& {Shields}}]{mccall85}
{McCall}, M.~L., {Rybski}, P.~M., \& {Shields}, G.~A. 1985, \apjs, 57, 1

\bibitem[{{McQuade} {et~al.}(1995){McQuade}, {Calzetti}, \&
  {Kinney}}]{mcquade95}
{McQuade}, K., {Calzetti}, D., \& {Kinney}, A.~L. 1995, \apjs, 97, 331

\bibitem[{{Moshir} {et~al.}(1990){Moshir}, {Kopan}, {Conrow}, {McCallon},
  {Hacking}, {Gregorich}, {Rohrbach}, {Melnyk}, {et~al.}}]{moshir90}
{Moshir}, M., {Kopan}, G., {Conrow}, T., {McCallon}, H., {Hacking}, P.,
  {Gregorich}, D., {Rohrbach}, G., {Melnyk}, M., {et~al.} 1990, \baas, 22, 1325

\bibitem[{{Mould} {et~al.}(2000){Mould}, {Huchra}, {Freedman}, {Kennicutt},
  {Ferrarese}, {Ford}, {Gibson}, {Graham}, {et~al.}}]{mould00}
{Mould}, J.~R., {Huchra}, J.~P., {Freedman}, W.~L., {Kennicutt}, R.~C.,
  {Ferrarese}, L., {Ford}, H.~C., {Gibson}, B.~K., {Graham}, J.~A., {et~al.}
  2000, \apj, 529, 786

\bibitem[{{Moustakas} {et~al.}(2005){Moustakas}, {Kennicutt}, \&
  {Tremonti}}]{moustakas05b}
{Moustakas}, J., {Kennicutt}, R.~C., \& {Tremonti}, C. 2005, ApJ, submitted

\bibitem[{{Nilson}(1973)}]{nilson73}
{Nilson}, P. 1973, Nova Acta Regiae Soc.~Sci.~Upsaliensis Ser.~V, 0

\bibitem[{{O'Donnell}(1994)}]{odonnell94}
{O'Donnell}, J.~E. 1994, \apj, 422, 158

\bibitem[{{Panter} {et~al.}(2003){Panter}, {Heavens}, \& {Jimenez}}]{panter03}
{Panter}, B., {Heavens}, A.~F., \& {Jimenez}, R. 2003, \mnras, 343, 1145

\bibitem[{{Prugniel} \& {Heraudeau}(1998)}]{prugniel98}
{Prugniel}, P., \& {Heraudeau}, P. 1998, \aaps, 128, 299

\bibitem[{{Reichardt} {et~al.}(2001){Reichardt}, {Jimenez}, \&
  {Heavens}}]{reichardt01}
{Reichardt}, C., {Jimenez}, R., \& {Heavens}, A.~F. 2001, \mnras, 327, 849

\bibitem[{{Rice} {et~al.}(1988){Rice}, {Lonsdale}, {Soifer}, {Neugebauer},
  {Koplan}, {Lloyd}, {de Jong}, \& {Habing}}]{rice88}
{Rice}, W., {Lonsdale}, C.~J., {Soifer}, B.~T., {Neugebauer}, G., {Koplan},
  E.~L., {Lloyd}, L.~A., {de Jong}, T., \& {Habing}, H.~J. 1988, \apjs, 68, 91

\bibitem[{{Rosa-Gonz{\' a}lez} {et~al.}(2002){Rosa-Gonz{\' a}lez}, {Terlevich},
  \& {Terlevich}}]{rosa-gonzalez02}
{Rosa-Gonz{\' a}lez}, D., {Terlevich}, E., \& {Terlevich}, R. 2002, \mnras,
  332, 283

\bibitem[{{Sakai} {et~al.}(2004){Sakai}, {Ferrarese}, {Kennicutt}, \&
  {Saha}}]{sakai04}
{Sakai}, S., {Ferrarese}, L., {Kennicutt}, R.~C., \& {Saha}, A. 2004, \apj,
  608, 42

\bibitem[{{Salpeter}(1955)}]{salpeter55}
{Salpeter}, E.~E. 1955, \apj, 121, 161

\bibitem[{{Sanders} \& {Mirabel}(1996)}]{sanders96}
{Sanders}, D.~B., \& {Mirabel}, I.~F. 1996, \araa, 34, 749

\bibitem[{{Savaglio} {et~al.}(2005){Savaglio}, {Glazebrook}, {Le Borgne},
  {Juneau}, {Abraham}, {Chen}, {Crampton}, {McCarthy}, {Carlberg}, {Marzke},
  {Roth}, {Jorgensen}, \& {Murowinski}}]{savaglio05}
{Savaglio}, S., {Glazebrook}, K., {Le Borgne}, D., {Juneau}, S., {Abraham}, R.,
  {Chen}, H.~., {Crampton}, D., {McCarthy}, P., {Carlberg}, R., {Marzke}, R.,
  {Roth}, K., {Jorgensen}, I., \& {Murowinski}, R. 2005, ArXiv Astrophysics
  e-prints

\bibitem[{{Schlegel} {et~al.}(1998){Schlegel}, {Finkbeiner}, \&
  {Davis}}]{schlegel98}
{Schlegel}, D.~J., {Finkbeiner}, D.~P., \& {Davis}, M. 1998, \apj, 500, 525

\bibitem[{{Shapley} {et~al.}(2001){Shapley}, {Fabbiano}, \&
  {Eskridge}}]{shapley01b}
{Shapley}, A., {Fabbiano}, G., \& {Eskridge}, P.~B. 2001, \apjs, 137, 139

\bibitem[{{Soifer} {et~al.}(1989){Soifer}, {Boehmer}, {Neugebauer}, \&
  {Sanders}}]{soifer89}
{Soifer}, B.~T., {Boehmer}, L., {Neugebauer}, G., \& {Sanders}, D.~B. 1989,
  \aj, 98, 766

\bibitem[{{Spergel} {et~al.}(2003){Spergel}, {Verde}, {Peiris}, {Komatsu},
  {Nolta}, {Bennett}, {Halpern}, {Hinshaw}, {Jarosik}, {Kogut}, {Limon},
  {Meyer}, {Page}, {Tucker}, {Weiland}, {Wollack}, \& {Wright}}]{spergel03}
{Spergel}, D.~N., {Verde}, L., {Peiris}, H.~V., {Komatsu}, E., {Nolta}, M.~R.,
  {Bennett}, C.~L., {Halpern}, M., {Hinshaw}, G., {Jarosik}, N., {Kogut}, A.,
  {Limon}, M., {Meyer}, S.~S., {Page}, L., {Tucker}, G.~S., {Weiland}, J.~L.,
  {Wollack}, E., \& {Wright}, E.~L. 2003, \apjs, 148, 175

\bibitem[{{Storchi-Bergmann} {et~al.}(1994){Storchi-Bergmann}, {Calzetti}, \&
  {Kinney}}]{storchi94}
{Storchi-Bergmann}, T., {Calzetti}, D., \& {Kinney}, A.~L. 1994, \apj, 429, 572

\bibitem[{{Storchi-Bergmann} {et~al.}(1995){Storchi-Bergmann}, {Kinney}, \&
  {Challis}}]{storchi95}
{Storchi-Bergmann}, T., {Kinney}, A.~L., \& {Challis}, P. 1995, \apjs, 98, 103

\bibitem[{{Storey} \& {Zeippen}(2000)}]{storey00}
{Storey}, P.~J., \& {Zeippen}, C.~J. 2000, \mnras, 312, 813

\bibitem[{{Tonry} {et~al.}(2001){Tonry}, {Dressler}, {Blakeslee}, {Ajhar},
  {Fletcher}, {Luppino}, {Metzger}, \& {Moore}}]{tonry01}
{Tonry}, J.~L., {Dressler}, A., {Blakeslee}, J.~P., {Ajhar}, E.~A., {Fletcher},
  A.~B., {Luppino}, G.~A., {Metzger}, M.~R., \& {Moore}, C.~B. 2001, \apj, 546,
  681

\bibitem[{{Trager} {et~al.}(1998){Trager}, {Worthey}, {Faber}, {Burstein}, \&
  {Gonzalez}}]{trager98}
{Trager}, S.~C., {Worthey}, G., {Faber}, S.~M., {Burstein}, D., \& {Gonzalez},
  J.~J. 1998, \apjs, 116, 1

\bibitem[{{Tremonti} {et~al.}(2004){Tremonti}, {Heckman}, {Kauffmann},
  {Brinchmann}, {Charlot}, {White}, {Seibert}, {Peng}, {et~al.}}]{tremonti04}
{Tremonti}, C.~A., {Heckman}, T.~M., {Kauffmann}, G., {Brinchmann}, J.,
  {Charlot}, S., {White}, S.~D.~M., {Seibert}, M., {Peng}, E.~W., {et~al.}
  2004, \apj, 613, 898

\bibitem[{{Tully}(1988)}]{tully88}
{Tully}, R.~B. 1988, {Nearby galaxies catalog} (Cambridge and New York,
  Cambridge University Press, 1988, 221 p.)

\bibitem[{{Tully} {et~al.}(1996){Tully}, {Verheijen}, {Pierce}, {Huang}, \&
  {Wainscoat}}]{tully96}
{Tully}, R.~B., {Verheijen}, M.~A.~W., {Pierce}, M.~J., {Huang}, J., \&
  {Wainscoat}, R.~J. 1996, \aj, 112, 2471

\bibitem[{{Turner}(1998)}]{turner98}
{Turner}, A.~M. 1998, Ph.D.~Thesis

\bibitem[{{van Dokkum}(2001)}]{dokkum01}
{van Dokkum}, P.~G. 2001, \pasp, 113, 1420

\bibitem[{{Veilleux} {et~al.}(1999){Veilleux}, {Kim}, \&
  {Sanders}}]{veilleux99}
{Veilleux}, S., {Kim}, D.-C., \& {Sanders}, D.~B. 1999, \apj, 522, 113

\bibitem[{{Veilleux} {et~al.}(1995){Veilleux}, {Kim}, {Sanders}, {Mazzarella},
  \& {Soifer}}]{veilleux95}
{Veilleux}, S., {Kim}, D.-C., {Sanders}, D.~B., {Mazzarella}, J.~M., \&
  {Soifer}, B.~T. 1995, \apjs, 98, 171

\bibitem[{{Vorontsov-Vel'Yaminov} \& {Arkhipova}(1962)}]{vorontsov62}
{Vorontsov-Vel'Yaminov}, B.~A., \& {Arkhipova}, V.~P. 1962, in Morphological
  catalogue of galaxies., 1 (1962), 0--+

\bibitem[{{Witt} \& {Gordon}(2000)}]{witt00}
{Witt}, A.~N., \& {Gordon}, K.~D. 2000, \apj, 528, 799

\bibitem[{{Worthey} {et~al.}(1994){Worthey}, {Faber}, {Gonzalez}, \&
  {Burstein}}]{worthey94a}
{Worthey}, G., {Faber}, S.~M., {Gonzalez}, J.~J., \& {Burstein}, D. 1994,
  \apjs, 94, 687

\bibitem[{{Worthey} \& {Ottaviani}(1997)}]{worthey97}
{Worthey}, G., \& {Ottaviani}, D.~L. 1997, \apjs, 111, 377

\bibitem[{{Wu} {et~al.}(2002){Wu}, {Clayton}, {Gordon}, {Misselt}, {Smith}, \&
  {Calzetti}}]{wu02}
{Wu}, W., {Clayton}, G.~C., {Gordon}, K.~D., {Misselt}, K.~A., {Smith}, T.~L.,
  \& {Calzetti}, D. 2002, \apjs, 143, 377

\bibitem[{{York} {et~al.}(2000)}]{york00}
{York}, D.~G., {et~al.} 2000, \aj, 120, 1579

\bibitem[{{Zaritsky} {et~al.}(2002){Zaritsky}, {Harris}, {Thompson}, {Grebel},
  \& {Massey}}]{zaritsky02}
{Zaritsky}, D., {Harris}, J., {Thompson}, I.~B., {Grebel}, E.~K., \& {Massey},
  P. 2002, \aj, 123, 855

\end{thebibliography}
\end{document}